\documentclass[12pt,a4paper]{article}

\usepackage[margin=1in]{geometry}  
\usepackage{amsmath,amsfonts,amssymb,latexsym,dsfont}
\usepackage{hhline}
\usepackage{amsthm}
\newtheorem{theorem}{Theorem}[subsection]

\usepackage{graphicx}
\usepackage[all]{xy}
\usepackage{tikz}
\usepackage{tikz-cd}
\definecolor{hanpurple}{rgb}{0.32, 0.09, 0.98}
\usepackage{tabu}
\usepackage{authblk}
\usepackage{longtable}

\usepackage[numbers]{natbib}
\usepackage[colorlinks]{hyperref}
\hypersetup{citecolor={kwcolor},linkcolor={hanpurple},urlcolor=hanpurple}

\usepackage{chngcntr}
\counterwithin{table}{subsection}
\counterwithin{figure}{subsection}

\usetikzlibrary{positioning,arrows}
\usetikzlibrary{decorations.pathmorphing}
\usetikzlibrary{decorations.markings}

\usetikzlibrary{decorations.pathreplacing,calc}

\tikzset{brace/.style={decorate, decoration={brace}},
 brace mirrored/.style={decorate, decoration={brace,mirror}},
}

\newcounter{brace}
\newcounter{arrow}
\newcommand{\tp}{\otimes}

\newcommand{\ra}{\rightarrow}
\newcommand{\unit}{\mathds{1}}
\newcommand{\zz}{\mathbb{Z}}
\newcommand{\mce}{\mathcal{E}}
\newcommand{\mcb}{\mathcal{B}}
\newcommand{\cc}{\mathbb{C}}
\newcommand{\rr}{\mathbb{R}}
\newcommand{\mcr}{\mathcal{R}}
\newcommand{\mcz}{\mathcal{Z}}
\newcommand{\mca}{\mathcal{A}}
\newcommand{\mcd}{\mathcal{D}}
\newcommand{\mcg}{\mathcal{G}}
\newcommand{\mct}{\mathcal{T}}

\newcommand{\mch}{\mathcal{H}}
\newcommand{\mcl}{\mathcal{L}}
\newcommand{\mcc}{\mathcal{C}}

\newcommand{\mcv}{\mathcal{V}}

\newcommand{\zt}{\mathbb{Z}_2}

\newcommand\be            {\begin{equation}}
\newcommand\ee            {\end{equation}}
\newcommand\ba            {\begin{aligned}}
\newcommand\ea            {\end{aligned}}
\newcommand{\mcf}{\mathcal{F}}

\newcommand{\id}{\text{id}}
\newcommand{\Hom}{\text{Hom}}
\newcommand{\mor}{\text{mor}}

\newcommand{\End}{\text{End}}
\newcommand{\tr}{\text{tr}}
\newcommand{\str}{\text{str}}

\usepackage{verbatim}
\newcommand{\cliff}{\mathbb{C}\ell}
\newcommand{\cl}{\text{cl}}

\newcommand{\spin}{\text{Spin}}
\newcommand{\pin}{\text{Pin}}	
\newcommand{\tube}{\textbf{Tube}}

\newcommand{\du}{\sqcup}

\newcommand{\sob}{\text{sob}_r}
\newcommand{\sobm}{\text{sob}_r^m}
\newcommand{\sobq}{\text{sob}_r^q}

\newcommand{\ot}{\otimes}
\newcommand{\bd}{\partial}
\DeclareMathOperator{\Arf}{Arf}
\definecolor{kwcolor}{rgb}{0.2, 0.5, 0.85}
\definecolor{kwcolorx}{rgb}{0.5, 0.75, 0.95}
\newcommand{\kw}[1]{{\color{kwcolor}\footnotesize{(KW) #1}}}

\newcommand{\ket}[1]{\ensuremath{\left|#1\right\rangle}}

\definecolor{ao(english)}{rgb}{0.0, 0.5, 0.0}
\definecolor{orange}{HTML}{FFA500}
\definecolor{americanrose}{rgb}{1.0, 0.01, 0.24}
\definecolor{amber(sae/ece)}{rgb}{1.0, 0.49, 0.0}

\newcommand{\dave}[1]{{\color{ao(english)}\footnotesize{(DA) #1}}}

\newcommand{\CapDotLeft}{\mathord{\vcenter{\hbox{\includegraphics[scale=1]{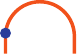}}}}}
\newcommand{\CapDotRight}{\mathord{\vcenter{\hbox{\includegraphics[scale=1]{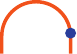}}}}}
\newcommand{\CupDotLeft}{\mathord{\vcenter{\hbox{\includegraphics[scale=1,angle=180,origin=c]{CapDotRight.pdf}}}}}
\newcommand{\CupDotRight}{\mathord{\vcenter{\hbox{\includegraphics[scale=1,angle=180,origin=c]{CapDotLeft.pdf}}}}}

\newcommand{\CupCap}{\mathord{\vcenter{\hbox{\includegraphics[scale=1]{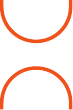}}}}}
\newcommand{\CupCapDots}{\mathord{\vcenter{\hbox{\includegraphics[scale=1]{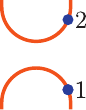}}}}}

\newcommand{\dbeta}{\mathord{\vcenter{\hbox{\includegraphics[scale=1]{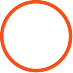}}}}}
\newcommand{\dpsi}{\mathord{\vcenter{\hbox{\includegraphics[scale=1]{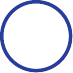}}}}}

\newcommand{\SigmaDotDot}{\mathord{\vcenter{\hbox{\includegraphics[scale=1]{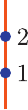}}}}}
\newcommand{\SigmaDotDotExchange}{\mathord{\vcenter{\hbox{\includegraphics[scale=1]{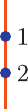}}}}}
\newcommand{\TwoLine}{\mathord{\vcenter{\hbox{\includegraphics[scale=1]{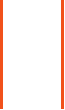}}}}}
\newcommand{\TwoLineDots}{\mathord{\vcenter{\hbox{\includegraphics[scale=1]{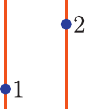}}}}}

\newcommand{\FubeXss}{\mathord{\vcenter{\hbox{\includegraphics[scale=1]{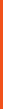}}}}}

\newcommand{\DiskLarge}{\mathord{\vcenter{\hbox{\includegraphics[scale=1]{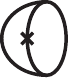}}}}}
\newcommand{\TubeIdempotentBasistrace}{\mathord{\vcenter{\hbox{\includegraphics[scale=1]{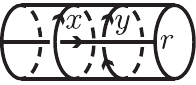}}}}}

\newcommand{\FubeXXXA}{\mathord{\vcenter{\hbox{\includegraphics[scale=1]{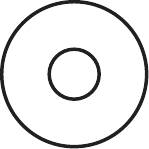}}}}}
\newcommand{\FubeXssA}{\mathord{\vcenter{\hbox{\includegraphics[scale=1]{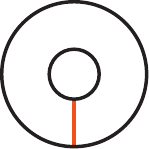}}}}}
\newcommand{\FubeXsdsA}{\mathord{\vcenter{\hbox{\includegraphics[scale=1]{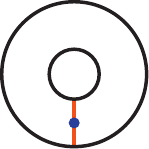}}}}}

\newcommand{\FubesddXsA}{\mathord{\vcenter{\hbox{\includegraphics[scale=1]{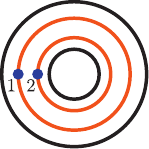}}}}}

\newcommand{\RDotTwobA}{\mathord{\vcenter{\hbox{\includegraphics[scale=1]{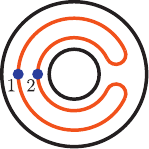}}}}}\newcommand{\RDotTwocA}{\mathord{\vcenter{\hbox{\includegraphics[scale=1]{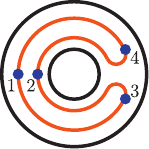}}}}}

\newcommand{\Fubex}[2]{{\mathord{\ooalign{ \vphantom{$\Big|^2$}\cr\hidewidth\ensuremath{\scriptstyle{#2}}\hidewidth\cr$\vcenter{\hbox{$#1$}}$\cr
  \hidewidth\raise0ex\hbox{$\scale{1.2}{\VerticalSpace}$}\cr
  }}}}

\newcommand{\TubeBC}{\mathord{\vcenter{\hbox{\includegraphics[scale=1]{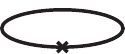}}}}}

\newcommand{\TubeBCx}[1]{{\mathord{\ooalign{ \vphantom{$\Big|^2$}\cr\hidewidth\ensuremath{\scriptstyle{#1}}\hidewidth\cr$\vcenter{\hbox{$\scale{1}{\TubeBC}$}}$\cr
  \hidewidth\raise0ex\hbox{$\scale{.25}{\VerticalSpace}$}\cr
  }}}}

\newcommand{\AnnularTubeNoIndex}{\mathord{\vcenter{\hbox{\includegraphics[scale=1]{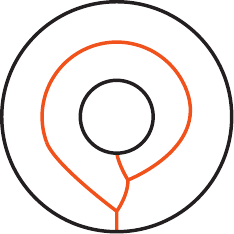}}}}}
\newcommand{\AnnulusTubeTube}{\mathord{\vcenter{\hbox{\includegraphics[scale=1]{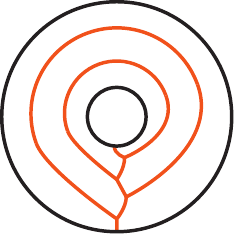}}}}}

\newcommand{\SAnnulusNoLabel}{\mathord{\vcenter{\hbox{\includegraphics[scale=1]{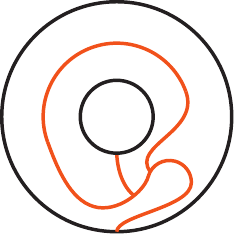}}}}}
\newcommand{\TAnnulusNoLabel}{\mathord{\vcenter{\hbox{\includegraphics[scale=1]{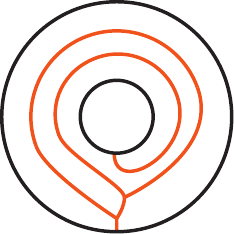}}}}}

\newcommand{\OneHandlePrime}{\mathord{\vcenter{\hbox{\includegraphics[scale=1]{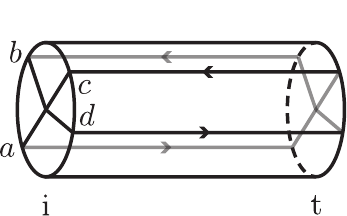}}}}}
\newcommand{\CellDecompNearOneHandle}{\mathord{\vcenter{\hbox{\includegraphics[scale=1]{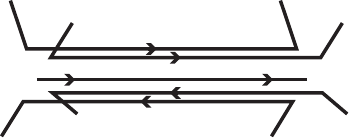}}}}}

\newcommand{\Bilineara}{\mathord{\vcenter{\hbox{\includegraphics[scale=0.85]{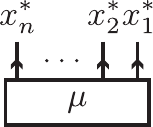}}}}}
\newcommand{\Bilinearb}{\mathord{\vcenter{\hbox{\includegraphics[scale=0.85]{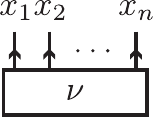}}}}}
\newcommand{\Bilinearc}{\mathord{\vcenter{\hbox{\includegraphics[scale=0.85]{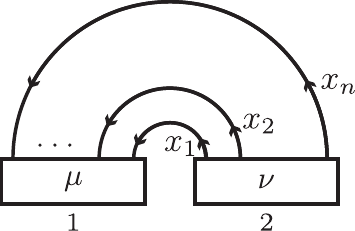}}}}}

\newcommand{\TetrahedronPrime}{\mathord{\vcenter{\hbox{\includegraphics[scale=1]{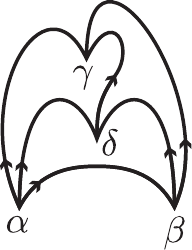}}}}}
\newcommand{\Tetrahedron}{\mathord{\vcenter{\hbox{\includegraphics[scale=1]{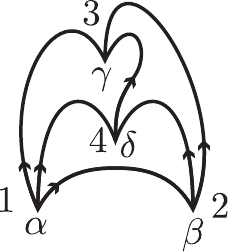}}}}}

\newcommand{\HorshoeIdentity}{\mathord{\vcenter{\hbox{\includegraphics[scale=1]{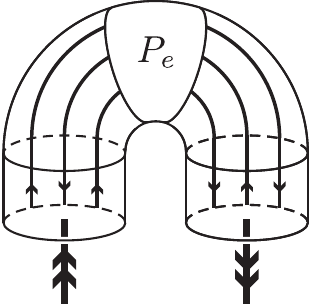}}}}}

\newcommand{\Pitchforkabc}{\mathord{\vcenter{\hbox{\includegraphics[scale=1.3]{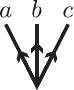}}}}}
\newcommand{\Pitchforkabcrot}{\mathord{\vcenter{\hbox{\includegraphics[scale=1.3]{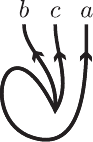}}}}}
\newcommand{\PivotThreeTimes}{\mathord{\vcenter{\hbox{\includegraphics[scale=1.3]{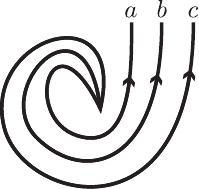}}}}}

\newcommand{\PitchforkLarge}{\mathord{\vcenter{\hbox{\includegraphics[scale=1.3]{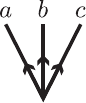}}}}}
\newcommand{\Pitchforkdotthree}{\mathord{\vcenter{\hbox{\includegraphics[scale=1.3]{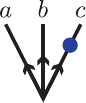}}}}}
\newcommand{\Pitchforkdottwo}{\mathord{\vcenter{\hbox{\includegraphics[scale=1.3]{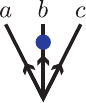}}}}}
\newcommand{\Pitchforkdotone}{\mathord{\vcenter{\hbox{\includegraphics[scale=1.3]{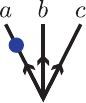}}}}}

\newcommand*{\Annulus}[2]{{ #1 }
\kern-3.5em\raisebox{1ex}{ $\scriptstyle{#2}$} \kern2.6em} 

\newcommand*{\AnnulusP}[3]{{{\Annulus{#1}{#2}} }
\kern-.5em\raisebox{3.5ex}{ $\scriptstyle{#3}$} \kern.3em} 

\newcommand*{\AnnulusPx}[4]{\AnnulusP{#1}{#2}{#3}
\kern-3.6em\raisebox{-4.5ex}{ $\scriptstyle{#4}$} \kern2.6em} 

\newcommand*{\AnularTubex}[6]{\AnnulusPx{#1}{#2}{#3}{#4}
\kern-4em\raisebox{-4.5ex}{ $\overset{#6}{\underset{#5}{\vphantom{\Big|^2}}}$} \kern3em} 

\newcommand*{\TorusTubex}[6]{\AnnulusPx{#1}{#2}{#3}{#4}
\kern-4em\raisebox{-4.5ex}{ $\overset{#6}{\underset{#5}{\vphantom{\Big|^2}}}$} \kern3em} 

\newcommand{\SAnnulusx}[3]{\mathrel{\ooalign{$\SAnnulusNoLabel$\cr
  \hidewidth\raise5ex\hbox{$\scriptstyle{#3}\mkern1mu$}\cr
  \hidewidth\raise.8ex\hbox{$\scriptstyle{#2}\mkern63mu$}\cr
    \hidewidth\raise-3.8ex\hbox{$\scriptstyle{#1}\mkern39mu$}\cr
  }}}
  
  \newcommand{\TAnnulusx}[3]{\mathrel{\ooalign{$\TAnnulusNoLabel$\cr
  \hidewidth\raise5ex\hbox{$\scriptstyle{#3}\mkern1mu$}\cr
  \hidewidth\raise.8ex\hbox{$\scriptstyle{#2}\mkern63mu$}\cr
    \hidewidth\raise-4.9ex\hbox{$\scriptstyle{#1}\mkern51mu$}\cr
  }}}

\newcommand{\AnnularTubex}[6]{\mathrel{\ooalign{$#1$\cr
  \hidewidth\raise5ex\hbox{$\scriptstyle{#6}\mkern1mu$}\cr
  \hidewidth\raise.8ex\hbox{$\scriptstyle{#5}\mkern63mu$}\cr
  \hidewidth\raise-.9ex\hbox{$\scriptstyle{#3}\mkern68mu$}\cr
    \hidewidth\raise-4.5ex\hbox{$\scriptstyle{#4} \mkern50mu$}\cr
  \hidewidth\raise-7ex\hbox{$\scriptstyle{#2}\mkern68mu$}\cr
  }}}

\newcommand{\AnnularTubexp}[9]{\mathrel{\ooalign{$#1$\cr
  \hidewidth\raise5ex\hbox{$\scriptstyle{#9}\mkern1mu$}\cr
 \hidewidth\raise.8ex\hbox{$\scriptstyle{#8}\mkern63mu$}\cr
  \hidewidth\raise-3.7ex\hbox{$\scriptstyle{#7}\mkern50mu$}\cr
    \hidewidth\raise-5.1ex\hbox{$\scriptstyle{#6}\mkern57mu$}\cr
  \hidewidth\raise-2.2ex\hbox{$\scriptstyle{#5}\mkern90mu$}\cr
  \hidewidth\raise-.9ex\hbox{$\scriptstyle{#4}\mkern68mu$}\cr
    \hidewidth\raise-3.8ex\hbox{$\scriptstyle{#3} \mkern69mu$}\cr
  \hidewidth\raise-7ex\hbox{$\scriptstyle{#2}\mkern68mu$}\cr
  }}}

\newcommand{\SmallTorus}[3]{\mathrel{\ooalign{$#1$\cr
  \hidewidth\raise0ex\hbox{$\scriptstyle{#2}\mkern36mu$}\cr
    \hidewidth\raise-1.5ex\hbox{$\scriptstyle{#3}\mkern10mu$}\cr
  }}}

  \newcommand{\AnnulusTubeTubex}[6]{\mathrel{\ooalign{$#1$\cr
  \hidewidth\raise.8ex\hbox{$\scriptstyle{#6}\mkern65mu$}\cr
  \hidewidth\raise-.5ex\hbox{$\scriptstyle{#3}\mkern68mu$}\cr
    \hidewidth\raise-4.9ex\hbox{$\scriptstyle{#4} \mkern50mu$}\cr
     \hidewidth\raise-2.7ex\hbox{$\scriptstyle{#5} \mkern54mu$}\cr
  \hidewidth\raise-7.3ex\hbox{$\scriptstyle{#2}\mkern68mu$}\cr
  }}}

\newcommand{\TorusLocalRelationc}{\mathord{\vcenter{\hbox{\includegraphics[scale=1]{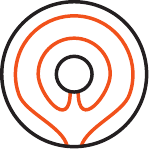}}}}}
\newcommand{\TorusLocalRelationb}{\mathord{\vcenter{\hbox{\includegraphics[scale=1]{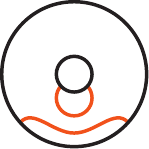}}}}}
\newcommand{\TorusLocalRelationa}{\mathord{\vcenter{\hbox{\includegraphics[scale=1]{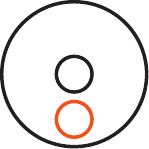}}}}}

\newcommand{\Dv}{\mathord{\vcenter{\hbox{\includegraphics[scale=1]{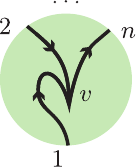}}}}}
\newcommand{\Dfv}{\mathord{\vcenter{\hbox{\includegraphics[scale=1]{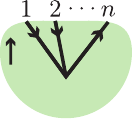}}}}}

\newcommand{\Horseshoe}{\mathord{\vcenter{\hbox{\includegraphics[scale=1]{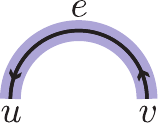}}}}}

\newcommand{\TubeElement}{\mathord{\vcenter{\hbox{\includegraphics[scale=1]{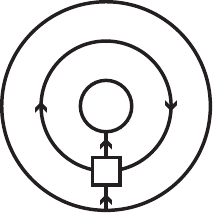}}}}}

\newcommand{\LoopOverId}{\mathord{\vcenter{\hbox{\includegraphics[scale=1]{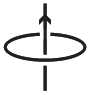}}}}}
\newcommand{\Ida}{\mathord{\vcenter{\hbox{\includegraphics[scale=1]{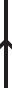}}}}}
\newcommand{\OmegaSLoop}{\mathord{\vcenter{\hbox{\includegraphics[scale=1]{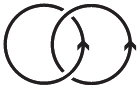}}}}}

\newcommand{\IdxOmegaLoopa}{\mathord{\vcenter{\hbox{\includegraphics[scale=1]{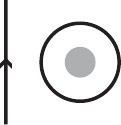}}}}}
\newcommand{\IdxOmegaLoopb}{\mathord{\vcenter{\hbox{\includegraphics[scale=1]{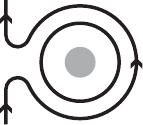}}}}}

\newcommand{\HandleSlidea}{\mathord{\vcenter{\hbox{\includegraphics[scale=1]{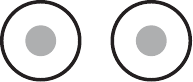}}}}}
\newcommand{\HandleSlideb}{\mathord{\vcenter{\hbox{\includegraphics[scale=1]{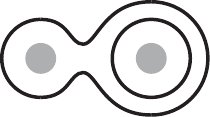}}}}}

\newcommand{\TubeBasisa}{\mathord{\vcenter{\hbox{\includegraphics[scale=1]{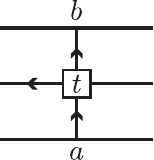}}}}}
\newcommand{\TubeBasisb}{\mathord{\vcenter{\hbox{\includegraphics[scale=1]{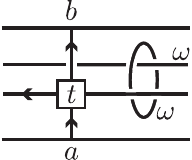}}}}}
\newcommand{\TubeBasisc}{\mathord{\vcenter{\hbox{\includegraphics[scale=1]{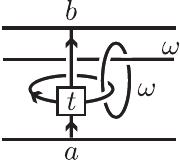}}}}}
\newcommand{\TubeBasisd}{\mathord{\vcenter{\hbox{\includegraphics[scale=1]{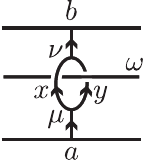}}}}}
\newcommand{\TubeBasisdprime}{\mathord{\vcenter{\hbox{\includegraphics[scale=1]{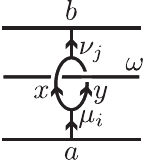}}}}}

\newcommand{\TubeBasisaF}{\mathord{\vcenter{\hbox{\includegraphics[scale=1]{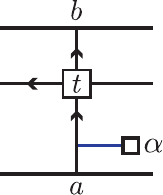}}}}}

\newcommand{\TubeBasisdF}{\mathord{\vcenter{\hbox{\includegraphics[scale=1]{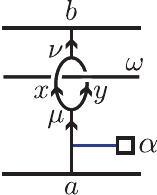}}}}}

\newcommand{\TubeFiga}{\mathord{\vcenter{\hbox{\includegraphics[scale=1]{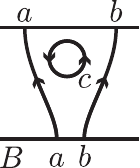}}}}}
\newcommand{\TubeFigb}{\mathord{\vcenter{\hbox{\includegraphics[scale=1]{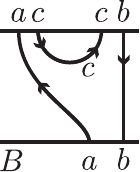}}}}}
\newcommand{\TubeFigc}{\mathord{\vcenter{\hbox{\includegraphics[scale=1]{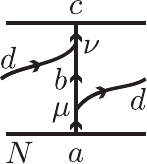}}}}}

\newcommand{\OneStrandIdempotentMTCpsi}{\mathord{\vcenter{\hbox{\includegraphics[scale=1]{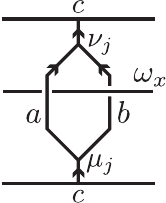}}}}}
\newcommand{\gxyaj}{\mathord{\vcenter{\hbox{\includegraphics[scale=1]{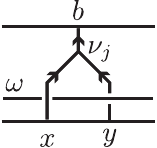}}}}}
\newcommand{\hxyai}{\mathord{\vcenter{\hbox{\includegraphics[scale=1]{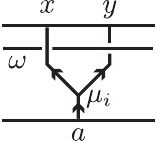}}}}}

\newcommand{\VxyaoVxya}{\mathord{\vcenter{\hbox{\includegraphics[scale=1]{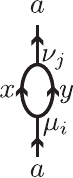}}}}}
\newcommand{\idaprime}{\mathord{\vcenter{\hbox{\includegraphics[scale=1]{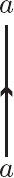}}}}}

\newcommand{\Vxyxy}{\mathord{\vcenter{\hbox{\includegraphics[scale=1]{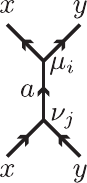}}}}}
\newcommand{\idxy}{\mathord{\vcenter{\hbox{\includegraphics[scale=1]{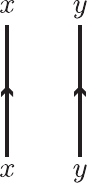}}}}}
\newcommand{\Vxyxyomega}{\mathord{\vcenter{\hbox{\includegraphics[scale=1]{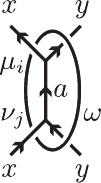}}}}}

\newcommand{\Tuberrprime}{\mathord{\vcenter{\hbox{\includegraphics[scale=1]{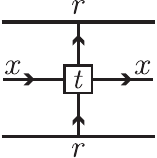}}}}}
\newcommand{\Tuberrprimecut}{\mathord{\vcenter{\hbox{\includegraphics[scale=1]{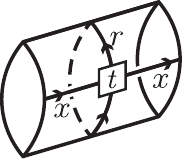}}}}}
\newcommand{\PinchedDisk}{\mathord{\vcenter{\hbox{\includegraphics[scale=1]{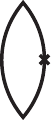}}}}}

\newcommand{\clx}{\mathord{\vcenter{\hbox{\includegraphics[scale=1]{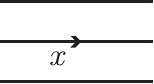}}}}}

\newcommand{\moronex}{\mathord{\vcenter{\hbox{\includegraphics[scale=1]{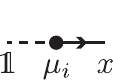}}}}}
\newcommand{\moronexj}{\mathord{\vcenter{\hbox{\includegraphics[scale=1]{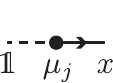}}}}}

\newcommand{\moronexcollar}{\mathord{\vcenter{\hbox{\includegraphics[scale=1]{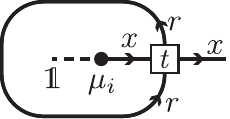}}}}}

\newcommand{\EFunctora}{\mathord{\vcenter{\hbox{\includegraphics[scale=1]{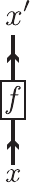}}}}}
\newcommand{\EFunctorb}{\mathord{\vcenter{\hbox{\includegraphics[scale=1]{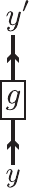}}}}}
\newcommand{\EFunctorc}{\mathord{\vcenter{\hbox{\includegraphics[scale=1]{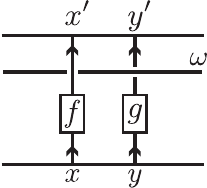}}}}}

\newcommand{\IdempotentBasis}{\mathord{\vcenter{\hbox{\includegraphics[scale=1]{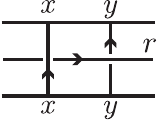}}}}}

\newcommand{\TubeIdempotentTwoStrand}{\mathord{\vcenter{\hbox{\includegraphics[scale=1]{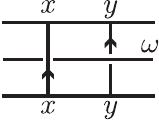}}}}}

\newcommand{\fTubefOdd}{\mathord{\vcenter{\hbox{\includegraphics[scale=1]{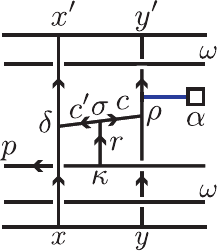}}}}}
\newcommand{\minimalBosonic}{\mathord{\vcenter{\hbox{\includegraphics[scale=1]{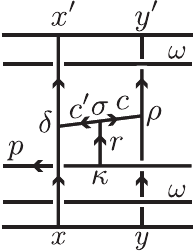}}}}}

\newcommand{\TubeIdempotentTwoStrandab}{\mathord{\vcenter{\hbox{\includegraphics[scale=1]{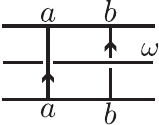}}}}}
\newcommand{\BasisForIdempotents}{\mathord{\vcenter{\hbox{\includegraphics[scale=1]{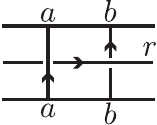}}}}}

\newcommand{\minimalBosonicRHS}{\mathord{\vcenter{\hbox{\includegraphics[scale=1]{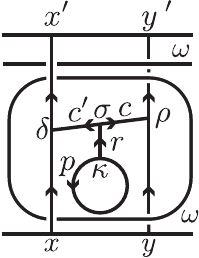}}}}}
\newcommand{\ftubefOddRHS}{\mathord{\vcenter{\hbox{\includegraphics[scale=1]{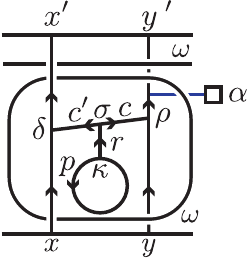}}}}}

\newcommand{\FusionIsomorphismprimereduced}{\mathord{\vcenter{\hbox{\includegraphics[scale=1]{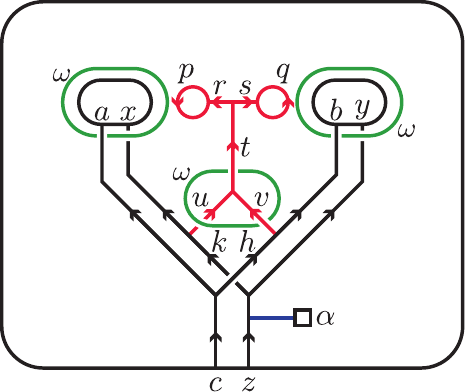}}}}}
\newcommand{\FusionIsomorphismprime}{\mathord{\vcenter{\hbox{\includegraphics[scale=1]{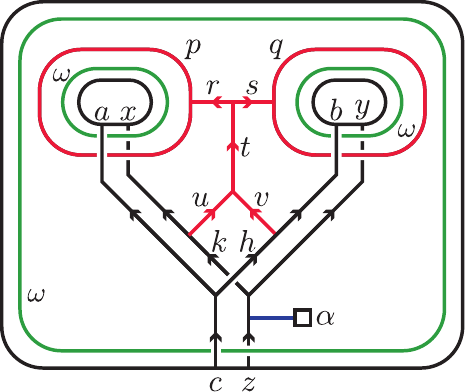}}}}}

\newcommand{\TubeCompletea}{\mathord{\vcenter{\hbox{\includegraphics[scale=1]{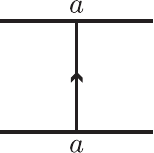}}}}}
\newcommand{\TubeCompleteb}{\mathord{\vcenter{\hbox{\includegraphics[scale=1]{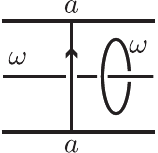}}}}}

\newcommand{\TubeCompletecprime}{\mathord{\vcenter{\hbox{\includegraphics[scale=1]{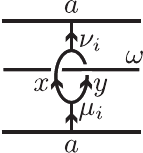}}}}}

\newcommand{\SMatrix}[2]{\mathrel{\ooalign{$\OmegaSLoop$\cr
  \hidewidth\raise0ex\hbox{$\scriptstyle{#1}\mkern18mu$}\cr
    \hidewidth\raise0ex\hbox{$\scriptstyle{#2}\mkern-10mu$}\cr
  }}}

  \newcommand{\SMatrixx}[2]{\mathrel{\ooalign{$\OmegaSLoop$\cr
  \hidewidth\raise0ex\hbox{$\scriptstyle{#1}\mkern12mu$}\cr
    \hidewidth\raise0ex\hbox{$\scriptstyle{#2}\mkern-10mu$}\cr
  }}}

  \newcommand{\SMatrixxx}[2]{\mathrel{\ooalign{$\OmegaSLoop$\cr
  \hidewidth\raise0ex\hbox{$\scriptstyle{#1}\mkern12mu$}\cr
    \hidewidth\raise0ex\hbox{$\scriptstyle{#2}\mkern-38mu$}\cr
  }}}

  \newcommand{\OmegaLoopDefectx}[2]{\mathrel{\ooalign{$\OmegaLoopDefect$\cr
  \hidewidth\raise0ex\hbox{$\scriptstyle{#1}\mkern8mu$}\cr
    \hidewidth\raise0ex\hbox{$\scriptstyle{#2}\mkern-22mu$}\cr
  }}}
  
  \newcommand{\OmegaLoopDefect}{\mathord{\vcenter{\hbox{\includegraphics[scale=1]{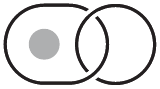}}}}}
    \newcommand{\DiscGray}{\mathord{\vcenter{\hbox{\includegraphics[scale=1]{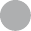}}}}}

\newcommand{\DCSmatrixa}{\mathord{\vcenter{\hbox{\includegraphics[scale=1]{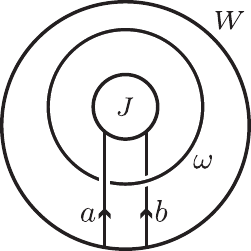}}}}}
\newcommand{\DCSmatrixb}{\mathord{\vcenter{\hbox{\includegraphics[scale=1]{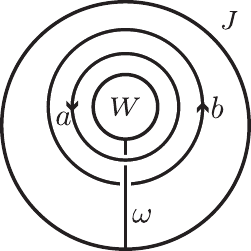}}}}}

\newcommand{\DCSmatrixh}{\mathord{\vcenter{\hbox{\includegraphics[scale=1]{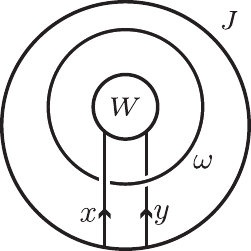}}}}}

\newcommand{\STorusBasisa}{\mathord{\vcenter{\hbox{\includegraphics[scale=1]{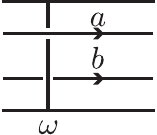}}}}}

\newcommand{\Scalcae}{\mathord{\vcenter{\hbox{\includegraphics[scale=1]{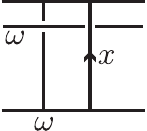}}}}}  
\newcommand{\Scalcad}{\mathord{\vcenter{\hbox{\includegraphics[scale=1]{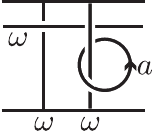}}}}}  
\newcommand{\Scalcac}{\mathord{\vcenter{\hbox{\includegraphics[scale=1]{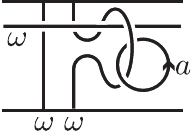}}}}}  
\newcommand{\Scalcab}{\mathord{\vcenter{\hbox{\includegraphics[scale=1]{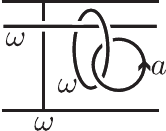}}}}}
\newcommand{\Scalcaa}{\mathord{\vcenter{\hbox{\includegraphics[scale=1]{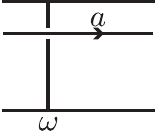}}}}}

\newcommand{\Scalcbe}{\mathord{\vcenter{\hbox{\includegraphics[scale=1]{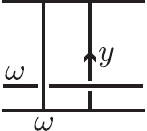}}}}}  
\newcommand{\Scalcbd}{\mathord{\vcenter{\hbox{\includegraphics[scale=1]{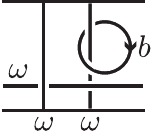}}}}}  
\newcommand{\Scalcbc}{\mathord{\vcenter{\hbox{\includegraphics[scale=1]{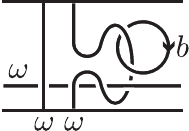}}}}}  
\newcommand{\Scalcbb}{\mathord{\vcenter{\hbox{\includegraphics[scale=1]{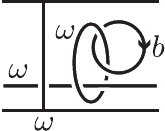}}}}}
\newcommand{\Scalcba}{\mathord{\vcenter{\hbox{\includegraphics[scale=1]{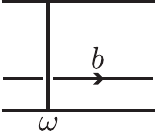}}}}}

\newcommand{\Scalcbedot}{\mathord{\vcenter{\hbox{\includegraphics[scale=1]{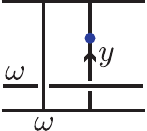}}}}}

\newcommand{\Scalcbddotprime}{\mathord{\vcenter{\hbox{\includegraphics[scale=1]{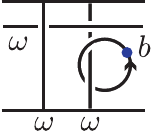}}}}}
\newcommand{\Scalcbadotprime}{\mathord{\vcenter{\hbox{\includegraphics[scale=1]{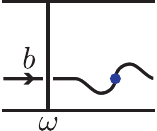}}}}}

\newcommand{\Szmatrix}{\mathord{\vcenter{\hbox{\includegraphics[scale=1]{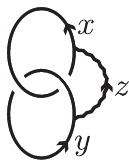}}}}}

\newcommand{\LoopArrow}{\mathord{\vcenter{\hbox{\includegraphics[scale=1]{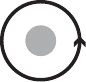}}}}}  
\newcommand{\LoopArrowx}[1]{\mathrel{\ooalign{$\LoopArrow_{\mathrel{\raisebox{4pt}{$\scriptstyle{#1}$}}}$
  }}}

\newcommand{\OmegaLoop}{\mathord{\vcenter{\hbox{\includegraphics[scale=1]{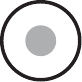}}}}}  
\newcommand{\OmegaLoopx}[1]{\mathrel{\ooalign{$\OmegaLoop_{\mathrel{\raisebox{4pt}{$\scriptstyle{#1}$}}}$
  }}}

\newcommand{\TubeElementx}[3]{\mathrel{\ooalign{$\TubeElement$\cr
  \hidewidth\raise-3.55ex\hbox{$\scriptstyle{#1}\mkern62mu$}\cr
        \hidewidth\raise-6.3ex\hbox{$\scriptstyle{#2}\mkern61mu$}\cr
                \hidewidth\raise-0.3ex\hbox{$\scriptstyle{#3}\mkern60mu$}\cr
          \hidewidth\raise0ex\hbox{$\scale{1.5}{\VerticalSpace}$}\cr
  }}}

\newcommand{\IdempotentMTCNoLabel}{\mathord{\vcenter{\hbox{\includegraphics[scale=1]{IdempotentMTC.pdf}}}}}
\newcommand{\IdempBraid}[5]{\mathrel{\ooalign{$\IdempotentMTCNoLabel$\cr
  \hidewidth\raise-4.3ex\hbox{$\scriptstyle{#1}\mkern78mu$}\cr
    \hidewidth\raise-4.3ex\hbox{$\scriptstyle{#2}\mkern43mu$}\cr
        \hidewidth\raise-6.3ex\hbox{$\scriptstyle{#3}\mkern61mu$}\cr
        \hidewidth\raise4.1ex\hbox{$\scriptstyle{#4}\mkern25mu$}\cr
                \hidewidth\raise0ex\hbox{$\scriptstyle{#5}\mkern60mu$}\cr
          \hidewidth\raise0ex\hbox{$\scale{1.5}{\VerticalSpace}$}\cr
  }}}
  
\newcommand{\TorusBasisMTCNoLabel}{\mathord{\vcenter{\hbox{\includegraphics[scale=1]{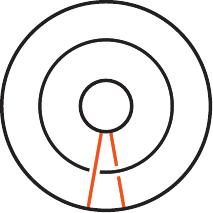}}}}}
\newcommand{\TorusBraidBasis}[6]{\mathrel{\ooalign{$\TorusBasisMTCNoLabel$\cr
  \hidewidth\raise-4.3ex\hbox{$\scriptstyle{#1}\mkern78mu$}\cr
    \hidewidth\raise-4.3ex\hbox{$\scriptstyle{#2}\mkern40mu$}\cr
        \hidewidth\raise-6.3ex\hbox{$\scriptstyle{#3}\mkern61mu$}\cr
        \hidewidth\raise4.1ex\hbox{$\scriptstyle{#4}\mkern25mu$}\cr
                \hidewidth\raise0ex\hbox{$\scriptstyle{#5}\mkern58mu$}\cr
                                \hidewidth\raise4.7ex\hbox{$\scriptstyle{#6}\mkern0mu$}\cr
          \hidewidth\raise0ex\hbox{$\scale{1.5}{\VerticalSpace}$}\cr
  }}}

\newcommand{\TorusBasisMTCdl}{\mathord{\vcenter{\hbox{\includegraphics[scale=1]{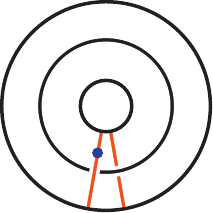}}}}}
\newcommand{\TorusBasisMTCdr}{\mathord{\vcenter{\hbox{\includegraphics[scale=1]{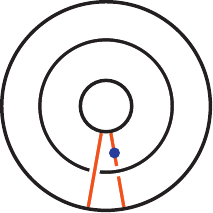}}}}}
\newcommand{\TorusBasisMTCdd}{\mathord{\vcenter{\hbox{\includegraphics[scale=1]{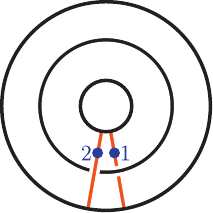}}}}}
 
 \newcommand{\TorusBraidBasisd}[7]{\mathrel{\ooalign{$#7$\cr
  \hidewidth\raise-4.3ex\hbox{$\scriptstyle{#1}\mkern78mu$}\cr
    \hidewidth\raise-4.3ex\hbox{$\scriptstyle{#2}\mkern40mu$}\cr
        \hidewidth\raise-6.3ex\hbox{$\scriptstyle{#3}\mkern61mu$}\cr
        \hidewidth\raise4.1ex\hbox{$\scriptstyle{#4}\mkern25mu$}\cr
                \hidewidth\raise0ex\hbox{$\scriptstyle{#5}\mkern58mu$}\cr
                                \hidewidth\raise4.7ex\hbox{$\scriptstyle{#6}\mkern0mu$}\cr
          \hidewidth\raise0ex\hbox{$\scale{1.5}{\VerticalSpace}$}\cr
  }}}

\newcommand{\TaddownTubeNoLabel}{\mathord{\vcenter{\hbox{\includegraphics[scale=1]{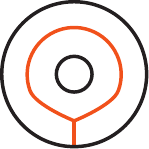}}}}}
\newcommand{\TadupTubeNoLabel}{\mathord{\vcenter{\hbox{\includegraphics[scale=1]{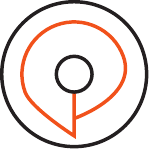}}}}}
\newcommand{\hTube}{\mathord{\vcenter{\hbox{\includegraphics[scale=1]{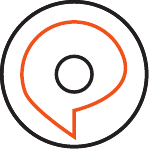}}}}}
\newcommand{\tTube}{\mathord{\vcenter{\hbox{\includegraphics[scale=1]{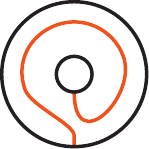}}}}}
\newcommand{\eTube}{\mathord{\vcenter{\hbox{\includegraphics[scale=1]{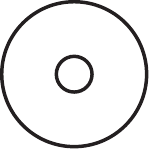}}}}}
\newcommand{\vTube}{\mathord{\vcenter{\hbox{\includegraphics[scale=1]{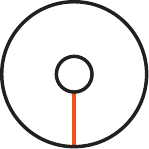}}}}}

\newcommand{\dota}{\mathord{\vcenter{\hbox{\includegraphics[scale=1]{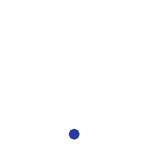}}}}}
\newcommand{\dotb}{\mathord{\vcenter{\hbox{\includegraphics[scale=1]{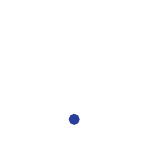}}}}}
\newcommand{\dotc}{\mathord{\vcenter{\hbox{\includegraphics[scale=1]{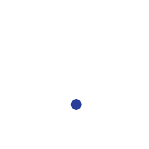}}}}}

\newcommand{\XTubeNoLabel}{\mathord{\vcenter{\hbox{\includegraphics[scale=1]{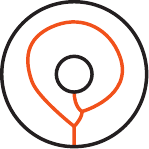}}}}}
\newcommand{\XTube}[2]{\mathrel{\ooalign{$\XTubeNoLabel$\cr
  \hidewidth\raise-2.5ex\hbox{$\scriptstyle{#2}\mkern28mu$}\cr
    \hidewidth\raise-2.9ex\hbox{$\scriptstyle{#1}\mkern50mu$}\cr
  }}}

\newcommand{\TaddownTube}[1]{\mathrel{\ooalign{$\TaddownTubeNoLabel$\cr
    \hidewidth\raise-2.8ex\hbox{$\scriptstyle{#1}\mkern30mu$}\cr
  }}}

  \newcommand{\TaddownTubeprime}[1]{\mathrel{\ooalign{$\TaddownTubeNoLabel$\cr
    \hidewidth\raise-3.0ex\hbox{$\scriptstyle{#1}\mkern32mu$}\cr
  }}}
  
  \newcommand{\TadupTube}[1]{\mathrel{\ooalign{$\TadupTubeNoLabel$\cr
    \hidewidth\raise-2.8ex\hbox{$\scriptstyle{#1}\mkern30mu$}\cr
  }}}
  
  \newcommand{\TorusNoLabels}{\mathord{\vcenter{\hbox{\includegraphics[scale=1]{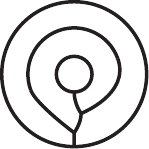}}}}}
\newcommand{\TorusNoLabelsx}[1]{\mathrel{\ooalign{$\TorusNoLabels$\cr
    \hidewidth\raise-2.8ex\hbox{$\scriptstyle{#1}\mkern30mu$}\cr
  }}}

\newcommand{\VerticalSpace}{\mathord{\vcenter{\hbox{\includegraphics[scale=1]{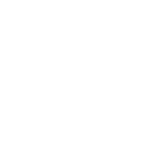}}}}}

\newcommand{\AddDat}[3]{\mathrel{\ooalign{  $#1$\cr
  \hidewidth\raise0ex\hbox{$\scriptstyle{#2}\mkern38mu$}\cr
  \hidewidth\raise0ex\hbox{${#3}\mkern0mu$}\cr
  \hidewidth\raise0ex\hbox{$\VerticalSpace$}\cr
  }}}
  
  \newcommand{\AddDatTorus}[3]{\mathrel{\ooalign{  $#1$\cr
  \hidewidth\raise0ex\hbox{$\scriptstyle{#2}\mkern38mu$}\cr
  \hidewidth\raise3.8ex\hbox{$\scriptstyle{#3}\mkern0mu$}\cr
  \hidewidth\raise0ex\hbox{$\VerticalSpace$}\cr
  }}}
  
    \newcommand{\AddDatTorusDot}[4]{\mathrel{\ooalign{  $#1$\cr
  \hidewidth\raise0ex\hbox{$\scriptstyle{#2}\mkern38mu$}\cr
  \hidewidth\raise3.8ex\hbox{$\scriptstyle{#3}\mkern0mu$}\cr
  \hidewidth\raise0ex\hbox{${#4}\mkern0mu$}\cr
  \hidewidth\raise0ex\hbox{$\VerticalSpace$}\cr
  }}}

  \newcommand{\TubeProductCoefficienta}{\mathord{\vcenter{\hbox{\includegraphics[scale=1]{TubeProductCoefficienta.pdf}}}}}
  \newcommand{\TubeProductCoefficientb}{\mathord{\vcenter{\hbox{\includegraphics[scale=1]{TubeProductCoefficientb.pdf}}}}}
  
    \newcommand{\ThetaSymbol}{\mathord{\vcenter{\hbox{\includegraphics[scale=1]{Theta.pdf}}}}}
 
   \newcommand{\ThetaSymbolx}[1]{\mathrel{\ooalign{$\ThetaSymbol$\cr
      \hidewidth\raise-1.3ex\hbox{$\scriptstyle{#1} \mkern0mu$}\cr
  }}}

  \newcommand{\TubeProductCoefficientbx}[2]{\mathrel{\ooalign{$\TubeProductCoefficientb \;\;\; $\cr
      \hidewidth\raise-.8ex\hbox{$\scriptstyle{#1} \mkern54mu$}\cr
      \hidewidth\raise-.8ex\hbox{$\scriptstyle{#2} \mkern6mu$}\cr
  }}}
  
  \newcommand{\TubeProductCoefficientax}[5]{\mathrel{\ooalign{$\TubeProductCoefficienta$\cr
      \hidewidth\raise-4.2ex\hbox{$\scriptstyle{#1} \mkern30mu$}\cr
      \hidewidth\raise-1.8ex\hbox{$\scriptstyle{#2} \mkern30mu$}\cr
  \hidewidth\raise6.1ex\hbox{$\scriptstyle{#3}\mkern44mu$}\cr
  \hidewidth\raise2.05ex\hbox{$\scriptstyle{#4}\mkern82mu$}\cr
    \hidewidth\raise3.2ex\hbox{$\scriptstyle{#5}\mkern3mu$}\cr
  }}}

\newcommand{\FubesdXsA}{\mathord{\vcenter{\hbox{\includegraphics[scale=1]{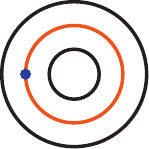}}}}}
\newcommand{\FubessXA}{\mathord{\vcenter{\hbox{\includegraphics[scale=1]{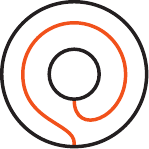}}}}}
\newcommand{\FubessdXA}{\mathord{\vcenter{\hbox{\includegraphics[scale=1]{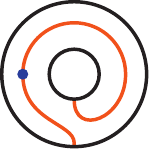}}}}}

\newcommand{\FubesXsA}{\mathord{\vcenter{\hbox{\includegraphics[scale=1]{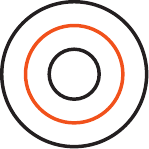}}}}}

\newcommand{\FubesXscA}{\mathord{\vcenter{\hbox{\includegraphics[scale=1]{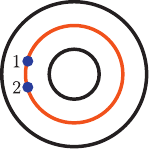}}}}}

\newcommand{\FubesXsaA}{\mathord{\vcenter{\hbox{\includegraphics[scale=1]{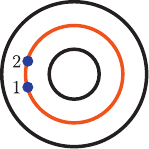}}}}}

\newcommand{\PantssvXsvdA}{\mathord{\vcenter{\hbox{\includegraphics[scale=.8,angle=0,origin=c]{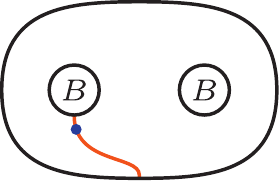}}}}}
\newcommand{\PantssvtXsvdA}{\mathord{\vcenter{\hbox{\includegraphics[scale=.8,angle=0,origin=c]{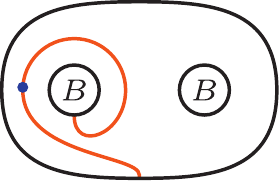}}}}}
\newcommand{\PantssvtshsvdA}{\mathord{\vcenter{\hbox{\includegraphics[scale=.8,angle=0,origin=c]{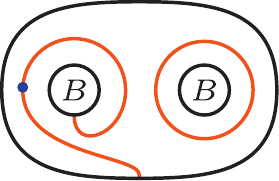}}}}}
\newcommand{\PantssvshsvdA}{\mathord{\vcenter{\hbox{\includegraphics[scale=.8,angle=0,origin=c]{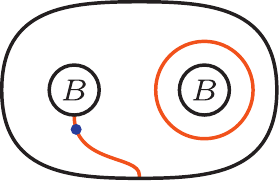}}}}}
\newcommand{\PantsPsdAsPA}{\mathord{\vcenter{\hbox{\includegraphics[scale=.8,angle=0,origin=c]{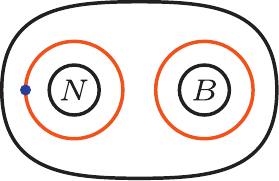}}}}}
\newcommand{\PantsPsdAPA}{\mathord{\vcenter{\hbox{\includegraphics[scale=.8,angle=0,origin=c]{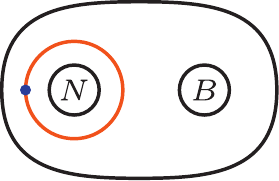}}}}}
\newcommand{\PantsPPAsA}{\mathord{\vcenter{\hbox{\includegraphics[scale=.8,angle=0,origin=c]{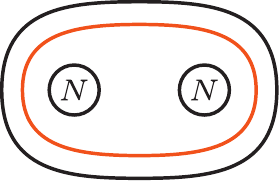}}}}}
\newcommand{\PantsPPAA}{\mathord{\vcenter{\hbox{\includegraphics[scale=.8,angle=0,origin=c]{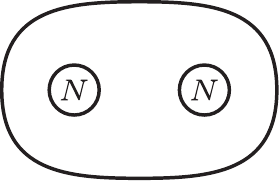}}}}}
\newcommand{\PantsPAsPA}{\mathord{\vcenter{\hbox{\includegraphics[scale=.8,angle=0,origin=c]{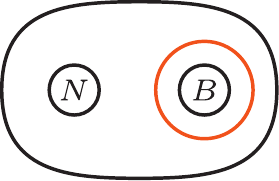}}}}}
\newcommand{\PantsPAPA}{\mathord{\vcenter{\hbox{\includegraphics[scale=.8,angle=0,origin=c]{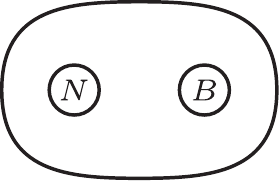}}}}}
\newcommand{\PantsAstAAsA}{\mathord{\vcenter{\hbox{\includegraphics[scale=.8,angle=0,origin=c]{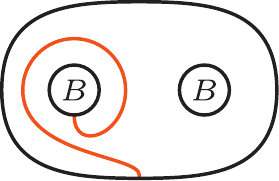}}}}}
\newcommand{\PantsAsAshAsvtA}{\mathord{\vcenter{\hbox{\includegraphics[scale=.8,angle=0,origin=c]{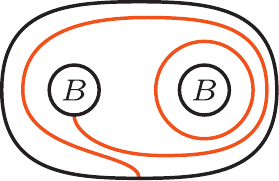}}}}}
\newcommand{\PantsNNda}{\mathord{\vcenter{\hbox{\includegraphics[scale=.8,angle=0,origin=c]{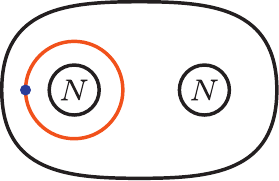}}}}}
\newcommand{\PantsNNd}{\mathord{\vcenter{\hbox{\includegraphics[scale=.8,angle=0,origin=c]{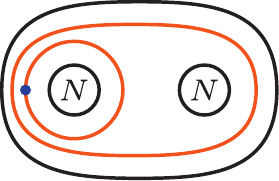}}}}}

\newcommand{\Pantsmsixa}{\mathord{\vcenter{\hbox{\includegraphics[scale=.8,angle=0,origin=c]{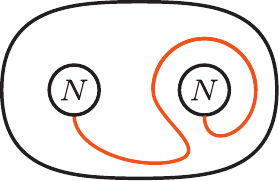}}}}}
\newcommand{\Pantsmsixb}{\mathord{\vcenter{\hbox{\includegraphics[scale=.8,angle=0,origin=c]{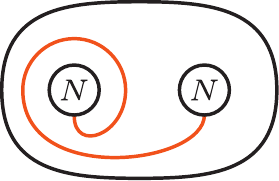}}}}}
\newcommand{\Pantsmsixc}{\mathord{\vcenter{\hbox{\includegraphics[scale=.8,angle=0,origin=c]{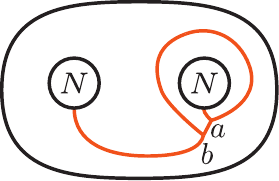}}}}}
\newcommand{\Pantsmsixd}{\mathord{\vcenter{\hbox{\includegraphics[scale=.8,angle=0,origin=c]{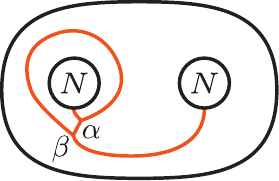}}}}}

\newcommand{\Pantsmsixe}{\mathord{\vcenter{\hbox{\includegraphics[scale=.8,angle=0,origin=c]{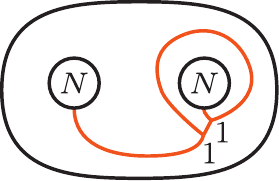}}}}}
\newcommand{\Pantsmsixf}{\mathord{\vcenter{\hbox{\includegraphics[scale=.8,angle=0,origin=c]{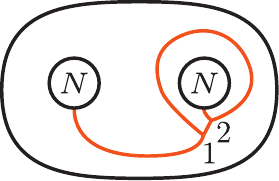}}}}}
\newcommand{\Pantsmsixg}{\mathord{\vcenter{\hbox{\includegraphics[scale=.8,angle=0,origin=c]{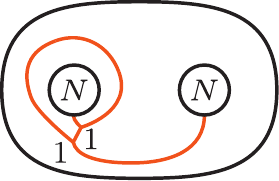}}}}}
\newcommand{\Pantsmsixh}{\mathord{\vcenter{\hbox{\includegraphics[scale=.8,angle=0,origin=c]{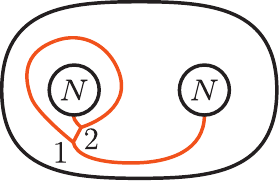}}}}}

\newcommand{\Vmsixa}{\mathord{\vcenter{\hbox{\includegraphics[scale=.8,angle=0,origin=c]{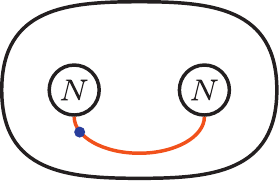}}}}}
\newcommand{\Vmsixb}{\mathord{\vcenter{\hbox{\includegraphics[scale=.8,angle=0,origin=c]{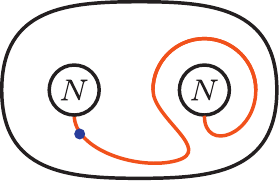}}}}}
\newcommand{\Vmsixc}{\mathord{\vcenter{\hbox{\includegraphics[scale=.8,angle=0,origin=c]{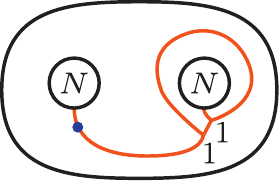}}}}}
\newcommand{\Vmsixd}{\mathord{\vcenter{\hbox{\includegraphics[scale=.8,angle=0,origin=c]{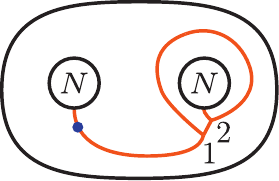}}}}}

\newcommand{\TwoLinedotdot}{\mathord{\vcenter{\hbox{\includegraphics[scale=1.5,angle=0,origin=c]{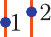}}}}}

\newcommand{\Id}{\mathord{\vcenter{\hbox{\includegraphics[scale=1.5,angle=0,origin=c]{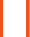}}}}}

\newcommand{\StaggaredGSOdd}{\mathord{\vcenter{\hbox{\includegraphics[scale=1.5,angle=0,origin=c]{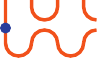}}}}}
\newcommand{\StaggaredGSEven}{\mathord{\vcenter{\hbox{\includegraphics[scale=1.5,angle=0,origin=c]{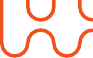}}}}}

\newcommand{\StaggaredGSEvenR}{\mathord{\vcenter{\hbox{\reflectbox{\includegraphics[scale=1.5,angle=0,origin=c]{StaggeredGSEven.pdf}}}}}}

\newcommand{\HambdRdot}{\mathord{\vcenter{\hbox{\includegraphics[scale=1.5,angle=0,origin=c]{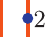}}}}}
\newcommand{\HambdR}{\mathord{\vcenter{\hbox{\includegraphics[scale=1.5,angle=0,origin=c]{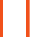}}}}}
\newcommand{\HambdLdot}{\mathord{\vcenter{\hbox{\includegraphics[scale=1.5,angle=0,origin=c]{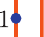}}}}}
\newcommand{\HambdL}{\mathord{\vcenter{\hbox{\includegraphics[scale=1.5,angle=0,origin=c]{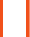}}}}}

\newcommand{\PsiFermion}{\mathord{\vcenter{\hbox{\includegraphics[scale=1.5]{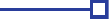}}}}}
\newcommand{\PsiFermionTwist}{\mathord{\vcenter{\hbox{\includegraphics[scale=1.5]{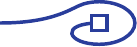}}}}}
\newcommand{\halfbentfermionline}{\mathord{\vcenter{\hbox{\includegraphics[scale=1.5]{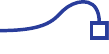}}}}}
\newcommand{\kinkedfermionline}{\mathord{\vcenter{\hbox{\includegraphics[scale=1.5]{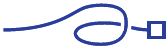}}}}}

\newcommand{\alphaboxbetabraidinga}{\mathord{\vcenter{\hbox{\includegraphics[scale=1.5]{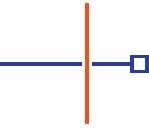}}}}}
\newcommand{\alphaboxbetabraidingb}{\mathord{\vcenter{\hbox{\includegraphics[scale=1.5]{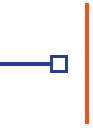}}}}}
\newcommand{\alphaboxbetabraidingc}{\mathord{\vcenter{\hbox{\includegraphics[scale=1.5]{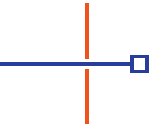}}}}}

\newcommand{\TwoFermionNoLabels}{\mathord{\vcenter{\hbox{\includegraphics[scale=1.5]{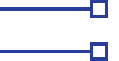}}}}}
\newcommand{\TwoFermionExchangeNoLabels}{\mathord{\vcenter{\hbox{\includegraphics[scale=1.5]{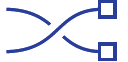}}}}}

\newcommand{\PsiEnd}{\mathord{\vcenter{\hbox{\includegraphics[scale=1.5]{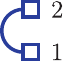}}}}}

\newcommand{\ebox}{\mathord{\vcenter{\hbox{\includegraphics[scale=1.3]{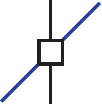}}}}}
\newcommand{\ex}{\mathord{\vcenter{\hbox{\includegraphics[scale=1.3]{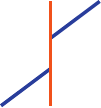}}}}}
\newcommand{\ey}{\mathord{\vcenter{\hbox{\includegraphics[scale=1.3]{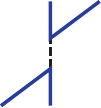}}}}}
\newcommand{\eone}{\mathord{\vcenter{\hbox{\includegraphics[scale=1.3]{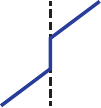}}}}}
\newcommand{\egeneral}{\mathord{\vcenter{\hbox{\includegraphics[scale=1.3]{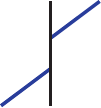}}}}}

\usepackage{leftidx}
\newcommand{\overunderset}[3]{\overset{#3}{\underset{#2}{#1}}}

  \newcommand{\halfbraid}[4]{\mathrel{\ooalign{
  \vphantom{$\Big|^2$}  
  $\leftidx{_{#2}}{\overunderset{#1}{#3}{#3}}{^{#2}}$\cr
   \hidewidth \hbox{$\scriptstyle{#4}\mkern32mu$}\cr
     \hidewidth\raise0ex\hbox{$\scale{1.5}{\VerticalSpace}$}\cr
 }}}

  \newcommand{\halfbraidHex}[5]{\mathrel{\ooalign{
  \vphantom{$\Big|^2$}  
  $\leftidx{_{#2}}{\overunderset{#1}{#3}{#4 \quad \; \; \; #5}}{^{#2}}$\cr
     \hidewidth\raise0ex\hbox{$\scale{2.5}{\VerticalSpace}$}\cr
 }}}

\newcommand{\HalfBraidHexa}{\mathord{\vcenter{\hbox{\includegraphics[scale=1.3]{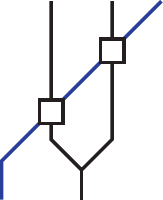}}}}}
\newcommand{\HalfBraidHexb}{\mathord{\vcenter{\hbox{\includegraphics[scale=1.3]{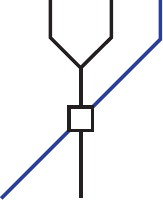}}}}}

\newcommand{\scale}[2]{\scalebox{#1}{$#2$}}

\newcommand{\VSa}{\mathord{\vcenter{\hbox{\includegraphics[scale=1.3]{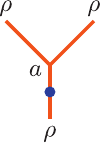}}}}}
\newcommand{\VSb}{\mathord{\vcenter{\hbox{\includegraphics[scale=1.3]{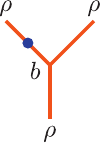}}}}}
\newcommand{\VSc}{\mathord{\vcenter{\hbox{\includegraphics[scale=1.3]{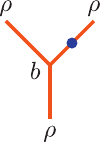}}}}}
\newcommand{\VSd}{\mathord{\vcenter{\hbox{\includegraphics[scale=1.3]{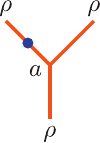}}}}}

\newcommand{\VNoLabel}{\mathord{\vcenter{\hbox{\includegraphics[scale=1.3]{VNoLabel.pdf}}}}}
\newcommand{\VVNoLabel}{\mathord{\vcenter{\hbox{\includegraphics[scale=1.3]{VVNolabel.pdf}}}}}

\newcommand{\PitchForkNoLabel}{\mathord{\vcenter{\hbox{\includegraphics[scale=1.3]{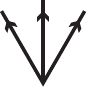}}}}}
\newcommand{\PitchForkNoLabelas}{\mathord{\vcenter{\hbox{\includegraphics[scale=1.3]{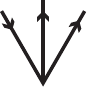}}}}}

\newcommand{\PitchForkWithEdge}{\mathord{\vcenter{\hbox{\includegraphics[scale=1.3]{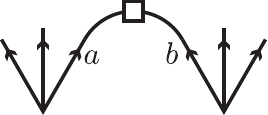}}}}}

\newcommand{\NonpitchforkNoLabel}{\mathord{\vcenter{\hbox{\includegraphics[scale=1.3]{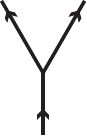}}}}}
\newcommand{\NonpitchforkNoVertex}[3]{\overset{#1 \quad \quad  \;\; #2}{\underset{#3}{\NonpitchforkNoLabel}}}

  \newcommand{\Nonpitchfork}[4]{\mathrel{\ooalign{$\NonpitchforkNoVertex{#1}{#2}{#3}$\cr
      \hidewidth\raise0ex\hbox{$\scriptstyle{#4} \mkern18mu$}\cr
  }}}

\newcommand{\PitchFork}[4]{\overset{#2}{\leftidx{_{}^{#1}}{\underset{#4}{\PitchForkNoLabel}}{^{#3}}}}
\newcommand{\PitchForkas}[4]{\underset{#4}{\overset{#2}{\leftidx{_{}^{#1}}{\PitchForkNoLabelas}{^{#3}}}}}

\newcommand{\NonPitchforkDotDisplacedNoLabel}{\mathord{\vcenter{\hbox{\includegraphics[scale=1.3]{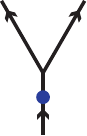}}}}}
\newcommand{\NonPitchforkDotDisplacedNoVertex}[3]{\overset{#1 \quad \quad  \;\; #2}{\underset{#3}{\NonPitchforkDotDisplacedNoLabel}}}

  \newcommand{\NonPitchforkDotDisplaced}[4]{\mathrel{\ooalign{$\NonPitchforkDotDisplacedNoVertex{#1}{#2}{#3}$\cr
      \hidewidth\raise0ex\hbox{$\scriptstyle{#4} \mkern13mu$}\cr
      \hidewidth\raise0ex\hbox{$\scale{1.5}{\VerticalSpace}$}\cr
  }}}

\newcommand{\NonPitchforkDotNoLabel}{\mathord{\vcenter{\hbox{\includegraphics[scale=1.3]{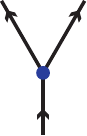}}}}}
\newcommand{\NonPitchforkDotNoVertex}[3]{\overset{#1 \quad \quad  \;\; #2}{\underset{#3}{\NonPitchforkDotNoLabel}}}

  \newcommand{\NonPitchforkDot}[4]{\mathrel{\ooalign{$\NonPitchforkDotNoVertex{#1}{#2}{#3}$\cr
      \hidewidth\raise0ex\hbox{$\scriptstyle{#4} \mkern13mu$}\cr
            \hidewidth\raise0ex\hbox{$\scale{1.5}{\VerticalSpace}$}\cr
  }}}

\newcommand{\fPLPR}{\mathord{\vcenter{\hbox{\includegraphics[scale=1.3]{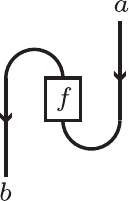}}}}}
\newcommand{\fstar}{\mathord{\vcenter{\hbox{\includegraphics[scale=1.3]{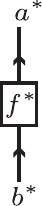}}}}}

\newcommand{\Tensora}{\mathord{\vcenter{\hbox{\includegraphics[scale=.7]{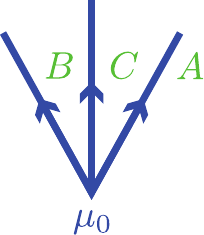}}}}}
\newcommand{\Tensorb}{\mathord{\vcenter{\hbox{\includegraphics[scale=.7]{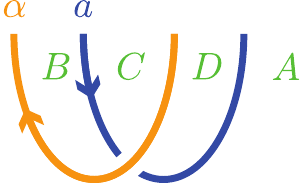}}}}}
\newcommand{\Tensorc}{\mathord{\vcenter{\hbox{\includegraphics[scale=.7]{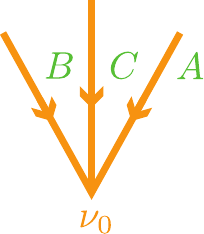}}}}}

\newcommand{\Tensoraa}{\mathord{\vcenter{\hbox{\includegraphics[scale=.7]{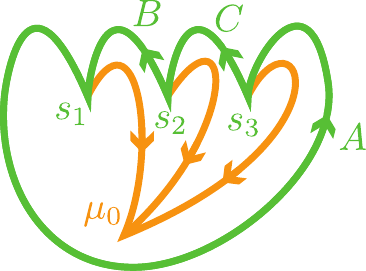}}}}}
\newcommand{\Tensorbb}{\mathord{\vcenter{\hbox{\includegraphics[scale=.7]{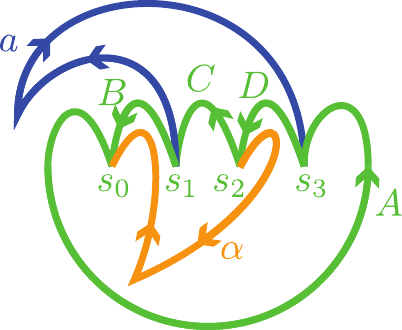}}}}}
\newcommand{\Tensorcc}{\mathord{\vcenter{\hbox{\includegraphics[scale=.7]{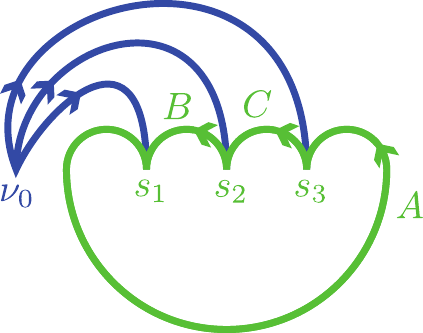}}}}}

\newcommand{\PivotYa}{\mathord{\vcenter{\hbox{\includegraphics[scale=1.3]{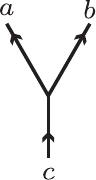}}}}}
\newcommand{\PivotYb}{\mathord{\vcenter{\hbox{\includegraphics[scale=1.3]{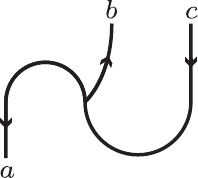}}}}}

\newcommand{\Fusionspacea}{\mathord{\vcenter{\hbox{\includegraphics[scale=1]{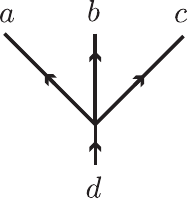}}}}}
\newcommand{\Fusionspaceb}{\mathord{\vcenter{\hbox{\includegraphics[scale=1]{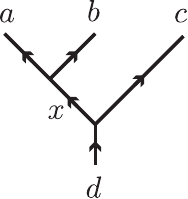}}}}}
\newcommand{\Fusionspacec}{\mathord{\vcenter{\hbox{\includegraphics[scale=1]{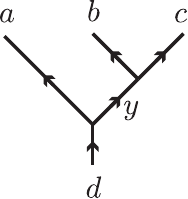}}}}}

\newcommand{\TensorProductaprime}{\mathord{\vcenter{\hbox{\includegraphics[scale=1]{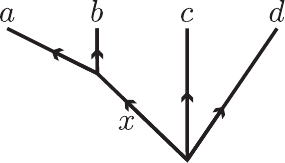}}}}}
\newcommand{\TensorProductbprime}{\mathord{\vcenter{\hbox{\includegraphics[scale=1]{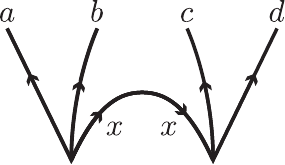}}}}}
\newcommand{\TensorProductcprime}{\mathord{\vcenter{\hbox{\includegraphics[scale=1]{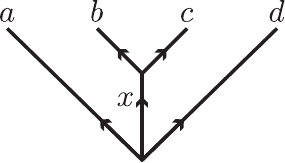}}}}}
\newcommand{\TensorProductdprime}{\mathord{\vcenter{\hbox{\includegraphics[scale=1]{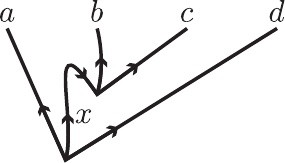}}}}}
\newcommand{\TensorProducteprime}{\mathord{\vcenter{\hbox{\includegraphics[scale=1]{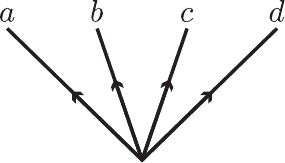}}}}}

\newcommand{\Idabba}{\mathord{\vcenter{\hbox{\includegraphics[scale=1.3]{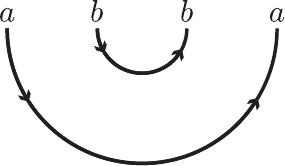}}}}}
\newcommand{\IdPitchfork}{\mathord{\vcenter{\hbox{\includegraphics[scale=1.3]{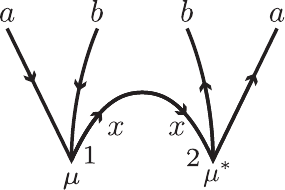}}}}}

     \newcommand{\VNoLabelDot}[1]{\mathrel{\ooalign{  $\VNoLabel$\cr
  \hidewidth\raise0ex\hbox{${#1}\mkern0mu$}\cr 
  }}}
  
       \newcommand{\VVNoLabelDot}[1]{\mathrel{\ooalign{  $\VVNoLabel$\cr
  \hidewidth\raise0ex\hbox{${#1}\mkern0mu$}\cr 
  }}}
 
    \newcommand{\Vx}[4]{\mathrel{\ooalign{  $\overset{#1\kern2.8em #2}{\underset{#3}{\VNoLabel}}$\cr
  \hidewidth\raise0ex\hbox{${#4}\mkern12mu$}\cr 
  }}}

    \newcommand{\VDotx}[5]{\mathrel{\ooalign{  $\overset{#1\kern2.8em #2}{\underset{#3}{\VNoLabelDot{#5}}}$\cr
  \hidewidth\raise0ex\hbox{${#4}\mkern12mu$}\cr 
       \hidewidth\raise0ex\hbox{$\scale{1.1}{\VerticalSpace}$}\cr
  }}}
  
      \newcommand{\VVDotx}[3]{\mathrel{\ooalign{  $\VVNoLabelDot{#3}$\cr
  \hidewidth\raise-.7ex\hbox{${#2}\mkern-8mu$}\cr 
    \hidewidth\raise-.7ex\hbox{${#1}\mkern65mu$}\cr 
       \hidewidth\raise0ex\hbox{$\scale{1.1}{\VerticalSpace}$}\cr
  }}}

\newcommand{\Vxyxxa}{\mathord{\vcenter{\hbox{\includegraphics[scale=1.3]{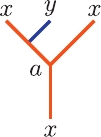}}}}}
\newcommand{\Vxyxxb}{\mathord{\vcenter{\hbox{\includegraphics[scale=1.3]{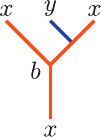}}}}}
\newcommand{\Vyxxxa}{\mathord{\vcenter{\hbox{\includegraphics[scale=1.3]{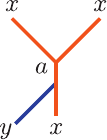}}}}}
\newcommand{\Vyxxxb}{\mathord{\vcenter{\hbox{\includegraphics[scale=1.3]{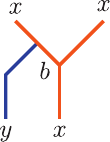}}}}}
\newcommand{\Vxxyxa}{\mathord{\vcenter{\hbox{\includegraphics[scale=1.3]{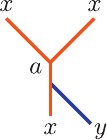}}}}}
\newcommand{\Vxxyxb}{\mathord{\vcenter{\hbox{\includegraphics[scale=1.3]{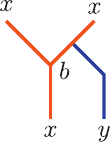}}}}}

\newcommand{\Vxxxa}{\mathord{\vcenter{\hbox{\includegraphics[scale=1.3]{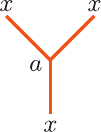}}}}}

\newcommand{\Vxxxya}{\mathord{\vcenter{\hbox{\includegraphics[scale=1.3]{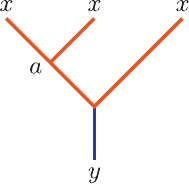}}}}}
\newcommand{\Vxxxyb}{\mathord{\vcenter{\hbox{\includegraphics[scale=1.3]{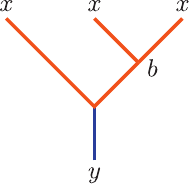}}}}}

\newcommand{\Vxxxva}{\mathord{\vcenter{\hbox{\includegraphics[scale=1.3]{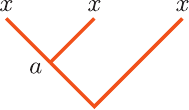}}}}}
\newcommand{\Vxxxvb}{\mathord{\vcenter{\hbox{\includegraphics[scale=1.3]{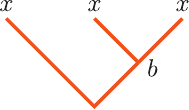}}}}}

\newcommand{\VSeven}{\mathord{\vcenter{\hbox{\includegraphics[scale=1.3]{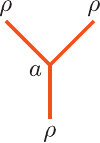}}}}}

\newcommand{\Vrhorhorhoodd}{\mathord{\vcenter{\hbox{\includegraphics[scale=1.3]{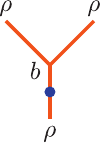}}}}}
\newcommand{\Vrhorhorho}{\mathord{\vcenter{\hbox{\includegraphics[scale=1.3]{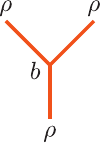}}}}}
\newcommand{\PivotEsixOdd}{\mathord{\vcenter{\hbox{\includegraphics[scale=1.3]{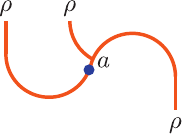}}}}}
\newcommand{\PivotEsixEven}{\mathord{\vcenter{\hbox{\includegraphics[scale=1.3]{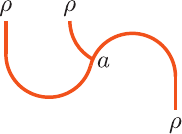}}}}}
 
 \newcommand{\EsixDynkin}{\mathord{\vcenter{\hbox{\includegraphics[scale=1]{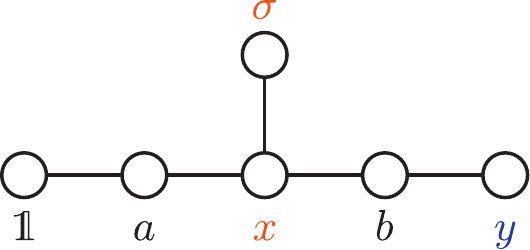}}}}}
 \newcommand{\EsixCondensePsi}{\mathord{\vcenter{\hbox{\includegraphics[scale=1]{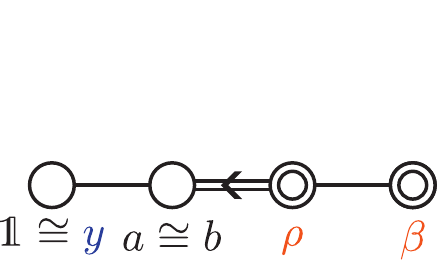}}}}}
 
 \newcommand{\HalfEsixDynkin}{\mathord{\vcenter{\hbox{\includegraphics[scale=1]{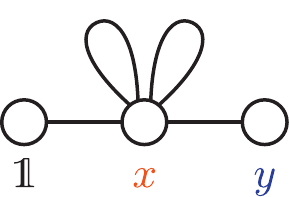}}}}}
 \newcommand{\HalfEsixDynkinCondensed}{\mathord{\vcenter{\hbox{\includegraphics[scale=1]{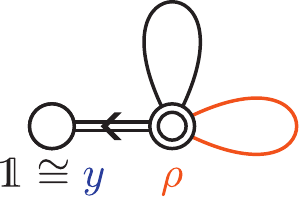}}}}}

 \newcommand{\AThreeDynkin}{\mathord{\vcenter{\hbox{\includegraphics[scale=1]{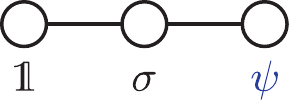}}}}}
  \newcommand{\CTwoDynkin}{\mathord{\vcenter{\hbox{\includegraphics[scale=1]{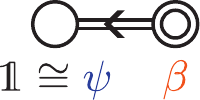}}}}}

 \newcommand{\halfesix}{\frac{1}{2}\text{E}_6}

\newcommand{\Vxxy}{\mathord{\vcenter{\hbox{\includegraphics[scale=1.3]{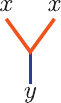}}}}}
\newcommand{\Vxxydual}{\mathord{\vcenter{\hbox{\includegraphics[scale=1.3]{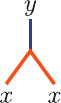}}}}}

\newcommand{\xxypivota}{\mathord{\vcenter{\hbox{\includegraphics[scale=1.3]{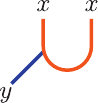}}}}}
\newcommand{\xxypivotb}{\mathord{\vcenter{\hbox{\includegraphics[scale=1.3]{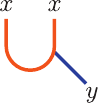}}}}}
\newcommand{\xxypivotadual}{\mathord{\vcenter{\hbox{\includegraphics[scale=1.3]{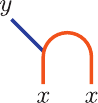}}}}}
\newcommand{\xxypivotbdual}{\mathord{\vcenter{\hbox{\includegraphics[scale=1.3]{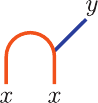}}}}}

\newcommand{\Vyxxdual}{\mathord{\vcenter{\hbox{\includegraphics[scale=1.3]{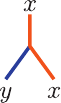}}}}}
\newcommand{\yxxpivotbdual}{\mathord{\vcenter{\hbox{\includegraphics[scale=1.3]{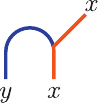}}}}}
\newcommand{\yxxpivotadual}{\mathord{\vcenter{\hbox{\includegraphics[scale=1.3]{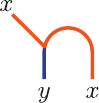}}}}}
\newcommand{\Vxyxdual}{\mathord{\vcenter{\hbox{\includegraphics[scale=1.3]{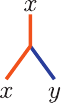}}}}}
\newcommand{\xyxpivotbdual}{\mathord{\vcenter{\hbox{\includegraphics[scale=1.3]{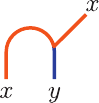}}}}}
\newcommand{\xyxpivotadual}{\mathord{\vcenter{\hbox{\includegraphics[scale=1.3]{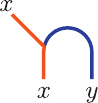}}}}}
\newcommand{\Vyxx}{\mathord{\vcenter{\hbox{\includegraphics[scale=1.3]{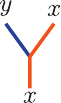}}}}}
\newcommand{\yxxpivotb}{\mathord{\vcenter{\hbox{\includegraphics[scale=1.3]{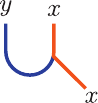}}}}}
\newcommand{\yxxpivota}{\mathord{\vcenter{\hbox{\includegraphics[scale=1.3]{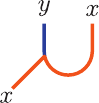}}}}}
\newcommand{\Vxyx}{\mathord{\vcenter{\hbox{\includegraphics[scale=1.3]{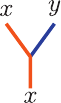}}}}}
\newcommand{\xyxpivotb}{\mathord{\vcenter{\hbox{\includegraphics[scale=1.3]{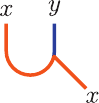}}}}}
\newcommand{\xyxpivota}{\mathord{\vcenter{\hbox{\includegraphics[scale=1.3]{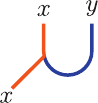}}}}}

\newcommand{\cupleftdot}{\mathord{\vcenter{\hbox{\includegraphics{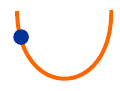}}}}}
\newcommand{\cuprightdot}{\mathord{\vcenter{\hbox{\includegraphics{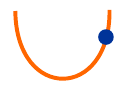}}}}}
\newcommand{\capleftdot}{\mathord{\vcenter{\hbox{\includegraphics{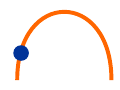}}}}}
\newcommand{\caprightdot}{\mathord{\vcenter{\hbox{\includegraphics{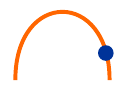}}}}}

\newcommand{\doublebeta}{\mathord{\vcenter{\hbox{\includegraphics{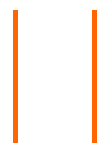}}}}}
\newcommand{\doublebetadots}{\mathord{\vcenter{\hbox{\includegraphics{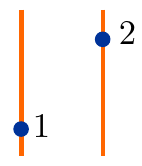}}}}}
\newcommand{\doublecups}{\mathord{\vcenter{\hbox{\includegraphics{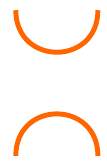}}}}}
\newcommand{\doublecupdots}{\mathord{\vcenter{\hbox{\includegraphics{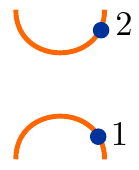}}}}}
\newcommand{\doublecuppsi}{\mathord{\vcenter{\hbox{\includegraphics{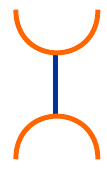}}}}}

\newcommand{\Vertexa}{\mathord{\vcenter{\hbox{\includegraphics[scale=1]{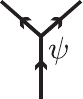}}}}}
\newcommand{\Vertexb}{\mathord{\vcenter{\hbox{\includegraphics[scale=1]{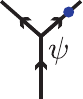}}}}}
\newcommand{\Vertexc}{\mathord{\vcenter{\hbox{\includegraphics[scale=1]{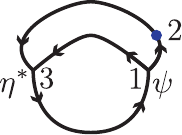}}}}}
\newcommand{\Vertexd}{\mathord{\vcenter{\hbox{\includegraphics[scale=1]{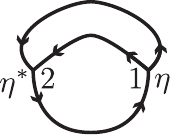}}}}}
\newcommand{\Vertexe}{\mathord{\vcenter{\hbox{\includegraphics[scale=1]{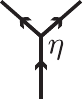}}}}}

\newcommand{\IsingDat}[5]{\underset{{\scriptstyle{#4} \quad\; \; \; \; \scriptstyle{#5}}}{\overset{\scriptstyle{#2}  \quad\; \; \; \; \scriptstyle{#3}}{#1}}}

\newcommand{\ScriptOverSymbol}[2]{{\mathord{\ooalign{ \vphantom{$\idpsishort$}\cr\hidewidth\ensuremath{\scriptstyle{#2}}\hidewidth\cr$\vcenter{\hbox{$\scale{1}{#1}$}}$\cr
  }}}}

\newcommand{\idorange}{\mathord{\vcenter{\hbox{\includegraphics[scale=1.3]{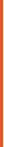}}}}}
\newcommand{\idblue}{\mathord{\vcenter{\hbox{\includegraphics[scale=1.3]{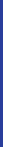}}}}}
\newcommand{\idblack}{\mathord{\vcenter{\hbox{\includegraphics[scale=1.3]{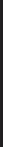}}}}}
\newcommand{\Bubblessp}{\mathord{\vcenter{\hbox{\includegraphics[scale=1.3]{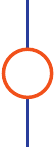}}}}}
\newcommand{\Bubblesps}{\mathord{\vcenter{\hbox{\includegraphics[scale=1.3]{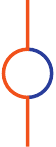}}}}}
\newcommand{\Bubbleabc}{\mathord{\vcenter{\hbox{\includegraphics[scale=1.3]{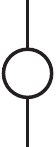}}}}}

\newcommand{\Vssp}{\mathord{\vcenter{\hbox{\includegraphics[scale=1.3]{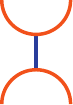}}}}}
\newcommand{\VppI}{\mathord{\vcenter{\hbox{\includegraphics[scale=1.3]{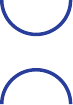}}}}}
\newcommand{\Vsps}{\mathord{\vcenter{\hbox{\includegraphics[scale=1.3]{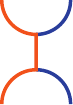}}}}}
\newcommand{\VssI}{\mathord{\vcenter{\hbox{\includegraphics[scale=1.3]{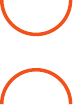}}}}}
\newcommand{\Hssp}{\mathord{\vcenter{\hbox{\includegraphics[scale=1.3]{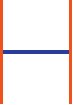}}}}}
\newcommand{\HssI}{\mathord{\vcenter{\hbox{\includegraphics[scale=1.3]{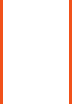}}}}}
\newcommand{\HspI}{\mathord{\vcenter{\hbox{\includegraphics[scale=1.3]{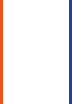}}}}}
\newcommand{\HppI}{\mathord{\vcenter{\hbox{\includegraphics[scale=1.3]{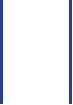}}}}}

\newcommand{\Braidpsconj}{\mathord{\vcenter{\hbox{\includegraphics[scale=1.3]{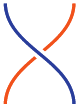}}}}}
\newcommand{\Braidsp}{\mathord{\vcenter{\hbox{\includegraphics[scale=1.3]{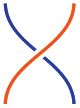}}}}}
\newcommand{\Braidpp}{\mathord{\vcenter{\hbox{\includegraphics[scale=1.3]{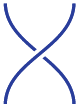}}}}}
\newcommand{\Braidss}{\mathord{\vcenter{\hbox{\includegraphics[scale=1.3]{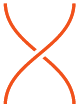}}}}}

\newcommand{\Twistp}{\mathord{\vcenter{\hbox{\includegraphics[scale=1.3]{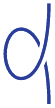}}}}}
\newcommand{\Twists}{\mathord{\vcenter{\hbox{\includegraphics[scale=1.3]{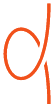}}}}}

\newcommand{\idsigmashort}{\mathord{\vcenter{\hbox{\includegraphics[scale=1.3]{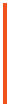}}}}}
\newcommand{\idsigmashortdot}{\mathord{\vcenter{\hbox{\includegraphics[scale=1.3]{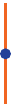}}}}}
\newcommand{\idsigmashortbox}{\mathord{\vcenter{\hbox{\includegraphics[scale=1.3]{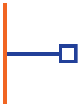}}}}}
\newcommand{\idpsishort}{\mathord{\vcenter{\hbox{\includegraphics[scale=1.3]{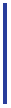}}}}}

\newcommand{\Rpss}{\mathord{\vcenter{\hbox{\includegraphics[scale=1.3]{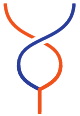}}}}}
\newcommand{\Vsigmapsisigma}{\mathord{\vcenter{\hbox{\includegraphics[scale=1.3]{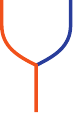}}}}}

\newcommand{\FusionSpaceRightOrdered}{\mathord{\vcenter{\hbox{\includegraphics[scale=1.3]{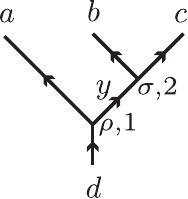}}}}}
\newcommand{\FusionSpaceLeftOrdered}{\mathord{\vcenter{\hbox{\includegraphics[scale=1.3]{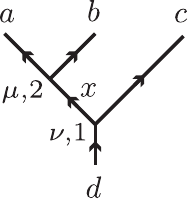}}}}}

\newcommand{\PitchforkFLeftprime}{\mathord{\vcenter{\hbox{\includegraphics[scale=1.3]{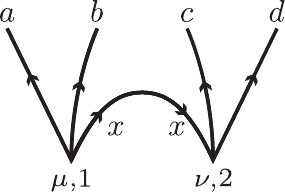}}}}}
\newcommand{\PitchforkFRightprime}{\mathord{\vcenter{\hbox{\includegraphics[scale=1.3]{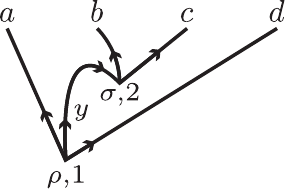}}}}}

\newcommand{\PitchforkFLeftprimesmall}{\mathord{\vcenter{\hbox{\includegraphics[scale=0.8]{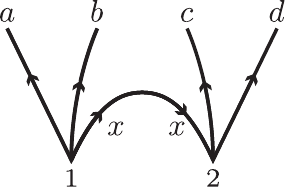}}}}}
\newcommand{\PitchforkFRightprimesmall}{\mathord{\vcenter{\hbox{\includegraphics[scale=0.8]{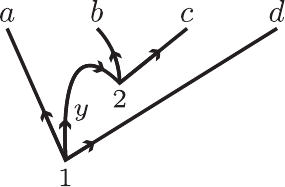}}}}}

\newcommand{\Edgea}{\mathord{\vcenter{\hbox{\includegraphics[scale=1.3]{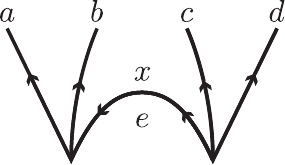}}}}}
\newcommand{\Edgeb}{\mathord{\vcenter{\hbox{\includegraphics[scale=1.3]{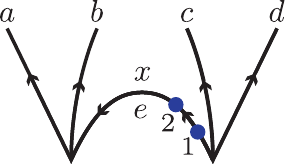}}}}}
\newcommand{\Edgec}{\mathord{\vcenter{\hbox{\includegraphics[scale=1.3]{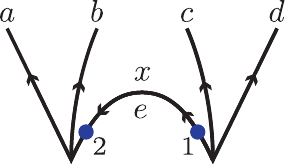}}}}}

\newcommand{\PivotCoherencea}{\mathord{\vcenter{\hbox{\includegraphics[scale=0.8]{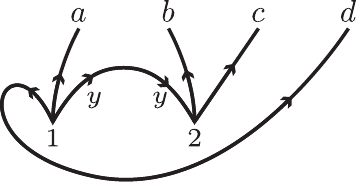}}}}}
\newcommand{\PivotCoherenceb}{\mathord{\vcenter{\hbox{\includegraphics[scale=0.8]{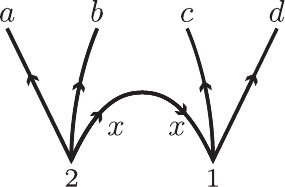}}}}}
\newcommand{\PivotCoherencec}{\mathord{\vcenter{\hbox{\includegraphics[scale=0.8]{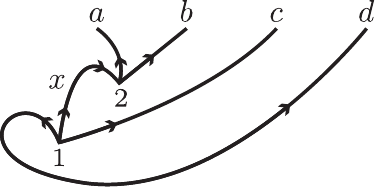}}}}}

\newcommand{\Penta}{\mathord{\vcenter{\hbox{\includegraphics[scale=0.7]{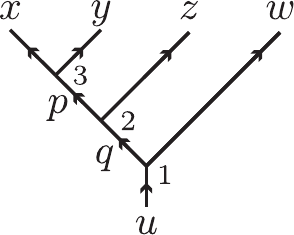}}}}}
\newcommand{\Pentb}{\mathord{\vcenter{\hbox{\includegraphics[scale=0.7]{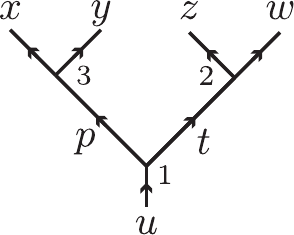}}}}}
\newcommand{\Pentc}{\mathord{\vcenter{\hbox{\includegraphics[scale=0.7]{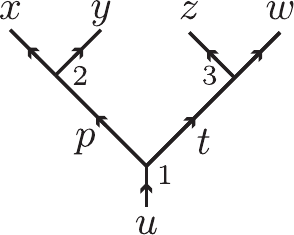}}}}}
\newcommand{\Pentd}{\mathord{\vcenter{\hbox{\includegraphics[scale=0.7]{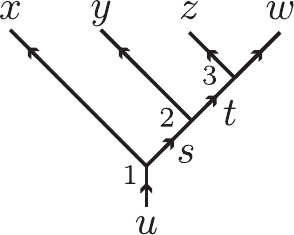}}}}}
\newcommand{\Pente}{\mathord{\vcenter{\hbox{\includegraphics[scale=0.7]{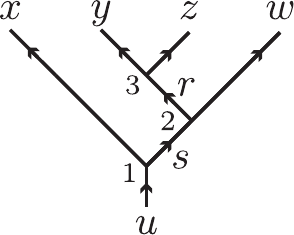}}}}}
\newcommand{\Pentf}{\mathord{\vcenter{\hbox{\includegraphics[scale=0.7]{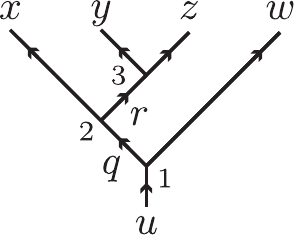}}}}}

\newcommand{\Bananafourmu}{\mathord{\vcenter{\hbox{\includegraphics[scale=1]{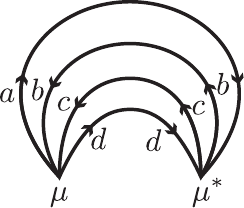}}}}}

\newcommand{\Plaquettea}{\mathord{\vcenter{\hbox{\includegraphics[scale=1]{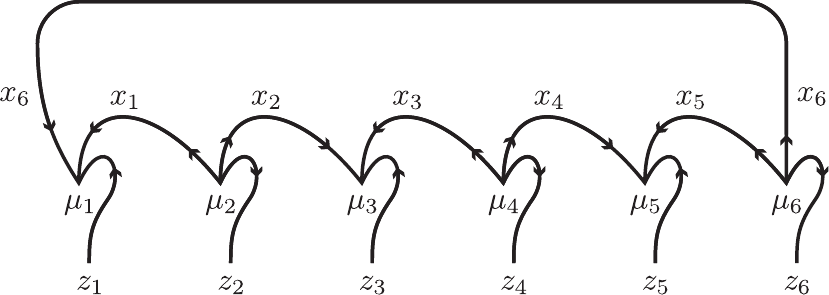}}}}}
\newcommand{\Plaquetteb}{\mathord{\vcenter{\hbox{\includegraphics[scale=1]{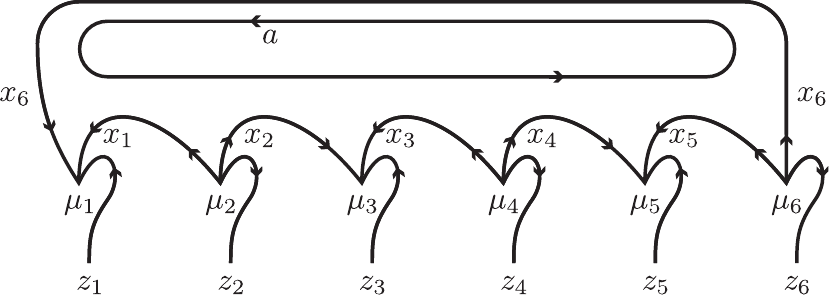}}}}}
\newcommand{\Plaquettec}{\mathord{\vcenter{\hbox{\includegraphics[scale=1]{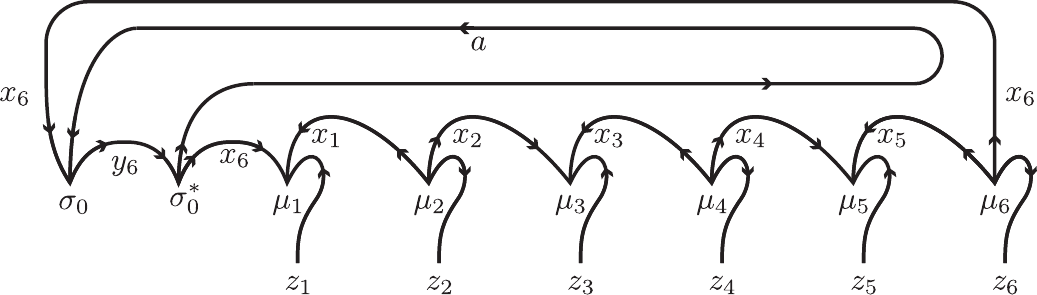}}}}}
\newcommand{\Plaquetted}{\mathord{\vcenter{\hbox{\includegraphics[scale=1]{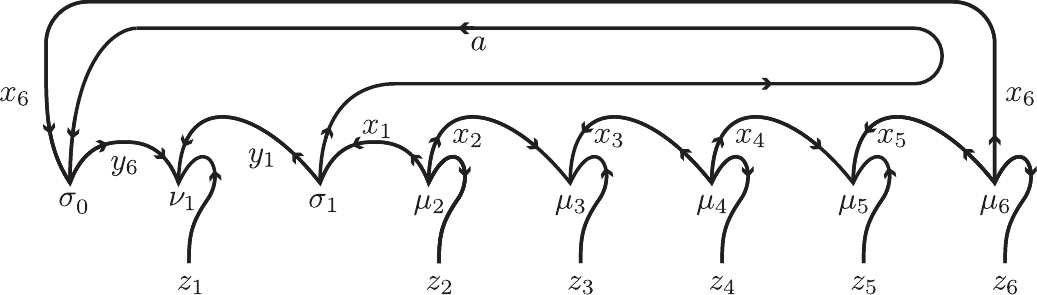}}}}}
\newcommand{\Plaquettee}{\mathord{\vcenter{\hbox{\includegraphics[scale=1]{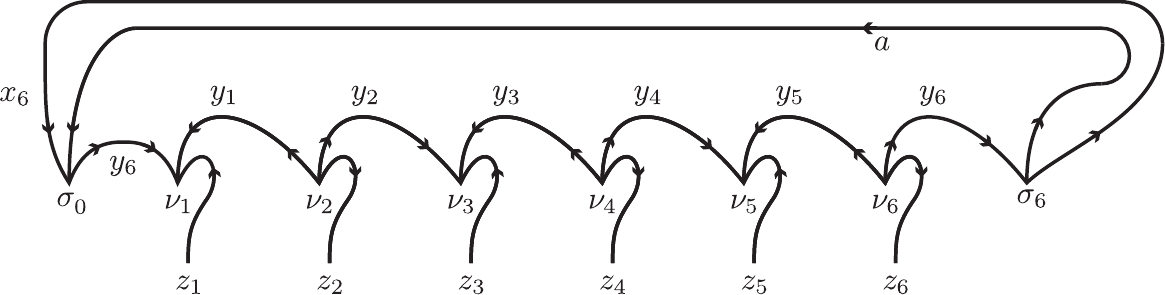}}}}}
\newcommand{\Plaquettef}{\mathord{\vcenter{\hbox{\includegraphics[scale=1]{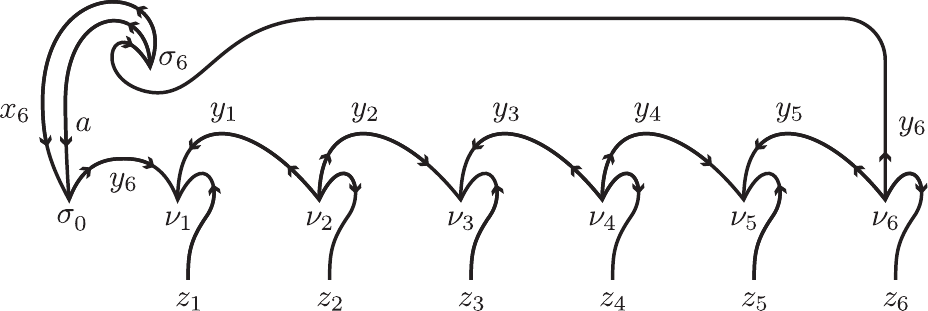}}}}}
\newcommand{\Plaquetteg}{\mathord{\vcenter{\hbox{\includegraphics[scale=1]{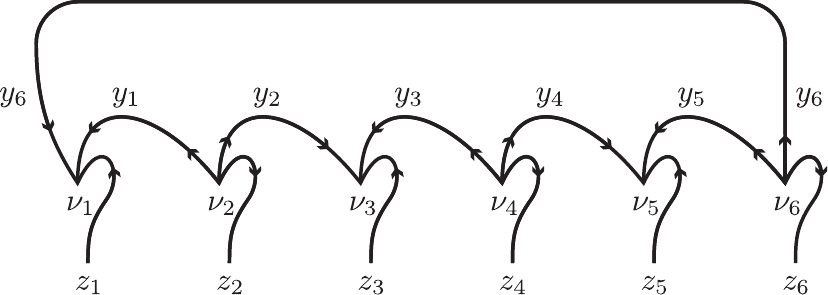}}}}}

\newcommand{\HexPlaquette}{\mathord{\vcenter{\hbox{\includegraphics[scale=1]{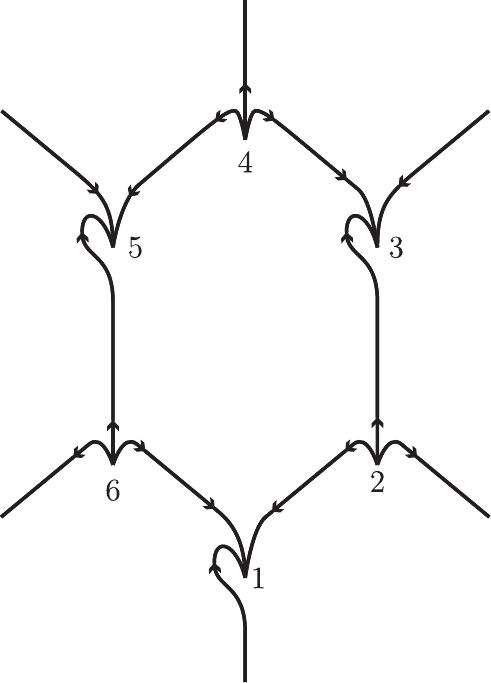}}}}}

\newcommand{\HexPlaquetteStretchedPrime}{\mathord{\vcenter{\hbox{\includegraphics[scale=1]{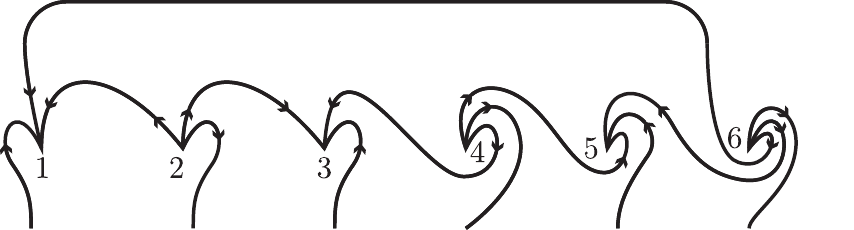}}}}}

\newcommand{\ZCPsiParent}{\mathord{\vcenter{\hbox{\includegraphics[scale=1]{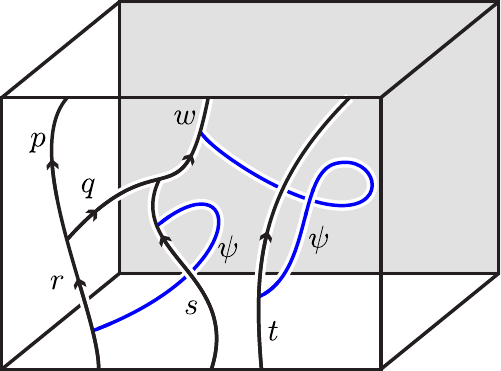}}}}}

\newcommand{\ZCPsiCondensed}{\mathord{\vcenter{\hbox{\includegraphics[scale=1]{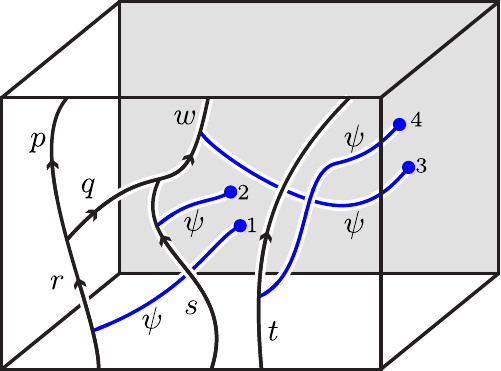}}}}}

\newcommand{\BackwallFigureprimeb}{\mathord{\vcenter{\hbox{\includegraphics[scale=1]{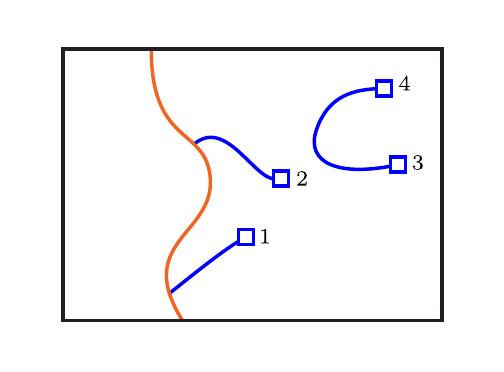}}}}}
\newcommand{\BackwallFigureprimea}{\mathord{\vcenter{\hbox{\includegraphics[scale=1]{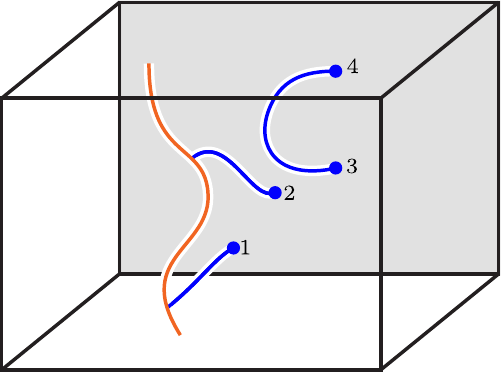}}}}}

\newcommand{\refcoha}{\mathord{\vcenter{\hbox{\includegraphics[scale=1]{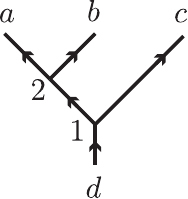}}}}}
\newcommand{\refcohb}{\mathord{\vcenter{\hbox{\includegraphics[scale=1]{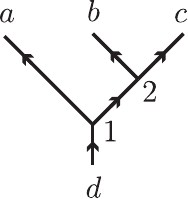}}}}}
\newcommand{\refcohc}{\mathord{\vcenter{\hbox{\includegraphics[scale=1]{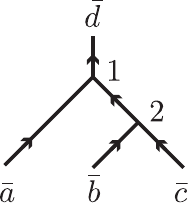}}}}}
\newcommand{\refcohd}{\mathord{\vcenter{\hbox{\includegraphics[scale=1]{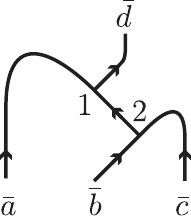}}}}}
\newcommand{\refcohe}{\mathord{\vcenter{\hbox{\includegraphics[scale=1]{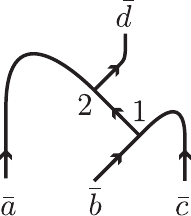}}}}}
\newcommand{\refcohf}{\mathord{\vcenter{\hbox{\includegraphics[scale=1]{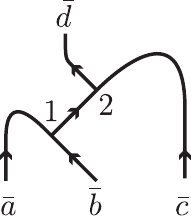}}}}}
\newcommand{\refcohg}{\mathord{\vcenter{\hbox{\includegraphics[scale=1]{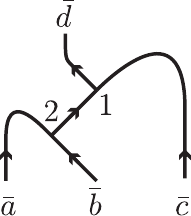}}}}}
\newcommand{\refcohh}{\mathord{\vcenter{\hbox{\includegraphics[scale=1]{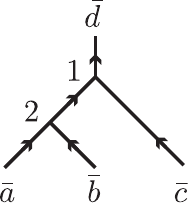}}}}}

\newcommand{\refa}{\mathord{\vcenter{\hbox{\includegraphics[scale=1]{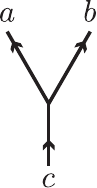}}}}}
\newcommand{\refb}{\mathord{\vcenter{\hbox{\includegraphics[scale=1]{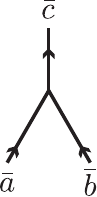}}}}}
\newcommand{\refc}{\mathord{\vcenter{\hbox{\includegraphics[scale=1]{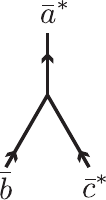}}}}}
\newcommand{\refd}{\mathord{\vcenter{\hbox{\includegraphics[scale=1]{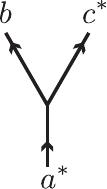}}}}}

\newcommand{\GraphG}{\mathord{\vcenter{\hbox{\includegraphics[scale=1]{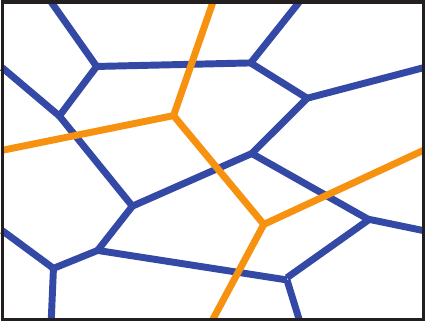}}}}}
\newcommand{\GraphGprimeprime}{\mathord{\vcenter{\hbox{\includegraphics[scale=1]{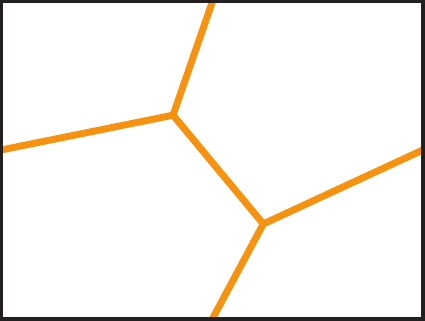}}}}}
\newcommand{\GraphGprime}{\mathord{\vcenter{\hbox{\includegraphics[scale=1]{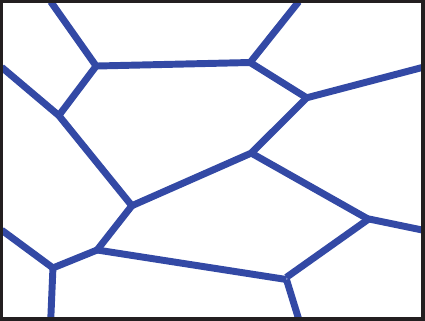}}}}}

\newcommand{\LocalHilba}{\mathord{\vcenter{\hbox{\includegraphics[scale=1.3]{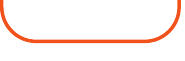}}}}}
\newcommand{\LocalHilbb}{\mathord{\vcenter{\hbox{\includegraphics[scale=1.3]{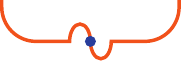}}}}}

\newcommand{\halfchain}{\mathord{\vcenter{\hbox{\includegraphics[scale=1.3]{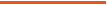}}}}}
\newcommand{\halfchaindot}{\mathord{\vcenter{\hbox{\includegraphics[scale=1.3]{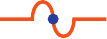}}}}}

\newcommand{\quarterpiper}{\mathord{\vcenter{\hbox{\includegraphics[scale=1.3]{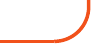}}}}}
\newcommand{\quarterpiperdot}{\mathord{\vcenter{\hbox{\includegraphics[scale=1.3]{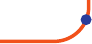}}}}}

\newcommand{\quarterpipel}{\mathord{\vcenter{\hbox{\includegraphics[scale=1.3]{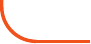}}}}}
\newcommand{\quarterpipeldot}{\mathord{\vcenter{\hbox{\includegraphics[scale=1.3]{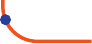}}}}}

\newcommand{\quarterpiperprime}{\mathord{\vcenter{\hbox{\includegraphics[scale=1.3]{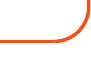}}}}}
\newcommand{\quarterpiperdotprime}{\mathord{\vcenter{\hbox{\includegraphics[scale=1.3]{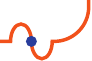}}}}}

\newcommand{\quarterpipelprime}{\mathord{\vcenter{\hbox{\includegraphics[scale=1.3]{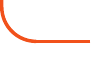}}}}}

\newcommand{\StaggaredGSOddprime}{\mathord{\vcenter{\hbox{\includegraphics[scale=1.3,angle=0,origin=c]{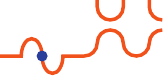}}}}}
\newcommand{\StaggaredGSEvenprime}{\mathord{\vcenter{\hbox{\includegraphics[scale=1.3,angle=0,origin=c]{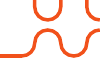}}}}}

\newcommand{\StaggaredGSEvenRprime}{\mathord{\vcenter{\hbox{\reflectbox{\includegraphics[scale=1.3,angle=0,origin=c]{StaggeredGSEvenprime.pdf}}}}}}

\newcommand{\scat}{\mathcal{T}}
\newcommand{\spc}{\mathcal{S}}

\begin{document}


\title{Fermion condensation and super pivotal categories}

\author{David Aasen$^1$$^2$, Ethan Lake$^{3}$, and Kevin Walker$^4$}
\affil{
$^1$ Department of Physics and Institute for Quantum Information and Matter, California Institute of Technology, Pasadena, CA 91125, USA}
\affil{$^2$Kavli Institute for Theoretical Physics, University of California, Santa Barbara, CA 93106, USA}
\affil{
$^3$ Department of Physics and Astronomy, University of Utah, Salt Lake City, UT 84112, USA}
\affil{
$^4$ Station Q, Microsoft Research, Santa Barbara, California 93106-6105, USA}


\date{\today}

\maketitle


\begin{abstract}
We study fermionic topological phases using the technique of fermion condensation. 
We give a prescription for performing fermion condensation in 
bosonic topological phases which contain a fermion.
Our approach to fermion condensation can roughly be understood 
as coupling the parent bosonic topological phase to a phase of physical fermions, and condensing 
pairs of physical and emergent fermions. 
There are two distinct types of objects in the resulting fermionic fusion categories, which we call ``m-type'' and ``q-type'' objects. 
The endomorphism algebras of q-type objects are complex Clifford algebras, and they
have no analogues in bosonic theories. 
We construct a fermionic generalization of the tube category, which allows us to compute
the quasiparticle excitations arising from the condensed theories.
We prove a series of results relating data in fermionic theories to data in their parent bosonic theories;
for example, if $\mathcal{C}$ is a modular tensor category containing a fermion, then the tube category 
constructed from the condensed theory satisfies 
$\textbf{Tube}(\mathcal{C}/\psi) \cong \mathcal{C} \times (\mathcal{C}/\psi)$.  
We also study how modular transformations, fusion rules, and coherence relations 
are modified in the fermionic setting, 
prove a fermionic version of the Verlinde dimension formula,
construct a commuting projector lattice Hamiltonian for fermionic theories, and 
write down a fermionic version of the Turaev-Viro-Barrett-Westbury state sum.  
A large portion of this work is devoted to three detailed examples of performing fermion condensation to 
produce fermionic topological phases: we condense fermions 
in the Ising theory, the $SO(3)_6$ theory, and the $\frac{1}{2}\text{E}_6$ theory, and compute the 
quasiparticle excitation spectrum in each of the condensed theories. 
\end{abstract}

\tableofcontents


\section{Introduction}

A large program in condensed matter physics in recent years has been to classify 
topological phases of matter which host emergent quasiparticle excitations 
with topological properties, such as exotic braiding statistics. 
The mathematical framework of category theory has proven to provide the right 
framework needed to formally develop classification efforts. 
Thus far, the bulk of these efforts have focused on understanding bosonic topological phases.
These phases are ``bosonic'' because the topological excitations emerge from bosonic 
local degrees of freedom.

By contrast, much less is known about how to complete a classification program for {\it fermionic} topological phases, in which fermions constitute 
the underlying degrees of freedom. 
Early work in this direction was presented by one of the authors in a series of talks \cite{Walker2013,Walker2014,Walker2015}.
Progress on understanding the coherence relations used to classify fermionic topological phases and the identification of a class of examples of such phases was made in \cite{gu2015,gu2014,Lan2016b}. 
Recently, Majorana dimer lattice models have appeared \cite{Freedman2011a,tarantino2016,ware2016}, 
which give an explicit Hamiltonian construction of a 
non-trivial fermionic phase closely related to the Ising theory. 
There has also been some recent work in the math community \cite{usher2016,brundan2016,bruillard2017,bonderson2017} 
devoted to studying the formal category-theoretic description of fermionic topological phases and spin-TQFTs. 

Intimately related to the description of fermionic topological phases is the concept of fermion condensation, 
whereby one passes to a phase in which the emergent (anyonic) fermions have become 
local particles (meaning that they can be created and destroyed locally). 
There have been several recent approaches related to understanding 
fermionic topological phases with field theoretic methods, fermion condensation, and bosonization \cite{Barkeshli2014b,barkeshli2015,gaiotto2016, bhardwaj2016, bhardwaj2016b,kapustin2017,putrov2016}, 
with a more algebraic take on fermion condensation given in \cite{wan2016}. 

In this work, we perform a systematic study of fermionic topological phases and fermion condensation from a category-theoretic point of view. 
The strategy we will use to construct examples of fermionic topological phases will be to start with 
a bosonic phase described by a tensor category $\mcc$
which contains an emergent fermion $\psi$.
To obtain a fermionic theory, we will ``condense'' $\psi$.
The resulting fermionic fusion category (or fermionic fusion theory) is described by a super pivotal category, which we denote
as $\mcc / \psi$.
$\mcc/\psi$ is intrinsically fermionic because in order for a phase to support local fermionic excitations, 
its underlying degrees of freedom must be fermionic. 

Due to their nontrivial spin and statistics, condensing fermions is not a straightforward business. 
From a mathematical perspective, in order to perform the condensation it is necessary to equip the configuration 
space of $\psi$ worldline endpoints with a certain complex line bundle. 
Physically, the construction of this bundle amounts to attaching a phase of physical (not emergent) fermions $f$ 
to the parent bosonic theory.
Fermion condensation then heuristically proceeds by coupling the $\psi$ fermions to 
the $f$ fermions, and condensing $\psi f$ bound states. 
In most cases (when $\psi$ is not transparent in $\mcc$), we will also need to perform 
the condensation with a construction we call the ``back wall'', which is a codimension-1 surface 
on which $\psi$ worldlines are allowed to end. 
Because of this construction, the resulting 
condensed theory does not inherit the braiding of the parent theory, and thus the condensed theory only describes 
fusion data, not braiding data. 
(In mathematical language, the result of condensation is merely a tensor category, not a braided category.)
The condensed theory can then be used as input data to a generalized 
string-net model which produces a braided fermionic topological phase.
(If $\psi$ is transparent, 
or if $\psi$ is non-transparent but we are willing to countenance vortices in the condensed theory,
then we can retain the braiding; see \ref{condense_transparent_fermion} and \ref{spin_defects_condensation}.)

An important difference between fermionic topological phases and their bosonic counterparts is that 
the former possesses two distinct classes of simple objects (or anyons). 
One class of simple objects, which we refer to as ``m-type'' objects, are identical in character
to the simple objects found in bosonic theories. 
The other class of objects, which we call ``q-type'' objects, have no bosonic analogues. 
From a formal perspective, they are distinguished by their nontrivial endomorphism algebras -- if 
$a$ is a q-type simple object, then $\End(a) \cong \cliff_1$, the first complex Clifford algebra. 
From a physical point of view, these can be thought of as ``Majorana objects'', which have the ability 
to ``absorb'' fermions. 
In string-net constructions, strings labeled by q-type objects are ``Kitaev strings'', in that they behave like Kitaev chains in the 
topological phase: the fermion parity of a closed loop of such a string is determined by 
the spin structure inherited by the string, and is delocalized in the sense that fermions living on the q-type string are allowed to fluctuate freely along its length.

One rather trivial way to produce a fermionic phase is to simply form a non-interacting stack of a phase of physical fermions 
and a known bosonic topological phase. 
To obtain examples of fermionic phases that do not arise in this way (which are called ``primitive'' in \cite{Lan2016b}),
it is essential to examine theories that contain q-type objects, which are fundamentally fermionic in nature and which are impossible to obtain in theories constructed through this stacking procedure. 

A categorical description of the condensed phase $\mcc / \psi$ cannot be obtained within the framework 
of regular tensor categories, and one needs to instead adopt the framework of so-called {\it super pivotal tensor categories}. 
The fermionic nature of $\mcc / \psi$ means that the fusion spaces of the theory become super vector spaces 
(as opposed to normal vector spaces), 
tensor products of morphisms and coherence relations like the pentagon identity are modified to account for Koszul signs, 
the pivotal structure of the theory is changed to account for the presence of fermions, and so on.

We should stress that all of the constructions we employ in this paper are mathematically well-defined and 
self-contained, independent of their physical interpretations. 
Most of the paper uses techniques from category theory, TQFTs and string nets to construct what one
might call fermionic Turaev-Viro theory.
This paper is mostly about fermionic TQFTs from this perspective, 
rather than ground states of Hamiltonians. 
Indeed, not until near the end of the paper do we define a fermionic Levin-Wen style Hamiltonian whose ground states coincide with 
the Hilbert spaces of this TQFT.
Foreshadowing the introduction of this Hamiltonian, we will, throughout the paper, talk about physical notions like ``excitations", ``ground state degeneracies", 
etc, even though
strictly speaking this point of view does not make sense until after we have introduced the Hamiltonian.
We emphasize that the Hamiltonian point of view is optional; most of the paper can be viewed as taking place
in the self-contained world of (fermionic) string-net TQFTs.

In order to examine the quasiparticle excitation spectrum of fermionic phases, 
we use the tube category construction \cite{ocneanu1994},\footnote{
Again, it is technically imprecise to call the objects of the tube category ``excitations'' until after we have carefully defined a Hamiltonian.}
which is closely related to the Drinfeld center.
This allows us to use the fermionic fusion categories 
produced by the fermion condensation procedure to construct examples of 
braided fermionic topological phases.
In the fermionic context, we must modify the tube category construction to take into account different spin structures on the circle.
For bosonic theories we have the isomorphism $\tube(\mcc)\cong\mcc\times\overline{\mcc}$ \cite{muger2003b} if $\mcc$ is modular, where $\overline{\mcc}$ denotes the opposite category of $\mcc$.
We prove a fermionic version of this isomorphism, namely that 
if $\mcc$ is modular then 
\be 
	\tube(\mcc / \psi) \cong \mcc \times (\overline{\mcc / \psi}).
\ee

Because quasiparticle excitations are associated to circular boundary components of the ambient 
manifold on which the theory is defined, there are two distinct classes of quasiparticles (a.k.a.\ representations of the tube category) in 
fermionic theories: those with anti-periodic fermion boundary conditions around the circle, and 
those with periodic boundary conditions. 
The tubes in the tube category come equipped with spin structures, which gives the tube category a $\zt$ grading
and separates the quasiparticle spectrum in to vortex quasiparticles (those which bind spin structure defects) 
and non-vortex quasiparticles (which are similar in character to the quasiparticles present in bosonic theories). 

The analysis of things related to the tube category, 
like modular transformations, braiding statistics, and the computation of ground state degeneracies on various spin surfaces,
is also modified in the fermionic setting.  
The behavior of the $S$ and $T$ modular transformations on the torus depends on the spin structure of the torus in question, and 
certain relations like $(ST)^3 = \id$ are modified in the fermionic setting, becoming 
$(ST)^3=(-1)^F\id$ where $(-1)^F$ is the fermion parity operator.

We will also show how to construct a commuting projector lattice Hamiltonian 
for fermionic phases, whose excitation spectrum is given by the objects in the fermionic tube category.  
It is similar in character to the Levin-Wen Hamiltonian \cite{levin2005}, 
except that it includes an extra edge term which is responsible for moving fermions along
edges labeled by q-type objects. 
Spin structure defect excitations 
are realized as violations of this edge term, and are linearly confined. 

Finally, we show how to construct a tensor network which produces the ground state 
wavefunction of our lattice Hamiltonian. We do this using a cut-and-glue approach to 
the construction of the Turaev-Viro-Barrett-Westbury state sum, which allows us to write the bosonic partition function
as a tensor contraction. 
We then show how to modify this construction in order to obtain a fermionic version of the TVBW state sum
and a corresponding fermionic tensor network.  

A series of in-depth examples occupies a large portion of this paper.
Our presentation is characterized by an emphasis on examples, 
and we often save formal definitions and more general statements 
for later sections after relevant examples have been presented.  
Many general observations are mixed in with the example sections,
and we stress that the more general sections are intended as a supplement
to the example sections; they are not stand-alone.

In several places in the paper we make use of a fermionic version of a Turaev-Viro type 2+1-dimensional TQFT.
Constructing this TQFT using the techniques of \cite{Walker2006} is straightforward -- one simply replaces oriented manifolds with
spin manifolds and allows for the possibility of non-trivial endomorphism algebras of minimal idempotents.
In the interest of not further increasing the length of this paper, we have not repeated details found in
\cite{Walker2006} here.

\medskip

We now give a quick section-by-section summary of the paper.
Much of the content of Section \ref{C2_condense_sect} was presented at \cite{Walker2013} in July 2013,
and many of the ideas contained in Sections \ref{C2excitations}, \ref{C2_fusion_rules}, and \ref{Super_pivotal_Hamiltonian} were presented at \cite{Walker2014} and again
at \cite{Walker2015}.

Sections \ref{C2_condense_sect} and \ref{C2_quasiparticles} are devoted to a detailed study of 
fermion condensation in the Ising TQFT, which is a simple case study that allows us to build intuition for the condensation procedure.
This condensed theory has been examined before in \cite{bhardwaj2016, kapustin2017}; 
here we examine it in greater detail. 
In Section \ref{C2_condense_sect} we review the Ising TQFT, describe how to perform the 
fermion condensation, and write down the local string-net rules in the condensed theory, 
which we call the $C_2$ theory. 
In Section \ref{C2_quasiparticles} we compute the simple objects of the tube category of the $C_2$ theory,
determine their fusion rules,
compute the ground state degeneracy on the torus, 
and study the modular transformations 
of the theory. 

In Section \ref{generalities}, we present a collection of more general results
on the machinery of fermionic theories. 
We show how to perform fermion condensation in more general settings
and discuss the fermionic version 
of the tube category construction in more generality.
We also discuss how to compute quantum dimensions and fusion rules,
and we prove some general
relations between the total quantum dimensions of condensed theories, their 
parent theories, and their tube categories.
Additionally, we present a Verlinde-type
dimension formula for fermionic theories.
The general dimension formula is given at \eqref{PplusQ} and \eqref{PminusQ}.
For 3-punctured spheres the general formula specializes to
\begin{samepage}
\begin{align}
	\dim_{\text{even}}(V^{abc}) = \frac{\sqrt{n_a n_b n_c}}{2}\left(
			\sum_{x\in B_m} \frac{S_{ax}S_{bx}S_{cx}}{S_{1x}} +
			\sum_{x\in N_m \cup N_q} \frac{S_{ax}S_{bx}S_{cx}}{S_{1x}}
	\right)\\
\intertext{and}
	\dim_{\text{odd}}(V^{abc}) = \frac{\sqrt{n_a n_b n_c}}{2}\left(
			\sum_{x\in B_m} \frac{S_{ax}S_{bx}S_{cx}}{S_{1x}} -
			\sum_{x\in N_m \cup N_q} \frac{S_{ax}S_{bx}S_{cx}}{S_{1x}}
	\right) .
\end{align}
\end{samepage}Here $n_x = 1$ if $x$ is m-type and $n_x = 2$ if $x$ is q-type, $S_{ij}$ denotes the modular $S$-matrix
with unitary normalization, $B_m$ denotes bounding (and therefore m-type; see Section \ref{ground_states_on_torus}) 
simple objects of the tube category, 
$N_m$ denotes non-bounding (i.e.\ vortex) simple m-type objects of the tube category, and
$N_q$ denotes non-bounding simple q-type objects of the tube category.

We devote Section \ref{more_on_tubes} to a detailed study of the tube category 
for fermionic theories which result from fermion condensation in a modular tensor category (MTC). 
We introduce several tools for performing calculations in tube categories, 
and use them to prove that if $\mcc$ is an MTC containing a fermion $\psi$, then
$\tube(\mcc/\psi)\cong \mcc\times (\overline{\mcc/\psi})$ as tensor 
categories. We also describe a way of easily computing the modular data for 
$\tube(\mcc/\psi)$. 

Section \ref{so36} is devoted to the example of performing condensation in the $SO(3)_6$ 
and $SU(2)_6$ theories, while Section \ref{halfesix} discusses fermion condensation in the $
\halfesix$ theory. 
Both of these examples are more involved than the $C_2$ theory, 
and illustrate some of the more interesting features of general fermionic topological phases. 

In Section \ref{def_sect} we discuss super pivotal categories from a more formal point of view. 
We show how fermionic fusion spaces are constructed and tensored together, how 
coherence relations like the pentagon identity are modified in the fermionic case, 
and 
explain various other modifications that are necessary in the fermionic case.

Section \ref{Super_pivotal_Hamiltonian} is devoted to a discussion of a lattice Hamiltonian for fermionic theories, which is an extension of the Levin-Wen Hamiltonian. 
It differs most significantly from the Levin-Wen Hamiltonian due to the presence of an edge term, which 
is responsible for allowing fermions to fluctuate across q-type strings.

We show how to construct a tensor network realizing the ground state of our lattice Hamiltonian 
in Section \ref{state_sums}. We first review how to write the partition function of a bosonic theory 
as a tensor contraction in a way amenable to generalization, and then show how our construction 
can be extended to cover the fermionic case. 

In Section \ref{kitaev_wire}, we discuss the Kitaev chain within the framework of super pivotal 
categories. 
We show how the Kitaev chain Hamiltonian and ground-state wavefunctions can be 
succinctly written down using the diagrammatic calculus developed earlier in the paper, and 
discuss connections between the Kitaev chain and the $C_2$ theory. 
This section also provides a connection between our work 
and recent work on fermionic topological phases in the physics community \cite{ware2016,tarantino2016,turzillo2016}. 
The only prerequisite for this section is a brief reading of Section \ref{C2_condense_sect}, and physically-inclined readers less interested in more general mathematical frameworks
are invited to read this section after reading Section \ref{C2_condense_sect}. 

We end with a conclusion and discussion in Section \ref{discussion}. 
Several appendices contain miscellaneous results and mathematical background information.

\subsection{Table of notation}

In partial compensation for the unwieldy length of this paper, we give a table summarizing some of the 
most frequently used and/or least standard notation found herein.

\medskip

\tabulinesep = 2mm
\begin{longtabu}{|X[1,$c]|X[4,l]|}
	\hline
    \mcc 	& tensor category \\ \hline
    \mcc/\psi 	& fermionic quotient of $\mcc$ \\ \hline
    \spc 	& super pivotal tensor category, e.g.~$\mcc/\psi$ \\ \hline
    \scat 	& super linear category \\ \hline
    \sob(\mcc) 	& complete set of representatives of simple objects for $\mcc$ \\ \hline
	\mor(x \to y)	& morphisms from $x$ to $y$ \\ \hline
    \End(x)	& endomorphism algebra of the object $x$; same as $\mor(x\to x)$ \\ \hline
    \cliff_1	& first complex Clifford algebra \\ \hline
    \cc^{r|s} & Super vector space with even dimension $r$ and odd dimension $s$ \\ \hline
    \text{m-type}	& simple object with $\End(x)\cong \cc$ \\ \hline
    \text{q-type}	& simple object with $\End(x)\cong \cliff_1$ \\ \hline
    n_x		& $\dim\End(x)$, 1 if $x$ is m-type and 2 if $x$ is q-type \\ \hline
	\cl(x) & closure of $x$ in either an annulus or torus depending on context \\ \hline
	\cl_W(x) & closure of $x$ with spin structure $W$ assigned to new cycle \\ \hline
	f\cdot g	& the composition of two morphisms $x \xrightarrow{f} y \xrightarrow{g} z$ (arrow order) \\ \hline
	g \circ f	& the composition of two morphisms $x \xrightarrow{f} y \xrightarrow{g} z$ (function order) \\ \hline
	A(Y)	& string nets modulo local relations on a 2-manifold $Y$; predual Hilbert space \\ \hline
	A(Y; c)	& string nets, with fixed boundary condition $c$ on $\bd Y$, modulo local relations \\ \hline
	Z(Y)	& functions on string nets invariant under local relations on a 2-manifold $Y$; dual space $A(Y)^*$; Hilbert space \\ \hline
	Z(Y; c)	& functions on string nets, with fixed boundary condition $c$ on $\bd Y$, invariant under local relations \\ \hline
	Z(M)	& path integral of a 3-manifold $M$ \\ \hline
	Z(M)(c)	& path integral of a 3-manifold $M$, evaluated on a boundary condition $c\in A(\bd M)$ \\ \hline
    \tube(\mcc)	& tube category of $\mcc$ \\ \hline
    \tube^B(\mcc)	& bounding tube category \\ \hline
    \tube^N(\mcc)	& non-bounding tube category \\ \hline
	\tube_{x \to y}	& morphisms from $x$ to $y$ in the tube category \\ \hline
	B, N	& bounding, nonbounding spin structures on the circle \\ \hline
	S^1_B	& spin circle with bounding (anti-periodic) spin structure \\ \hline
	S^1_N	& spin circle with nonbounding (periodic) spin structure \\ \hline
	\unit	& tensor unit in a tensor category; trivial object \\ \hline
	d_a		& quantum dimension of the object $a$; loop value of $a$ \\ \hline
	\theta_a		& twist eigenvalue of the simple object $a$ \\ \hline
	\mcd^2	& $\sum_{a\in \sob(\mcc)} d_a^2$ in bosonic case; $\sum_{a\in \sob(\spc)} d_a^2/n_a$ in fermionic case \\ \hline
\end{longtabu}


\section{Fermion condensation in the Ising TQFT}  \label{C2_condense_sect}

Before discussing super pivotal categories in the abstract and general techniques for constructing examples thereof,
we will give a detailed account of one of the simplest examples:
the $C_2$ super pivotal category.
This theory 
is obtained from the Ising TQFT by condensing the emergent fermion $\psi$, and provides a good demonstration of the 
qualitatively new features that occur as a result of fermion condensation. 
This section (and the next) is organized as follows: in \ref{Ising_review} we briefly review the aspects of the 
Ising TQFT we will need in later sections. 
In \ref{general_condensation} we comment on the general procedure of anyon condensation, and in \ref{condensing_psi} 
we show how to condense $\psi$ in the Ising theory, obtaining the $C_2$ super pivotal category.\footnote{
	As mentioned in the introduction, the $C_2$ category lacks a braiding, and so does not directly
	describe an anyonic topological phase.
	The $C_2$ category can be used as input for a fermionic string-net model, and the tube category (or Drinfeld center)
	of $C_2$ describes the resulting anyonic topological phase.}
Section \ref{C2_local_relns} details the diagrammatic properties of the $C_2$ theory, and in \ref{C2excitations} we compute 
the quasiparticle excitations of the theory. 
In \ref{C2_fusion_rules} we determine the fusion rules of these quasiparticles, and in \ref{modulartforms} we compute 
the modular $S$ and $T$ matrices of the theory.

\subsection{Ising TQFT}  \label{Ising_review}

Here we provide a brief review the Ising TQFT (see e.g. \cite{Lins1994}). 
There are three particles in the theory, which we label as $\unit$ (the trivial particle), $\sigma$ 
(the non-abelian Ising anyon), and $\psi$ (the emergent fermion).
The nontrivial fusion rules of the theory are as follows:
\be 
	\sigma\tp\sigma\cong\unit\oplus\psi,\quad \sigma\tp\psi\cong\sigma,\quad\psi\tp\psi\cong\unit.
\ee

The quantum dimensions of the particles are 
\be
\label{QuantumDimensions}
d_\unit = 1 \quad d:=d_\sigma = -A^2 - A^{-2} \quad d_\psi =1,
\ee
where $A$ is a primitive 16th root of unity. Graphically, this means that
\begin{align}
\dpsi_\psi = \text{(vaccuum)} \qquad \qquad \dbeta_\sigma = d \times \text{(vaccuum)}
\end{align}
where the blue (orange) circle denotes a circular $\psi$ ($\sigma$) worldline.
$\unit$ worldlines, being identified with the vaccuum, are not drawn in diagrams. 

Out of the eight possible choices for $A$, the four different choices $A = ie^{\pm i\pi/8}, -ie^{\pm i\pi/8}$ 
all give a positive quantum dimension for the $\sigma$ particle of $d = \sqrt{2}$.
The other four choices of $A$ give $d=-\sqrt{2}$, which can alternatively be defined
with $d=\sqrt{2}$ but with a negative Frobenius-Schur indicator of $\kappa_\sigma=-1$. 
In what follows, we will specify to $A = ie^{\pm i\pi/8}, -ie^{\pm i\pi/8}$ so that $d_\sigma = \sqrt{2},\kappa_\sigma=1$.

We now turn our attention to the graphical calculus of the Ising TQFT. 
We pick a normalization more common in the physics literature:
bubbles in diagrams can be eliminated by using the rule
\be
\overset{\;c'}{\underset{c}{{\scriptstyle{a}}\Bubbleabc {\scriptstyle{b}}}} = \delta_{c c'} \sqrt{\frac{d_a d_b}{d_c}}\times \overset{c}{\underset{c}{\idblack}}\,.\ee
In particular, we have 
\be
\overset{\sigma}{\underset{\sigma}{{\scriptstyle{\sigma}}\Bubblesps {\scriptstyle{\psi}}}} = \overset{\sigma}{\underset{\sigma}{\idorange}}\;, \quad \quad 
\overset{\psi}{\underset{\psi}{{\scriptstyle{\sigma}}\Bubblessp {\scriptstyle{\sigma}}}}
 = d \times \overset{\psi}{\underset{\psi}{\idblue}}\; ,
  \label{Ising_bubble_relns} 
\ee
where again, we are marking $\psi$ worldlines in dark blue and $\sigma$ worldlines in orange.

The non-trivial $F$-moves in the theory are as follows:
\begin{align}	
& \IsingDat{\VssI}{\sigma}{\sigma}{\sigma}{\sigma}  =\frac{1}{d} \left(   \IsingDat{\HssI}{\sigma}{\sigma}{\sigma}{\sigma} + \IsingDat{\ScriptOverSymbol{\Hssp}{\substack{\; \\ \; \\ \psi }}}{\sigma}{\sigma}{\sigma}{\sigma} \right) \qquad \qquad
\IsingDat{\ScriptOverSymbol{\Vsps}{\;\; \;\; \sigma }}{\sigma}{\psi}{\sigma}{\psi} = 
 \IsingDat{\HspI}{\sigma}{\psi}{\sigma}{\psi}  \\[4ex]
& \IsingDat{\ScriptOverSymbol{\Vssp}{\;\; \;\; \psi }}{\sigma}{\sigma}{\sigma}{\sigma}  = \frac{1}{d} \left(   \IsingDat{\HssI}{\sigma}{\sigma}{\sigma}{\sigma} - \IsingDat{\ScriptOverSymbol{\Hssp}{\substack{\; \\ \; \\ \psi }}}{\sigma}{\sigma}{\sigma}{\sigma} \right) \qquad \qquad 
 \IsingDat{\VppI}{\psi}{\psi}{\psi}{\psi} =  \IsingDat{\HppI}{\psi}{\psi}{\psi}{\psi} 	
 \end{align} 
The twist and braiding of the $\sigma$ particle are given by 
\begin{align}
\underset{\;\;\; \; \;\;\;  \sigma }{\overset{\;\;\; \; \;\;\;  \sigma }{\Twists} } = -A^3\; \underset{\sigma}{\overset{\sigma}{\idsigmashort}}  \qquad \qquad 
\IsingDat{\Braidss}{\sigma}{\sigma}{\sigma}{\sigma} = A \: \IsingDat{\HssI}{\sigma}{\sigma}{\sigma}{\sigma} + A^{-1} \IsingDat{\VssI}{\sigma}{\sigma}{\sigma}{\sigma} 
\label{sigma_braiding}
\end{align}
The formula for the topological twist on the left follows from the relation given on the right.
Using that $\sigma \tp \sigma = \unit \oplus \psi$ and \eqref{sigma_braiding} 
we can derive the braiding data for the $\psi$ particle. 
Most importantly for us, the
$\psi$ particle is fermionic in both its topological twist 
and statistics, regardless of the choice of $A$:
\begin{align}
  \underset{\;\;\; \; \;\;\;  \psi }{\overset{\;\;\; \; \;\;\;  \psi }{\Twistp} }= -\;  \underset{\psi}{\overset{\psi}{\idpsishort}} \qquad \qquad \IsingDat{\Braidpp}{\psi}{\psi}{\psi}{\psi}  = (-1) \IsingDat{\HppI}{\psi}{\psi}{\psi}{\psi} 
\label{psi_a_fermion}
\end{align}
Additionally, $\psi$ is not transparent, as it braids nontrivially with $\sigma$:
\begin{align}
 \overset{\sigma \quad \; \; \; \; \psi }{\underset{\sigma}{\Rpss}}
 =A^4 \;  \overset{\sigma \quad \; \; \; \; \psi }{\underset{\sigma}{\Vsigmapsisigma}}, \qquad \qquad \IsingDat{\Braidsp}{\psi}{\sigma}{\sigma}{\psi}  = - \IsingDat{\Braidpsconj}{\psi}{\sigma}{\sigma}{\psi}. 
  \label{sig_psi_Rsymbol}
\end{align}

Our goal in what follows is to describe how to condense the $\psi$ particle in the Ising TQFT. 
To put this in context, we will first make a few remarks on the more familiar bosonic condensation.

\subsection{Condensation of transparent bosons}\label{general_condensation}

We now briefly review how to perform condensation with transparent bosons (see e.g. \cite{eliens2014}).  
Let $\mcc$ be a ribbon category and let $\alpha$ be a particle (simple object) of $\mcc$ which we hope to condense.
In category theoretic terms, we want to add morphisms to $\mcc$ so that $\alpha$ becomes
isomorphic to the trivial particle $\unit$ (or, more generally, to a direct sum of several copies of $\unit$). 
We can think of this as a categorical quotient, denoted $\mcc/\langle \alpha \cong \unit \rangle$ or
more simply as $\mcc/\alpha$.
Physically, this amounts to turning the anyon $\alpha$ into a {\it local} particle (one which 
can be created locally).

In our graphical calculus, condensing $\alpha$ means that $\alpha$ worldlines are allowed to 
have endpoints at locations where they are ``absorbed'' into the condensate (allowing them 
to be created locally).
We will mark the locations where $\alpha$ particles are absorbed into the condensate with boxes
\begin{align}
\label{box_def}
\alpha\; \PsiFermion\; , 
\end{align}
where the horizontal blue line is an $\alpha$ worldline. 
We can also think of these boxes as morphisms from $\alpha$ to the trivial particle.
For simplicity, we will assume that $\alpha\otimes\alpha\cong \unit$ (which will be true in the examples considered in this paper), 
although most of the following discussion can be made to work more generally.

In order for this condensation procedure to not cause unintended collapse 
in $\mcc$ (e.g.\ confine other particles of $\mcc$, or result in a trivial theory),
$\alpha$ must satisfy three conditions:

\begin{itemize}

\item First, the twist of $\alpha$ must be 1.
This is because
\begin{align}
\label{twist_inconsistency}
\PsiFermion =\halfbentfermionline = \PsiFermionTwist = 
\kinkedfermionline=
\theta_\alpha \ \PsiFermion\;,
\end{align}
and so if $\theta_\alpha \neq 1$, diagrams in which $\alpha$ worldlines are absorbed into the 
condensate are identically zero, and condensation is impossible. 

\item Secondly, $\alpha$ must be statistically bosonic, i.e.\ it must braid trivially with itself.  
This is because
\be \label{statistics_inconsistency}
\mathord{\vcenter{\hbox{\includegraphics[scale=1.5]{TwoFermion_nolabels.pdf}}}} = \ \mathord{\vcenter{\hbox{\includegraphics[scale=1.5]{TwoFermionExchange_nolabels.pdf}}}} = \theta_{\alpha,\alpha} \mathord{\vcenter{\hbox{\includegraphics[scale=1.5]{TwoFermion_nolabels.pdf}}}},\ee
where $\theta_{\alpha,\alpha}$ is the self-statistics of $\alpha$. 
(By our assumption that $\alpha\otimes\alpha \cong \unit$, the last equality
must hold with $\theta_{\alpha,\alpha} = \pm 1$.)
By the spin-statistics relation this condition is not independent of the previous one, but it will be 
useful to regard them as separate constraints for the purpose of the fermion condensation 
procedure described in the next section.\footnote{If we drop positivity requirements, we can 
violate the spin-statistics theorem and have particles which are fermionic in spin but not statistics 
or vice-versa; see \cite{MorWalker} for a discussion of this.}

\item Finally, 
$\alpha$ must braid trivially with every particle in $\mcc$. In category-theoretic language, 
this means that $\alpha$ must lie in the transparent subcategory of $\mcc$.
If a particle $\beta$ braids non-trivially with $\alpha$ (i.e.\ if the left and right braidings are not equal),
then any string diagram which includes a $\beta$ particle must be zero: (we assume the quantum 
dimension of alpha is one for convenience)
\begin{align}
\label{transparency_inconsistency}
\underset{\;\; \;\;\beta}{\alphaboxbetabraidinga} \; = \; \underset{\quad \quad \quad\beta}{\alphaboxbetabraidingb} \; = \;  \underset{\;\; \;\;\beta}{\alphaboxbetabraidingc} \; = \; \theta_{\alpha \beta} \; \underset{\;\; \;\;\beta}{\alphaboxbetabraidinga} \;,
\end{align}
where the orange line is a $\beta$ worldline and $\theta_{\alpha,\beta}$ is the mutual statistics of $\alpha$ and $\beta$. 
The first two equalities follow from the fact that the location of a particle being absorbed into the 
condensate is not physically significant: no operator can distinguish between states that differ only 
by the location of an $\alpha$ endpoint. 
Therefore, if $\theta_{\alpha,\beta}\neq1$, condensing $\alpha$ causes unintended collapse in $\mcc$, since it confines $\beta$.
\end{itemize}

To summarize, in order to condense $\alpha$, $\alpha$ must have a twist of 1, have bosonic self-statistics, 
and braid trivially with every other particle in the theory. 
If any of these conditions are violated, we
will have to work harder to construct $\mcc/\alpha$.

\subsection{Condensing $\psi$ in Ising} \label{condensing_psi}

In this subsection we describe our procedure for condensing $\psi$ in the Ising theory.
Again, the motivation for doing this is to produce an example of a fermionic fusion category, which we can later
feed into a fermionic string-net model to construct an example of an anyonic topological phase.
While we will focus on the Ising theory in this section, our discussion will be fairly general, and can be 
applied to perform fermion condensation in more general scenarios. 

We would like to ``condense" the $\psi$ particle by constructing the quotient theory $\mcc / \psi$. 
We will denote the condensed theory by $C_2$, since the fusion rules are described by the $C_2$ Dynkin diagram; see below.
We will denote the image of $\sigma$ in the condensed theory $C_2$ as $\beta$.

First, from \eqref{psi_a_fermion}, we recall that $\psi$ is fermionic in both spin and statistics,
Additionally, we recall that $\psi$ has nontrivial braiding with $\sigma$, and as such $\psi$ is not transparent. 
Thus, $\psi$ violates all three of the conditions that particles in a condensate must satisfy! 
To condense $\psi$, we will clearly need some new tricks. 

\medskip

We will first examine how to address the non-transparency of $\psi$. 
As we saw above in \eqref{transparency_inconsistency}, 
we can't allow world lines of $\psi$ to disappear at arbitrary points in the 3-dimensional spacetime,
since this would confine $\beta$ (a.k.a.\ $\sigma$).
\begin{align}
\label{box_beta_nowall_braiding}
\underset{\;\; \;\;\beta}{\alphaboxbetabraidinga} \; = \; \underset{\quad \quad \quad\beta}{\alphaboxbetabraidingb} \; = \;  \underset{\;\; \;\;\beta}{\alphaboxbetabraidingc} \; = \; (-1)\; \underset{\;\; \;\;\beta}{\alphaboxbetabraidinga} \;,
\end{align}
However, if we restrict the $\psi$ worldline endpoints to lie on a 2-dimensional 
subset of the boundary of the ambient 3-dimensional spacetime,
we {\it can} obtain a consistent graphical calculus.

In this paper, we will adopt the convention that $\psi$ world lines are allowed to terminate 
on a codimension-1 ``back wall", located on a boundary of the system that is positioned ``behind'' all 
other world lines drawn in our graphical calculus. 

\begin{figure}
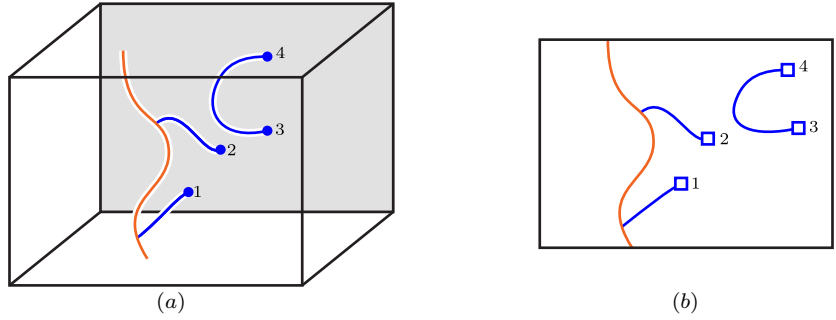

  \centering
\begin{align}
\nonumber
\underset{(a)\quad \quad }{\BackwallFigureprimea} \quad \quad \quad \underset{(b)}{\BackwallFigureprimeb}
\end{align}
      \caption{\label{backwall} (a) The ``back wall'' picture. 
      The box represents a section of a (2+1)D Ising TQFT, with the codimension-1 back wall 
      indicated by the gray back side of the box. 
$\psi$ worldlines are absorbed into (or emitted from) the back wall at marked points labelled by $1,2,3,4$. 
Free $\psi$ endpoints which do not terminate on the back wall are not allowed. 
(b) Our way of representing the picture (a) in a (1+1)D graphical calculus. 
We have squashed the box down to a two dimensional plane, with the blue boxes representing the 
points at which the $\psi$ lines ``go straight back and hit the back wall''.}
\end{figure}

Figure \ref{backwall}(a) demonstrates this graphically, 
with the gray back section of the box denoting the back wall, 
on which the $\psi$ worldlines can be absorbed or emitted. 
String-net graphs in our (2+1)D spacetime can be reduced to string-net graphs 
in (1+1)D spacetime by introducing a shorthand notation for $\psi$-lines terminating on the back wall. 
This notation is shown in Figure \ref{backwall}(b), 
where we use boxes to denote places where $\psi$ worldlines head straight back into the back wall and terminate.

Even though we use the same box-at-the-end-of-the-string graphical convention to denote ordinary condensation
and back wall condensation, there is an important difference.
In both cases, we can slide a box behind another strand, with the before and after pictures differing by an isotopy:
\begin{align}
\underset{\quad \quad \quad\beta}{\alphaboxbetabraidingb} \; = \;  \underset{\;\; \;\;\beta}{\alphaboxbetabraidinga}\;.
\end{align}
However, in the back wall case, we cannot slide a box in front of another strand
\begin{align}
\underset{\quad \quad \quad\beta}{\alphaboxbetabraidingb} \; \neq \;  \underset{\;\; \;\;\beta}{\alphaboxbetabraidingc}\; .
\end{align}
This is because doing so would involve the other strand crossing the $\psi$ strand as it
heads into the page en route to the back wall; the before and after pictures are not isotopic.
Thus restricting $\psi$ emission/absorption to the back wall
disallows the series of diagram equalities in Figure \ref{box_beta_nowall_braiding}.

An important consequence of the existence of the back wall this is that the quotient category $C_2$ 
is not braided (although there is a way to perform the condensation so that the resulting theory is 
braided, which we mention briefly in \ref{spin_defects_condensation}).
String diagrams for braided categories (such as Ising) can be glued together in three independent 
dimensions: right/left, top/bottom, and back/front.
In more formal terms, braided categories are (special cases of) 3-categories.
Because of the back wall, it does not make sense to glue string diagrams for the $C_2$ category in the back/front
dimension.
In more formal terms, $C_2$ is a mere 2-category, or, more specifically, a (super) tensor category.
We can, however, glue an Ising string diagram to the front (but not back) of a $C_2$ diagram.
In category theoretic language, $C_2$ is a module 2-category for the Ising 3-category; equivalently, $C_2$ is a codimension 1 defect
connecting Ising to the vacuum.

\medskip

The back wall construction fixes the problems caused by $\psi$ not being transparent.
However, we have not yet addressed the spin and statistics 
inconsistencies \eqref{twist_inconsistency} and \eqref{statistics_inconsistency}.
To fix these inconsistencies, we will couple the boxes marking the $\psi$ endpoints to a 
complex line bundle associated to a spin structure. 
(When we later consider reflections, this spin structure will be promoted to a $\mbox{pin}_+$ structure.)
Readers unfamiliar with spin and pin structures are referred to Appendix \ref{spin_and_pin}. 
The spin structure will enable us to cancel out the factors of $-1$ 
arising from the endpoints of $\psi$ worldlines twisting through $2\pi$ as in \eqref{twist_inconsistency}.
This ``bosonizes'' the $\psi$ endpoints and allowing the condensation process to go through.

\medskip

To make this precise, let $Y$ be an oriented 2-manifold, let $M = Y\times [0,1]$, and let $B = Y\times \{1\}$.
(Everything we do in this subsection will work more generally for $M$ any oriented 
3-manifold and $B$ some codimension 0 submanifold
of $\bd M$.)
In addition, choose a spin structure on $B$.
Our goal is to associate a Hilbert space to $Y$ based on string nets in $M = Y\times I$
and some extra data on the back wall $B$.

Consider the configuration space $\mcr(B)$ of all $\psi$ ribbon endpoints on $B$.
Let $\mcr(B)_k$ denote the subspace of $\mcr(B)$ corresponding to configurations with exactly $k$ endpoints.
(If $B$ is connected, then these are the connected components of $\mcr(B)$.)
We can think of the ribbon endpoint as a point $p$ of $B$ plus a tangent vector at $p$ which
points in the direction of the ``front" of the ribbon.\footnote{Ribbons 
have a distinguished front and back; if they did not, then we would have to assign
phases $\theta_a^{1/2}$ corresponding to half twists of ribbons.}
This means that $\mcr(B)_1$ is diffeomorphic to the unit tangent bundle of $B$.
We also stipulate that $\mcr(B)_0$ consists of a single point.

In the next subsection we will construct a complex line bundle $F(B)$, with flat connection, over $\mcr(B)$, 
which will provide us with the additional structure we need to perform the condensation.
The extra data alluded to above will be a vector in this vector bundle.
More specifically, for each string net $S$ in $M$, we have an endpoint configuration $e(S) \in \mcr(B)$.
Our Hilbert space will be generated by pairs $(S, v)$, where $v \in F(B)_{e(S)}$, the fiber of the bundle $F(B)$
at the configuration $e(S)$.

We impose the usual local string net relations in the interior of $M$.
(Ribbon end points are fixed for these relations.)
We also impose the following additional relation:
Let $\{S_t\}$, with $0 \le t \le 1$, be a 1-parameter family of string nets.
The $\psi$ endpoints on $B$ are allowed to move, but any other ribbon endpoints must remain fixed.
Let $v \in F(B)_{e(S_0)}$.
Using parallel translation for the connection on $F(B)$ and the path of configurations $\{e(S_t)\}$, 
we obtain $v' \in F(B)_{e(S_1)}$.
We now identify these configurations by imposing the relation
\be \label{bw_rel}
	(S_0, v) = (S_1, v') .
\ee

\medskip

Let us now see how coupling the system to $F(B)$ fixes the inconsistencies 
\eqref{twist_inconsistency} and \eqref{statistics_inconsistency}.
The crucial property of the flat connection on $F(B)$ is that the holonomies around loops
corresponding to \eqref{twist_inconsistency} and \eqref{statistics_inconsistency} are $-1$.
In other words, in \eqref{bw_rel}, we have $v = -v'$ if $\{S_t\}$ is the family of string nets
depicted in \eqref{twist_inconsistency} or \eqref{statistics_inconsistency}.

So, \eqref{twist_inconsistency} now becomes
\begin{align} \label{boxspin}
	\PsiFermion = (-1)\ \PsiFermionTwist = (-1)^2\ \PsiFermion\;. 
\end{align}
and \eqref{statistics_inconsistency} becomes
\begin{align} \label{boxbraid}
\TwoFermionNoLabels =(-1)\ \TwoFermionExchangeNoLabels = (-1)^2\ \TwoFermionNoLabels\;.
\end{align}

\medskip

Is the choice of $F(B)$ together with its flat connection uniquely determined by the above holonomy requirements?
No: in fact, there is a natural bijection between the set of spin structures on $B$ and flat connections fulfilling
the above holonomy requirements.
In other words, in order to perform this sort of fermionic condensation, we 
we need to choose a spin structure on $B$ (or on $Y$, in the main case where $M = Y\times I$ and $B \cong Y$).

\medskip

The details of the construction of the bundle-with-connection $F(B)$ are somewhat technical, 
and can be found in in Appendix \ref{flb_appendix}.
Some of the key properties of $F(B)$ are:
\begin{itemize}
\item In order to specify an element $v\in F(B)_{e(S)}$, it suffices to (a) assign an ordering to the $\psi$-endpoints,
and (b) choose a spin-framing at each such endpoint.
Recall that a spin structure can be thought of as a double-covering of the unit tangent bundle of $B$.
By ``spin framing" we mean a choice of lift to this double cover from the unit tangent vector determined by the ribbon orientation.
\item The most general way of specifying a spin framing is via a ``Dirac belt" connecting a base framing 
on $B$ to the tangent vector in question.
For manifolds equipped with global framings there are no ambiguities, and we can choose
a ``gauge'' in which the orientation of the belt is consistent
with the global framing, allowing us to drop the belts from the figures.
The standard Euclidean structure on a page of this paper determines a global framing (the ``blackboard framing"),
and unless stated otherwise we implicitly equip each 
$\psi$ endpoint with a spin framing determined by this choice of global framing.
In some figures we have in mind a spin structure with a different global framing,
and in these cases we will draw dashed ``branch cut" lines to indicate how the framing
differs from the standard blackboard framing. 
The branch cuts will be chosen such that 
the spin framing rotates by $2\pi$ (switches sheets on the double cover) across a branch cut. 
\item Another important property of $F(B)$ is locality: The bundles behave well with respect to gluing surfaces.
By ``behave well'', we roughly mean that for $B = B' \cup B''$, 
there is an isomorphism between $F(B)$ and $F(B')\tp F(B'')$.
\item $F(B)$ comes equipped with a way to cancel pairs of $\psi$ endpoints, which is required since the fermion 
$\psi$ we will be condensing satisfies the fusion rule $\psi \tp \psi \cong \unit$.
This requires us to supplement the parallel transport of the flat connection with additional isomorphisms of
fibers of $F(B)$ connecting points of $\mcr(B)_k$ to $\mcr(B)_{k-2}$.
\item Finally, in order to define Hermitian inner products on our Hilbert spaces, $F(B)$ comes equipped with 
an antilinear bundle map $F(B) \to F(B')$ for all orientation-reversing maps $B \to B'$.
\end{itemize}

\medskip

Our spin-structure-equipped back wall construction admits a simple physical interpretation: coupling the 
theory to the bundle $F(B)$ is equivalent to adding a phase of {\it physical} (not emergent) fermions to the theory, 
and binding single physical fermions to each $\psi$ endpoint.
The physical fermion attached to each $\psi$ endpoint compensates for the fermionic nature of the $\psi$ particle 
in the condensate: the factors of $-1$ that relate two different orderings of the $\psi$ endpoints are the Koszul 
signs associated with the physical fermions, and the factors of $-1$ that we pick up when rotating the $\psi$ 
endpoint framing by $2\pi$ come from the spin $1/2$ of the physical fermions. 
Thus, attaching physical fermions to the $\psi$ endpoints transforms them into bosonic objects, which are then 
allowed to undergo normal boson condensation. 
Therefore, although we are indeed identifying emergent fermions with the vacuum, the term ``fermion condensation'' 
is a bit misleading, as what we are actually doing is closer to {\it boson} condensation of bound states of $\psi$ 
and a physical fermion. 
However, we reiterate that the $\psi$ endpoints are not technically bosons, since they see the background spin structure: 
the emergent $\psi$ fermions do not see the spin structure, but the physical fermions attached to their endpoints do. 

\medskip

To summarize, we have seen that by introducing the back wall and equipping it with a spin structure, all three 
conditions necessary for $\psi$ endpoints to condense are satisfied.
We emphasize that we have utilized the back wall and the spin structure for two {\it independent} reasons 
(the non-transparency of $\psi$ and $\psi$'s fermionic spin and statistics, respectively). 
For example, if $\psi$ were non-transparent but bosonic, we would still need a back wall, but the back wall 
would not need a spin structure.
Conversely, if $\psi$ were transparent but fermionic, then we would not need a back wall, but we would still 
need to introduce spin
structures.

\subsection{Local relations in the $C_2$ theory} \label{C2_local_relns}

Now that we have worked out how to condense $\psi$, we can determine the graphical rules that govern the condensed theory. 

We have already seen that in order for the condensation procedure to go through, 
the manifold on which we define the $C_2$ theory must be equipped with a spin structure. 
The interaction of fermionic morphisms with this spin structure leads to
relations in the diagrammatic calculus that are not present in bosonic theories.  

First, we first observe that any $\psi$ strands can be absorbed into the 
condensate at the expense of phase factors, 
and so in the condensed theory the only object remaining will be the image of $\sigma$ under condensation, 
which will denote by $\beta$. 
$\beta$ lines will be drawn in orange, and $\psi$ lines will be drawn in blue (and can also be distinguished 
from $\beta$ lines through their termination points). 
In the condensed theory, the $\beta$ line may or may not have $\psi$ fermions attached to them. 
We will introduce a blue dot as a compact notation for a $\psi$ worldline that terminates on a $\beta$ line:
\begin{align} 
\idsigmashortdot \quad = \quad \idsigmashortbox
\end{align}
Note that in order for the dot notation to be unambiguous, we must specify
a spin-framing at the dot.
We will usually do this implicitly as follows.
The dots are only allowed to occur on vertical strands,
and the spin framing at the dot is obtained from the base framing of the manifold
by rigidly translating with respect to the ``blackboard framing" of the page.
For diagrams which do not inherit their spin structure from the blackboard,
we will have to use other means to specify the spin-framing.

Because $\beta$ lines in the condensed theory can have fermionic dots on them, 
there are several new diagrammatic rules involving them that need to be included in our graphical calculus. 
One of the most important local relations is 
the addition and removal of an even number of fermions on $\beta$ lines. 
The process of removing two fermions can be done at the cost of a phase factor: 
\begin{align}
\mathord{\vcenter{\hbox{\includegraphics[scale=1]{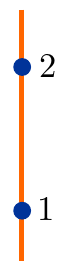}}}} \quad =  \quad \mathord{\vcenter{\hbox{\includegraphics[scale=1]{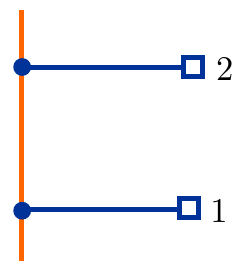}}}} \quad = \quad 
\mathord{\vcenter{\hbox{\includegraphics[scale=1]{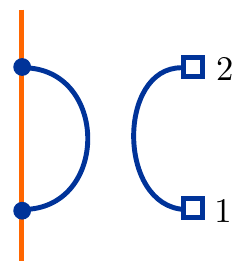}}}} \quad =\quad  
\lambda\; \mathord{\vcenter{\hbox{\includegraphics[scale=1]{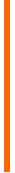}}}}\;,
\end{align}
where the labels $1$ and $2$ denote the 
ordering of the fermions in question.
We have performed an $F$-move on the $\psi$ worldlines to get the second equality, 
used \eqref{Ising_bubble_relns} 
to remove the $\psi$ line attached to the $\sigma$ line in the third diagram, 
and have used  
\be \label{eval_psi_semicirc}
   \PsiEnd  = \lambda \times \text{(vaccuum)},
\ee
in the last step to remove the semicircular $\psi$ line, where $\lambda\in \cc$ is some complex number. 
We show in Appendix \ref{flb_appendix} that we must have $\lambda=\pm i$, and since $A^4=\pm i$ in the Ising theory, 
we may choose either $\lambda = A^4$ or $\lambda = -A^{4}$ (the choice of $A$ does not constrain the choice of $\lambda$).  
The choice of $\lambda = \pm A^4$ affects the $F$-symbols in the condensed theory, 
but does not affect observable quantities like the twists or mutual statistics of the 
quasiparticles in the theory.\footnote{
This is because the choice of $\lambda=\pm A^4$ doesn't 
change the ``multiplication table'' of the tube category, which we will discuss below. 
Essentially, changing $\lambda$ just performs a scaling on the $\psi$ endpoints in the figures.}
Therefore without loss of generality we may choose a gauge in which $\lambda=A^4$, and so 
\begin{align}  \label{removing_fermions}
\mathord{\vcenter{\hbox{\includegraphics[scale=1]{beta_dotdot.pdf}}}} \quad = A^4\; \mathord{\vcenter{\hbox{\includegraphics[scale=1]{straight_beta.pdf}}}}\;,
\end{align}

Since $\psi$ is statistically fermionic, exchanging two fermions on a $\beta$ line results in a minus sign:
\begin{align}
\mathord{\vcenter{\hbox{\includegraphics[scale=1]{beta_dotdot.pdf}}}} \quad = \quad (-1)\times
\mathord{\vcenter{\hbox{\includegraphics[scale=1]{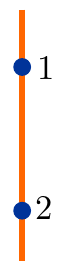}}}}\;.
\end{align}

The braiding in the parent Ising theory means that we pick up non-trivial phase 
factors when sliding fermion dots over and around $\beta$ caps and cups. 
For example, we can compute
\begin{align} \label{capslide}
\capleftdot =
\mathord{\vcenter{\hbox{\includegraphics[scale=1]{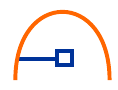}}}} =
\mathord{\vcenter{\hbox{\includegraphics[scale=1]{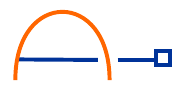}}}} =
\mathord{\vcenter{\hbox{\includegraphics[scale=1]{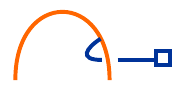}}}}
= A^4 \caprightdot,
\end{align}
where in the last step we have used \eqref{sig_psi_Rsymbol}. 
Since in the Ising theory $A^4=\pm i$, we can invert \eqref{capslide} to find 
\begin{align} \label{cap_slide_back}
\caprightdot\; &= - A^4 \; \capleftdot.
\end{align}
By a similar argument we find
\be  \cuprightdot\; =  A^4 \; \cupleftdot\;. \ee
Because of the nontrivial rules for sliding dots through cups and caps, dots which live 
on the apex of a cup or the bottom of a cap (where the $\beta$ line is horizontal) are ambiguous, 
and we will not allow them to be drawn in our fusion diagrams.
We stress that these phases are derived completely from the {\it braiding data} of the Ising theory.

Note that the above relations imply that even-parity $\beta$ loops are non-zero, while odd-parity $\beta$ loops vanish:
\be \mathord{\vcenter{\hbox{\includegraphics[scale=1]{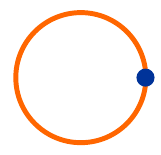}}}}=0 .\ee
This is as expected -- dragging the fermionic dot around the circle rotates it through $2\pi$.

The final local relations that will be important in what follows are the $F$-moves, 
which provide linear relations between different isotopy classes of diagrams.
The $F$-symbols in the condensed $C_2$ theory can be worked out using our rules 
for manipulating condensed $\psi$ worldlines and our knowledge of the $F$-symbols 
in the parent Ising theory.
We begin with a diagram in the Ising theory, apply an $F$-move, 
and then remove all $\psi$ lines through condensation to evaluate the $F$-move in the $C_2$ theory. 
For example, in the parent Ising theory we have
\begin{align}
\doublebeta = \frac{1}{d} \left(\doublecups + \doublecuppsi\right).
\end{align}
When we condense $\psi$, the second diagram on the left hand side becomes 
\be \label{fmove_isingtoC2}
\doublecuppsi = A^{-4}\; \mathord{\vcenter{\hbox{\includegraphics[scale=1]{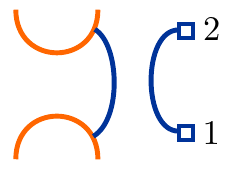}}}}\; \mapsto A^{-4}\;
\doublecupdots\;. \ee
Note that in the first step of \eqref{fmove_isingtoC2} we have displaced the 
vertical $\psi$ line to the right, so that it never intersects the $\beta$ line at 
the apex of a cap or the bottom of a cup.
We do this to avoid ambiguities in the fermion framing, 
which as discussed earlier always points ``to the left'', 
meaning that dots living on horizontal $\beta$ lines are not well-defined. 
While we choose to displace the $\psi$ line to the right in \eqref{fmove_isingtoC2}, 
this is merely a gauge choice: we could have equally well chosen it to be displaced to the left. 
Recapitulating, we see that in the $C_2$ theory we have the $F$-move
\be \doublebeta\; = \frac{1}{d} \left( \doublecups \; + \;A^{-4} \doublecupdots\right).
\label{Fmove}\ee
By similar reasoning we can derive the other nontrivial $F$-move in the $C_2$ theory, which is
\begin{align} \label{cupcap_fmove}
\doublecups\; &= \frac{1}{d} \left(\doublebeta \; + \; \doublebetadots \right).
\end{align}

The fact that $\beta$ lines can host dots means that $\beta$ has an endomorphism which 
is not a multiple of the identity, 
which is a hallmark of physics that cannot be found in bosonic topological phases. 
Indeed, any section of a given $\beta$ worldline may look like
\begin{align} \label{betaendos}
\mathord{\vcenter{\hbox{\includegraphics[scale=1]{straight_beta.pdf}}}} \qquad {\rm or} \qquad \mathord{\vcenter{\hbox{\includegraphics[scale=1]{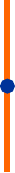}}}}.
\end{align}
The two diagrams in \eqref{betaendos} are the generators of the endorphism algebra of $\beta$.
Since we have one even generator and one odd generator we see that $\text{End}(\beta) \cong \mathbb{C}\ell_1$, 
where $\cliff_1$ is the first complex Clifford algebra (generated by $1$ and a single odd-parity generator). 
More generally, the vector space assigned to a disk with $2n$ $\beta$ strings ending on its boundary is 
has dimension $2^n$, and the endomorphism algebra of $\beta^{\tp n}$ (i.e.\ $n$ copies of $\beta$ on a line) is
isomorphic to $\cliff_n$.

In more general contexts, we will refer to simple objects whose endomorphism algebras are isomorphic 
to $\cliff_1$ as ``q-type objects'', and those whose endomorphism algebras are isomorphic to $\cc$ 
as ``m-type objects''. 
From the above information, we can infer the fusion rule
\be \beta\tp\beta \cong \cc^{1|1} \cdot \unit, \ee
where $\cc^{1|1}$ is the complex vector space with a single even generator and a single odd generator, 
corresponding to the even and odd channels of the fusion product $\beta\tp\beta$.

In bosonic theories, having $\End(\beta)\cong\cliff_1$ would imply that $\beta$ is not a simple object, 
since by Shur's lemma the endomorphism algebra of any simple object must be a division algebra, and
since $\cc$ is the only ungraded division algebra.
Nevertheless, $\beta$ {\it is} a simple object.\footnote{Indeed, if we tried to decompose $\beta$ 
into two idempotents, they would each be linear combinations of the two pictures in \eqref{betaendos}, 
which have different fermion parity.
But idempotents must have even parity.}
This is possible because in fermionic theories, the Hilbert spaces we use are supervector spaces, 
compared to the regular vector spaces of bosonic theories. 
Unlike in the bosonic case, there are {\it two} super division algebras, $\cc$ and $\cliff_1$.
This means that simple objects in the fermionic setting can have endomorphism algebras of either $\cc$ or $\cliff_1$.  
Later on, we will see that the existence of simple objects with $\cliff_1$ endomorphism 
algebras is responsible for a large part of the novel physics that occurs as a result of fermion condensation. 

Finally, we mention a higher-level way of understanding the content of the $C_2$ theory from the
parent Ising theory. We begin by noting that the principle graph for the Ising theory 
is given by the $A_3$ Dynkin diagram (see the top left of Table \ref{C2_data_table}). 
Condensing $\psi$ means establishing an isomorphism between $\psi$ and $\unit$, 
so that in the condensed theory $\unit$ and $\psi$ correspond to the same node of the 
principal graph. This identification can be done by ``folding'' the $A_3$ principal graph 
about the central node so that the $\unit$ and $\psi$ nodes are identified. 
Note that $\sigma$ is preserved under the folding, which translates into the fact that the 
image of $\sigma$ in the condensed theory (namely $\beta$) has a nontrivial endomorphism 
algebra. 
The resulting folded principal graph is shown in the top right of Table \ref{C2_data_table}: the double 
line indicates the two fusion channels in $\beta\tp\beta \cong\cc^{1|1}\cdot \unit$, the 
double circle indicates that $\beta$ has a two-dimensional endomorphism algebra, and 
the arrows have been chosen to point away from the object with larger endomorphism 
algebra. The resulting principal graph is precisely the $C_2$ Dynkin diagram, 
which is why we call the condensed theory the $C_2$ theory. 
This idea can be applied to perform fermion condensation in many other theories (most straightforwardly 
the other theories in the $A_n$ series which contain a fermion). 
We will explore several examples along these lines in later sections.

\begin{table}
	\fbox{\begin{minipage}{148mm}
	\vspace{3mm}\begin{center}{\begin{minipage}{128mm}
		\setlength{\parskip}{0ex}
		\begin{align*}
		\xymatrix @!0 @M=4mm @R=6mm @C=60mm {
		 \AThreeDynkin \ar@<7pt>[r]^{\text{condense $\psi$}}&   \CTwoDynkin
		 }
		\end{align*}
		
		\textbf{Basic data:}\\[-4ex]
\begin{flalign*} & \begin{array}{c @{\quad \quad  } c @{\quad \quad } c @{\quad \quad } c}
			\text{simple objects}	&	\text{fusion rules} &   \text{dimensions} & \text{Koszul signs}
		\\[.5ex]
			\text{$\unit$ and $\beta$}
			&	 \beta \tp \beta \cong \mathbb{C}^{1|1}\cdot \mathds{1}  & \dbeta_\beta  = d &  \SigmaDotDot = - \SigmaDotDotExchange
		\end{array} & \end{flalign*}
\vspace{3mm}
		\textbf{Linear relations:}\\[-4ex]
		\begin{flalign*} & \begin{array}{r @{\quad \quad} l @{\quad \quad} l @{\quad \quad} l}
			\text{F-symbols}
			&	d\;  \TwoLine =  \CupCap  +A^{12}\CupCapDots  & d\; \CupCap =\TwoLine + \TwoLineDots
				&	\eqref{Fmove}
		\\[4ex]
			\begin{tabular}{c}Fermion\\ pairing\end{tabular}
			&	\;\;  \SigmaDotDot = A^4\; \FubeXss \quad  \eqref{removing_fermions} & \; \begin{tabular}{l}$\CapDotLeft =  A^4 \CapDotRight$ \\ \\$ \CupDotRight = A^{4} \CupDotLeft $ \end{tabular}
				&	\eqref{cap_slide_back}
			\end{array} & \end{flalign*}
			 $A^2 = - e^{\pm i \pi /4}$ \quad \quad $d  = -A^2 - A^{-2}$ \quad  \eqref{QuantumDimensions}
	\end{minipage}}\end{center}\vspace{1mm}
	\end{minipage}}
	\caption{Summary of $C_2$ data 
	}  \label{C2_data_table}
\end{table}


\section{Quasiparticle excitations and the tube category of $C_2$} \label{C2_quasiparticles}

In this section, we identify the quasiparticle excitations 
in the $C_2$ theory.
We will discuss the quasiparticles in the theory, their fusion rules, their statistics, and the 
modular transformations of the ground states on the torus.
We will identify the excitations using a fermionic generalization of a device known as the {\it tube category}. 
We will briefly review the tube category as applied to the $C_2$ theory below; 
for a more detailed overview and for an explanation of why the fermionic version of the tube category computes the excitations, 
we refer the reader to Section \ref{more_on_tubes}, where we discuss the construction in full generality. 

For the benefit of more physically-inclined readers 
we will use the ``Hamiltonian/ground-state/excitation'' terminology 
in this section, 
even though (as discussed in the introduction) we will not define the relevant Hamiltonian until Section \ref{Super_pivotal_Hamiltonian}.
The constructions in this section all take place within the self-contained world of TQFTs defined via string nets;
the Hamiltonian/ground-state/excitation interpretation is optional.

\subsection{Finding the quasiparticle excitations}   \label{C2excitations}

We now turn to a detailed study of the tube category for $C_2$. 
The objects of $\tube(C_2)$ are given by spin circles with a finite number of marked points 
labeled by simple objects of $C_2$.
There are two spin structures on the circle: bounding (anti-periodic boundary conditions; non-vortex) and 
non-bounding (periodic boundary conditions; vortex).
Each object in $\tube(C_2)$ thus determines a choice of spin structure of the underlying circle. 

It suffices to consider only objects with at most one labeled point.
This is because any other object is isomorphic to a direct sum of objects with at most one labeled point.
Note also that an object with a point labeled by $\mathds{1}$ (the trivial object of $C_2$)
is isomorphic to the object obtained by erasing that marked point.
In particular, a circle with no marked points is isomorphic to a circle with a single point labeled by $\mathds{1}$.

Since there are two simple objects in $C_2$, 
for a fixed spin structure there are only two possible labels that can be assigned to a marked point,
\begin{align}
\underset{\mathds{1}}{\TubeBCx{B}} \quad \quad \quad \underset{\beta}{\TubeBCx{B}} \quad \quad \quad \quad \quad \quad \underset{\mathds{1}}{\TubeBCx{N}} \quad \quad \quad \underset{\beta}{\TubeBCx{N}}
\end{align}
where $B$, $N$ denote bounding and non-bounding spin structures respectively. 

Morphisms of $\tube(C_2)$ are 
defined to be cylinders (which we will draw as annuli) decorated by $C_2$ string-nets.
A given morphism $a\ra b$ is thus an annulus whose outer (inner) boundary conditions are determined by the object $a$ ($b$). 

After applying local relations (F-moves, dot-cancellations, removing trivial loops), 
an arbitrary string-net diagram on any tube can be reduced to a linear 
combination of the following diagrams:
\be
\begin{aligned}
\label{mortube}
\text{mor}(\underset{\mathds{1}}{\TubeBCx{J}} \rightarrow \underset{\mathds{1}}{\TubeBCx{J}}) &= \mathbb{C}\left[ \Fubex{\FubeXXXA}{J} ,  \Fubex{\FubesXsA}{J}, \Fubex{\FubesdXsA}{J} \right] \\
\text{mor}(\underset{\beta}{\TubeBCx{J}} \rightarrow \underset{\beta}{\TubeBCx{J}}) &= \mathbb{C}\left[   \Fubex{\FubeXssA}{J} ,  \Fubex{\FubessXA}{J}, \Fubex{\FubeXsdsA}{J} , \Fubex{\FubessdXA}{J} \right] \\
\\
\text{mor}(\underset{\mathds{1}}{\TubeBCx{B}} \rightarrow \underset{\beta}{\TubeBCx{N}}) &= 0 \quad \quad \quad \quad \text{mor}(\underset{\beta}{\TubeBCx{N}} \rightarrow \underset{\mathds{1}}{\TubeBCx{B}}) = 0
\end{aligned}
\ee
with $J =B$ for bounding spin structure, and $J =N$ for non-bounding spin structure. 
In the first row we have listed all possible tubes which take the trivial boundary condition back to itself, 
and in the second row the tubes which take the $\beta$ marked point back to itself.
Notice that all non-zero tubes for $C_2$ have the same label at the two marked points.
For general input categories this won't happen; see Sections \ref{so36} and \ref{halfesix} for examples.

Depending on the spin structure on the annulus, some of the above diagrams may be zero.
With a bounding spin structure, a fermionic dot picks up a factor of $-1$ if it moves around the annulus.
On the other hand a fermionic dot picks up a factor of $+1$ if it moves around the annulus with non-bounding spin structure.

As we said above, not every string diagram in the annulus is consistent with a given spin structure.
For example,
\begin{align}
\label{BoundingNullVector}
\Fubex{\FubesdXsA}{B}\;  = - \; \Fubex{\FubesdXsA}{B} \implies \Fubex{\FubesdXsA}{B} = 0,
\end{align}
where we have simply pulled the fermion around the non-contractible loop on the tube. 
A non-bounding tube with a horizontal $\beta$ line is also zero, since
\begin{align}
\label{non-boundingNullVector}
\Fubex{\FubesXsA}{N}\; =\; A^{-4} \Fubex{\FubesXsaA}{N} \; = A^{-4} \Fubex{\FubesXscA}{N} \; = \; -\Fubex{\FubesXsA}{N}
\end{align}
Where in the second step we have dragged one of the two fermions around the annulus.
All other tubes are nonzero for both spin structures, and so a complete basis for tubes in the bounding sector is given by
\begin{align}
 \label{atubes}
\tube^B_{\unit \rightarrow \unit} &= \mathbb{C} \left[\Fubex{\FubeXXXA}{B}  ,\Fubex{\FubesXsA}{B}\right]\\
\tube^B_{\beta \rightarrow \beta}  &=  \mathbb{C} \left[\Fubex{\FubeXssA}{B}  ,\Fubex{\FubessXA}{B},\Fubex{\FubeXsdsA}{B},\Fubex{\FubessdXA}{B}\right]
\end{align}
while a basis for the non-bounding sector is given by
\begin{align}
\label{ptubes}
\tube_{\unit \rightarrow \unit}^N &= \mathbb{C}\left[ \Fubex{\FubeXXXA}{N}  ,\Fubex{\FubesdXsA}{N} \right]\\
\tube_{\beta \rightarrow \beta}^N &= \mathbb{C}\left[ \Fubex{\FubeXssA}{N}  ,\Fubex{\FubessXA}{N} ,\Fubex{\FubeXsdsA}{N} ,\Fubex{\FubessdXA}{N}  \right]
\end{align}

The multiplication operation in the tube category is given by stacking tubes on top of one another 
and simplifying the resulting tube using local relations.  
For example, in the non-bounding sector we have
\begin{align}
\label{OddOddC2Stack}
\Fubex{\FubesdXsA}{N} \;  \cdot  \; \Fubex{\FubesdXsA}{N}  \;=\; \Fubex{\FubesddXsA}{N} \;=\; \frac{1}{d} \left(
\Fubex{\RDotTwobA}{N}
+ A^{-4} \Fubex{\RDotTwocA}{N}  \right) \; = \; 2 \Fubex{\FubeXXXA}{N}
\end{align}

Note that since the spin structures on two tubes being fused must agree on the boundary 
at which they are fused, non-bounding tubes can only be stacked on top of non-bounding tubes, 
and similarly for bounding tubes.

Relations like the ones above allow us to find the (isomorphism classes of) minimal idempotents of the tube category. 
First, we turn to an analysis of tubes with bounding spin structure.

\subsubsection{Non-vortex spin structure}

Let us first examine the tubes with no charge,\footnote{
We define the ``charge'' of a tube to be the label of the 
marked point of the upper boundary of the tube, so that tubes in $\tube_{a \rightarrow b}$ have charge $b$. 
This is slightly misleading however, because charge is not a good quantum number: in more general theories, 
$\tube_{a \ra b}$ will be nonzero even when $a\neq b$. 
}
that is, cylinders 
with bounding spin structures and empty boundary conditions on both their top and bottom, which are the cylinders in $\tube^B_{\unit \rightarrow \unit}$. 
We see that this algeba \eqref{atubes} has two even generators, and so as a vector space
\begin{align}
\label{C2etoe}
\tube^B_{\unit \rightarrow \unit} \cong \mathbb{C}^{2|0}
\end{align}
There is only one possible super algebra structure on $\mathbb{C}^{2|0}$; it is the sum of two trivial 1-dimensional algebras
$\End(\cc)\oplus \End(\cc)$ (or $\cc \oplus \cc$ for short, where here $\cc$ denotes a 1-dimensional 
algebra rather than a 1-dimensional vector space).
This sector therefore contains two minimal idempotents, which we will call $m_\unit$ and $m_\psi$.
Explicitly, they are 
\be
m_\unit = \frac{1}{2}\left( \Fubex{\FubeXXXA}{B} \; +\; \frac{1}{d}\Fubex{\FubesXsA}{B}\right),\qquad 
m_\psi = \frac{1}{2} \left( \Fubex{\FubeXXXA}{B} \; -\; \frac{1}{d} \Fubex{\FubesXsA}{B}\right).\ee
One can check that the action of any element from $\tube_{\unit \rightarrow \unit}^B$
on both $m_\unit$ and $m_\psi$ is simply scalar multiplication.

Now we turn to the endomorphism algebra $\tube^B_{\beta \rightarrow \beta}$, defined in \eqref{atubes}, 
of charged tubes: those whose top and bottom 
boundary conditions consist of a single marked $\beta$ point. 
There are two non-zero even tubes and two non-zero odd tubes, hence as a vector space we have,
\begin{align}
\tube^B_{\beta \rightarrow \beta} \cong \cc^{2|2}.
\end{align}
This means that as an algebra $\tube^B_{\beta \rightarrow \beta}$ is either 
$\cliff_2$ (a.k.a.\ $\End(\cc^{1|1})$) or $\cliff_1\oplus \cliff_1$. 

To figure out which case we have, we begin by writing down the multiplication rules. 
By using the local relations in the $C_2$ theory we can work out the multiplication table, 
which is presented in the following table. 
In the table, $A\times B$ means ``stack $A$ (left most column) on top of $B$ (top most row)''. 
For multiplications involving odd tubes, we always take fermions in the $A$ tube to have a 
higher ordering than the fermions in the $B$ tube. 

\be
\renewcommand{\arraystretch}{3}
\centering
\begin{tabular}{c | r r r r r}
$\times$ in $\tube^B_{\beta\rightarrow \beta} \vphantom{\scale{1.3}{\VerticalSpace}}$          & $\FubeXssA $ & $\FubeXsdsA $ & $\FubessXA $&$ \FubessdXA$  \\
\hline
$\FubeXssA$ & $\FubeXssA$ & $\ \FubeXsdsA$  & $\FubessXA$ & $\FubessdXA$  \\

$\FubeXsdsA$ & $\FubeXsdsA $& $A^{4}\FubeXssA$ & $-\FubessdXA $& $-A^{4}\FubessXA  $\\

$\FubessXA$   & $\FubessXA $&$ \FubessdXA  $&$A^{10}\FubeXssA $&$-A^{2}\FubeXsdsA $\\

$\FubessdXA    $&$ \FubessdXA $&$ A^{4}\FubessXA $&$ A^2 \FubeXsdsA $&$ A^6\FubeXssA$   \\
\end{tabular}
\ee

Since the multiplication table for $\tube_\beta^B$ is non-abelian, 
as an algebra it must be $\cliff_2$, as the other possibility (namely 
$\cliff_1\oplus\cliff_1$) is abelian. 
In order to show that the previous table is indeed the multiplication table of $\cliff_2$, 
one can identify,
\begin{align}
1 = \Fubex{\FubeXssA}{B} \qquad \gamma_1 = A^6 \Fubex{\FubeXsdsA}{B} \qquad \gamma_2 = A^5  \Fubex{\FubessdXA}{B} 
\end{align}
and check that the odd generators $\gamma_1$ and $\gamma_2$ satisfy $\gamma_1^2 = \gamma_2^2 = 1$, and $\{ \gamma_i, \gamma_j \} = 2 \delta_{ij}$. 
These are precisely the defining relations of $\cliff_2 \cong \langle 1, \gamma_1,\gamma_2 \rangle$, and so we have $\tube_{\beta \rightarrow \beta}^B \cong \cliff_2$.

The super algebra $\cliff_2$ contains exactly two minimal idempotents, which we will call $m_\sigma^+$ and $m_\sigma^-$.
Explicitly,
\begin{align}
m_\sigma^{\pm} = \frac{1}{2}\left( \Fubex{\FubeXssA}{B} \pm  A^{3}\Fubex{\FubessXA}{B} \right) .
\end{align}
These two idempotents are isomorphic, in the sense that there exist 
endomorphisms (tubes) $u$ and $v$ such that $uv = m_\sigma^+$
and $vu = m_\sigma^-$.\footnote{Indeed this must be the case, since $\cliff_2$ is Morita equivalent to $\cc$, 
which has only one minimal idempotent.}
The existence of this isomorphism means that $m_\sigma^+$ and $m_\sigma^-$ correspond to isomorphic simple 
modules and so represent the same
quasiparticle type.
Note that in this case $u$ and $v$ are necessarily odd; we say that $m_\sigma^+$ and $m_\sigma^-$ are {\it oddly} isomorphic.
When doing calculations we fix a particular representative of the $m_\sigma^\pm$ 
equivalence class, which we will choose to be $m_\sigma^+$.

\subsubsection{Vortex spin structure} 

As in the last section, we first examine the endomorphism algebra $\tube^N_{\unit \rightarrow \unit}$, 
consisting of tubes with no charge and non-bounding spin structure.
As we saw in \eqref{ptubes}, this algebra is two dimensional, and generated by a single even vector 
and a single odd vector, so as a vector space we have
\be \tube^N_{\unit \rightarrow \unit } \cong \cc^{1|1}
\ee
The only possible algebra structure on $\cc^{1|1}$ is $\cliff_1$. 
$\cliff_1 = \langle 1,\gamma\rangle$ has only one simple module, namely $\cc^{1|1}$ with the matrix representation 
$\rho(1) = \sigma^0,\rho(\gamma)=\sigma^x$. 
Therefore, this endomorphism algebra will support only one quasiparticle. Since idempotents must always be even, 
the explicit presentation of this quasiparticle is simply the empty tube. We will denote this quasiparticle by $q_\sigma$:
\be q_\sigma = \Fubex{\FubeXXXA}{N} \ee

Now we examine the charge sector, corresponding to the algebra $\tube_{\beta \rightarrow \beta}^N$ 
of vortex tubes with nontrivial charge. 
As we have seen in \eqref{ptubes} this subalgebra again has two even generators and two odd generators, 
and so as a vector space:
\be \tube_{\beta \rightarrow \beta}^N \cong \cc^{2|2} \ee
Therefore, as an algebra, we must have $\tube_{\beta \rightarrow \beta}^N \cong \cliff_2$ or $\tube_{\beta \rightarrow \beta}^N \cong \cliff_1\oplus\cliff_1$. 
To determine which choice is correct, we work out the multiplication table, which is
\be
\renewcommand{\arraystretch}{3}
\centering
\begin{tabular}{c | r r r r r}
$\times$ in $\tube^N_{\beta\rightarrow \beta} \vphantom{\scale{1.3}{\VerticalSpace}}$          & $\FubeXssA $ & $\FubeXsdsA $ & $\FubessXA $&$ \FubessdXA$  \\
\hline
$\FubeXssA$ & $\FubeXssA$ & $\ \FubeXsdsA$  & $\FubessXA$ & $\FubessdXA$  \\

$\FubeXsdsA$ & $\FubeXsdsA $& $A^{4}\FubeXssA$ & $\FubessdXA $& $A^{4}\FubessXA  $\\

$\FubessXA$   & $\FubessXA $&$ \FubessdXA  $&$A^{6}\FubeXssA $&$A^{6}\FubeXsdsA $\\

$\FubessdXA    $&$ \FubessdXA $&$ A^{4}\FubessXA $&$ A^6 \FubeXsdsA $&$ A^{10}\FubeXssA$   \\
\end{tabular}
\ee
Since the multiplication table is abelian, we must have $\tube_\beta^N \cong \cliff_1\oplus\cliff_1$ (as the other choice, $\cliff_2$, is non-abelian). 
To see this explicitly, we make the identifications
\be \unit^\pm = \frac{1}{2}\left ( \Fubex{\FubeXssA}{N}  \pm A^5\Fubex{\FubessXA}{N} \right),\qquad   \gamma^\pm = \frac{A^6}{2}\left ( \Fubex{\FubeXsdsA}{N}  \pm A^5\Fubex{\FubessdXA}{N} \right).\ee
If we then re-write the multiplication table for $\tube_{\beta \rightarrow \beta}^N$ in terms of these generators, 
we see that it is indeed isomorphic to $\langle \unit^+ ,\gamma^+\rangle \oplus \langle \unit^- , \gamma^-\rangle =\cliff_1\oplus\cliff_1$, with $(\gamma^\pm)^2 = \unit^\pm$. 
We therefore have two (non-isomorphic) idempotents, $q_\unit = \unit^+$ and $q_\psi = \unit^-$.
Thus, the $\tube_{\beta \rightarrow \beta}^N$ 
endomorphism algebra gives rise to two q-type quasiparticles. (Recall that a q-type object is a simple
object whose endomorphism algebra is $\cc^{1|1}$. Simple 
objects whose endomorphism algebras are isomorphic to $\cc$ are referred to as m-type.)

\medskip

To summarize, we have found six types of quasiparticles in the theory: three non-vortex quasiparticles 
associated with tubes possessing bounding spin structures, and three vortex quasiparticles associated 
with tubes possessing non-bounding spin structures. 
They are displayed in Table \ref{CTwoParticles}.
The quantum dimension of these excitations can be computed 
by tracing out the 
idempotents associated with each excitation, which we elaborate on in Section \ref{traces_and_innerproducts}. 
The quantum dimensions are displayed in Table \ref{C2Data}. 
\begin{table}
  \centering
    \begin{align}
\nonumber
\begin{array}{l l l}
\multicolumn{1}{c}{\text{bounding}} &{\quad \quad \quad}& \multicolumn{1}{c}{\text{non-bounding}} \\
\cline{1-1}\cline{3-3}&\qquad&\\
m_\unit = \frac{1}{2}\left( \Fubex{\FubeXXXA}{B} \; +\; \frac{1}{d}\Fubex{\FubesXsA}{B}\right) & & 
q_\unit = \frac{1}{2}\left ( \Fubex{\FubeXssA}{N}  + A^5\Fubex{\FubessXA}{N} \right)\\
&&\\
m_\sigma^{+} = \frac{1}{2}\left( \Fubex{\FubeXssA}{B} +  A^{3}\Fubex{\FubessXA}{B} \right) & &
q_\sigma = \Fubex{\FubeXXXA}{N} \\
&&\\
m_\psi = \frac{1}{2} \left( \Fubex{\FubeXXXA}{B} \; -\; \frac{1}{d} \Fubex{\FubesXsA}{B}\right) & & 
q_\psi = \frac{1}{2}\left ( \Fubex{\FubeXssA}{N}  - A^5\Fubex{\FubessXA}{N} \right)\\
&&\\
\end{array}
\end{align}
\caption{\label{CTwoParticles} Representative idempotents for the six quasiparticles in the $C_2$ theory.
The non-bounding (i.e., vortices) are all q-type particles, while the bounding (i.e., non-vortex) particles are all m-type.
We have chosen $m_\sigma^+$ as the representative of the isomorphism class given by $m_{\sigma}^+$ and $m_\sigma^-$. }
\end{table}

\begin{table}
\begin{flalign*} & \begin{array}{r@{ \quad \quad \quad}  c @{\quad \quad} c @{\quad \quad} c @{\quad \quad \quad \quad } c  @{\quad \quad} c @{\quad \quad} c  }
			\text{particle}				&m_\unit		&m_\sigma^+		&m_\psi	&q_\unit	&q_\sigma	&q_\psi \\[.5ex] \hline \\ [-2ex]
			\text{quantum dimension}		&1			&\sqrt{2}		&1		&\sqrt{2}	&2		&\sqrt{2} \\ [.5ex]
						\end{array} & \end{flalign*}
	\caption{\label{C2Data} $\tube(C_2)$ quantum dimensions. The total quantum dimension is $\mcd = \sqrt{8}$.
	The quantum dimensions above have been normalized so that the trivial idempotent $m_\unit$ has unit quantum dimension.
}
	
\end{table}

\subsection{Quasiparticle fusion rules} \label{C2_fusion_rules}

With all quasiparticles in hand, we are ready to compute their fusion rules. 
We postpone a more general discussion of how to compute fusion rules in fermionic theories to Section \ref{fusion_rules}, 
and in this section restrict ourselves to working out examples for the $C_2$ theory. 

Recall that to each minimal idempotent $e$ we can associate an irreducible module 
(a.k.a.\ irreducible representation) $M_e$ as follows.
Let $\mct$ denote the tube category, and let $x$ be object
of $\mct$ which hosts $e$ (i.e.\ $e\in \End(x)$).
To each object $y$ of $\mct$, the module $M_e$ associates the subspace of $\mor(x\to y)$ consisting of morphisms
of the form $ef$, where $e$ is our chosen idempotent and $f$ is an arbitrary morphism from $x$ to $y$.
We can express this compactly as
\be
	M_e = e\mct .
\ee
It is easy to check that if $e$ and $e'$ are isomorphic idempotents, then $M_e$ and $M_{e'}$ are isomorphic modules, and conversely.
Geometrically, the module $M_e$ consists of tubes with $e$ fixed at one end and arbitrary string nets ($f$ above)
at the other end.

We will frequently simplify notation and denote both the idempotent $e$ and the corresponding module $M_e$ as simply $e$.

It is important to note that there are many idempotents within a given equivalence class,
and these idempotents might be hosted at different objects.
Furthermore, the isomorphism relating two idempotents might be odd
(i.e.\ it reverses fermion parity).
Despite these differences, all of the idempotents within an equivalence class should be thought of as representing the same
anyon type.
To do calculations, we must choose a particular idempotent within the equivalence class.
This is analogous to a gauge choice.
For example, the idempotent given by ${m_\sigma^+}$ constitutes a choice of representative of the equivalence 
class containing $m_\sigma^+$ and $m_\sigma^-$, but we could have just as easily chosen ${m_\sigma^-}$ as a representative.

Given two modules $a$ and $b$ of $\mct$, we can construct a tensor product module $a\tp b$.
Intuitively, forming $a\tp b$ amounts to fusing the quasiparticles $a$ and $b$ together by bringing them close 
to one another, and ``zooming out'' to view $a$ and $b$ as a single composite quasiparticle. 
To make this precise, we can impost the idempotent versions of $a$ and $b$ as boundary conditions on the two inner boundary components
of a twice-punctured disk $P$ (a.k.a.\ pair of pants).
Adding tubes to the outer boundary component of $P$ gives a module for the tube category, and this
module is, by definition $a\tp b$.
We will discuss this in more detail in \ref{fusion_rules}. 

We define the fusion space $V^{ab}_c$ as the space
\be V^{ab}_c \equiv {\rm mor}(c \ra a \tp b).\ee
Geometrically, $V^{ab}_c$ corresponds to the space of all string-net configurations (modulo local relations) 
on a pair of pants whose outgoing legs are labeled by the quasiparticles $a$ and $b$, and whose incoming leg is labelled by $c$. 

One subtle property of fermion theories is that the vector spaces $V^{ab}_c$ are {\it not} 
the vector spaces which appear in the direct sum decomposition of $a\tp b$. 
Instead, we define the fusion rule coefficients $\Delta^{ab}_c$ via\footnote{The $\Delta^{ab}_c$ are 
complex super vector spaces, not numbers.
They are a categorified version of the coefficients used to write a general vector as a linear combination
of basis vectors.}
\be \label{fusion_coeffs_defn} 
	a \tp b \cong \bigoplus_c \Delta^{ab}_ c c,
\ee
where the sum runs over a set of representatives for the equivalence classes of irreducible 
representations (equivalently, of minimal idempotents). 
In bosonic theories $V^{ab}_c = \Delta^{ab}_c$, but in fermionic theories the fusion spaces 
can be larger than the fusion coefficients:
\be V^{ab}_c \cong \Delta^{ab}_c \tp \End(c).\ee
We demonstrate and elaborate on this in Section \ref{fusion_rules_and_fusion_spaces}. 

\medskip

We will now illustrate how to compute the fusion spaces with simple examples in the $C_2$ theory. 
Suppose we want to find the fusion rule for $q_\sigma \otimes m_\unit$.
We first note that spin structure considerations on the pair of pants require that any quasiparticle 
appearing in $q_\sigma \tp m_\unit$ be a vortex-type quasiparticle (one with a non-bounding spin structure). 
Furthermore, since both $m_\unit$ and $q_\sigma$ have no charge (no $\beta$ lines fixed to the boundaries of their tubes), 
we know that their fusion products cannot have any charge.
Since $q_\sigma$ is the only vortex-type quasi particle with no charge we know that it is the only particle 
which can appear in the tensor product of $q_\sigma$ and $m_\unit$.
By searching for pants invariant under the applications of the appropriate idempotents, we see that the super vector space 
$V^{q_\sigma m_\unit}_{q_\sigma}$ is isomorphic to $\cc^{1|1}$, with the even subspace generated by a single even vector
\begin{align}
[V^{q_\sigma m_1}_{q_\sigma}]^0 = \left\langle \PantsPAPA \; +\;  \frac{1}{d} \PantsPAsPA \right \rangle
\end{align}
and the odd subspace generated by a single odd vector
\begin{align}
[V^{q_\sigma m_1}_{q_\sigma}]^1 = \left\langle \PantsPsdAPA \; +\;  \frac{1}{d} \PantsPsdAsPA \right \rangle.
\label{apply_odd_endo}
\end{align}
Notice that to find the odd generator from the even one, we apply the odd endomorphism of $q_\sigma$ to the ``exterior'' 
boundary of the pair of pants.
Graphically, this corresponds to taking a $\beta$ loop with a single dot on it and inserting it in a position parallel 
to the outer boundary of the pair of pants.

A more nontrivial fusion rule is $q_\sigma \tp q_\sigma$.
Since $q_\sigma$ has no charge, any quasiparticles appearing in $q_\sigma \tp q_\sigma$ must also carry no charge, 
since for the $C_2$ theory charge is a good quantum number. 
Additionally, any quasiparticles in $q_\sigma \tp q_\sigma$ must have non-vortex spin structures, so we know that 
only $m_\unit$ and $m_\psi$ can appear in $q_\sigma \tp q_\sigma$. This lets us work out the fusion spaces explicitly. 
We first work out the fusion space for $V^{q_\sigma q_\sigma}_{m_\unit}$.
The even part is generated by a single vector:
\begin{align}
[V^{q_\sigma  q_\sigma}_{m_\unit}]^0 = \left\langle \PantsPPAA  + \frac{1}{\sqrt{2}} \PantsPPAsA \right \rangle.
\end{align}
As with \eqref{apply_odd_endo}, we find the odd generator by using the odd endomorphism coming from $q_\sigma$:
\begin{align}
[V^{q_\sigma  q_\sigma}_{m_\unit}]^1 =\left \langle \PantsNNda  + \frac{1}{\sqrt{2}} \PantsNNd \right \rangle.
\end{align}
Therefore, $V^{q_\sigma q_\sigma}_{m_\unit} =  [V^{q_\sigma\tp q_\sigma}_{m_\unit}]^0 \oplus [V^{q_\sigma\tp q_\sigma}_{m_\unit}]^1 \cong \cc^{1|1}$.
An analogous calculation shows that $V^{q_\sigma q_\sigma}_{m_\psi} \cong \cc^{1|1}$,
and is generated by the two vectors
\begin{align}
V^{q_\sigma q_\sigma}_{m_\psi} =
\left \langle \PantsPPAA  - \frac{1}{\sqrt{2}} \PantsPPAsA \right \rangle 
\oplus \left\langle \PantsNNda  - \frac{1}{\sqrt{2}} \PantsNNd \right \rangle.
\end{align}
Summarizing, we have $q_\sigma \tp q_\sigma \cong \mathbb{C}^{1|1}m_\unit \oplus \mathbb{C}^{1|1} m_\psi$, with $V^{q_\sigma q_\sigma}_{m_\unit} \cong V^{q_\sigma q_\sigma}_{m_\psi}\cong\cc^{1|1}$.

As a final example, we will examine the fusion channel $m_\sigma^{+} \tp m_\psi$. 
Since $m_\sigma^{+}$ has nonzero charge while $m_\psi$ has no charge, anything appearing in $m_\sigma^+ \tp m_\psi$ 
must carry nonzero charge. Additionally, since both $m_\sigma^+$ and $m_\psi$ have non-vortex spin structures, 
their fusion products must possess non-vortex spin structures as well. Therefore, they must fuse to $m_\sigma^\pm$. 
Determining the fusion space $V^{m_\sigma^+ m_\psi}_{m_\sigma^\pm}$ thus amounts to identifying string-nets 
on the pair of pants which are invariant under the 
application of $m_\sigma^+$ and $m_\psi$ on the inner legs of the pants and invariant under 
the application of $m_\sigma^\pm$ on the outer leg of the pants. 

First apply $m_{\sigma}^+$ and $m_\psi$ on the inner legs, 
and $m_{\sigma}^+$ on the outer leg to a generic linear combination of string nets on the pants. 
By using the linear relation,
\begin{align}
\PantsAstAAsA  = \frac{1}{\sqrt{2}} \;  \PantsAsAshAsvtA
\end{align}
one can check that the resulting vector space will be zero dimensional if the pair of pants has even parity, 
and will be one dimensional if the pair of pants has odd parity, generated as follows:
\begin{align}
V^{m_\sigma^+ m_\psi}_{m_\sigma^+} = \left\langle \PantssvXsvdA -\frac{A^3}{\sqrt{2}}\PantssvtshsvdA + A^3\PantssvtXsvdA  - \frac{1}{\sqrt{2}} \PantssvshsvdA   \right \rangle.
\end{align}
Hence $m_\sigma^+ \tp m_\psi \cong \cc^{0|1} m_\sigma^+$.
Repeating the calculation for $m_\sigma^-$ on the outer leg results in a one dimensional vector 
space if the pair of pants has even parity, and zero otherwise;
$m_\sigma^+ \tp m_\psi \cong \cc^{1|0} m_\sigma^-$.
This reflects the fact that $m_\sigma^-$ is oddly isomorphic to $m_\sigma^+$.
We emphasize that to write the fusion rules, we need to work with actual idempotents, not merely equivalence classes of idempotents.

\medskip

By following the approach outlined in these examples, it is straightforward to write down the table of fusion 
rules in the theory, which we present in Table \ref{fusiontable}. In the table, we list the fusion rule 
coefficients from \eqref{fusion_coeffs_defn}, with the bullets ($\bullet$) representing cases 
where $\Delta^{ab}_c \cong \cc^{1|1}$, and the appearance of $\cc^{0|1}$ indicating a purely odd fusion channel. 
The fusion spaces $V^{ab}_c$ can then be obtained through the use of $V^{ab}_c \cong \Delta^{ab}_c \tp \text{End}(c)$. 

\begin{table} 
\begin{flalign*} 
 \begin{array}{c@{ \;}  | c@{\quad } c @{\quad } c }
			\mca \tp \mca 		&m_\unit		&m_\sigma^+		&m_\psi		\\[.5ex] \hline \\ [-2ex]
			m_\unit		 	&m_\unit		&m_\sigma^+		&m_\psi		\\
			m_\sigma^+		&m_\sigma^+	&m_\unit \oplus \cc^{0|1} m_\psi &\cc^{0|1} m_\sigma^+		\\
			m_\psi		 	&m_\psi		&\cc^{0|1} m_{\sigma}^+		&m_\unit		\\
			\multicolumn{1}{c}{} \\ [1ex]
			\mca \tp \mcv 		&q_\unit			&q_\sigma						&q_\psi		\\[.5ex] \hline \\ [-2ex]
			m_\unit		 	& q_\unit		& q_\sigma					& q_\psi \\
			m_\sigma^+	 	& q_\sigma	& q_\unit \oplus  q_\psi			&  q_\sigma	\\
			m_\psi		 	& q_\psi		& q_\sigma					& q_\unit		\\
			\end{array} 	
& \quad \quad \quad 
\begin{array}{c@{ \;}  | c @{ \quad} c @{\quad } c }
			\mcv \tp \mca 		&m_\unit			&m_\sigma^+				&m_\psi		\\[.5ex] \hline \\ [-2ex]
			q_\unit		 	& q_\unit		& q_\sigma					& q_\psi \\
			q_\sigma	 		& q_\sigma	& q_\unit \oplus  q_\psi			&  q_\sigma	\\
			q_\psi		 	& q_\psi		& q_\sigma					& q_\unit		\\
			\multicolumn{1}{c}{} \\ [1ex]
			\mcv \tp \mcv 			&q_\unit				&q_\sigma 						&q_\psi		\\[.5ex] \hline \\ [-2ex]
			q_\unit		 		&\bullet m_\unit			&\bullet m_\sigma^+						&\bullet m_\psi 	\\
			q_\sigma		 		&\bullet m_\sigma^+		&\bullet (m_\unit \oplus  m_\psi )		&\bullet m_\sigma^+		\\
			q_\psi		 		&\bullet m_\psi			&\bullet m_\sigma^+ 					&\bullet m_\unit		\\
			\end{array}
 \end{flalign*}

\caption{ \label{fusiontable} The fusion rules in $\tube(C_2)$. 
	We have defined $\mca = \{ m_\unit, m_\sigma^+, m_\psi \} $ and $\mcv = \{ q_\unit, q_\sigma, q_\psi \}$ as the set of anyons and set of vortices, respectively.
	The ($a$-$b$)th entry in each table is the sum $\oplus_c \Delta_c^{ab} c$, where we have omitted any $\Delta^{ab}_c$ that is equal to $\cc$ and used $\bullet = \mathbb{C}^{1|1}$ to signify that the associated $\Delta^{ab}_c$ is isomorphic to $\cc^{1|1}$.
		Entries with $\cc^{0|1}$ indicate that the fusion channel is purely odd. 
	The fusion spaces can be obtained from this table according to $V^{ab}_c \cong \Delta^{ab}_c \tp \text{End}(c)$.
	For example, $V^{m_\psi q_\sigma}_{q_\sigma} \cong \cc \tp \cc^{1|1} = \cc^{1|1}$.
	}
\end{table}

\subsection{Modular transformations and ground states on the torus} \label{modulartforms}

\subsubsection{$C_2$ string-nets on the torus} \label{c2_stringnets_torus}

In this section, we compute a standard basis for $C_2$ string-nets-modulo-local-relations on spin tori as well 
as the action of the mapping class groupoid (i.e. the modular $S$ and $T$ matrices).

\begin{figure}
  \centering
    \begin{align}
\nonumber
\newcommand{\AnnulusOption}{\mathord{\vcenter{\hbox{\includegraphics[scale=1]{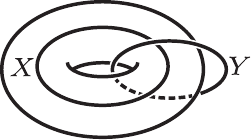}}}}}
\AnnulusOption
\quad \quad \leftrightarrow \quad \quad 
\AddDatTorus{\FubeXXXA}{X}{Y}\quad \quad \quad X,Y \in\{ B, N \}
\end{align}
\caption{\label{TorusNotation} 
To go from the annulus to the torus we identify the inner and outer circles, and label the different spin structures by $X$ and $Y$ for the longitudinal and meridional cycles on the torus, respectively.
}
\end{figure}

Because the $C_2$ theory is dependent on the existence of a spin structure, 
in order to talk about ground states on a torus we must first specify the spin structure on the torus.
There are $|H^1(T^2;\zt)|=4$ different spin structures on the torus, obtained by choosing either bounding or non-bounding spin 
structures for two of the torus's three non-contractible cycles 
(with the spin structure along the third non-contractible cycle determined by those of the other two). 
To go from the annulus to the torus we identify the inner and outer circles, and label the different spin structures by 
\begin{align}
\AddDatTorus{\FubeXXXA}{X}{Y}\quad \quad \quad X,Y \in\{ B, N \}
\end{align}
where $X$ and $Y$ specify the spin structure along the longitudinal and meridional cycles, respectively (see Figure \ref{TorusNotation}).

If there was no interplay between a chosen spin structure and the string-net pictures drawn on the torus, 
and if the fermion parity of a ground state is fixed by the spin structure 
and the string-net picture drawn on the torus, we would expect $|H_1(T^2;\zt)|=4$ 
degenerate ground states 
for each choice of spin structure, since the $\beta$ lines obey $\zt$ fusion rules. 
We will see that naive guess is incorrect: instead, for each spin structure, 
one of the four putative states is a null vector, meaning that there is only a 
3-dimensional groundstate for each choice spin structure.

We will first use elementary arguments to find a spanning set for each of the four spin tori.
Later, using more sophisticated techniques, we will prove that these spanning sets are in fact bases.

Since $\beta$ lines obey $\zz/2$ fusion rules (see \ref{C2_data_table}), it is easy to see that any string net on a spin torus
is a linear combination of the following seven diagrams:
\be \label{seven_tubes} \Fubex{\FubeXXXA}{},\;  \Fubex{\FubesXsA}{},\;     \Fubex{\FubesdXsA}{},\; 
\Fubex{\FubeXssA}{},\; \Fubex{\FubeXsdsA}{},\; \Fubex{\FubessXA}{}, \; \Fubex{\FubessdXA}{}.
\ee
Indeed, using \eqref{Fmove} and the ability to remove topologically trivial $\beta$ loops, 
we can transform the $\beta$ loops into a standard representative of their
$\zz/2$ homology class.
The resulting loop will be decorated by some number of fermionic dots; using dot cancellation we can assume that the
loop will contain either 0 or 1 dot.
The seven diagrams listed above are the only independent diagrams that remain after applying these relations.

However, for a given spin structure, some of the diagrams in \eqref{seven_tubes} will be zero.
To determine which of the diagrams are zero, we use the following three obervations:
\begin{enumerate}
	\item If one of the non-trivial cycles of the torus has an antiperiodic spin structure, then any odd diagram is zero.
	This is because we can translate the entire string net in the direction of the antiperiodic cycle.
	When we arrive back at our starting point, we have picked up a factor of $-1$ from the antiperiodicity.
	Thus odd diagrams are equal to $-1$ times themselves and are therefore zero.
	In particular, the three odd diagrams of \eqref{seven_tubes} are zero in the $BB$, $BN$ and $NB$ spin structures.
	\item If a diagram contains a $\beta$ loop without a dot (even fermion parity), and if that loop inherits the periodic spin structure,
	then the diagram is zero.
	To see this, we create two dots on the loop, slide one around the loop, and then cancel the dots again.
	Because of the Koszul sign, we see that the diagram is equal to $-1$ times itself.
	(If the spin structure along the cycle were antiperiodic, there would be an additional sign to cancel the Koszul sign.)
	In particular, in the $NN$ spin structure all three of the even $\beta$ loops are zero.
	For the $BB$, $BN$ and $NB$ spin structures, exactly one of the three even $\beta$ loops fails this test and is zero.
	\item Finally, the empty diagram with $NN$ spin structure is zero. To show this, 
	we first create a topologically trivial circular $\beta$ loop. We then wrap it around the tube and fuse it 
	with itself to get the sum of two diagrams with parallel $\beta$ loops wrapping around the tube, one of 
	which has no fermion dots (the even channel of $\beta\tp \beta$) and one which has one dot on each $
	\beta$ loop (the odd fusion channel of $\beta\tp\beta$). 
	The diagram corresponding to the even fusion channel 
	is zero by observation 2 above, while the diagram with two odd loops is zero by an argument similar to observation 2: we slide one of the loops in the direction
	orthogonal to itself and pick up a Koszul sign, showing that the diagram must be zero.
\end{enumerate}

\begin{figure}
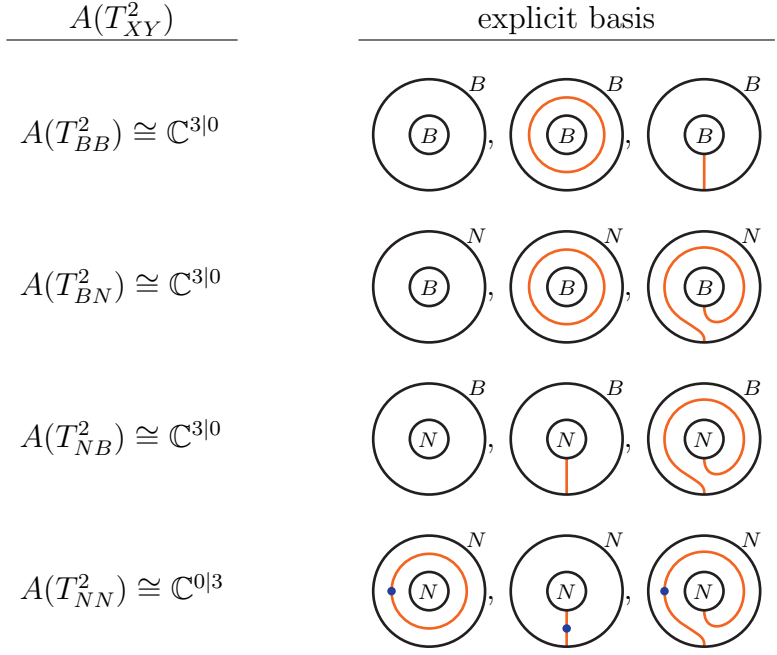

  \centering
    \begin{align}
\nonumber
\begin{array}{ l l l l }
\multicolumn{1}{c}{A(T^2_{XY})} &{\quad \quad \quad}& \multicolumn{1}{c}{\text{explicit basis}} & \\ \cline{1-1}\cline{3-3} && &\\
A(T^2_{BB}) \cong \mathbb{C}^{3|0} & & \AddDatTorus{\FubeXXXA}{B}{B} , \; \AddDatTorus{\FubesXsA}{B}{B}, \;  \AddDatTorus{\FubeXssA}{B}{B} &\\   
&&&\\
A(T^2_{BN}) \cong \mathbb{C}^{3|0} & & \AddDatTorus{\FubeXXXA}{B}{N} , \; \AddDatTorus{\FubesXsA}{B}{N}, \;  \AddDatTorus{\FubessXA}{B}{N} &\\  
&&&\\
A(T^2_{NB}) \cong \mathbb{C}^{3|0} & & \AddDatTorus{\FubeXXXA}{N}{B} , \; \AddDatTorus{\FubeXssA}{N}{B}, \;  \AddDatTorus{\FubessXA}{N}{B} &  \\
&&&\\
A(T^2_{NN}) \cong \mathbb{C}^{0|3} & & \AddDatTorus{\FubesdXsA}{N}{N} , \; \AddDatTorus{\FubeXsdsA}{N}{N}, \;  \AddDatTorus{\FubessdXA}{N}{N} &\\  
&&&\\ 
\end{array}
\end{align}
\caption{\label{TorusBasisC2} The ground states on the four different spin tori. 
Notice that the non-bounding torus (with $NN$ spin structure) has only odd fermion parity ground states.}
\end{figure}

Figure \ref{TorusBasisC2} shows the remaining non-zero diagrams for each spin structure.
It is easy to see that these diagrams are linearly independent if they are non-zero, but we have not yet proved that they are in fact non-zero.
To do this, we will employ a fancier argument relating a basis of the torus Hilbert space to minimal idempotents of the tube category.

Let $\tube^B$ denote the bounding tube category and $\tube^N$ denote the non-bounding tube category.
We can make use of the following results, which we prove in a more general context in Section \ref{torus_basis_theorem}:
The ground state spaces of the ${BB}, {NB},$ and ${BN}$ tori are purely even, with an orthogonal basis given by closed-up minimal idempotents of $\tube^B$.
Additionally, the ground state space of the $NN$ torus is isomorphic to $\cc^{p|q}$, where $p$ is the number of m-type idempotents of $\tube^N$
and $q$ is the number of q-type idempotents of $\tube^N$.
An orthogonal basis is given by
closures of
a set of representatives of the minimal m-type idempotents of $\tube^N$,
union $\{\cl(\gamma_j)\}$,
where $\gamma_j$ runs through a set of representatives of odd endomorphisms of the minimal q-type idempotents of $\tube^N$.
We use $\cl$ to denote the closure of an annular diagram on the torus.

For the $C_2$ theory, we have shown that $\tube^B$ has three m-type minimal idempotents, 
and $\tube^N$ has three q-type minimal idempotents.
Letting $A(T^2_{JK})$ denote the Hilbert space on the torus with $JK$ spin structure, 
it follows that $A(T^2_{BB}) \cong A(T^2_{NB}) \cong A(T^2_{BN}) \cong \cc^{3|0}$ and
$A(T^2_{NN}) \cong \cc^{0|3}$,
and so the diagrams of Figure \ref{TorusBasisC2} must indeed all be non-zero.

For future reference we also tabulate the change of basis between the ground-state tori in Figure \ref{TorusBasisC2} 
and the closed-up minimal idempotents.
To form string-net pictures drawn on tori from the idempotents, 
we close up the idempotents along the longitudinal direction by identifying the inner boundary 
of the annulus on which the idempotent lives with the outer boundary, 
specifying a choice of spin structure along the newly made cycle. 
We then express the result as a linear combination of the tori in Figure \ref{TorusBasisC2}.
For simplicity of notation, we will define 
\begin{align}
e=\; {\FubeXXXA} \quad h = \;{\FubesXsA}  \quad v =\; {\FubeXssA} \quad t = \; {\FubessXA}
\end{align}
and append subscripts to denote a particular spin structure. 
We will also use an overscript $\bullet$
if we are closing up an odd endomorphism rather than the idempotent itself.
For example,
\begin{align}
h_{NB}= \; \AddDatTorus{\FubesXsA}{N}{B} \quad \quad \text{and} \quad \quad \overset{\bullet}{t}_{NN} = \AddDatTorus{\FubessdXA}{N}{N} \; .
\end{align}
We can then compute the change of basis shown in Figure \ref{C2Change_of_Basis}.
\begin{figure}
  \centering
\begin{align}
\nonumber
\left( \begin{matrix}
m_\unit \\
m_\sigma^+\\
m_\psi \\
\end{matrix} \right)_{BB} 
&= \frac{1}{2}\left( \begin{matrix}
1&1/d&0\\
0&0&1\\
1&-1/d&0\\
\end{matrix} \right)
\left( \begin{matrix}
e\\
h\\
v\\
\end{matrix} \right)_{BB}
\quad \quad \quad
\left( \begin{matrix}
m_\unit \\
m_\sigma^+\\
m_\psi \\
\end{matrix} \right)_{BN} 
= \frac{1}{2}\left( \begin{matrix}
1&1/d&0\\
0&0&A^3\\
1&-1/d&0\\
\end{matrix} \right)
\left( \begin{matrix}
e\\
h\\
t\\
\end{matrix} \right)_{BN}\\
\nonumber
\left( \begin{matrix}
q_\unit \\
q_\sigma\\
q_\psi \\
\end{matrix} \right)_{NB} 
&= \frac{1}{2}\left( \begin{matrix}
0&1&A^5\\
2&0&0\\
0&1&-A^5\\
\end{matrix} \right)
\left( \begin{matrix}
e\\
v\\
t\\
\end{matrix} \right)_{NB}
\quad \quad \quad
\left( \begin{matrix}
\overset{\bullet}{q}_\unit \\
\overset{\bullet}{q}_\sigma\\
\overset{\bullet}{q}_\psi \\
\end{matrix} \right)_{NN} 
= \frac{1}{2}\left( \begin{matrix}
0&A^6&-A^{3}\\
\sqrt{2}&0&0\\
0&A^6&A^{3}\\
\end{matrix} \right)
\left( \begin{matrix}
\overset{\bullet}{h}\\
\overset{\bullet}{v}\\
\overset{\bullet}{t}\\
\end{matrix} \right)_{NN}
\end{align}
\caption{Change of basis between the quasiparticle (idempotent) basis given in Table \eqref{CTwoParticles} 
and the topological bases in Figure \eqref{TorusBasisC2} for the torus.
These are simply given by expressing the idempotents in the topological bases. 
Note that the odd torus has a sign ambiguity on each of the of the idempotents. 
We can require that $(\overset{\bullet}{q})^2 = q$, but that leaves $\overset{\bullet}{q}$ ambiguous up to a $\pm$ sign. 
This ambiguity can lead to different $S$ matrices, see \eqref{NNSmatrix} and surrounding text for more details.}
\label{C2Change_of_Basis}
\end{figure}

\subsubsection{The modular $S$ and $T$ matrices} \label{C2_modular_mats}

\begin{figure} 
\newcommand{\Space}{\; \; \; \; \; \; }
\newcommand{\Spacep}{ \; \; \; \mathop{\vphantom{\int}} \; \; \;    } 
\begin{align}
\nonumber
\xymatrix@!0 @M=1mm @R=7mm @C=30mm{
\AddDatTorus{\FubeXXXA}{B}{B}\ar @`{p+(-28.28,7.07),p+(-7.07,28.28)}^{S}  \ar@<1ex>[r]^T  & \AddDatTorus{\FubeXXXA}{B}{N} \ar@<1ex>[l]^T   \ar@<1ex>[r]^S&  \AddDatTorus{\FubeXXXA}{N}{B}  \ar@<1ex>[l]^S  \ar @`{p+(7.07,28.28),p+(28.28,7.07)}^{T } &\AddDatTorus{\FubeXXXA}{N}{N}  \ar @`{p+(7.07,28.28),p+(28.28,7.07)}^{S,T } 
}
\end{align} \nonumber
\caption{\label{spin_str_mapping_class_group}The action of the mapping class group on the four spin tori. 
}
\end{figure}
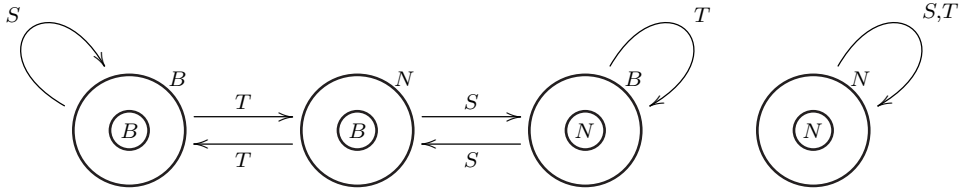

In this section, we will compute the representation of the modular $S$ and $T$ matrices in the $C_2$ theory, which together generate the modular groupoid. 
The modular $S$-matrix acts on states on the torus by interchanging the meridional and longitudinal cycles 
of the torus.
If we draw the torus as a square with opposite sides identified, then $S$ acts by rotating the square by $\pi/2$ clockwise. 
The modular $T$ matrix represents the action of the Dehn twist on the torus, 
which corresponds to cutting the torus along a meridional cycle, 
twisting the boundary conditions at the cut by $2\pi$ relative to one another, and gluing the torus back together.
In terms of the annular pictures we have been drawing of the tubes, the twist is accomplished by 
twisting the inner boundary of the annulus by $2\pi$ counterclockwise relative to the outer boundary. 

Importantly, the $S$ and $T$ modular transformations do {\it not} always preserve the spin structure of the torus they act on. 
Figure \ref{spin_str_mapping_class_group} shows how the different possible spin structures are permuted under $S$ and $T$. 
Since $T$ does not preserve the spin structures, it cannot have well-defined eigenstates with a definite spin structure,
meaning that it will not be diagonal in the idempotent (quasiparticle) basis. 
This means that the topological spins of bounding (non-vortex) quasiparticles, defined as their eigenvalues under $T$, will not be well-defined.
In contrast with $T$, the action of $T^2$ preserves spin structures, and so we are still able to associate 
definite $T^2$ 
eigenvalues to the quasiparticles in the theory.
Putting aside the issue of spin structures, the twist of an idempotent (defined as the phase acquired when performing a $2\pi$ twist on the tubes in a given idempotent) 
is in general only defined 
up to a $\pm$ sign (which has been discussed in e.g. \cite{cano2014,bruillard2017,gu2014}).
More precisely, the twist of a given simple object in the tube category is ambiguous across the isomorphism 
class of that object. 
Indeed, we will see that the twists of the two idempotents in the $m_\sigma$ isomorphism class (namely $m_\sigma^+$ and 
$m_\sigma^-$) have twists that differ by a factor of $-1$. 

We now proceed to examine the action of the $S$ and $T$ modular transformations on the four spin tori which compose 
the 12-dimensional space listed in Figure \ref{TorusBasisC2}.
Although the calculations are most easily performed in the topological basis in Figure \ref{TorusBasisC2}, 
it is more natural to analyze the resulting transformations in the idempotent basis given in Table \ref{CTwoParticles}. 
As mentioned earlier, the basis vectors in the idempotent basis are constructed by taking the idempotents associated with the 
quasiparticles identified in the previous section and closing them up (with different choices of spin structure) 
along the longitudinal direction. 
The change of basis between the topological and idempotent bases are written explicitly in Figure \ref{C2Change_of_Basis}.

We'll start with the $BB$ spin structure, which is preserved under the action of $S$. 
In the topological basis $(e,v,h)^T$ we find
\begin{align}
\left( \begin{matrix}
e\\
v\\
h\\
\end{matrix} \right)_{BB} 
\xrightarrow{S^{BB \rightarrow BB}} & \left( \begin{matrix}
1&0&0\\
0&0&1\\
0&1&0\\
\end{matrix} \right)
\left( \begin{matrix}
e\\
v\\
h\\
\end{matrix} \right)_{BB}.
\end{align}
To transform $S^{BB\rightarrow BB}$ into the quasiparticle basis we use Figure \ref{C2Change_of_Basis}.
After the change of basis we find the familiar matrix
\begin{align}
\left( \begin{matrix}
m_\unit\\
m_\sigma^+\\
m_\psi \\
\end{matrix} \right)_{BB} 
\xrightarrow{S^{BB \rightarrow BB}} &\frac{1}{2} \left( \begin{matrix}
1&d&1\\
d&0&-d\\
1&-d&1\\
\end{matrix} \right)
\left( \begin{matrix}
m_\unit\\
m_\sigma^+\\
m_\psi \\
\end{matrix} \right)_{BB}
\end{align}
which is identical to the $S$-matrix for the Ising TQFT. 

Now for the $BN$ and $NB$ spin structures.
These are interchanged by the $S$-matrix, as indicated in Figure \ref{spin_str_mapping_class_group}.
In the idempotent bases these induce transformations between the non-vortex and vortex quasi-particles.
We obtain 
\begin{align}
\left( \begin{matrix}
e\\
h\\
t\\
\end{matrix} \right)_{BN} 
\xrightarrow{S^{BN \rightarrow NB}} & \left( \begin{matrix}
1&0&0\\
0&1&0\\
0&0&A^{10}\\
\end{matrix} \right)
\left( \begin{matrix}
e\\
v\\
t\\
\end{matrix} \right)_{NB}
\quad \text{and} \quad 
\left( \begin{matrix}
e\\
v\\
t\\
\end{matrix} \right)_{NB} 
\xrightarrow{S^{NB \rightarrow BN}} & \left( \begin{matrix}
1&0&0\\
0&1&0\\
0&0&A^{6}\\
\end{matrix} \right)
\left( \begin{matrix}
e\\
h\\
t\\
\end{matrix} \right)_{BN}.
\end{align}
Notice that if we compose both transformations we get the identity. 
These can each be transformed into the idempotent bases using \eqref{C2Change_of_Basis}, 
where one finds,
\begin{align}
\left( \begin{matrix}
m_\unit\\
m_\sigma^+\\
m_\psi\\
\end{matrix} \right)_{BN} 
\xrightarrow{S^{BN \rightarrow NB}} &\frac{1}{2} \left( \begin{matrix}
1&d&1\\
-d&0&d\\ 
-1&d&-1\\
\end{matrix} \right)
\left( \begin{matrix}
\hat{q}_\unit\\
\hat{q}_\sigma^+\\
\hat{q}_\psi\\
\end{matrix} \right)_{NB}
\end{align}
and
\begin{align}
\label{closing_q_type_C2}
\left( \begin{matrix}
\hat{q}_\unit\\
\hat{q}_\sigma^+\\
\hat{q}_\psi\\
\end{matrix} \right)_{NB} 
\xrightarrow{S^{NB \rightarrow BN}} & \frac{1}{2}\left( \begin{matrix}
1&-d&-1\\ 
d&0&d\\
1&d&-1\\
\end{matrix} \right)
\left( \begin{matrix}
m_\unit\\
m_\sigma^+\\
m_\psi\\
\end{matrix} \right)_{BN}.
\end{align}
In order to make the matrix unitary, we have defined $\hat{q} = q/\sqrt{2}$ so that each $\hat{q}$ idempotent has unit norm (how to compute the norms of idempotents will be discussed in Section \ref{traces_and_innerproducts}). 
To collect the results we've arrived at so far, we define
\begin{align}
{\bf M} = [ m_\unit \;  m_\sigma^+\;  m_\psi]^T \qquad \widehat{{\bf Q} }= [\hat{q}_\unit \; \hat{q}_\sigma \; \hat{q}_\psi ]^T
\end{align}
Then we have,
\begin{align}
\left( \begin{matrix}
{\bf M}_{BN}\\
\widehat{{\bf Q} }_{NB}\\
{\bf M}_{BB}\\
\end{matrix} \right)
\xrightarrow{\;\; S \; \; } & \left( \begin{matrix}
&S^{BN \rightarrow NB}&\\
S^{NB \rightarrow BN}&&\\
&&S^{BB \rightarrow BB}\\
\end{matrix} \right)
\left( \begin{matrix}
{\bf M}_{BN}\\
\widehat{{\bf Q} }_{NB}\\
{\bf M}_{BB}\\
\end{matrix} \right).
\end{align}

Similarly we can compute the modular $T$-matrix, the action of which twists the inner boundary of an annulus by an angle of $2\pi$ counterclockwise with respect to its outer boundary. 
This definition ensures that the $T$-matrix acts as the identity on tubes with no charge line 
(i.e. with no strings ending on their inner annular boundaries).
Within each spin structure sector the $T$-matrix is diagonal in the idempotent basis. 
We can read off the structure of the $T$-matrix with the help of Figure \ref{spin_str_mapping_class_group} to find
\begin{align}
\left( \begin{matrix}
{\bf M}_{BN}\\
\widehat{{\bf Q} }_{NB}\\
{\bf M}_{BB}\\
\end{matrix} \right)
\xrightarrow{\;\; T \; \; } & \left( \begin{matrix}
&&T^{BN \rightarrow BB}\\
&T^{NB \rightarrow NB}&\\
T^{BB \rightarrow BN}&&\\
\end{matrix} \right)
\left( \begin{matrix}
{\bf M}_{BN}\\
\widehat{{\bf Q} }_{NB}\\
{\bf M}_{BB}\\
\end{matrix} \right).
\end{align}
With 
\begin{align}
T^{BN \rightarrow BB} =  & \left( \begin{matrix}
1&&\\
&A^3&\\
&&1\\
\end{matrix} \right)
\quad 
T^{NB \rightarrow NB}=  & \left( \begin{matrix}
A^5&&\\
&1&\\
&&-A^5\\
\end{matrix} \right)
\quad 
T^{BB \rightarrow BN}=  & \left( \begin{matrix}
1&&\\
&A^3&\\
&&1\\
\end{matrix} \right)
\end{align}
One can check that the usual modular group relation $(ST)^3 = \unit$ holds as expected.

More interesting is the $NN$ torus, whose spin structure is preserved under both $S$ and $T$. 
This has been investigated before in \cite{ware2016}, and our results agree with theirs in this case. 
According to Figure \ref{TorusBasisC2}, the Hilbert space is spanned by $\overset{\bullet}{h}$, $\overset{\bullet}{v}$, and $\overset{\bullet}{t}$,
where as before the $\bullet$ means that the associated tubes have odd fermion parity.
In the topological basis, we obtain
\begin{align}
\left( \begin{matrix}
\overset{\bullet}{h}\\
\overset{\bullet}{v}\\
\overset{\bullet}{t}\\
\end{matrix} \right)_{NN} 
\xrightarrow{S^{NN \rightarrow NN}} & \left( \begin{matrix}
0&1&0\\
 A^4&0&0\\
0&0&A^{10}\\
\end{matrix} \right)
\left( \begin{matrix}
\overset{\bullet}{h}\\
\overset{\bullet}{v}\\
\overset{\bullet}{t}\\
\end{matrix} \right)_{NN}
\end{align}
Now we need to transform into the idempotent (quasiparticle) basis. 
With the choice of basis in Figure \ref{C2Change_of_Basis} we find
\begin{align}
\left( \begin{matrix}
\overset{\bullet}{q}_\unit\\
\overset{\bullet}{q}_\sigma\\
\overset{\bullet}{q}_\psi\\
\end{matrix} \right)_{NN} 
\xrightarrow{S^{NN \rightarrow NN}} &\frac{-A^2}{2} \left( \begin{matrix}
1& d&-1\\
d&0&d\\
-1& d&1\\
\end{matrix} \right)
\left( \begin{matrix}
\overset{\bullet}{q}_\unit\\
\overset{\bullet}{q}_\sigma\\
\overset{\bullet}{q}_\psi\\
\end{matrix} \right)_{NN}.
\label{NNSmatrix}
\end{align}
Note that the requirement $\overset{\bullet}{q}^2 =q$ only determines $\overset{\bullet}{q}$ up to a $\pm$ sign. 
Consequently the off-diagonal matrix elements $(S^{NN \ra NN})_{ij}$ between q-type idempotents are only determined up to a sign 
(indeed, sending $\overset{\bullet}{q}_j \ra s_j \overset{\bullet}{q}_j$, $s_j =\pm1$ conjugates the S-matrix by a diagonal matrix of $\pm1$'s).
Similarly, we can compute the modular $T$-matrix, 
\begin{align}
\left( \begin{matrix}
\overset{\bullet}{q}_\unit\\
\overset{\bullet}{q}_\sigma\\
\overset{\bullet}{q}_\psi\\
\end{matrix} \right)_{NN} 
\xrightarrow{T^{NN \rightarrow NN}} &\left( \begin{matrix}
A^5& &\\
&1&\\
&&-A^5\\
\end{matrix} \right)
\left( \begin{matrix}
\overset{\bullet}{q}_\unit\\
\overset{\bullet}{q}_\sigma\\
\overset{\bullet}{q}_\psi\\
\end{matrix} \right)_{NN}
\end{align}
One can check that we have the modular relations
\begin{align}
 (S^{NN\ra NN}T^{NN\ra NN})^3 = (S^{NN \ra NN})^4= A^{8}\text{id} = -\text{id}
\end{align}
The minus sign comes from the fact that acting by $S^4$ or $(ST)^3$ performs a $2\pi$ rotation 
of the fermion framing, resulting in a phase of $-1$, since all states on the $NN$ torus have odd fermion parity. 
For general theories, these relations become $(ST)^3=S^4=(-1)^F$, where $(-1)^F$ is the fermion 
parity operator.
See Sections \ref{SO36ModularTransformations} and \ref{E6S_matrix_section} for examples of this more general scenario.

\medskip

Collecting these results, we can now write down the complete modular $S$ and $T$ matrices in the $C_2$ theory, 
which act across all spin structures. 
In the quasiparticle 
basis $[(m_\unit,m_\sigma^+,m_\psi)_{BB},(m_\unit,m_\sigma^+,m_\psi)_{BN},(q_\unit,q_\sigma,q_\psi)_{NB},(\overset{\bullet}{q}_\unit,\overset{\bullet}{q}_\sigma,\overset{\bullet}{q}_\psi)_{NN}]^T$, 
we have
\be \label{modularS}
S = \frac{1}{2}\begin{pmatrix} 1 & d & 1 &			&&&			&&&			&& \\ 
					      d & 0 &-d &			&&&			&&&			&&\\
					      1&-d&1 & 			&&&			&&&			&&\\
						&&&				&&&			1&d&1&		&& \\
						&&&				&&&			-d&0&d&		&&\\ 
						&&&				&&&			-1&d&-1&		&&\\
						&&&				1&-d&-1&		&&&			&&\\ 
						&&&				d&0&d&		&&&			&&\\
						&&&				1&d&-1&		&&&			&&\\ 
						&&&				&&&			&&&			-A^{2} & -A^{2}d & A^{2}\\
						&&&				&&&			&&&			-A^{2}d & 0 & -A^{2}d \\ 
						&&&				&&&			&&&			A^{2} & -A^{2}d & -A^{-2} \end{pmatrix}\ee		
where we have only listed the non-zero entries. 
$S$ has two different direct-sum decompositions. First, essentially by construction, 
it splits into a direct sum over blocks according to spin structures 
under the $S$ modular transformation. 
Additionally, we have $S = S_{even} \oplus S_{odd}$, where $S_{odd}$ is the $S$-matrix 
acting on ground states with odd fermion parity.
This decomposition is always possible, but it will not always match a decomposition based 
on spin structures. 
That is, while $S_{odd} = S^{NN\ra NN}$ in this theory, spin structure blocks of $S^{NN\ra NN}$ 
will not have a definite fermion parity in general; see Sections \ref{so36} and \ref{halfesix} for examples. 
Also note that $S^4 = \unit_{9\times9}\oplus(-\unit_{3\times3})$ in accordance with $S = S_{even} \oplus S_{odd}$ and $S^4=(-1)^F\unit$. 

Now for the $T$-matrix. 
In the same quasiparticle basis as before, the $T$-matrix is
\be 
T = \begin{pmatrix}   			&&&				1&0&0&		&&&			&& \\ 
					        &&&				0&A^3&0&	&&&			&&\\
					        &&& 				0&0&1&		&&&			&&\\
						1&0&0&			&&&			&&&			&& \\
						0&A^3&0&		&&&			&&&			&&\\
						0&0&1&			&&&			&&&			&&\\
						&&&				&&&			A^{5}&0&0&	&&\\
						&&&				&&&			0&1&0&		&&\\
						&&&				&&&			0&0&-A^{5}&	&&\\
						&&&				&&&			&&&			A^{5}&0&0\\
						&&&				&&&			&&&			0&1&0 \\ 
						&&&				&&&			&&&			0&0&-A^{5} \end{pmatrix}\ee	
The $T$-matrix is not completely diagonalized in the idempotent basis, since it acts nontrivially on the spin structures (although it is diagonalized within each spin structure block). 


\section{Generalities on fermion condensation and tube categories} \label{generalities} 

The techniques used in Section \ref{C2_condense_sect} work more generally.
In this section we discuss the general case and some variants thereof.

\subsection{General comments on fermion condensation}
\label{General_comments_on_fermion_condensation}

In this section we will explore fermion condensation in more general settings. 
In Section \ref{gntf_condense} we make some general remarks on fermion condensation in a
generic unitary braided fusion category $\mcc$. 
In Section \ref{condense_transparent_fermion} we add the stipulation that the fermion we aim 
to condense is {\it transparent}, in that it braids trivially with the other particles in the theory. 
Section \ref{lift_and_condense} examines the more general case where the parent category $\mcc$ is not braided,
but the object we want to condense lifts to a fermionic object in the Drinfeld center $\mcz(\mcc)$.
Finally, Section \ref{spin_defects_condensation} sketches a construction in which the braiding 
of the parent theory is retained after condensation, and spin structure defects are bound to 
particles that the fermion braids nontrivially with. 

First, some preliminary remarks. 
In what follows we will work with a 
unitary braided fusion category $\mathcal{C}$ which contains a fermionic object $\psi$ that we aim to condense. 
We require that $\psi$ satisfy the following conditions:
\begin{itemize} 
	\item $\psi\ot\psi\cong\unit$
	\item The Frobenius-Schur indicator of $\psi$ is 1
	\item The topological twist of $\psi$ is fermionic, i.e. $\theta_\psi =-1$
	\item The braiding of $\psi$ with itself (see \eqref{statistics_inconsistency}) is equal to $-1$ times the identity
	\item The quantum dimension $d_\psi$ of $\psi$ is 1
\end{itemize}
Note that there is some redundancy in this list.
For example, 
in unitary theories we have the spin-statistics relation, and $d_\psi=1$ can be inferred from $\psi\tp\psi \cong \unit$.

\medskip

We remark that the above assumptions allow us to endow the object $\unit\oplus\psi$ 
with the structure of a {\it fermionic commutative algebra object} in $\mcc$.
We can more generally condense any fermionic commutative algebra object $\varphi$.
In the quotient category $\mcc/\varphi$, any simple summand of $\varphi$ becomes isomorphic to $\unit$,
or perhaps isomorphic to a direct sum of copies of $\unit$.

\subsubsection{Condensing non-transparent fermions in a braided category}  \label{gntf_condense}

In this subsection we will assume $\mcc$ is a unitary braided fusion category (UBFC).
We will also assume that $\psi$ is non-transparent, meaning that it braids non-trivially with at least one non-trivial object in $\mcc$. 
We will examine the case where $\psi$ is transparent in a later subsection.

We will proceed 
as in Section \ref{C2_condense_sect} and define a super pivotal category $\mcc/\psi$. 
Since $\psi$ is fermionic in spin and statistics, and since it braids non-trivially with 
at least one other non-trivial object in $\mcc$ (from its assumed non-transparency), we
must utilize both spin structures and the back wall construction in the condensation 
procedure. 

The objects of $\mcc/\psi$ are the same as the objects of $\mcc$.
Similarly, morphisms of the condensed theory are defined via morphisms in the parent theory:
the even and odd parts of the morphism super vector space are given (up to isomorphism) by
\begin{align}
\mor_{\mcc/\varphi}(a \to b)^0 & \cong \mor_\mcc(a \to b) \\
\mor_{\mcc/\varphi}(a \to b)^1 & \cong \mor_\mcc(a \to b \tp \psi).
\end{align}


To be more precise, the morphism space assigned to a disk with a boundary condition is the space of all back wall diagrams
modulo local relations.
(More concretely, the even part of this morphism space is the corresponding morphism space in the parent category $\mcc$,
and the odd part of this morphism space is (up to isomorphism) the morphism space of $\mcc$ 
obtained by adding $\psi$ to the boundary condition.)
In order for these relations to make sense, the disk must be equipped with a spin structure.
This morphism space is a super vector space, with the $\zt$-grading given by the number of $\psi$ endpoints in the diagram modulo 2.
Composition of morphisms is given by gluing diagrams together.
We can take the domain of the composition map to be the unordered tensor product (see \ref{koszul_signs})
of the two morphism spaces we are combining.
When doing computations it is necessary to choose an ordering and to take Koszul signs into account.

\medskip

\begin{figure}
\begin{center}
\includegraphics{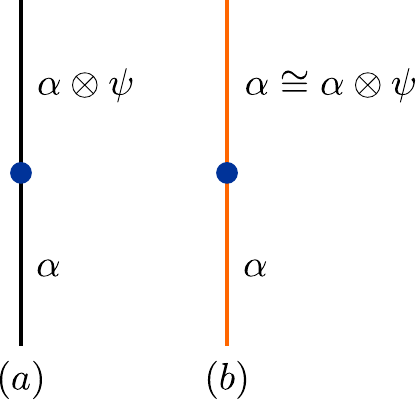}
\caption{ \label{mvsqtype} (a) An m-type particle $\alpha$. The fermion $\psi$ (blue dot) is 
an odd map from $\alpha$ to $\alpha\tp \psi$, with $\alpha\not\cong\alpha\tp\psi$. (b) A q-type particle $\alpha$ with $\alpha\cong\alpha\tp\psi$, where $\psi$ now acts as an odd 
endomorphism.
}
\end{center}
\end{figure} 

It follows from our assumptions about $\psi$ that if $\alpha$ is a simple object of $\mcc$, then
$\alpha\ot\psi$ is also a simple object of $\mcc$.
There are two cases:
\begin{itemize}
	\item If $\alpha\ot\psi$ is not isomorphic to $\alpha$, then $\alpha$ and $\alpha\ot\psi$ are both m-type
	objects of $\mcc/\psi$.
	While $\alpha$ and $\alpha\ot\psi$ represent distinct equivalence classes of simple objects in $\mcc$,
	they belong to the same equivalence class in $\mcc/
	\psi$.
	More specifically, $\alpha$ and $\alpha\ot\psi$ are oddly isomorphic in $\mcc/\psi$, 
	via a digram with a single $\psi$ dot (see Figure \ref{mvsqtype} (a)).
	\item If $\alpha\ot\psi \cong \alpha$, then $\alpha$ becomes a q-type simple object in $\mcc/\psi$.
	The odd endomorphism from $\alpha$ to itself is as shown in Figure \ref{mvsqtype} (b).
\end{itemize}

We will sometimes use the notation
\be
	\sob(\mcc/\psi) = \sobm(\mcc/\psi) \du \sobq(\mcc/\psi) ,
\ee
with $\sobm(\mcc/\psi)$ a complete set of representatives for m-type simple objects, and
$\sobq(\mcc/\psi)$ a complete set of representatives for q-type simple objects.

\medskip

$\mcc/\psi$ is not a braided category, since the back wall used in the condensation procedure 
prevents us from braiding particles completely around one another. 
However, it does have the structure of a front-braiding by $\mcc$, since we are allowed to 
pass $\psi$ worldlines between other particle worldlines and the back wall. 
(Another way of saying this is that $\mcc/\psi$ is a (fermionic) module 2-category for the 3-category $\mcc$.
Yet another way of putting it is that $\mcc/\psi$ is a codimension-1 defect between $\mcc$ and the vacuum.)

\medskip

A particularly simple class of condensed theories obtained from condensing a non-
transparent fermion in a braided theory are provided by the $C_{2(n+1)} = A_{4n+3} / \psi$ series, 
where $A_k$ is the category whose principal graph is the Dynkin diagram of the Lie algebra $\mathfrak{sl}_{k+1}$, 
and $C_k$ is the category whose principal graph is the Dynkin 
diagram of the Lie algebra $\mathfrak{sp}_{2k}$.
The choice $n=0$ 
corresponds to the Ising example considered in Section \ref{C2_condense_sect}.

\subsubsection{Condensing transparent fermions in a braided category}
\label{condense_transparent_fermion}

In this subsection we make the same assumptions about $\psi$ and $\mcc$ as in \ref{gntf_condense}, 
but we replace the assumption of non-transparency of $\psi$ with the 
assumption that $\psi$ is transparent in $\mcc$, meaning that it braids trivially with every other particle in $\mcc$.

Since $\psi$ is transparent, we do not need the back wall when performing the condensation.
We still need a spin structure, however (since $\theta_\psi = -1$), and we also need to keep track of 
Koszul signs (since $\psi$ has fermionic statistics).

The resulting category $\mcc/\psi$ is a fermionic braided pivotal category.
The absence of the back wall is what allows for $\mcc / \psi$ to possess a full braiding (rather than the 
``front-braiding'' forced on condensed theories where back walls are needed). 
We will construct an example of such a theory in Section \ref{so36}, where we condense a transparent fermion 
in the $SO(3)_6$ theory.

We note that q-type particles can never arise in the condensed theory if $\psi$ is transparent.
To see this, we note that if $\alpha\ot\psi = \beta$ for $\alpha$ and $\beta$ simple objects of $\mcc$, it follows that 
\be\theta_\beta = \theta_\alpha \theta_\psi = -\theta_\alpha,
\label{noq-type}\ee
where the first equality is because $\psi$ is transparent.
It follows that $\alpha \not\cong \beta$ in $\mcc$, and hence $\alpha$ descends to an m-type 
simple object in $\mcc/\psi$, since if $\alpha$ were q-type we would have $\alpha \tp \psi \cong \alpha$.

\subsubsection{Condensing objects that lift to fermions in the Drinfeld Center}
\label{lift_and_condense}

In this section we drop the assumption that $\mcc$ has a braiding.
We will describe how to condense an object $y$ of $\mcc$ which lifts to a fermion in the Drinfeld center $\mcz(\mcc)$. 
An instance of this construction is the $\halfesix$ theory we study in detail in Section \ref{halfesix}.

\medskip

\begin{figure}
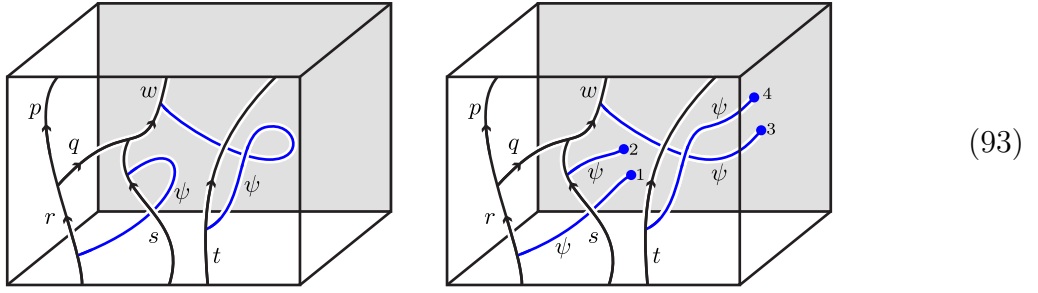

\begin{centering}
\begin{align}
\begin{array}{ccc}
\ZCPsiParent &&\ZCPsiCondensed  
\end{array}
\end{align}
\end{centering} 
\caption{\label{ZCPsiCondensed_fig}
On the left we have a 3d box. 
The front of the box is a 2d boundary on which $\mcc$ string nets are drawn (viewed as a module for $\mcz(\mcc)$). 
The bulk of the box contains braided nets from $\mcz(\mcc)$, which we restrict to $\psi$.
The back of the box can be viewed as a trivial boundary to the vacuum.
The right picture shows the box after performing back wall condensation on $\psi \in \mcz(\mcc)$.
The $\psi$ lines terminate on the back boundary at marked points $1,2,3,4$. 
}
\end{figure}

The basic idea is as follows.
The tensor category $\mcc$ can be thought of as a module for the braided category $\mcz(\mcc)$.
In terms of string net pictures, this means that $\mcc$ string nets can be thought of as living on the 2d boundary of a 3d bulk
of $\mcz(\mcc)$ string nets.
We can, if we like, restrict the bulk to only contain strings from a subcategory of $\mcz(\mcc)$, in particular the subcategory
generated by $\unit$ and the lift of $y$.
We can now do the back wall construction on the opposite side of the bulk and proceed as before.
See Figure \ref{ZCPsiCondensed_fig}.

\medskip

Now for a few more details.

To condense $y$, we first need to lift $y$ to $\mcz(\mcc)$, which means defining half-braidings for $y$. 
For a formal definition of half braid see 
\cite{muger2003b} (Definition 3.1 and Lemma 3.3).
A half-braid on an object $y \in \mcc $ is an isomorphism from $y \tp r \rightarrow r \tp y$ for each $r \in \mcc$,
the isomorphism being the half-braiding of $y$ with $r$. 
We will denote these isomorphisms as $e_y(r)$, and write them diagrammatically as
\begin{align} 
e_y(r)  = \halfbraid{\ebox}{y}{r}{}{}.
\label{halfbraid}
\end{align}
One should think of the box braiding $y$ under $r$. 
We could equally well think of this as braiding $y$ over $r$, but since we will use the ``back wall condensation" procedure to condense $y$, 
it is natural to choose $y$ braiding under $r$. 

Using the semisimplicity of $\mcc$ we can write $e_y(r)$ as
\begin{align}
e_y(r)  = \halfbraid{\ebox}{y}{r}{} = \sum_{w \in V^{ry}_{yr}} \left[ e_y(r)  \right]_{w} \halfbraid{\egeneral}{y}{r}{w},
\label{halfbraid_resolution}
\end{align}

Not any isomorphism $e_y(r)$ is allowed; $e_y(r)$ must satisfy some consistency conditions.
For example, braiding with the identity object should be trivial (up to unitors):
\begin{align}
\halfbraid{\ebox}{y}{\mathds{1}}{} = \text{id}_y.
\label{identityBraid}
\end{align} 
The most important property of the braiding isomorphisms is that they commute with fusion, meaning that they can freely slide past the fusion spaces $V^{ab}_c$:
\begin{align}
\halfbraidHex{\HalfBraidHexa}{y}{a}{b}{c} = \halfbraidHex{\HalfBraidHexb}{y}{a}{b}{c}.
\label{junctionBraid}
\end{align}
Taking \eqref{halfbraid_resolution} and inserting it into \ref{identityBraid} and \ref{junctionBraid} allows one to find the $\left[ e_y(r)  \right]_{w}$ defined in \eqref{halfbraid_resolution}. 

In order to do fermion condensation we require $y$ to lift to a fermionic object.
Fermionicity under exchanges means that
\begin{align}
e_y(y) =  -\; \text{id}_y \tp \text{id}_y.
\end{align}
If the quantum dimension of $y$ is 1, then the spin-statistics theorem will imply that the lift of $y$
is also fermionic with respect to twists.

Once a fermion has been defined one can proceed with the techniques of Sections \ref{gntf_condense} and \ref{condense_transparent_fermion}. 
We will provide an example of such a condensation in Section \ref{halfesix}.

\subsubsection{Condensing non-transparent fermions using spin defects instead of a back wall} \label{spin_defects_condensation}

In this section, we briefly comment on another way to condense non-transparent fermions through a type of flux attachement, 
which doesn't make use of the back wall construction and allows the condensed theory to remain braided 
(for a braided input category). 
For related ideas, see \cite{kapustin2017,thorngren2015}. 

The construction proceeds schematically as follows.
We allow the $\psi$ worldlines to be absorbed into the vacuum at any point.
In this picture the $\psi$ lines end anywhere in 3-space, they are not restricted to terminating on a back wall. 
In order to resolve problems with the twist and self-statistics of $\psi$, 
we must couple the $\psi$ endpoints to a spin structure and introduce Koszul signs, as before.
This yields an inconsistent theory if $\psi$ is not transparent (see \eqref{transparency_inconsistency}).

However, this inconsistency is rather mild. 
A natural way to distinguish the simple objects in $\mcc$ to use the $\mathbb{Z}_2$ 
grading inherited from the full braid of objects with $\psi$. 
This can be defined by the indicator 
\begin{align} 
(-1)^{\nu_x} := S_{a \psi}/S_{a0} \in \{+1, -1 \}
\label{grading}
\end{align}
It is easy to check that this grading is preserved under fusion.
Hence we can partition the simple objects in $\mcc$ as 
\begin{align} \label{braiding_indicator}
\text{sobj}\; \mathcal{C}  = I_0 \cup I_1,
\end{align}
where $I_k$ is the set of simple objects $x$ with $\nu_x = k$.
Since $I_a \tp I_b = I_{a+b\; \text{mod} \; 2}$, this is a $\zt$ grading of the simple objects. 
It can be shown that in any UBFC, $\nu_x = 1$ for all q-type objects (those for which $x\tp \psi \cong x$).
Indeed, if $x$ braided trivially with $\psi$ and $x\tp \psi \cong x$, then it would follow that
$\theta_x = \theta_x \theta_\psi = -\theta_x$, a contradiction.
Therefore, for a given object $a\in\mcc$, $\psi$ must either be transparent with respect to $a$, or be non-transparent only by a minus sign. 

We now observe that the problem in \eqref{transparency_inconsistency} would be 
surmounted if the ``box'' (the physical fermion attached to the $\psi$ endpoint) had a $-1$ 
braiding phase when taken around any anyon with which $\psi$ braids nontrivially. 
This extra $-1$ braiding phase cannot be due to the presence of any additional anyonic 
degrees of freedom, since the physical fermions braid trivially with any emergent anyon. 
However, there's actually a very natural way to do this: bind spin structure defects to the 
worldlines of anyons $x$ with which $\nu_x=1$. 
If the $x$ worldlines get bound to spin structure defects during the condensation 
process, then a box will pick up a factor of $-1$ when traveling around a $x$ line 
(when it passes through the branch cut), which cancels the $-1$ sign from the braiding of 
$x$ and $\psi$, and solves the transparency inconsistency. 
The spin structure defects have $\zt$ fusion rules, and so this procedure is consistent since 
the $\zt$ grading of objects in \eqref{braiding_indicator} is preserved under fusion. 
This allows the condensation to go through without the use of the back wall construction, 
which allows us to perform fermion condensation without sacrificing the existence of a 
braided structure. 

This picture also gives us a schematic physical interpretation of how to invert the condensation 
procedure. 
To go the other way and un-condense $\psi$, we just decouple the $x$ lines from the spin structure defects.
This gives us a phase consisting of the MTC $\mcc$ together with a loop gas of spin structure defects.
The loop gas of spin structure defects confine free $\psi$ endpoints, forcing all $\psi$ 
worldlines to be closed and restoring the original phase.

\subsection{The tube category of $\mcc/\psi$}

The quasiparticle excitations in a bosonic topological phase obtained from a category $\mcc$ 
are given by the simple objects in the Drinfeld center $\mcz(\mcc)$ \cite{levin2005}.
These excitations are also naturally described by minimal idempotents of a category called the tube category of $\mcc$ (see e.g. \cite{ocneanu2001,evans1995,Izumi2000,muger2003b,Bultinck2017,Lan2014, Hu2015}, 
also variously referred to as the tube algebra and the Q-algebra), 
which we will write as $\tube(\mcc)$.

The tube category was first introduced by Ocneanu \cite{ocneanu1994} and has since 
been dubbed ``Ocneanu's tube algebra"; it is also referred to as the annular category $\text{Ann}(\mcc)$, 
or the categorified degree zero Hochschild homology of $\mcc$.
It is closely related to the Drinfeld center:
If $\mcc$ is pivotal then there is a natural isomorphism
$\text{Rep}(\text{Ann}(\mcc)) \cong \mcz(\text{Rep}(\mcc))$, and if $\mcc$ is semisimple then we can drop
the $\text{Rep}$s and obtain $\text{Ann}(\mcc) \cong \mcz(\mcc)$.
With appropriate modifications accounting for spin structure issues, a similar construction holds 
in the more general fermionic setting considered here.

\medskip
We will now define two categories $\tube^B(\spc)$ and $\tube^N(\spc)$ for a string-net TQFT derived from a given super pivotal category $\spc$. 
We will postpone the most general definition of a super pivotal category until Section \ref{def_sect}. 
However, categories
obtained from fermion condensation $\spc \cong \mcc / \psi$ all constitute examples of super pivotal categories (and provide all the examples of super pivotal categories discussed explicitly in this paper), 
and so the reader may substitute $\mcc/\psi$ for $\spc$ in what follows before reading the more 
general definitions in Section \ref{def_sect}. 

The objects of the two spin tube categories $\tube^B(\spc)$ and $\tube^N(\spc)$ are defined as isomorphism classes of 
string-net boundary conditions on spin circles with bounding 
and non-bounding spin structures, respectively. 
For each spin tube category, we will fix a representative spin circle with which to define its
objects. 
Other possible choices of spin circles are related to the chosen representative
by spin diffeomorphisms, and the exact specification of the particular representative will 
be unimportant in our analysis. 

\begin{figure}
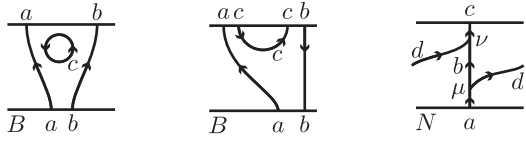

\begin{centering} 
\begin{align}
\nonumber
\TubeFiga \quad \quad \quad
\TubeFigb  \quad \quad \quad
\TubeFigc  \quad \quad \quad
\end{align}
\caption{\label{ExampleMorphismsTubeC}
Some examples of morphisms in $\tube(\spc)$,
with $\spc$ a super pivotal category obtained through a fermionic quotient ($\spc/\psi$) or more 
generally a category satisfying the assumptions laid out in Section \ref{def_sect}.
We have written the annulus as a square with (unmarked) left and right sides identified. 
At the bottom left of each annulus we have denoted the spin structure of the annulus by B (or N) for bounding (or non-bounding). 
The labels $a,b,c,d \in \sob(\spc)$, and $\nu \in \mor_{\spc}(d \tp b \ra c)$, and $\mu \in \mor_{\spc}(a \ra b \tp d)$.
}
\end{centering} 
\end{figure}

The morphism spaces of the tube category are finite linear combinations of
spin annuli decorated with different string-net configurations. 
Again, we will fix a particular representative spin annulus for each spin structure to use
when defining morphisms, with other choices being related by spin diffeomorphisms. 
Figure \ref{ExampleMorphismsTubeC}
shows some examples
of morphisms in the tube category.

Up to isomorphism, every object in the tube category is isomorphic to a direct sum of objects with a exactly one string net endpoint on the circle.
Put another way, the full subcategory spanned by such one-endpoint objects is Morita equivalent to the entire category.
This means that for purposes of, for example, enumerating equivalence classes of minimal idempotents or computing Hilbert spaces (ground states),
we can restrict our attention to the one-endpoint subcategory.
And indeed we will usually do so without comment.
(Note however that it is sometimes convenient to work in the larger category; see e.g.\ \ref{mtc_idem_subsect}.)

In the one-endpoint subcategory, morphism spaces can be presented as
\begin{align}
\text{mor}(\underset{a}{\TubeBCx{}} {\scriptstyle{X}} \rightarrow \underset{c}{\TubeBCx{}} {\scriptstyle{X}} )\; =\; \mathbb{C} \left[\AnnularTubexp{\AnnularTubeNoIndex}{a}{b}{c}{d}{\mu}{\nu}{X}{} \right] 
\end{align}
where $a,b,c,d$ are simple objects in $\spc$ and $X\in\{B,N\}$ denotes the spin structure of the annulus. 
The multiplicity indices $\mu$, and $\nu$ are 
collective indices that denote the vector in the fusion space $V^{db}_a$ and $V^{c \bar{d}}_b$, 
as well as the ordering of the tensor product ($V^{d b}_a \tp V^{c d^*}_b$ or $V^{c d^*}_b\tp V^{d b}_a$) which forms the Hilbert 
space of the tube.

We define the {\it full} tube category $\tube(\spc)$ to be the direct sum
\be \label{tube_spin_decomp} \tube(\spc) \cong \tube^B(\spc) \oplus \tube^N(\spc).\ee
In what follows we will mostly think of $\tube(\spc)$ as the fundamental category of interest. 
This is because in what follows we will be led to consider a $\tp$ structure on $\tube(\spc)$, which mixes the two spin components (we defer a discussion of this to 
Section \ref{fusion_rules}). 

Composition of morphisms in $\tube(\spc)$ is defined by stacking tubes together.
Graphically,
\begin{align}
\AnnularTubex{\AnnularTubeNoIndex}{a}{b}{\eta}{X}{} \cdot
\AnnularTubex{\AnnularTubeNoIndex}{c}{d}{\psi}{Y}{} \;
=\; \delta_{X,Y} \delta_{b,c}\AnnulusTubeTubex{\AnnulusTubeTube}{a}{d}{\psi}{\eta}{X}
\end{align}
The $\delta$ functions ensure that if the string labels of the two tubes don't agree on the 
boundary on which they are being glued (i.e. if $b\neq c$), or if the spin structures of the 
two tubes disagree, the two tubes compose (or multiply) to zero. 

\medskip

Before moving on, we briefly remark on the physical interpretation of the tube category. 
In Section \ref{Super_pivotal_Hamiltonian} we will define a Hamiltonian whose ground state wave functions assign amplitudes to string nets in such a way
that equivalent string nets receive equal amplitudes.
(In other words, the ground states are naturally identified with the string net TQFT Hilbert space $Z(Y; c)$,
where $Y$ is a spin surface and $c$ is a boundary condition.)
Let $S$ be a boundary component of $Y$, which we will think of as a puncture.
We can act on the space of string nets of of $Y$ (a.k.a.\ $A(Y; c)$) by gluing morphisms of the tube category 
(which are string-net-decorated spin annuli)
to $Y$ at $S$.
Dually, we get an action of the tube category on the ground state vector spaces $Z(Y; c)$ (for various values of $c$).
If we like, we can think of this action of the tube category as a scale transformation; see \cite{Lan2014}.
We can also think of it as a generalized symmetry of the Hamiltonian.
The collection of ground states $Z(Y; c)$ can be decomposed as a direct sum of irreducible representations of the tube category.
The irreducible representations of the tube category are thus identified with the elementary particles (i.e.\ anyons) of the theory.
Using the usual correspondence between minimal idempotents and irreducible representations, we can also identify
the minimal idempotents of the tube category with anyons.
See the end of Section \ref{ss_plaquette_term} for slightly more detail.

\subsubsection{Traces and inner products} \label{traces_and_innerproducts}


Recall that a trace on a super linear category $\scat$
is an even linear function from endomorphisms to $\cc$, satisfying
\be  \label{trace_cycle_rel}
	\tr(fg) = \tr(gf) ,
\ee
for all $f \in \mor(x\to y)$ and $g\in \mor(y\to x)$.
(Note that $\tr(f) = 0$ if $f$ is odd, since $\tr$ is an even function.
Note also that there is no Koszul sign in \eqref{trace_cycle_rel}.)

All of the categories we consider have a reflection structure\footnote{Usually called a $*$-structure,
but we are already committed to denoting the pivotal structure in tensor categories by $*$.}
(more specifically, a pin+ reflection structure),
which is an order 2 antilinear antiautomorphism of $\scat$
(i.e.\ $\mor(x\to y)$ is sent to $\mor(y\to x)$, for all objects $x$ and $y$).
We will denote the image of a morphism $f$ under reflection as $\bar f$.
The antiautomorphism property has the usual Koszul sign
\be
	\overline{fg} = (-1)^{|f||g|} \overline{g} \overline{f} .
\ee

A trace is equivalent to a collection of sesquilinear inner products on $\mor(x\to y)$ satisfying
\be
	\langle fg, h \rangle =(-1)^{|g||h|} \langle f, h \bar{g} \rangle \;\;\; {\rm and} \;\;\; \langle fg, h \rangle = (-1)^{|f||h|}\langle g,\bar{f}  h  \rangle 
\ee
The trace and inner products are related by
\be
\label{trace_to_innerproduct}
	\tr(f) = \langle f, \id_x \rangle   \;\;\; {\rm and} \;\;\;   \langle g,h \rangle = \tr(g\bar{h})
\ee
for $f\in\mor(x\to x)$ and  $g,h\in\mor(x\to y)$.
A trace is called nondegenerate if the corresponding inner product is nondegenerate in the usual sense (on each morphism space individually).

\medskip

We now recall two facts about TQFTs.
The first is that we can define a nondegenerate pairing on the predual Hilbert
space $A(Y; a_1,\ldots,a_k)$, where $Y$ is a spin surface with $k$ disjoint boundary components 
and the $a_i$ are tube category idempotents, satisfying
$\bar a_i = a^*_i$, which specify boundary conditions at each boundary component.
The nondegenerate pairing is defined via
\be
	\langle u, v\rangle = Z(Y\times I) (u\cup \bar v) ,
\ee
where the bar denotes the reflection map from 
$A(Y; a_1,\ldots,a_k)$ to $A(-Y; a_1,\ldots,a_k)$.
Here we are using the ``pinched boundary" condition for $Y\times I$, so that $\bd (Y\times I) = Y\cup -Y$.
In other words, we glue together the string nets $u$ and $\bar v$ to get a string net on $\bd (Y\times I)$, and then we evaluate
the path integral of $Y\times I$ with the $u\cup \bar v$ string net as the boundary condition.

The second fact concerns the path integral of 3-manifolds of the form $Y\times S^1_B$ or $Y\times S^1_N$.
Let $c$ be a string net on $(\bd Y)\times S^1$ (with either spin structure on $S^1$).
$(\bd Y)\times S^1$ is a disjoint union of tori, and we can cut these tori open into a disjoint union of annuli.
Let $c'$ denote the cut-open string nets on the annuli.
Note that $c'$ is not uniquely determined by $c$; $c'$ depends on where we cut the tori.
The string net $c'$ determines a linear map
\be
	g(c') : A(Y; a_i) \to A(Y; a_i) ,
\ee
given by gluing the $c'$ annuli on to the boundary of a string net on $Y$.
The gluing rules for the path integral imply that
\be
\label{bounding_trace}
	Z(Y\times S^1_B)(c) = \tr(g(c'))
\ee
and
\be
\label{nonbounding_trace}
	Z(Y\times S^1_N)(c) = \str(g(c')),
\ee
where $\str$ denotes the supertrace, which is the trace weighted by the fermion parity operator, i.e. $\str(f) = \tr((-1)^F f)$.
The association of the trace (supertrace) with the $B$ ($N$) spin structure was 
also noticed in \cite{turzillo2016}. 
Note that the partition functions above are independent of the details of the cutting procedure;
any choice of cutting curves and any choice of $c'$ will yield the same answer on the RHS above.


\medskip
We now apply the above to the case where $Y$ is an annulus to obtain a nondegenerate trace on the tube category.
In what follows we will continue to let $\spc$ be a super pivotal fusion category 
satisfying the assumptions of Section \ref{def_sect}
(e.g.~one coming from a fermionic quotient).
Letting $t \in \tube^W(\spc)$ and writing $\tr(t)$ for the trace of $t$, we can use \eqref{trace_to_innerproduct} to write
\begin{align}
\label{trace_t}
\tr (t)  = \langle t, \text{id} \rangle  = Z((S^1_W \times I) \times I ) (t \cup \overline{\text{id} })  = Z(S^1_W \times D^2) (t \cup \overline{\text{id}})
		= Z(S^1_W \times D^2) (\cl(t)) .
\end{align}
Diagrammatically we can write the evaluation of the partition function on the right hand side of \eqref{trace_t} as the trace of a matrix, through 
\begin{align}
\label{trace_formula}
\langle t,\text{id}\rangle = \left \langle\;\Tuberrprime_W,\text{id} \; \right \rangle =
 \tr_W \left \{ \Tuberrprimecut \; : A\left(  \PinchedDisk \scriptstyle{x\;} \right)  \xrightarrow{\quad \quad} A \left(  \PinchedDisk \scriptstyle{x\;} \right) 
 \right \}, 
\end{align}
where we have made use of the graphical representation of the endomorphism $t$ in the first equality. 
The solid cylinder in the third equality provides a linear map from the disk at the origin of the $x$ line to the disk at the end of the $x$ line, with each disk 
possessing a single marked point on its boundary labeled  $x$.
The subscript $W$ denotes that the solid cylinder was found by cutting open a cycle with spin structure $W$. 
From \eqref{bounding_trace} and \eqref{nonbounding_trace}, we see that $\tr_W$ will be a trace if $W$ is bounding, and a super trace if $W$ is non-bounding.
The source and target of this linear map is $\text{mor}(\unit \ra x)$, which we write as
the vector space assigned to a disc with one marked point labeled  $x$:
\begin{align}
A\left( \PinchedDisk  \scriptstyle{x\;} \right)  \cong \text{mor}(\unit \ra x) = \bigoplus_i \moronex,
\end{align}
with $\mu_i$ denoting a complete basis of morphisms for $\mor(\unit \to x)$ 
We allow for $x \cong a_1 \tp a_2 \tp \cdots \tp a_k$ so that a basis of $\mor(\unit \to x)$ could be relatively large.
Often $x$ will denote a simple object 
and so $\mor(\unit \to x)$ will be zero or one dimensional: 
\begin{align}
\label{Disk_Hilb}
A \left( \PinchedDisk \scriptstyle{x} \right)  = 
\begin{cases} 
\cc^{1|0} & \text{if $x$ is evenly isomorphic to $\unit$ } \\
\cc^{0|1} & \text{if $x $ is evenly isomorphic to $ \psi$ } \\ 
0 & \text{otherwise} \\
\end{cases}.
\end{align} 

The linear map induced by the cut on these basis elements is given by
\begin{align} 
\moronex \mapsto \moronexcollar = \sum_{j} g_{ij} \; \moronexj.
\end{align}
The coefficients $g_{ij}$ are found by reducing the middle diagram using local relations.
As usual, the trace is basis independent and any basis of $\text{mor}(\unit \ra x)$ will do.

Let us introduce the notation 
\begin{align} \label{sX_defn}
s(X) =\left\{
                \begin{array}{ll}
                  1 \quad &\text{if } X =B \quad \text{Bounding} \\
                  -1 \quad &\text{if } X = N \quad \text{Non-bounding}
                \end{array}
              \right.
\end{align}
Explicitly, \eqref{trace_formula} is then given by
\begin{align} 
\label{matrix_trace}
\tr_W(g) = \sum_{ j } g_{jj} s(W)^{|\mu_j|}.
\end{align} 
If $W$ is bounding then $s(W) = 1$ and \eqref{matrix_trace} is just $\tr(g)$. On the other hand, 
if $W$ is non-bounding then $s(W)  = -1$, and \eqref{matrix_trace} is the super trace $\str(g)$. 

This trace is defined for the tube category of any super pivotal category satisfying the assumptions of Section~\ref{def_sect}.
The inner product obtained from the trace is even, in the sense that $\langle v,w\rangle = 0$ if $|v|\neq|w|$. 

\medskip

In summary, the trace on the tube category is obtained as follows.
Start with a string net $t$ on the annulus $S^1_W\times I$.
Cut $t$ along an interval to obtain a string net on a square.
Rotate this square $\pi/2$ and then reglue (with bounding spin structure) to obtain a new annular string net $\text{rot}(t)$ on $S^1_B\times I$.
Gluing $\text{rot}(t)$ to the boundary of a disk induces a linear map $r_t$ on disk string nets.
If $W = B$ then $\tr(t) = \tr(r_t)$, where the $\tr$ on the RHS is the usual linear algebra trace.
If $W = N$ then $\tr(t) = \str(r_t)$, where the $\str$ on the RHS is the usual linear algebra super trace.
Note that $\text{rot}(t)$ and $r_t$ depend on the choice of initial cut, but $\tr(r_t)$ and $\str(r_t)$ are independent of this choice.

\medskip

We now compute some traces that illustrate the techniques 
described in this section, and which will also be of use to us later.

First we compute the norm squared of a single strand on an annulus with empty boundary conditions.
We denote by $\cl_B(x)$ the closure of $x \in \sob(\spc)$ on an annulus with bounding spin structure:
\begin{align}
\cl_B(x) \; = \; \underset{B}{\clx}
\end{align}
The norm squared of $\cl_B(x)$ is
\begin{align}
\langle \cl_B(x), \cl_B(x) \rangle = \tr_B \{ \id: \; \mor(\unit \to x \tp x^{*}) \ra \mor(\unit \to x \tp x^*)  \} .
\end{align}
Since $\mor(\unit\ra x\tp x^*) \cong \mor(x\ra x) \cong \End(x)$, we get
 \begin{align}
 \label{clxnorm}
 \langle \cl_B(x), \cl_B(x) \rangle = \dim \End(x) = 
\begin{cases} 
1 & \text{if $x$ is m-type} \\
2 & \text{if $x$ is q-type} 
\end{cases}.
\end{align}

We can also compute the quantum dimensions $d_{e_j}$ of the minimal idempotents in $\tube(\spc)$.
The un-normalized quantum dimension $\tilde d_{e_j}$ is defined by the trace of the idempotent: 
\begin{align}
\tilde d_{e_j} = \tr (e_j),
\end{align}
The normalized quantum dimension $d_{e_j}$ is then given by $\tilde d_{e_j}/\tilde d_{e_0}$, where $e_0$ is the trivial idempotent.
For example, in the $C_2$ theory, we can use this approach to obtain $\tilde d_{m_{\unit / \psi}} = 1/2, \tilde d_{m_\sigma^+} = \tilde d_{q_{\unit/\psi}} = 1/\sqrt{2}, \tilde d_{q_\sigma} = 1$. 
Normalizing so that $d_{m_\unit} = 1$, we obtain the normalized quantum dimensions listed in Table \ref{C2Data}. 

The total squared dimension $\mcd^2$ of the theory is defined to be 
$ \langle S^2_\phi, S^2_\phi \rangle$, the inner product of the empty diagram on the 2-sphere with itself.
We can then compute
\begin{align}
\langle S^2_\phi, S^2_\phi \rangle = Z(S^2 \times I)(\overline{S^2_\phi} \cup {S^2_\phi}) = \sum_{x \in \sob(\spc)} \frac{Z(B^3, \cl(x)) Z(B^3, \overline{\cl(x)}) }{\langle  \cl_B(x), \cl_B(x) \rangle},
\end{align}
where $B^3$ is the 3-ball and $Z(B^3,\cl(x))$ denotes the partition function 
of a 3-ball with a closed $x$ loop on its surface.
To obtain the second equality, we have written $S^2\times I$ 
as a union of two manifolds homeomorphic to $B^3$ glued along a bounding annulus and 
have made use of the gluing axioms of TQFT. 
We can compute $Z(B^3, {\cl(x)}) = Z(B^3, \overline{\cl(x)}) = d_x$ by definition of the quantum dimension, and hence the total squared dimension is given by
\begin{align}
\label{total_qdim_defn}
\mcd^2 = \sum_{x\in \sob(\spc)} \frac{d_x^2}{\dim \End(x)}.
\end{align} 

A similar derivation recovers \eqref{non-boundingNullVector}, 
where we pointed out that $\cl_N(\beta)$ is zero.
Let us see how this works out in the tube category $\tube(\spc)$.
Let $x \in \sob(\spc)$, it follows that
\begin{align}
\langle \cl_N(x),\cl_N(x) \rangle  = \tr_N\{ \id: \; \mor(\unit \to x \tp x^*) \ra  \mor(\unit \to x \tp x^*)  \}
= \begin{cases} 
1 & \text{if $x$ is m-type} \\
0 & \text{if $x$ is q-type} 
\end{cases}
\end{align} 
The last equality follows from $\mor(\unit \to x \tp x^*)  \cong \End(x)$ 
and that $\tr_N$ is the super trace of $\id: \; \End(x) \to \End(x)$.
More explicitly,
if $x$ is m-type then as a vector space $\End(x) \cong \cc$ and $\str\{ \id: \; \cc \ra \cc \} = 1$. 
On the other hand, if $x$ is q-type then as a vector space $\End(x) \cong \cc^{1|1}$ and
\begin{align} 
\str \{\id: \; \cc^{1|1} \ra \cc^{1|1} \} = \tr \{(-1)^F: \;  \cc^{1|1} \ra \cc^{1|1} \} =  1-1 = 0
\end{align}
Hence q-type idempotent closed up to an annulus with non-bounding spin structure has norm zero. 
Similarly, one can show that $\langle \cl_N(\gamma), \cl_N(\gamma) \rangle = 2$ 
when $x$ is q-type and gamma is an odd endomorphism such that $\gamma^2 = \id$.

We now show that closing up a q-type idempotent of $\tube(\spc)$ onto a torus results in a state with norm $\sqrt{2}$.
Let $x$ be a minimal idempotent of $\tube(\spc)$, 
and $\cl_W(x) \in A(T^2)$ be the string net found by closing $x$ onto a torus with spin structure $W$ 
along the newly closed cycle. 
First consider the case where $W = B$.
The norm squared of $\cl_B(x)$ is
\begin{align}
\langle \cl_B(x), \cl_B(x) \rangle = Z(T^2 \times I)(\overline{\cl_B(x)} \cup \cl_B(x)) = Z((S^1 \times I)\times S^1_B)(\overline{\cl_B(x)} \cup \cl_B(x)).
\end{align}
In the third equality we have rewritten the torus in a form where we can readily apply equation \eqref{bounding_trace} or \eqref{nonbounding_trace}.
The role of $Y$ is played by $S^1 \times I$, an annulus with spin structure determined by $x$.
Further, we can assume a boundary condition given by the idempotent $x$ on each boundary component of the annulus.
We will also assume that $\bar x = x$.
The linear map we need to take the matrix (super) trace of is 
just the identity map, so
\begin{align}
\langle \cl_B(x), \cl_B(x) \rangle = \dim \End(x)
\end{align}
A similar calculation for the non-bounding torus yields the same answer, 
however the q-type idempotents need to be closed with an odd endomorphism.
We can now justify the mysterious normalization factor introduced 
in \eqref{closing_q_type_C2} when closing up q-type idempotents on the torus.
A complete orthogonal basis for the torus is given by closing up a complete set of representatives of minimal idempotents.
To find a unitary S-matrix we require each of the basis states to have unit norm,
hence we divide closed up q-type idempotents by $\sqrt{2}$.

Lastly we point out a useful relation between the total dimension of a pivotal fusion category $\mcc$ and its fermionic quotient $\mcc/\psi$.
Let $\mcc$ be a pivotal fusion category, $\psi \in \mcz(\mcc)$ a fermion with $\psi \tp \psi \cong \unit$,  and $\mcc/\psi$ the fermionic quotient of $\mcc$,
then
\begin{align}
\label{DC_and_DCpsi}
\mcd_\mcc^2 = 2 \mcd^2_{\mcc/\psi}
\end{align}
if in addition we assume that $\mcc$ is a modular tensor category we also have,
\begin{align}
\label{TubeDim_and_dimCCpsi}
\mcd_{\tube(\mcc/\psi)}^2 = \mcd_\mcc^2 \mcd_{\mcc/\psi}^2
\end{align}
To show \eqref{DC_and_DCpsi}, we simply note 
\begin{align}
\mcd^2_\mcc = \sum_{x\in \sob(\mcc)} d_x^2 = \sum_{x \in \sobm(\mcc/\psi)} d_x^2 + d_{x \tp \psi}^2  + \sum_{q \in \sobq(\mcc/\psi)} d_x^2.
\end{align}
Since $d_{x \tp \psi}  = d_x d_\psi = d_x$ we can write
\begin{align} 
\label{dimCtoDimCpsi}
\mcd_\mcc^2 =2 \sum_{ x \in \sobm(\mcc/\psi)} d_x^2 + \sum_{x \in \sobq(\mcc/\psi)} d_x^2 = 2 \sum_{x \in \sob(\mcc/\psi)} \frac{d_x^2}{\dim \End(x)} = 2 \mcd_{\mcc/\psi}^2
\end{align} 
hence, \eqref{DC_and_DCpsi}. 
For the $C_2$ theory, this works out as 
$\mcd_{C_2} = \sqrt{2} = \mcd_{\rm Ising} / \sqrt{2}$.
Using \eqref{DC_and_DCpsi} and specializing to the case where $\mcc$ is a modular tensor category then $\tube(\mcc/\psi) \cong \mcc \times (\mcc/\psi) $ 
(a result that we prove in Section \ref{double_fermionic_quotient}), and we have 
\be \mcd_{\tube(\mcc/\psi)} = \mcd_{\mcc} \mcd_{\mcc/\psi}.\ee
For example, in the $C_2$ theory this is verified by $\mcd_{\tube(C_2)} = \sqrt{8} = \sqrt{2} (\sqrt{2})^2 = \sqrt{2} \mcd^2_{C_2}.$

\subsubsection{The sum-of-squares formula}

When analyzing tube categories, we are often presented with a collection of non-simple objects $x_1, x_2, \ldots$
(e.g.~single string net endpoints), together with the super dimensions of the vector spaces $\mor(x_i\to x_j)$.
From this we want to deduce a complete set of minimal idempotents $\{e_\alpha\}$, together with isomorphisms
\be \label{xi_decomp}
	x_i \cong \bigoplus_\alpha W_{i\alpha} \cdot e_\alpha 
\ee
where the $\{W_{i\alpha}\}$ are supervector spaces. 
We will show in \eqref{amordef} how to compute morphism spaces between objects that are 
of the form of those in the RHS of \eqref{xi_decomp}. 
Applying \eqref{amordef} yields
\be  \label{coldotprod}
	\mor(x_i\to x_j) \cong \bigoplus_\alpha \Hom(W_{i\alpha} \to W_{j\alpha})\otimes_\cc \End(e_\alpha) .
\ee
When $i\ne j$, this is merely an isomorphism of supervector spaces, but when $i = j$ this is an isomorphism of super algebras.

It is frequently possible, given the left hand side of \eqref{coldotprod} and a small amount of additional information, 
to solve for the things on the right hand side: the idempotents $\{e_\alpha\}$, their types, and the coefficients $W_{i\alpha}$.
This is useful since the morphisms that constitute the left hand side are often very easy to enumerate: they are simply the different tubes in $\tube_{x_i\ra x_j}$. 
In terms of super dimensions, setting $i=j$ we have 
\be \label{sum_of_squares} \dim(\mor(x_i\ra x_i)) = \sum_\alpha \dim(\End(W_{i\alpha}))\dim(\End(e_\alpha)).\ee
This is a fermionic ``sum-of-squares'' formula since $\dim\,\End(W_{i\alpha})$ will always be the square of 
an integer. 

For example, we can consider the space $\mor^X(\unit\ra\unit)$ in the $C_2$ theory.
Letting $X=B$ we see immediately from \eqref{mortube} that $\mor(x_i\ra x_i)$ has super dimension $2|0$, meaning that there must be two summands on the RHS of \eqref{sum_of_squares}, and hence 
two minimal idempotents in $\tube^B_{\unit\ra\unit}$. Letting $X=N$ we read off a super dimension of $1|1$, which implies that there is only one q-type idempotent in $\tube^N_{\unit\ra\unit}$. 
Similarly from \eqref{mortube} one verifies that the super dimensions of $\mor^X(\beta\ra\beta)$ 
are each $2|2$, meaning that each sector must contain either one m-type idempotent with $\dim\End(W_{\alpha})=2^2$ or two q-type idempotents $e_\alpha$ with 
$\dim\End(W_{\alpha})=1$ (although the latter is ruled out if $X=B$ since $\tube^B$ can 
host no q-type idempotents).
Thus, one can learn a good deal about the number of minimal idempotents and their type in the tube category 
simply from a list of the non-zero linearly independent tubes. 

In Section \ref{he6mp-tube-cat} we will see a further example of these techniques.

\subsection{Ground states on the torus} \label{ground_states_on_torus}

In what follows, we will let $T_{XY}$ with $X,Y\in\{B,N\}$ denote the torus with spin structure $XY$
(with spin structure $X$ along the meridional cycle and $Y$ along the longitudinal cycle),
and $A(T_{XY})$ 
the Hilbert space of ground-state string-net configurations on $T_{XY}$.  
As before, we will also let $\tube^B$ denote the bounding tube category and $\tube^N$ denote the non-bounding tube category. 

We have the following theorem, valid for any super pivotal category, which allows us to determine 
the ground states on a torus from the tube category:

\begin{theorem} \label{torus_basis_theorem}
Let $X$ equal $B$ or $N$.
The Hilbert space $A(T_{XB})$ is purely even, with an orthogonal basis given by closed-up idempotents $\{\cl(e_i)\}$,
where $e_i$ runs through a set of representatives of the minimal idempotents of $\tube^X$.
The Hilbert space of $T_{XN}$ is isomorphic to $\cc^{p|q}$, where $p$ is the number of m-type idempotents of $\tube^X$
and $q$ is the number of q-type idempotents of $\tube^X$.
An orthogonal basis is given by $\{\cl(e_i)\}$, where $e_i$ runs through a set of 
representatives of the minimal m-type idempotents of $\tube^X$,
union $\{\cl(\gamma_j)\}$,
where $\gamma_j$ runs through a set of representatives of odd endomorphisms of the minimal q-type idempotents of $\tube^X$.
\end{theorem}

The proof of the above claim consists of a sequence of fairly simple observations:
\begin{enumerate}

\item Let $c$ be an object of $\tube^X$ and let $f\in\End(c)$.
Recall that $f$ is a linear combination of string nets on the annulus (with spin structure $X$), with boundary conditions
$c$ on both boundary components of the annulus.
We can therefore glue up $f$ to obtain a new string net $\cl(f)$ (the closure of $f$) on either $T_{XB}$ or $T_{XN}$.
This map is clearly linear, so we have a linear map $\cl_c : \End(c) \to A(T_{XY})$, where $Y$ is either $B$ or $N$.
(For future reference, we will call the image of the boundary of the annulus in the torus $K$.)

\item Taking (finite) sums over different boundary conditions, we have a linear map
\be
	\cl : \bigoplus_c \End(c) \to A(T_{XY}) ,
\ee
where $c$ ranges over all possible boundary conditions.
(Note that even though $c$ ranges over an uncountable set, the direct sum consists only of finite linear combinations; 
there are no convergence issues.)

\item The map $\cl$ is surjective.
This is because any string net on the torus is, after a small isotopy, 
transverse to the gluing locus $K$.
But $\cl$ is very far from being injective, so our next task is to characterize the kernel of $\cl$.

\item One way of constructing elements in the kernel of $\cl$ is as follows.
Let $c$ and $d$ be two objects of $\tube^X$.
Let $g: c\to d$ and $h: d\to c$.
Then $gh \in \End(c)$ and $hg \in \End(d)$.
We have
\be
	\cl(gh) = (-1)^{|g| |h|} s(Y)^{|h|} \cl(hg) ,
\ee
where $s(B) = -1$ (antiperiodic) and $s(N) = 1$ (periodic).
This follows from the fact that the two string nets $\cl(gh)$ and $\cl(hg)$ are isotopic
via a ``shift" isotopy which pushes $h$ past the gluing locus $K$.
The factor of $(-1)^{|g| |h|}$ is the usual Koszul sign.
The factor of $s(Y)^{|h|}$ comes from sliding $h$ past the spin structure branch cut at $K$; see Figure~\ref{KernalCl}.
It follows that elements of the form
\be \label{cl_ker}
	gh - (-1)^{|g| |h|} s(Y)^{|h|} hg \in \bigoplus_x \End(x)
\ee
are in the kernel of $\cl$.
\begin{figure}
\begin{center}
\includegraphics{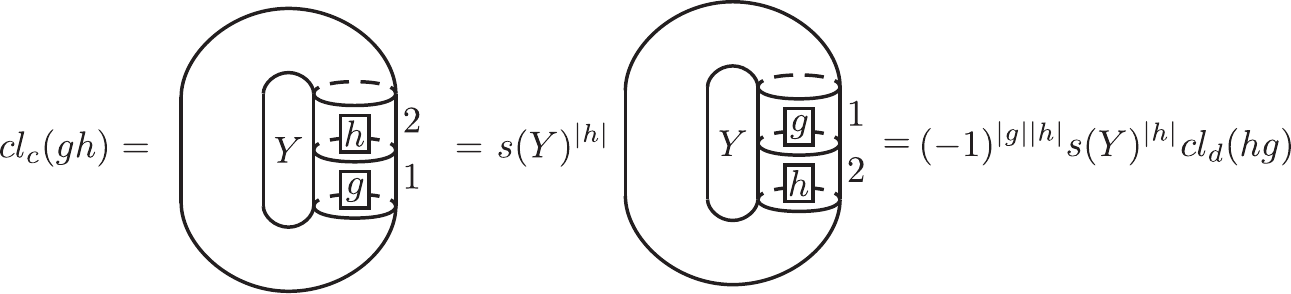}
\caption{ A graphical illustration of how to interchange morphisms in the tube category. 
The labels $1$ and $2$ denote the order in which the morphisms $g$ and $h$ appear in 
tensor products. The factor of $s(Y)$ comes from transporting $h$ around the torus, and the factor of 
$(-1)^{|g||h|}$ 
is the usual Koszul sign.
}
\label{KernalCl} 
\end{center}
\end{figure}

\item In fact, elements of the form \eqref{cl_ker} generate all of the kernel of $\cl$.
In the bosonic case, this is a standard fact (see \cite{Walker2006}).
The proof for the fermionic case is exactly the same, except that we have to keep track of
Koszul signs and signs coming from the spin structure.
The key idea of the proof is that any isotopy of the torus can be decomposed into (a) isotopies which
are fixed near $K$, and therefore can be lifted to isotopies of the annulus, and (b) a ``shift" isotopy
as described above.
In summary,
\be \label{torus_thm}
	A(T_{XY}) \cong \left( \bigoplus_x \End(x) \right) / \left\langle gh - (-1)^{|g| |h|} s(Y)^{|h|} hg \right\rangle .
\ee

\item In the semisimple case, the expression \eqref{torus_thm} can be greatly simplified.
Let $\{e_i\}$ be a complete set of minimal idempotents for $\tube^X$.
Any endomorphism $f$ can be written as a sum of endomorphisms of the form $f' e_i f''$.
Using \eqref{cl_ker}, we see that the subspace (of the big direct sum)
\be
	\bigoplus_i \End(e_i)
\ee
maps surjectively to $A(T_{XY})$.
Furthermore, because the minimal idempotents are orthogonal (in the sense that $e_i f e_j$ is zero for any $f$ unless $i=j$),
the only relations we have to consider are of the form
\be \label{t2_idem_rels}
	gh - (-1)^{|g| |h|} s(Y)^{|h|} hg
\ee
where both $g$ and $h$ are endomorphisms of $e_i$.
The theorem now follows.
\end{enumerate}

If $e_i$ is m-type, then \eqref{t2_idem_rels} is always zero and we get a summand of $\cc^{1|0}$ in $A(T_{XY})$.
If $e_i$ is q-type and $Y$ is $B$, then any odd endomorphism is of the form $\eqref{t2_idem_rels}$
and we get a summand of $\cc^{1|0}$.
If $e_i$ is q-type and $Y$ is $N$, then any even endomorphism is of the form $\eqref{t2_idem_rels}$
and we get a summand of $\cc^{0|1}$.

A useful corollary of Theorem \ref{torus_basis_theorem} is that all the idempotents of 
$\tube^B$ must be m-type.
Since $T_{BN}$ is spin diffeomorphic to $T_{NB}$, we can compute the Hilbert space for 
these spin surfaces in two different ways, one using
idempotents of $\tube^B$ and the other using idempotents of $\tube^N$.
By the first part of the theorem, the Hilbert space of $T_{NB}$ is purely even.
By the second part of theorem, the dimension of the odd part of the Hilbert space of $T_{BN}$
is given by the number of q-type idempotents of $\tube^B$.
Since the two Hilbert spaces are isomorphic, there can be no q-type idempotents in $\tube^B$.

A similar argument shows that the total number of (equivalence classes of) minimal idempotents of $\tube^B$
must equal that of $\tube^N$.

\subsection{Fusion rules} \label{fusion_rules}

In this subsection, we will define the fusion (tensor product) of representations of the tube category.

We begin with some general observations.
Let $Y$ be a spin surface with $k$ boundary components, denoted by $S_1,\ldots,S_k$.
Let $T_i$ denote the tube category corresponding to the circle $S_i$.
Each $T_i$ is (non-canonically) isomorphic to either $\tube^B$ or $\tube^N$.
(In order to make the isomorphisms canonical, we must choose a spin framing in each
boundary component.)
Given objects $c_i$ of $T_i$, for $1\le i \le k$, we have a super vector space $A(Y; c_1,\ldots,c_k)$ consisting of string nets
on $Y$, modulo local relations, with boundary conditions $c_i$ at $S_i$.

We can then glue morphisms of $T_i$ (tubes) onto $S_i$ to obtain a new string net with (possibly) different boundary condition.
A concise way to describe this algebraic structure is to say that the collection of super vector spaces $\{A(Y; c_1,\ldots,c_k)\}$
(for all possible values of $c_1,\ldots,c_k$)
affords a representation of the category $T_1\times\cdots\times T_k$.
We will denote this representation by $A(Y)$.

To define the fusion rules of excitations, we take $Y$ to be the 
pair of pants (a.k.a three-punctured sphere), which we will denote as $P$. 

There are four spin structures on $P$. 
In one of them, all three boundary components have a bounding spin structure, while in 
the remaining three, two out of the three boundary components have a non-bounding spin structure.
We will choose a standard representative for each of these spin pairs of pants, 
so that each boundary component is equipped with a spin diffeomorphism to a 
standard copy of $S^1_B$ or $S^1_N$.

Let $T_a$, $T_b$ and $T_c$ denote the three copies of the tube category associated to the boundary components of $P$.
Given representations $\rho_a$ and $\rho_b$ of $T_a$ and $T_b$, we can define a new representation of $T_c$,
denoted $\rho_a\tp\rho_b$, via
\be  \label{tctpdef}
	\rho_a\tp\rho_b = (\rho_a \boxtimes \rho_b) \tp_{T_a\times T_b} A(P) .
\ee
Here $\rho_a \boxtimes \rho_b$ denotes the ``outer" tensor product (so that $\rho_a \boxtimes \rho_b$ is a representation of $T_a\times T_b$), 
and $A(P)$ is the 
trimodule defined above, built out of
vector spaces of string-net configurations (modulo local relations) on $P$ with all possible boundary conditions. 
Informally, $\rho_a\tp \rho_b$ is found by taking a superposition of tubes carrying $\rho_a$ and $\rho_b$ 
(which is the outer product $\rho_a\boxtimes\rho_b$) and gluing them onto a pair of pants (given by $A(P)$). 
The algebraic implementation of gluing is the tensor product $\tp_{T_a\times T_b}$. 

If $\rho$ is the representation (i.e.\ module) determined by an idempotent $e$ (as described at the start of \ref{C2_fusion_rules}),
then the above association of $\rho$ to a boundary component is equivalent to imposing $e$ as a boundary condition
in a annular neighborhood of that boundary component.

Note that the spin structure grading of the tube category (and its modules) is respected by the above tensor product:
\begin{align}
B \tp B = B \quad \quad B \tp N = N \quad \quad N \tp N = B
\end{align}
where $B$ $(N)$ is shorthand for the bounding (non-bounding) sector of the tube category.

\subsection{Dimension formula}

In this subsection we give a Verlinde-type formula for the super dimension of the Hilbert space of a surface $Y$.

\subsubsection{The formula}   \label{dimformula_ss}

Let $Y$ be a spin surface with boundary components $U_1, \ldots, U_k$.
Each $U_i$ inherits either a bounding (a.k.a.\ antiperiodic or non-vortex) spin structure or a nonbounding 
(a.k.a.\ periodic or vortex) spin structure.

Let $a_1,\ldots, a_k$ be a set of labels for $\bd Y$.
Each $a_i$ is a minimal idempotent in the tube category $\tube^{U_i}(\mcc)$
(which is isomorphic to either $\tube^B(\mcc)$ or $\tube^N(\mcc)$); 
either a non-vortex anyon or a vortex anyon
(according to the spin structure on $U_i$).
The predual Hilbert space of string-net configurations on $Y$ with boundary conditions 
determined by the $a_i$ is $A(Y; a_1,\ldots, a_k) \cong \cc^{p|q}$ for some integers $p,q$.
Our goal is to compute $p$ and $q$.

\medskip

Let $S_{ab}$ denote the normalized, unitary $S$-matrix.
The indices $a$ and $b$ are closed-up idempotents on a spin torus.
They are specified by giving an idempotent (either bounding or nonbounding) together with
the way the annulus was glued 
to obtain the torus (again either bounding or nonbounding,
independent of the bounding/nonbounding status of the idempotent).
The idempotent $a$ is glued up in a (non)bounding way if $b$ is a (non)bounding idempotent, 
and vice versa.\footnote{
This is because spin structure on the cutting circle of one torus must
match the spin structure of the circle perpendicular to the cutting circle of the other torus.
}
If the idempotent is q-type, it is rescaled by $1/\sqrt 2$ to obtain a unit vector in $A(T^2)$, 
the vector space of string-net configurations modulo local relations on the torus
(see \ref{traces_and_innerproducts}).
(Note that there is still some ambiguity for entries in the $S$-matrix corresponding
to odd vectors in $T^2_{NN}$, but the formulas below will not use these $S$-matrix entries.)

Let $S'_{ab}$ be $S_{ab}$ if $a$ is m-type and $\sqrt 2 \cdot S_{ab}$ if $a$ is q-type.
Note that $S'_{ab}$ is asymmetric in $a$ and $b$;
we're undoing the normalization for $a$ but not for $b$.
The idempotents we will be summing over fall into three classes:
$B_m$ (bounding and m-type), $N_m$ (non-bounding and m-type), and $N_q$ (non-bounding and q-type).
Recall 
(from the discussion at the end of \ref{ground_states_on_torus}) 
that bounding idempotents are always m-type; in other words the potential fourth class $B_q$ is empty.

\medskip

We can now state the dimension formula:
We have
\be
	p + q = \sum_{x\in B_m} {S_{1x}}^{\chi(Y)} {\textstyle \prod_i} S'_{a_i x} 
	\label{PplusQ}
\ee
where $\chi(Y)$ is the Euler characteristic of $Y$.
If any of the $a_i$ are q-type, then we know that $p=q$ and we are done, since in that case 
$\cc^{p|q}$ has an odd isomorphism coming from the action of an odd element of $\End(a_i)$.

Now assume that none of the $a_i$ are q-type.
In this case we have
\be
	p - q = \sum_{x\in N_m} {S_{1x}}^{\chi(Y)} {\textstyle \prod_i} S'_{a_i x}
			\;\;+\;\; (-1)^{\Arf(Y)} \sum_{x\in N_q} {S_{1x}}^{\chi(Y)} {\textstyle \prod_i} S'_{a_i x} .
			\label{PminusQ}
\ee
Where $\Arf(Y)$ is the Arf invariant of $Y$.
If $Y$ is closed, then $\Arf(Y) = 0$ if $Y$ has a bounding spin structure and 
$\Arf(Y) = 1$ if $Y$ has a nonbounding spin structure.
For a torus, the $BB$, $BN$ and $NB$ spin structures are bounding and the $NN$ spin structure is nonbounding.
$\Arf(Y)$ for a higher genus spin surface can be determined by writing $Y$ as a connected sum of spin tori,
and using the fact that $\Arf(Y)$ is additive under connected sums.

If $Y$ has non-empty boundary and each boundary component has a bounding spin structure, 
we define $\Arf(Y)$ to be the Arf invariant of the closed 
surface obtained by capping each boundary component off with a disk.

If $Y$ has a non-bounding boundary component, say $U_1$, then by assumption
the label $a_1$ is m-type.
It follows that $S'_{a_1 x} = 0$ if $x$ is q-type, so there is no need to define $\Arf(Y)$ in this case.
We can see that $S'_{a_1 x} = 0$
because the torus basis vector corresponding to $x$ is odd (odd endomorphism of $x$
glued up periodically), while the basis vector corresponding to $a_1$ is even (the idempotent $a_1$
glued up periodically), and $S$ is an even operator.
This can be observed, for example, the $NN$ block of the $S$-matrix
for the $\halfesix/\psi$ theory, \eqref{hE6_S_NN}.

Note that the formula for $p+q$ above does not depend on the Arf invariant of $Y$, while the formula for $p-q$ does.
Thus the total dimension of the Hilbert space is not sensitive to the spin structure
of $Y$, but the even and odd Hilbert space dimensions do depend on the spin structure.

\subsubsection{Sketch of proof}

In this subsection we sketch the proof of the above dimension formula.

\newcommand{\ztt}{Z}
\newcommand{\zob}{Z_{S^1_B}}
\newcommand{\zon}{Z_{S^1_N}}

Let $\ztt$ denote the $2{+}1$-dimensional TQFT associated to the super pivotal category $\mcc$.
Let $Y$ be a closed spin surface with Hilbert space $\ztt(Y)$, and let $p|q = \dim(\ztt(Y))$.
Then we have
\be  \label{df1}
	p+q = \tr(\id: \ztt(Y) \to \ztt(Y)) = \ztt(Y\times S^1_B)
\ee
and
\be  \label{df2}
	p-q = \tr((-1)^F: \ztt(Y) \to \ztt(Y)) = \ztt(Y\times S^1_N) .
\ee
More generally, if $Y$ has nonempty boundary and the boundary components are labeled by $a_1,\ldots,a_k$ 
(minimal idempotents of the tube category), then
\begin{align}  \label{df3}
	p+q & = \tr(\id: \ztt(Y; a_1,\ldots,a_k) \to \ztt(Y; a_1,\ldots,a_k)) \\
		& = \ztt(Y\times S^1_B)(\cl_B(a_1)\du\cdots\du \cl_B(a_k))
\end{align}
and
\begin{align}  \label{df4}
	p-q & = \tr((-1)^F:  \ztt(Y; a_1,\ldots,a_k) \to \ztt(Y; a_1,\ldots,a_k)) \\
		& = \ztt(Y\times S^1_N)(\cl_N(a_1)\du\cdots\du \cl_N(a_k)) .
\end{align}
Here $\cl_X(a_i)$ denotes the element of $\ztt(T^2_{UX})$ obtained by closing up the idempotent $a_i$,
and $U$ is the spin structure ($B$ or $N$) on the $i$-th boundary component.
Note that if $a_i$ is q-type, then $\cl_N(a_i) \in \ztt(T^2_{NN})$ is zero.
But in this case we also know that $p-q = 0$, since the odd endomorphism of $a_i$ maps the even part of 
$\ztt(Y; a_1,\ldots,a_k)$ isomorphically to the odd part.

\medskip

Recall that we can define reduced $1{+}1$-dimensional TQFTs $\zob$ and $\zon$ via
\be
	\zob(M) = \ztt(M\times S^1_B) \quad\quad \zon(M) = \ztt(M\times S^1_N) ,
\ee
where $M$ is a manifold of dimension 0, 1 or 2.
Combining the above we have, for closed spin surfaces $Y$,
\be  \label{df5}
	p+q = \zob(Y)
\ee
and
\be  \label{df6}
	p-q = \zon(Y) ,
\ee
and there are similar formulas when $Y$ has boundary.

\medskip

We can now outline the remainder of the proof of the dimension formula.
We have just seen that the super dimension $p|q$ can be calculated entirely in terms of the reduced 
$1{+}1$-dimensional TQFTs $\zob$ and $\zon$.
Because 1 is a small number, we can completely classify $1{+}1$-dimensional spin TQFTs and give an explicit
expression for the path integral in terms of basic structure constants of the $1{+}1$-dimensional TQFT.
Then all that remains to be done is express the structure constants of the $1{+}1$-dimensional TQFTs in terms
of the structure constants of the original $2{+}1$-dimensional TQFT $\ztt$.
It turns out that the only structure constants we will need are the list of minimal idempotents of the tube category, 
their types (m or q), and the $S$-matrix.

\medskip

Spin $1{+}1$-dimensional TQFTs are determined by two pieces of data:
the cylinder category of a point, which is a linear super category $C$,
and a non-degenerate trace on $C$.
The trace is the path integral of the disk $D^2$; if $f$ is an endomorphism of $C$, then
\be
	\tr(f) = Z_{C,\tr}(D^2)(\cl(f)) ,
\ee
where, as usual, $\cl(f)$ denotes the closure of $f$, an element of the (pre-dual) Hilbert space associated to $S^1_B = \bd D^2$.
The non-degenerate trace implies that $C$ is semisimple.
Up to Morita equivalence, there are only two indecomposable semisimple super categories, the 
trivial algebra $\cc$ and the complex Clifford algebra $\cliff_1$.

We first consider the case $C = \cc$.
Let $e$ be the unique minimal idempotent of $\cc$ (i.e.\ $e = 1\in \cc$).
Let $\lambda = \tr(e)$.
Let $Y$ be a spin surface with $k$ boundary components, and let
$\cl(e)\du\cdots\du \cl(e)$ denote the boundary condition given by placing $\cl(e)$ on each boundary component of $Y$.
The path integral for the TQFT determined by $(\cc, \lambda)$ is
\be  \label{zc_ans}
	Z_{\cc,\lambda}(Y)(\cl(e)\du\cdots\du \cl(e)) = \lambda^{\chi(Y)} ,
\ee
where $\chi(Y)$ denotes the Euler characteristic of $Y$.

Next we consider the case $C = \cliff_1$.
Again let $e$ be the unique minimal idempotent of $\cliff_1$.
Let $\lambda = \tr(e)$.
Let $Y$ be a spin surface with $k$ boundary components.
We will assume that each boundary component of $Y$ has the bounding spin structure, since that is the only case
we will need for the dimension formula.
Let
\be
	\hat e = \frac{1}{\sqrt 2} \, e
\ee
be the normalized idempotent.
(The norm of $\cl(\hat e)$ in $A(S^1_B)$ is 1.)
The path integral for the TQFT determined by $(\cliff_1, \lambda)$ is
\be \label{zcl1_ans}
	Z_{\cliff_1,\lambda}(Y)(\cl(\hat e)\du\cdots\du \cl(\hat e)) = (-1)^{\Arf(Y)} \left(\frac{\lambda}{\sqrt 2}\right)^{\chi(Y)} ,
\ee
where $\chi(Y)$ denotes the Euler characteristic of $Y$,
and $\Arf(Y)$ is the Arf invariant of $Y$ with its boundary components capped off by disks.

A general $1{+}1$-dimensional spin TQFT is a direct sum of instances of the two theories described above.

\medskip

All that remains to be done is to write the reduced theories $\zob$ and $\zon$ as a direct sum of the $(\cc, \lambda)$ and
$(\cliff_1, \lambda)$ theories described above.

The first task is to obtain a list of the idempotents (and their type, m or q) of the minimal
idempotents of $\zob$ and $\zon$.
This is easily done: the category which $\zob$ assigns to a point is $\tube^B$, and the 
category which $\zon$ assigns to a point is $\tube^N$.
The m-type idempotents correspond to $(\cc,\lambda)$ theories, and the q-type idempotents
correspond to $(\cliff_1,\lambda)$ theories.

The second task is to determine, for each idempotent $a$ in $\tube^B$ and $\tube^N$,
the number $\lambda$ above (path integral of the disk evaluated on a closed-up idempotent).
This is exactly the $S$-matrix entry $S_{1a}$ if $a$ is m-type.
If $a$ is q-type, then (since we have normalized the $S$-matrix) $S_{1a}$ is equal to $\lambda/\sqrt 2$, but this
is the value we need for \eqref{zcl1_ans}.

The third and final task is to convert the $\cl(a_i)$ boundary conditions from the beginning of \ref{dimformula_ss}
to the $\cl(e)$ and $\cl(\hat e)$ boundary conditions appearing in \eqref{zc_ans} and \eqref{zcl1_ans}.
Both of these boundary conditions are (after undoing the dimensional reduction along $S^1_B$ or $S^1_N$) vectors
in $A(T^2)$, and both boundary conditions are closed-up idempotents (possibly normalized with a factor of $1/\sqrt 2$).
But the $\cl(a_i)$ boundary condition cuts the torus along a longitude, while the $\cl(e)$ and $\cl(\hat e)$ boundary conditions
cut the torus along a meridian.
So we need to apply the $S$-matrix (actually $S'$, because $\hat e$ is normalized while $a_i$ is not) to change basis:
\be  \label{df_cob}
	\cl(a_i) = \sum_x S'_{a_ix} \cl(\hat x),
\ee
where for convenience we have defined $\hat x = x$ if $x$ is m-type.

Combining \eqref{df5}, \eqref{df6}, \eqref{zc_ans}, \eqref{zcl1_ans}, and \eqref{df_cob}
yields the dimension formula.

\subsubsection{Sample calculations}

\begin{figure}\begin{center}  \tabulinesep = 1mm
\begin{tabu}{|X[$l]|X[$c]|X[$c]|X[$c]|}
	\hline
	& C_2 & SO(3)_6/\psi & \halfesix/y \\ \hline
	g=1, \Arf = 0 & 3 \;|\; 0 & 4 \;|\; 0 & 3 \;|\; 0  \\ 
	g=1, \Arf = 1 & 0 \;|\; 3 & 2 \;|\; 2 & 1 \;|\; 2  \\ \hline
	g=2, \Arf = 0 & 10 \;|\; 0 & 40 \;|\; 24 & 19 \;|\; 8  \\
	g=2, \Arf = 1 & 0 \;|\; 10 & 32 \;|\; 32 & 11 \;|\; 16  \\ \hline
	g=3, \Arf = 0 & 36 \;|\; 0 & 1184 \;|\; 1120 & 281 \;|\; 232  \\
	g=3, \Arf = 1 & 0 \;|\; 36 & 1152 \;|\; 1152 & 241 \;|\; 272  \\ \hline
	g=4, \Arf = 0 & 136 \;|\; 0 & 51328 \;|\; 51072 & 5755 \;|\; 5504  \\
	g=4, \Arf = 1 & 0 \;|\; 136 & 51200 \;|\; 51200 & 5531 \;|\; 5728  \\ \hline
	g=5, \Arf = 0 & 528 \;|\; 0 & 2368000 \;|\; 2366976 & 126449 \;|\; 125056  \\ 
	g=5, \Arf = 1 & 0 \;|\; 528 & 2367488 \;|\; 2367488 & 125137 \;|\; 126368  \\ \hline
\end{tabu}
\caption{Hilbert space dimensions for closed surfaces in various theories.} \label{dim_formula_fig1}
\end{center}\end{figure}

There are three specific $S$ matrices calculated in this paper, for the TQFTs based on the $C_2$, $SO(3)_6/\psi$, and $\halfesix/y$ theories.
Note that all three of these theories have just one non-trivial simple object.
Plugging the $S$-matrix entries into the above dimension formula, we find, for closed surfaces of genus $g$ and specified Arf invariant,
the results in Figure \ref{dim_formula_fig1}.

If we take $Y$ to be a 3-punctured sphere (with various spin structures),
then we can use the dimension formula to compute the fusion rules of the tube category.
For example, the fusion rules of Table \ref{TubehalfesixFusionRules} were computed using
the dimension formula.
Explicitly, we have
\begin{samepage}
\begin{align}
	\dim_{\text{even}}(V^{abc}) = \frac{\sqrt{n_a n_b n_c}}{2}\left(
			\sum_{x\in B_m} \frac{S_{ax}S_{bx}S_{cx}}{S_{1x}} +
			\sum_{x\in N_m \cup N_q} \frac{S_{ax}S_{bx}S_{cx}}{S_{1x}}
	\right)\\
\intertext{and}
	\dim_{\text{odd}}(V^{abc}) = \frac{\sqrt{n_a n_b n_c}}{2}\left(
			\sum_{x\in B_m} \frac{S_{ax}S_{bx}S_{cx}}{S_{1x}} -
			\sum_{x\in N_m \cup N_q} \frac{S_{ax}S_{bx}S_{cx}}{S_{1x}}
	\right) .
\end{align}
\end{samepage}
Recall that $n_a$ is defined to be $\dim(\End(a))$, i.e.\ 1 if $a$ is m-type and 2 if $a$ is q-type.

We remark that the dimension formula is a useful check on $S$-matrix accuracy.
Mistakes in calculating the $S$ matrix typically lead to non-integer outputs from the dimension formula.


\section{More on fermion condensation in modular tensor categories and the tube category} \label{more_on_tubes}

In this section we investigate $\tube(\mcc/\psi)$ when $\mcc$ is a modular tensor category.
If $\mcc$ is a MTC, it is a well known theorem that $\tube(\mcc) \cong \mcc \times \overline{\mcc}$ as braided tensor categories
(see for example Theorem 7.10 of \cite{muger2003b}). 
In the this section we will prove an analogous theorem for the super pivotal categories resulting from fermion condensation on MTCs.
Specifically, if $\mcc$ is a MTC we prove that 
\be \label{tube_theorem_teaser}
	\tube(\mcc/\psi) \cong \mcc \times \overline{\mcc/\psi}
\ee 
as tensor categories.
(Neither side of this equivalence is braided in the usual sense.)

The analogous result when $\psi$ is a boson is a special case of Corollary 4.8 of \cite{kawahigashi2001}
(see also the 1998 announcement by Ocneaunu referred to therein).

To begin, we remind the reader of this known result for $\tube(\mcc)$.
We then turn our attention to super pivotal categories of the form $\tube(\mcc/\psi)$ and make the necessary modifications.

\subsection{$\omega$ loops}

An essential tool in what follows will be the $\omega$ loop \cite{Lins1994}.
We take $\mcc$ to be a MTC and $\sob(\mcc)$ 
be the set of the simple objects of $\mcc$. 
The $\omega$ loop is defined by
\begin{align}
	\omega  = \frac{1}{\mathcal{D}^2}\sum_{x\in \sob(\mcc)} d_x \cdot \cl(x) ,
\label{omega_loop}
\end{align}
where as before, $\cl(x)$ denotes a closed loop labeled by $x$, i.e.\ the closure of $x$ inside a solid torus.

One way to think of the $\omega$ loop is as follows.
In any premodular category,
string nets in the solid torus (with the empty boundary condition) form a semisimple commutative algebra
(isomorphic to the fusion ring of the premodular category).
Therefore this vector space has a basis given by the minimal idempotents of the algebra structure.
The $S$-matrix gives a bijection between these idempotents and $\sob(\mcc)$.
The $\omega$ loop is the minimal idempotent in the solid torus corresponding to the trivial object of $\mcc$.

Diagrammatically, we will denote the $\omega$ loop embedded in an ambient 3-manifold by
\begin{align}
\OmegaLoopx{\omega} = \frac{1}{\mcd^2}\sum_{x\in \sob(\mcc)} d_x\; \LoopArrowx{x} ,
\end{align}
where the gray disk in the center 
indicates that this relation holds in the solid torus.
If the gray region is empty (i.e.\ if the solid torus is standardly embedded in the 3-ball), 
then the $\omega$ loop can be shrunk and evaluated using the rules of MTC (since in that case $\cl(x)$ is equal to $d_x$ times the empty diagram), 
and since $1 = \sum_x d_x^2/\mcd^2$, the $\omega$ loop simply acts as the identity. 
We summarize all the properties of the $\omega$ loop which we will make use of in Table~\ref{omega_loop_properties}.

For a modular theory, we can easily use part (a) of Table \ref{omega_loop_properties} to see that
\begin{align}
\OmegaLoopDefectx{\omega\;\;}{\omega\;\;} \quad \; =  \frac{1}{\mcd^2} \; \DiscGray.
\label{Omega_in_SLoop}
\end{align}
Note that this is true independent of what is inside the gray disc.
In the next section, we will see that this allows us to rewrite elements in the tube algebra in a particularly nice basis.
\begin{table}
\begin{center}{
\begin{flalign*} & \begin{array}{c @{\quad \quad  \quad  } c @{\quad \quad \quad } c }
			\text{(a)}	&	\text{(b)} &   \text{(c)}
		\\[1.0ex]
			\underset{x}{ \LoopOverId}{\scriptstyle{\omega}} = \delta_{x,\unit} \underset{x}{\Ida}
			&\scriptstyle{x}\;\IdxOmegaLoopa \;= \;\IdxOmegaLoopb\;\scriptstyle{x} 
			&	\HandleSlidea = \HandleSlideb
		\end{array} & \end{flalign*}
		}\end{center}
		\caption{\label{omega_loop_properties}
		All unlabeled lines in the above figures are $\omega$ loops as defined in \eqref{omega_loop}.
For a modular theory, the $\omega$ loop projects onto the vacuum as shown in part (a) (agreeing with the interpretation of $\omega$ as the minimal idempotent in the solid torus corresponding to the trivial object of $\mcc$).
Part (b) shows that arbitrary string-net lines can be deformed across any $\omega$ loop. 
Part(c) shows the same move as in part (b) but with an $\omega$ loop, 
rather than a single simple object $x$. }		
\end{table}

\subsection{Minimal idempotents of $\tube(\mcc)$, when $\mcc$ is a modular tensor category}   \label{mtc_idem_subsect}

The starting point for our proof of \eqref{tube_theorem_teaser} will be
a convenient set of minimal idempotents of $\tube(\mcc)$ \cite{ocneanu1994}. 
We will give two constructions for complete sets of minimal idempotents of $\tube(\mcc)$:
one set is more conventional (and appeared first historically), while the second 
is more suited to the proof of \eqref{tube_theorem_teaser}.

The first construction of a set of minimal idempotents utilizes annuli that possess only one marked point at each boundary.
A basis for the morphism space from $a$ to $b$ in 
the annular category $\tube(\mcc)$ is given by
\begin{align}
\text{mor}(\underset{a}{\TubeBCx{}} \rightarrow \underset{b}{\TubeBCx{}})=
\; \cc\left[ \TubeElementx{t}{a}{b} \right]\quad \quad \text{with} \quad \quad t \in \bigoplus_r V^{rb r^*}_a,
\end{align}
with $r$ in each summand labeling the string wrapping around the annulus. 
With the help of \eqref{Omega_in_SLoop} we can change to a much more convenient basis via
\begin{align}
\TubeBasisa \times \mcd^2 =  
\TubeBasisb   =
\TubeBasisc  = \sum_{\substack{x,y \in \sob(\mcc) \\
\mu =1,\cdots, N^{xy}_a \\
\nu =1,\cdots, N_{xy}^b}
}C_{t; xy\mu\nu} \times 
\TubeBasisd
\label{idempotent_complete}
\end{align}
where we have written the annulus as a rectangle with the left and right (blank) edges identified and
the indices $\mu$ and $\nu$ run over complete orthogonal bases of $V^{xy}_a$ and $V_{xy}^b$, respectively.
The constants $C_{t,xy\mu \nu}$ can be determined by fusing the $\omega$ loop into the strand labeled $b$ in the second to last diagram, 
and then using a series of $F$ and $R$ moves to reduce the diagram to the form of the final diagram on the right. 
Since $\mcc$ is assumed to be modular and all transformations shown 
above are invertible, we have shown that the morphism spaces can be alternatively presented as 
\begin{align}
\text{mor}(\underset{a}{\TubeBCx{}} \rightarrow \underset{b}{\TubeBCx{}})\cong \cc\left[ \TubeBasisd \right]. 
\label{AnnCatMTC}
\end{align}
Equivalently, we have shown that
\begin{align}
\text{mor}(\underset{a}{\TubeBCx{}} \rightarrow \underset{b}{\TubeBCx{}})\cong \bigoplus_{xy} V^{b}_{xy} \tp V^{xy}_a .
\label{morabMTC}
\end{align}
This basis for the morphism space of $\tube(\mcc)$ is rather special, and we will see that the diagonal elements (those with $a=b$) are proportional to the minimal idempotents.

Letting $\mu_i$, $i= 1, \cdots,N_{xy}^a$ be a basis of $V^{xy}_a$ and similarly letting $\nu_i$ be a basis for $V^{a}_{xy}$, we take the normalization convention 
\begin{align} \label{bubble_normalization} 
\VxyaoVxya = \delta_{ij} \frac{d_x d_y}{d_a} \idaprime.
\end{align}
It then follows that
\begin{align}
\Vxyxy\;  =\delta_{ij}\;  \idxy\; + \; \cdots, 
\end{align}
where the $+\dots$ represents diagrams that have a nontrivial string connecting the $x$ and $y$ strings (if admissible diagrams of such a form exist).
From the properties of the $\omega$ loop, we thus have 
\begin{align}
\Vxyxyomega \;=\delta_{ij}\; \idxy.
\label{vertex_normalization}
\end{align}
Our normalization is thus chosen so that there is no numerical prefactor in front of the right hand side 
of the above equality. 

With these conventions, we define a basis of morphisms by
\begin{align} \label{f_morphism_defn}
f_a^b(x,y,j,i)\; = \; \TubeBasisdprime,
\end{align}
where $\nu_j$ is a basis of $V_{xy}^b$ and $\mu_i$ a basis of $V^{xy}_a$, which are normalized according to \eqref{bubble_normalization}.
It follows that the $f_a^b$ morphisms compose as 
\begin{align}
f_a^b(x,y,j,i) \cdot f_b^c(x',y',j',i') = \delta_{xx'}\delta_{yy'} \delta_{ij'} f_a^c(x,y,j,i').
\label{matrix_units}
\end{align}
Therefore, the $f_a^b(x,y,j,i)$ is a basis of matrix units for $\tube(\mcc)$. 
Said another way,
$\tube(\mcc)$ 
(strictly speaking, the subcategory of $\tube(\mcc)$ spanned by objects with only a single marked point)
splits as a direct sum of full matrix categories\footnote{
Recall that a full matrix category is one in which each object is a finite-dimensional vector space,
and the morphisms between two objects are all linear maps.
A full matrix category with only one object is a full matrix algebra.
Full matrix categories are Morita trivial; all minimal idempotents within a full matrix category are equivalent to each other.
} 
labeled by pairs of simple objects in $\sob(\mcc) \times \sob(\mcc)$:
\begin{align}
\tube(\mcc) \cong \bigoplus_{xy} \text{Mat}(x,y),
\label{tube_matrix_algebra}
\end{align}
with the vector space associated to the object $a$ of $\tube(\mcc)$ at the $(x,y)$ summand being $V^a_{xy}$,
and 
\be
	\mor(a\to b) \cong \bigoplus_{x,y} \hom(V^a_{xy} \to V^b_{xy}) .
\ee
It follows from \eqref{matrix_units} that the ``diagonal'' morphisms 
\begin{align} 
e(a,x,y,j) = f_a^a(x,y,j,j)
\label{idempotent_one_strand}
\end{align}
are each a minimal idempotent. 
The idempotents $e(a,x,y,i)$ and $e(b,x',y',j)$ are equivalent\footnote{Two idempotents 
$e$ and $e'$ are equivalent if $e = uv$ and $e' = vu$ for some $u$ and $v$.}
if and only if $x \cong x'$ and $y \cong y'$, with the isomorphism given by 
$e(a,x,y,i) = f_a^b(x,y,i,j) \cdot f_b^a(x,y,j,i)$ and $e(b,x,y,j) = f_b^a(x,y,j,i) \cdot f_a^b(x,y,i,j)$.
This presentation of the minimal idempotents also appeared in \cite{ocneanu1994}.

It will be useful to have another complete set of minimal idempotents for $\tube(\mcc)$ at our disposal. 
These are the minimal idempotents that live in the annular category with two marked points on each of the circles bounding the annulus (rather than one marked point on each circle).
The idempotents are given by
\begin{align}
e_{xy} = \TubeIdempotentTwoStrand.
\label{Bosonic_twostrand_idempotent}
\end{align}

To show that the $e_{xy}$ are a complete set of minimal idempotents, 
we first show that $e_{xy} \tube(\mcc) e_{x'y'}$
is zero unless $x=x'$ and $y=y'$, in which case it is 1-dimensional.
(This implies that the $e_{xy}$ are minimal and pairwise orthogonal.)
Using the spine lemma,\footnote{
This is a well-known and easy-to-prove folk result which says that arbitrary string nets are 
equivalent to linear combinations of labeled spines.
We don't know a reference for this result. 
}
a basis for $e_{xy} \tube(\mcc) e_{x'y'}$ is spanned by
\begin{align}
\minimalBosonic \; = \; \minimalBosonicRHS.
\end{align}
The RHS is derived from the LHS by first sliding the $p$ loop over the lower $\omega$ loop, then sliding the lower $\omega$ loop
over the upper $\omega$ loop.
The ``no tadpole" axiom implies that the diagram is zero unless $r\cong\unit$,
and (a) of Table \ref{omega_loop_properties} implies that $c$ (and hence also $c'$) must be $\unit$ in any non-zero diagram.
This proves the claim.

Completeness of the idempotents follows from the resolution of the identity 
\begin{align}
\mcd^2 \TubeCompletea =  \TubeCompleteb = \frac{1}{\mcd^2}\sum_{x,y,i} \sqrt{\frac{d_xd_y}{d_a}} \;  \TubeCompletecprime .
\end{align}

It is easy to show directly that the idempotents $e_{xy}$ and $e(a,x,y,j)$ are equivalent.
Let
\begin{align}
\label{TwoStrandToOneStrand}
	g(b,x,y,j) =\; \gxyaj \quad \quad \text{and} \quad \quad h(a,x,y,i) = \; \hxyai .
\end{align}
Then we have
\be
	e_{xy} = g(a,x,y,j) \cdot h(a,x,y,j)
\ee
and
\be
	e(a,x,y,j) = h(a,x,y,j) \cdot g(a,x,y,j) .
\ee

The idempotents above can be used to show that $\tube(\mcc) \cong \mcc \times \overline{\mcc}$.
In the following subsection we will state and prove an analogous theorem for $\tube(\mcc/\psi)$.

\subsection{Double of the fermionic quotient}
\label{double_fermionic_quotient}

In this subsection we prove that $\tube(\mcc/\psi) \cong \mcc \times \overline{\mcc/\psi}$ as tensor categories
when $\psi$ is a fermion satisfying conditions of \ref{gntf_condense} and $\mcc$ is a modular tensor category.

To gain insights on the relation between $\tube(\mcc/\psi)$ and $\mcc$, 
we will first construct the minimal idempotents of the condensed theory, which are 
useful objects in their own right.
To facilitate this construction, we note that any string net configuration in the parent theory $\mcc$ descends to a string net configuration 
in the condensed theory $\mcc/\psi$.
Conversely, we can always take an even morphism in $\mcc/\psi$ and lift it to $\mcc$, 
giving us a way of lifting tubes in $\tube(\mcc/\psi)$ to those in $\tube(\mcc)$.
These two facts allow us to find the minimal idempotents of the quotient theory 
using knowledge of the minimal idempotents of the parent theory.
The details of the condensation functor $\tube(\mcc) \rightarrow \tube(\mcc/\psi)$ are important:
for example, the image of some of the idempotents may simply be zero, 
while the images of nonisomorphic idempotents of $\tube(\mcc)$ may map to the same isomorphism class in $\tube(\mcc/\psi)$. 
These details, as well as minimality and completeness of the idempotents, will have to be addressed carefully.
Once this is done, we arrive at the following theorem:

\begin{theorem}  \label{minimal_idempotents_modular_C/psi}
Let ${\mcc}$ be a modular tensor category and let $\psi$ be a fermion in ${\mcc}$ as in \ref{gntf_condense}.
Let $\mcc/\psi$ be the super pivotal category resulting from the fermionic quotient.
Let $\tube(\mcc/\psi) = \tube^B(\mcc/\psi) \cup \tube^N(\mcc/\psi)$
be the annular category of $\mcc/\psi$. 
Then as tensor categories,
\begin{align}
\tube(\mcc/\psi) \cong \mcc \times \overline{\mcc/\psi}.
\end{align}
In particular, $\sob(\tube(\mcc/\psi)) \cong \sob({\mcc}) \times \sob(\overline{\mcc/\psi})$. 
Let $a \in \sob(\mcc/\psi)$ and $\tilde{a} \in \sob({\mcc})$ be a lift of $a$.
If $\tilde a$ is transparent with respect to $\psi$, then $(x, a)$ is in the bounding sector $\tube^B(\mcc/\psi)$ of $\tube(\mcc/\psi)$
(for any $x\in\sob(\mcc)$).
If $\tilde a$ is not transparent with respect to $\psi$, then $(x, a)$ is in the non-bounding sector $\tube^N(\mcc/\psi)$.
\end{theorem}

The above result can also be written in terms of the tube category of the parent theory $\tube(\mcc)$. 
We first write 
$\mcc \times \bar\mcc/\psi \cong (\mcc/\unit)\times(\bar\mcc/\psi) \cong (\mcc\times\bar\mcc)/(\unit\times\psi)$.
Using the isomorphism $\tube(\mcc) \cong \mcc \times \bar\mcc$, we can embed 
$\psi \in \sob(\mcc)$ into $\tube(\mcc)$ by $\psi \mapsto \tilde\psi$, where $\tilde\psi \cong \unit \times \psi$. This means that as tensor categories, 
\be \tube(\mcc/\psi) \cong \tube(\mcc) / \tilde\psi,\ee
showing that fermion condensation commutes with constructing the tube category.

We prove the theorem by defining a tensor functor $E:\mcc \times \overline{\mcc/\psi} \rightarrow \tube(\mcc/\psi)$
and showing that it is an equivalence of tensor categories.
It is given by
\begin{align}
\xymatrix @!0 @M=2mm @R=15mm @C=48mm{
 &E: \; \; \mcc \times \overline{\mcc/\psi} \ar[r]            & \tube(\mcc/\psi) & \\
		  &\quad \quad  \;\; \left(\;  \EFunctora
\;,\;\EFunctorb \; \right)\;  \ar@{|->}[r] &\EFunctorc_{J(y)}&
		  },
		  \label{FunctorCxCtoTubeC}
\end{align}
where $J(y) = B$ if $\nu_y = 0$ and $J(y) = N$ if $\nu_y=1$, with $\nu_y$ the indicator 
defined in \eqref{grading}.
This is clearly a functor; it preserves composition of morphisms in an obvious way. 
To show that $E$ is a tensor isomorphism we need to show two things: 
\begin{enumerate} 
\item if $\{e_i \}$ and $\{f_j \}$ are a complete set of minimal idempotents for $\mcc $ and $ \overline{\mcc / \psi }$ respectively, 
then $\{ E(e_i, f_j) \}$ is a complete set of minimal 
idempotents for $\tube(\mcc/\psi)$,
\item  $E$ is a tensor functor. 
\end{enumerate} 

We first establish that $\{ E(e_i, f_j) \}$ are a complete set of minimal idempotents for $\tube(\mcc/\psi)$.
This is done in three parts, first we show completeness, then that the idempotents are non-zero, 
and finally that they are minimal and orthogonal. 

A complete basis of morphisms for $\tube(\mcc/\psi)$ is given by
\begin{align} 
 \TubeBasisaF_J, 
 \end{align}
with $t$ an even morphism of $\tube(\mcc/\psi)$ and $\alpha = \unit$ or $\psi$ denotes whether the 
morphism of $\tube(\mcc/\psi)$ has even fermion parity or odd fermion parity.
An even parity tube ($\alpha = \unit$) has $ t \in \bigoplus_x V^{x b x^*}_{a} $ 
while an odd parity tube ($\alpha = \psi$) has $ t \in \bigoplus_x V^{x b x^*}_{a\tp \psi} $.
Since $t$ is an even morphism in $\tube(\mcc/\psi)$, 
we can (trivially) lift it to $\tube(\mcc)$,
use the completeness relation in \eqref{idempotent_complete}, and then (trivially) include the morphism back into $\tube(\mcc/\psi)$. 
Hence we have
 \begin{align}
\TubeBasisaF_J
 =  
\frac{1}{\mcd^2}\sum_{\substack{x,y \in \sob(\mcc) \\
\nu \in V^b_{xy} \\
\mu \in V^{xy}_{a \tp \alpha}}}
C_{t; xy\mu\nu} \times 
\TubeBasisdF_J.
\end{align}
for some coefficients $C_{t; xy \mu \nu}$.
The morphism on the right hand side of the equation is isomorphic to $E(x,y)$.
Making use of the fact that $E(x,y) \cong E(x, y \tp \psi)$,\footnote{Note that $E(x,y)$ is not isomorphic to $E(x \tp \psi, y)$, see \eqref{Etilde} for more details.} 
allows us to replace the sum over $y \in \sob(\mcc)$ by a sum over $y \in \sob(\mcc/\psi)$.
Therefore, any morphism $t \in \tube(\mcc/\psi)$ can be written as
\begin{align} 
t = \sum_{k} x_k \cdot E(e_{i_k},f_{j_k} ) \cdot y_{k},
\end{align}
and hence the $\{ E(e_i, f_j) \}$ are complete. 

We now establish that the set of idempotents $\{ E(e_i, f_j) \} $ are non-zero.
We do this using the trace defined Section \ref{traces_and_innerproducts}.
We have
\begin{align}
\tr \left ( E(x,y) \right) &= \tr \left( \TubeIdempotentTwoStrand_{J} \right) \\
&= \frac{1}{\mcd^2} \sum_{r \in \mcc}  d_r\;   \tr \left( \IdempotentBasis_J \right)\\
&  =\frac{1}{\mcd^2} (1+s(J)(-1)^{\nu_y} )d_x d_y. 
\end{align}
The second line follows from linearity of the trace and the last line from,
\begin{align}
\tr \left( \IdempotentBasis_J \right) &=  \tr_J   \left\{ \TubeIdempotentBasistrace : A\left( \DiskLarge \right)  \ra A\left( \DiskLarge\right)  \right \}\\
&= (\delta_{r \unit} + s(J) (-1)^{\nu_y}\delta_{r \psi} )d_x d_y. 
 \end{align} 
We have used that when $\cl(\psi)$ is pushed past $y$ the trace picks up the phase $ (-1)^{ \nu_y}$;
see section \ref{traces_and_innerproducts} for more details on the trace.
Hence $\tr \left ( E(e_i,f_j) \right)$ is non-zero so long as $s(J) = (-1)^{\nu_{f_j}}$, 
which is true by definition of the $E$ idempotents (recall \eqref{FunctorCxCtoTubeC}).

Now we show that the $\{ E(e_i, f_j) \}$ are minimal and orthogonal. 
We do this by computing the dimension of $E(x,y) \cdot \tube(\mcc/\psi) \cdot E(x',y')$.
By the spine lemma, 
a generic element of $E(x,y) \cdot \tube(\mcc/\psi) \cdot E(x',y')$ can be written as the LHS of
\begin{align}
\fTubefOdd \; = \;  \ftubefOddRHS,
\label{MinimalProof}
\end{align}
with $\delta \in V^{x'}_{xc}, \; \rho \in V^{y'\tp \alpha}_{c y}, \; \sigma \in V^{c'c}_r, \; \kappa \in V^{pr}_p$.
We have suppressed the spin structure index (the spin structure is determined by $y$; recall \eqref{FunctorCxCtoTubeC}).
All vector spaces appearing are written in terms of the 
parent theory, $\mcc$, 
and $\alpha$ is either $\unit$ or $\psi$ denoting whether the tube is 
even or odd in $\tube^J(\mcc/\psi)$.

The LHS is equal to the RHS by sliding the $p$ strand over the lower $\omega$ loop 
then sliding the lower $\omega$ loop over the upper $\omega$ loop.
On the RHS,
the no tadpole axiom guaranties that $r \cong \unit$ and consequently that $c^* \cong c'$.
The remaining loop labeled $p$ can be removed at the expense of multiplying by its quantum dimension $d_p$.
Using property (a) of \ref{omega_loop_properties} we see that $c \cong \unit \cong c'$.
Using the orthogonality of the idempotents in the parent theory, we have
\begin{align}
\label{minimal_table}
E(x,y) \cdot \tube(\mcc/\psi) \cdot E(x',y') \cong
\begin{cases}
\cc^{1|0}& \text{if}\; x \cong x' \; \text{and}\; y \cong y' \; \text{and}\; y \not\cong y'\tp \psi \\
\cc^{0|1}& \text{if}\; x \cong x' \; \text{and}\; y \cong \psi \tp y' \; \text{and}\; y \not \cong y' \\
\cc^{1|1}& \text{if}\; x\cong x' \; \text{and}\; y \cong y'\; \text{and} \; y \cong y \tp \psi \\
0 & \text{otherwise}
\end{cases}
\end{align}
Taking $x' = x$ and $y' = y$, it follow that $E(x,y)$ is a minimal m-type idempotent if $y\not\cong y\tp\psi$
and is a minimal q-type idempotent if $y \cong y\tp\psi$. 
This also confirms that $E(x,y)$ and $E(x,y \tp \psi)$ are oddly isomorphic.
The orthogonality of the idempotents follows from the fourth line of \eqref{minimal_table}.
This completes the proof that $\{ E(e_i, f_j) \}$ is a complete set of minimal idempotents.

\medskip

The tensor structure on $\tube(\mcc/\psi)$ is initially defined on $\text{Rep}(\tube(\mcc/\psi))$ and 
then transferred to $\tube(\mcc/\psi)$ using semisimplicity ($\tube(\mcc) \cong \text{Rep}(\tube(\mcc))$ in the semisimple case). 
Consequently, we only need to show that $E$ induces a tensor functor from 
$\text{Rep}(\mcc \times \overline{\mcc/\psi})$ to $\text{Rep}(\tube(\mcc/\psi))$.
To establish this, we show that
\begin{align} \label{fusion_isomorphism}
V^{E(a,x),E(b,y)}_{E(c,z)} \cong V^{ab}_c(\mcc) \tp V^{xy}_z(\mcc/\psi),
\end{align}
where $V(\mcc)$ denotes the fusion space for $\mcc$ and $V(\mcc/\psi)$ denotes the fusion space for $\mcc/\psi$.
This isomorphism is established by the following figure:
\begin{align}
\FusionIsomorphismprime \; = \; 
\FusionIsomorphismprimereduced
\end{align}
with $\alpha = \unit$ for the even fusion space and $\alpha = \psi$ for the odd fusion space.
By the spine lemma, the internal lines (red), labeled $k,h,r,s,t,u,v,p,q$ (multiplicity indices suppressed) span the entire space of 
net configurations for $V(P)$ with marked points, $(a,x)$, $(b,y)$, 
and $(c,z)$ living at the boundary circles (as before, $P$ is the pair of pants). 
Near each boundary circle we have applied the corresponding minimal idempotent to each boundary condition. 
Using the arguments similar to those following \eqref{MinimalProof} we can simplify the diagram 
using the $\omega$ loop (green) relations of Table ~\ref{omega_loop_properties} to find the diagram on the right. 
One finds that $r \cong s \cong t \cong u \cong v \cong \unit$, $k\cong x$, $h\cong b$, 
and the left over $p$ and $q$ loops can be removed by multiplying the picture with their quantum dimensions. 
The span of the resulting simplified pictures is isomorphic to $V^{ab}_c(\mcc) \tp V^{xy}_z(\mcc/\psi)$.
Using semisimplicity, this implies that $E$ is a tensor functor.

\subsection{Modular transformations}

The explicit representation of the minimal idempotents allows us to compute the 
modular transformations for the condensed theory. 

We first examine the $S$ transformation on bounding spin tori (i.e.\ the three spin 
tori that have at least one bounding cycle). 
The $S$ transformation acts to interchange the longitudinal and meridional cycles of the torus, and so it acts as
\begin{align}
S:\;\; \DCSmatrixa\; \mapsto \; \DCSmatrixb \;=\; \STorusBasisa
\label{s_transformation}
\end{align}
In the first two pictures we have drawn 
the torus as an annulus with inner and outer boundaries identified, while in the last picture we have re-written 
the torus on the plane as a square with the top and bottom as well as left and right edges identified.
Additionally, recall that from the way we constructed the idempotents, 
if the spin structure along the azimuthal direction is bounding, then $b$
must be transparent with respect to $\psi$, and if the azimuthal spin structure is non-bounding, then $b$ must be non-transparent
with respect to $\psi$.
Since we are working with bounding spin tori, and since we always transform to the standard basis of idempotents, 
the spin structure can be inferred from context,
and so we will suppress the labels in some of the diagrams. 

We now need to perform a series of manipulations that returns the right hand side of \eqref{s_transformation} 
to a linear combination of pictures that are written in the standard basis (the same as the left 
hand side of \eqref{s_transformation} with the spin structures interchanged).
We first investigate the part of the diagram with the $a$ string and the $\omega$ loop:
\begin{align}
\frac{1}{\mcd^2} \; \Scalcaa 
\;=\; \Scalcab 
\;=\; \Scalcac
\;=\; \Scalcad
\;=\; \sum_{x\in \sob(\mcc)} \frac{1}{\mcd^2} \SMatrix{a}{x} \; \Scalcae
\label{Scalc_above}
\end{align}
We can now do the same for the $b$ loop,
\begin{align}
\label{ScalcBack}
\frac{1}{\mcd^2}\; \Scalcba 
\; = \; \Scalcbb
\; = \; \Scalcbc
\; = \; \Scalcbd
\; = \; \sum_{y \in \sob(\mcc)} \frac{1}{\mcd^2} \SMatrixx{b^{*}}{y} \;  \Scalcbe
\end{align}
where we have used that the $\omega$ loop is a projector onto the vacuum. 
In the last summation we need to replace $\sum_{y \in \sob(\mcc)}$ with $\sum_{y \in \sob(\mcc/\psi)}$:
\begin{align}
\label{interim_S_calc}
 \sum_{y \in \sob(\mcc)} \frac{1}{\mcd^2} \SMatrixx{b^{*}}{y} \;  \Scalcbe 
\; =\; \sum_{y \in \sob(\mcc/\psi)} \frac{1}{\mcd^2} \frac{1}{2^{n_y}}\left(  \SMatrixx{b^{*}}{y} \;  +s(J)\SMatrixxx{b^{*}}{y \tp \psi} \;  \quad  \right) \Scalcbe.
\end{align}
The factor of $s(J)$ appeared due to $\cl_J(\psi \tp y) = s(J) \cl_J (y)$, where the subscript $J$ means we close up $y$ around a cycle with spin structure $J$.
The normalization factor $2^{-n_y} = 1/\dim \End(y)$ is inserted so that we don't overcount 
the q-type simple objects from $\sob(\mcc)$ (recall for example, \eqref{dimCtoDimCpsi}). 
Using that $S_{b^* (y\tp \psi)} = (-1)^{\nu_b} S_{b^* y}$, and that $(-1)^{\nu_b} s(J) = 1$ by assumption, 
the right hand side of \eqref{interim_S_calc} can be simplified so that \eqref{ScalcBack} becomes
\begin{align}
\frac{1}{\mcd^2}\; \Scalcba\; = \; \sum_{y \in \sob(\mcc/\psi)} \frac{2}{2^{n_y}} \frac{1}{\mcd^2} \SMatrixx{b^{*}}{y} \;  \Scalcbe 
\end{align}
Putting all calculations together, and removing leftover $\omega$ loops (which provide an 
additional factor of $\mcd^{-2}$) we find that the matrix elements of the (un-normalized) $S$-matrix can be written as
\begin{align}
\label{Smatrix_final_step}
\DCSmatrixb =  \sum_{\substack{x \in \sob(\mcc) \\ y \in \sob(\mcc/\psi) }} \frac{2}{2^{n_y}}S_{ax} S_{b^*y}  \; \DCSmatrixh.
\end{align}
In the above formula, the $S_{ax}$ and $S_{b^*y}$ are matrix elements of the $S$-matrix in the original 
input theory $\mcc$ (which we assumed to be an MTC).
Note that $\nu_b$ must be $0$ if $J$ is bounding, $1$ if $J$ is non-bounding, and similarly for $\nu_y$.
 The simple object $y$ appearing in $S_{b^*y}$ on the right hand side of \eqref{Smatrix_final_step} is a trivial lift of the $y$ written in the closed up idempotent 
(recall that the first is a simple object of $\mcc$, while the latter is a simple object of $\mcc/\psi$).
One can change the representative of the isomorphism class of $y \in \mcc/\psi$ with an odd isomorphism $\text{mor}(y \ra \psi \tp y)$.
Under this odd isomorphism the right hand side of \eqref{Smatrix_final_step} picks up a factor of $s(J)(-1)^{\nu_b}$ which is equal to $1$ since $s(J(b)) = (-1)^{\nu_b}$.

In order for the $S$-matrix to be unitary, we need to normalize each q-type idempotent properly.
In the discussion following \eqref{minimal_table} we pointed out that $E(a,b)$ is q-type if $b$ is q-type.
Hence we can normalize our idempotents by re-scaling the q-type idempotents by a factor of $1/\sqrt{2}$. This results in the ``pseudo idempotents'' 
\begin{align} 
\widehat{E}(a,b) = E(a,b)/(\sqrt{2})^{n_b},
\end{align}
which have unit norm.
The resulting unitary $S$-matrix is given by
\begin{align}
\label{normalized_S_matrix}
\cl_W(\widehat{E}(a,b)) \xrightarrow{S^{JW \ra WJ}}  \sum_{\substack{x \in \sob(\mcc) \\ y \in \sob(\mcc/\psi) }} \frac{2}{(\sqrt{2})^{n_b+n_y}}S_{ax}  S_{b^*y}   \cl_J(\widehat{E}(x,y))
\end{align}
Note that $\cl_J(\widehat{E}(x,y))$ on the right hand side of \eqref{normalized_S_matrix} is zero unless $y$ is compatible with the spin structure inherited from the left hand side of \eqref{normalized_S_matrix};
explicitly $y$ must satisfy $s(W) = (-1)^{\nu_y}$.

The matrix elements of the $S$-matrix on the torus with non-bounding spin structure (periodic 
boundary conditions around both cycles) can be calculated in an analogous way.
The first half of the calculation remains the same as in \eqref{Scalc_above}.
The second half of the calculation changes only if $b$ is q-type, in which case the idempotent 
$E(a,b)$ is q-type, and has to be closed up on the torus with an odd endomorphism. 
As discussed in the caption of Figure \ref{C2Change_of_Basis}, closing an idempotent with an odd endomorphism always results in a sign ambiguity for the closed up idempotent.
In such a case we have:\footnote{
We have used $(S^z)_{xy} = \frac{1}{\mcd} \Szmatrix$.}
\begin{align}
\frac{1}{\mcd^2} \; \Scalcbadotprime  \; = \;  \Scalcbddotprime \; =\;  \frac{1}{\mcd} \sum_{y\in Q}  [ S^{\psi} ]_{yb} \; \Scalcbedot
\end{align}

This completes our calculation of the $S$-matrix of the condensed theory in 
terms of the modular data of the input theory.  

\medskip

The $T$-matrix is found by twisting one boundary of an idempotent by $2\pi$ before closing it up. 
For the annulus, the twisting is implemented by performing a $2\pi$ counterclockwise rotation of the inner $S^1$ with respect to the outer $S^1$.
The matrix elements are given by
\begin{align}
\cl_W(\widehat{E}(a,b)) \xrightarrow{\; \; \; T^{JW \ra J\widetilde{W}}\;\;\; } \theta_a \theta_b^* \cl_{\widetilde{W}}(\widehat{E}(a,b))
\end{align}
where again $J = J(b)$, and where $\widetilde{W}$ can be read off from Figure \ref{spin_str_mapping_class_group}.
The phases $\theta_a$ and $\theta_b$ are the twists of the lifts of $a$ and $b$ to the parent theory.
For example, in the $C_2$ theory one verifies that $\theta_{m_\unit}=\theta_\unit\theta^*_\unit,\theta_{m_\sigma^+}=\theta_\sigma\theta_\unit^*,\theta_{q_\unit} = \theta_\unit\theta^*_\sigma,\theta_{q_\sigma}=\theta_\sigma\theta^*_\sigma$ and $\theta_{\psi}=\theta_\psi\theta^*_\sigma$, where $\theta_\sigma = -A^3$. 

Note that if $J = B$ then replacing $b$ with $b\tp \psi$ changes the sign of the twist.
Since we can choose either $b$ or $b\tp\psi$ for the lift, this gives a sign ambiguity in the twist (for the 
$C_2$ theory, this is manifested by $\theta_{m_\sigma^+} = \theta_\sigma\theta^*_\unit,\theta_{m_\sigma^-}=\theta_\sigma\theta^*_\psi=-\theta_{m_\sigma^+}$).
This sign ambiguity is expected, since only $T^2$ has well-defined eigenvalues 
on idempotents (see the discussion near the beginning of Section \ref{C2_modular_mats}).


\section{Fermion condensation in $SO(3)_6$} \label{so36}

Here we provide more examples of fermion condensation in two theories which are closely related to each other:
$SU(2)_6$ and $SO(3)_6$. 
Each theory contains a fermion $\psi$, which we will condense. 
The main difference between these two theories is that in $SO(3)_6$ the fermion $\psi$ is transparent 
(i.e.\ it braids trivially with every other particle in the theory), while in $SU(2)_6$ it is not. 
This means that when condensing $\psi$ in $SO(3)_6$, we do not need to use the ``back wall'' construction employed earlier,
and the quotient theory will be braided. 
However, the transparency of $\psi$ also means that the $S$-matrix in the $SO(3)_6$ theory is degenerate, 
and hence the theory is not modular. 
In this case the lack of modularity is fairly benign, 
$SO(3)_6$ is a subcategory of the modular tensor category $SU(2)_6$.
This will allow us to infer the minimal idempotents of $SO(3)_6/\psi$ from the minimal idempotents of $SU(2)_6/\psi$.
From the minimal idempotents one can also compute the mapping class group action, which we will work out for the $SO(3)_6/\psi$ example.
First we will establish some notation for UBFC's with fermions.

\subsection{Fusion theory of $SU(2)_6/\psi$ and $SO(3)_6/\psi$}

\begin{figure} 
\centering
\includegraphics{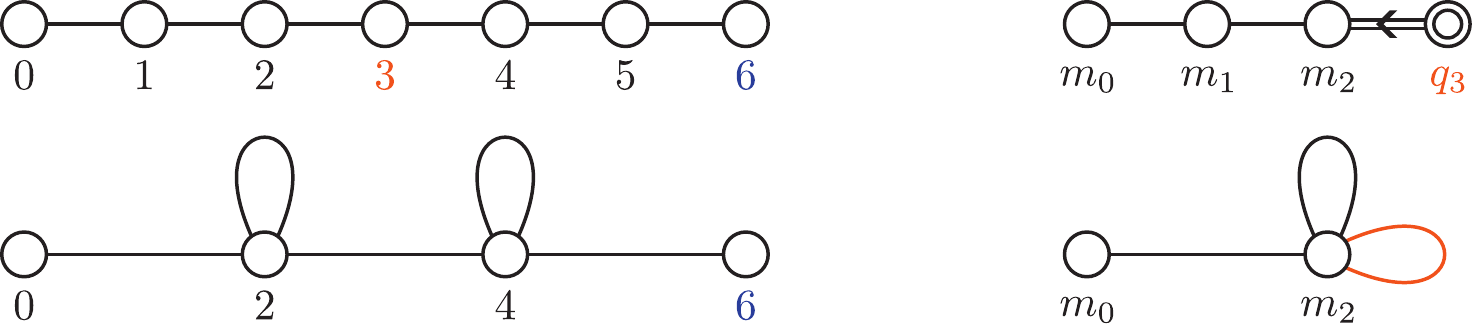}
\caption{\label{SUSOsix} The upper left diagram is the principle graph of $SU(2)_6$, and the 
lower left diagram is the principle graph for $SO(3)_6$. 
On the right we give the principle graphs of the condensed theories $SU(2)_6/\psi$ (top right) and 
$SO(3)_6/\psi$ (bottom right), both with the identification $\psi=6$. 
The naming convention of the condensed theories has been inherited from the parent theories, 
along with an $m$ or $q$ denoting whether the particle is m-type or q-type.
Black links denote even fusion channels, and the red link connecting $m_2$ to itself 
denotes an odd fusion channel in accordance with the rule $m_2\tp m_2 \cong m_0 \oplus \cc^{1|1}m_2$. 
}
\end{figure}

We will now briefly review $SU(2)_6$ and its connection with $SO(3)_6$.
Since these are well known theories we only list out some of their key properties and point the 
reader to some references for more details: see e.g. \cite{kirillow1989} and \cite{Bonderson2007}.
There are seven objects in $SU(2)_6$, labeled by $0,1,2,\cdots, 6$.
The principle graph for the theory is shown in the upper left of Fig.~\ref{SUSOsix}. 
The $0$ particle is the trivial object, $6$ is a fermion, and we have $6 \tp x = (6-x)$. 
Hence the particle $3$ is invariant under fusion with $6$, and so under condensation of the $6$ 
particle $3$ becomes a q-type simple object in $SU(2)_6/\psi$.
Since one m-type particle is always related to another by fusion with $\psi$ and 
there are six m-type particles, there are only three distinct equivalence classes of m-type
particles under fusion with $\psi$. 
We can take $\{0,1,2\}$ as the complete list of representatives.

We give the principle graph for $SU(2)_6/\psi$ in Fig.~\ref{SUSOsix} in the upper right, where $q_3$ is the q-type image of $3$ under condensation.
The particles $0,2,4,$ and $6$ form a closed sub-category of $SU(2)_6$. 
The principle graph of this theory is shown in the bottom left of Fig.~\ref{SUSOsix}. 
This is the subcategory known as $SO(3)_6$, it is a braided theory, with braiding and 
fusion inherited from $SU(2)_6$, however, it is not modular.
The $6$ particle braids trivially within this subcategory, and is therefore transparent, 
which breaks the modularity.

We now perform fermion condensation in $SU(2)_6$ and $SO(3)_6$ to obtain two super pivotal 
categories $SU(2)_6/\psi$ and $SO(3)_6/\psi$. 
Since $\psi$ is not transparent in $SU(2)_6$, we must perform the back-wall condensation 
process described earlier. 
However, since $\psi$ {\it is} transparent in $SO(3)_6$, condensation of $\psi$ is possible 
without employing a back-wall (although a spin structure is still needed). 
The principle graphs of the condensed theories are shown on the right of Fig.~\ref{SUSOsix}.
The simple objects of the two theories are given as follows:
\begin{align}
\xymatrix @!0 @M=1mm @C=10mm {
SU(2)_6/\psi:&&m_0 & m_1 & m_2 & q_3 \\
SO(3)_6/ \psi:&&m_0 && m_2& 
}
\end{align}
The particles have a natural grading given by (\ref{grading}) with the even set given by 
$I_0 = \{ m_0, m_2 \}$ and the odd set given by $I_1 = \{m_1, q_3\}$, with $I_a \tp I_b = I_{a+b \;  \text{mod} \; 2}$.
The closed sub-fusion algebra given by $I_0$ contains all of the objects in $SO(3)_6/\psi$, 
which occurs since $\psi$ is transparent in $SO(3)_6$. 
Note that there are no q-type objects in $SO(3)_6/\psi$. 

The non-trivial fusion rules of $SU(2)_6/\psi$ are given in Table \ref{SU2six_psi_fusion}.
\begin{table} 
\begin{align}
\xymatrix{
\text{
{\tabulinesep=1.2mm
\begin{tabu}{ c | c c  c c |c  c   }
$I_0 \tp I_0$&$m_0$ &$m_2$&$\quad\quad$&
$I_0 \tp I_1$&$m_1 $& $q_3$\\  
\cline{1-3}\cline{5-7}$m_0$&$m_0$ & $m_2$ &&
$m_0$& $m_1 $& $q_3$\\   
$m_2$ & $m_2$ & $m_0 \oplus \mathbb{C}^{1|1}m_2$&&
$m_2$ &$ m_1\oplus q_3 $ & $\mathbb{C}^{1|1} m_1 \oplus q_3$\\
\multicolumn{1}{r}{}&& &&\multicolumn{1}{r}{}&&\\
$I_1 \tp I_0$&$m_0 $& $m_2$ &&
$I_1 \tp I_1$ &$m_1 $& $q_3$\\  
\cline{1-3}\cline{5-7} $m_1$& $m_1 $& $m_1\oplus q_3$ &&
$m_1$ &$m_0 \oplus m_2$&$ \mathbb{C}^{1|1} m_2$ \\
$q_3$ &$ q_3  $ & $\mathbb{C}^{1|1} m_1 \oplus q_3$ &&
$q_3$ &$\mathbb{C}^{1|1} m_2 $&$ \mathbb{C}^{1|1}m_0 \oplus \mathbb{C}^{1|1}m_2$\\
\end{tabu}
}}
}
\end{align}
\caption{Fusion rules for $SU(2)_6/\psi$
\label{SU2six_psi_fusion}}
\end{table}

We note two features of these examples which were not present in our earlier $C_2$ example:
\begin{itemize}
\item Even though $m_2$ is an m-type particle, $\mathbb{C}^{1|1}m_2$ 
appears in the tensor product of $m_2$ with itself.
So $SO(3)_6/\psi$ provides us with an example of a theory which has 
no q-type objects, but which is still fermionic in the sense that its fusion spaces contain both even and odd elements. 
\item The q-type particle $q_3$ appears in the tensor product of two m-type 
particles, namely $m_2 \tp m_1$. 
Thus the classification of simple objects as m- or q-type should not be thought of as a $\zz/2$ grading, since the types of 
$a$ and $b$ in no way constrain the possible types of the simple objects appearing in $a\tp b$.
(In the next section we will also see an example of two q-type particles fusing to another q-type particle.)
\end{itemize}

The $F$-symbols of the condensed theories can be deduced from those of the parent theories, so we will not list them here.
We now compute the minimal idempotents of the tube category in the condensed theories.

\subsection{Minimal idempotents of $SU(2)_6/\psi$}
\label{SU2psiIdempotents}

The minimal idempotents of the tube category of $SU(2)_6/\psi$ can be computed directly 
using the techniques of Section \ref{more_on_tubes}. 
They are given by
\begin{align}
E(a,b) = \TubeIdempotentTwoStrandab_{J}, \quad \quad  (a,b) \in \sob(SU(2)_6) \times \sob(SU(2)_6/\psi),
\end{align}
where as before we require that $s(J) = (-1)^{\nu_b}$.
It will be useful to recast these idempotents in a form 
that can be easily used to compute the idempotents for $SO(3)_6$. 
We first define
\begin{align}
\widetilde{E}(a,b,\cl(r)) \; = \; \BasisForIdempotents
\end{align}
and for $x = \unit$ or $x = \psi$ we define the $\omega_x$ loop 
\begin{align}
 \omega_x = \frac{d_x}{\mathcal{D}_{\mcc/\psi}^2} \sum_{r\in \sob(\mcc/\psi)} \frac{d_r }{ \dim\End(r)} \left( \frac{S_{xr}}{S_{\unit r}} \right)   \cl(r).
\end{align}
The $S$-matrix above is that of the parent theory, and the labels are trivial lifts from $\mcc/\psi$.
When $x = \unit$, $ \frac{S_{x r}}{S_{\unit r}}=1$ this is just the standard $\omega$ loop, 
when $x =\psi$, $ \frac{S_{x r}}{S_{\unit r}}=(-1)^{\nu_r}$ and $\omega_\psi$ is a projector onto the $\psi$ strand.
The factor $\mcd_{\mcc/\psi}^2 = \sum_{x \in \mcc/\psi} d_x^2 / \dim \End(x)$ is the total quantum dimension of 
the condensed theory in the sense of \eqref{total_qdim_defn}. 
The minimal idempotents given by $E(a,b)$ in \eqref{FunctorCxCtoTubeC}
can be re-written as
\begin{align}
\label{Etilde}
E(a\tp x, b) \cong \widetilde{E}(a,b, \omega_x),
\end{align}
with $a,b \in \mcc/\psi$ and $x =\unit, \psi$.
The isomorphism relating the two idempotents is an odd isomorphism if $x = \psi$.
Notice that running over all all pairs $a,b \in \mcc/\psi$ and $x = \unit$ or $\psi$ runs over all possible $E(a,b)$.
Additionally if $a \tp \psi \cong a$ then $\widetilde{E}(a,b, \omega_\unit)$ and $\widetilde{E}(a,b, \omega_\psi)$ are equivalent. 
This presentation of the idempotents is more symmetric than those discussed in Section \ref{more_on_tubes}.

We now write these idempotents so that they have a single strand at the boundary rather than two.
We use the fermionic analogue of \eqref{TwoStrandToOneStrand}
to write down the single strand idempotents
\begin{align}
\label{OneStrandIdempotent}
\tilde{e}_{ab}(\omega_x,c,j) \; = \; \OneStrandIdempotentMTCpsi.
\end{align}
These idempotents are similar to the ones described earlier in \eqref{idempotent_one_strand}. 
The particular representative of the isomorphism class is given by choosing $c \in a\tp b$, 
and appropriately normalized vectors $\mu_j \in V^{ab}_c(\mcc/\psi)$, and $\nu_j \in V^{c}_{ab}(\mcc/\psi)$.
We will denote the parity of $\mu_j \in V^{ab}_c(\mcc/\psi)$ by $\sigma_j=1$ $(\sigma_j = -1)$ 
if the chosen basis vector $\mu_j$ in the fusion space $V^{ab}_c(\mcc/\psi)$ is even (odd).
The twists of the idempotents are given by 
\begin{align} 
\label{CmodPsiTwistsModular}
T\cdot \tilde{e}(a,b,\omega_x,c,j)  =
\begin{cases} 
  \frac{\theta_a}{\theta_b} (-1)^{x (\nu_a + \nu_b)}\sigma_j\;  \tilde{e}(a,b,\omega_x,c,j) & \text{if bounding}\\[.5em]
\frac{\theta_a}{\theta_b} (-1)^{x (\nu_a + \nu_b)} \; \tilde{e}(a,b,\omega_x,c,j) 
& \text{if non-bounding} \end{cases} 
\end{align} 
We will write $\tilde{e}(a,b,\omega_x,c,j)$ as $m_{ab}(\omega_x,c,j)$ 
if the idempotent is m-type and $q_{ab}(\omega_x,c,j)$ if the idempotent is q-type.
For $SU(2)_6/\psi$ we list a complete set of representatives of minimal idempotents.
We find 14 m-type idempotents for the bounding spin structure, 
and 14 idempotents for the non-bounding spin structure, 7 are m-type and 7 are q-type. 
Explicitly, the idempotents and corresponding twists are listed in tables \ref{bounding_SU26} and \ref{nonbounding_SU26}.
Note that if $a$ is q-type, then $\tilde{e}(a,b,\omega_\unit,j)$ is isomorphic to $\tilde{e}(a,b,\omega_\psi,j)$ and so we only list one of them.
We now turn our attention to the $SO(3)_6$ theory. 
\begin{table}
\begin{center}
\begin{align}
\nonumber
\xymatrix @M=0mm @R=5mm @C=5mm {
&\text{
\begin{tabu}{ c | c }
type& $\quad$twist $\quad$ \\ \hline
$m_{00}(\omega_0,c,j)$ &$1$ \\
$m_{10}(\omega_0,c,j)$ &$e^{3 i \pi /16}$ \\
$m_{20}(\omega_0,c,j)$ &$i$ \\
\end{tabu}
}
&
\text{
\begin{tabu}{ c | c }
type& $\quad$twist $\quad$  \\ \hline
$m_{00}(\omega_\psi,c,j)$ &$1$ \\
$m_{10}(\omega_\psi,c,j)$ &$-e^{3 i \pi/16}$ \\
$m_{20}(\omega_\psi,c,j)$ &$i$ \\
\end{tabu}
}
&
\text{
\begin{tabu}{ c | c }
type& twist $\times \; \sigma_j$ \\ \hline
$m_{30}(\omega_0,c,j)$ &$ e^{15i \pi /16}$ \\
$m_{32}(\omega_0,c,j)$ &$-ie^{15i \pi /16}$ \\
\end{tabu}
}
\\
&\text{
\begin{tabu}{ c | c }
type& twist $\times \; \sigma_j $ \\ \hline
$m_{02}(\omega_0,c,j)$ &$-i $ \\
$m_{12}(\omega_0,c,j)$ &$-i e^{3 i \pi/16}$ \\
$m_{22}(\omega_0,c,j)$ &$1$ \\
\end{tabu}
}
&
\text{
\begin{tabu}{ c | c }
type&twist $\times \; \sigma_j $ \\ \hline
$m_{02}(\omega_\psi,c,j)$ &$-i$ \\
$m_{12}(\omega_\psi,c,j)$ &$ie^{3 i \pi /16}$ \\
$m_{22}(\omega_\psi,c,j)$ &$1$ \\
\end{tabu}
}&}
\end{align}
\caption{ Bounding idempotents for $SU(2)_6/\psi$.
Where $c \in a \tp b$ and $j$ labels the choice of in the fusion space $V^{ab}_c (\mcc/\psi)$.
Some of these labels are determined, 
e.g., $m_{00}(\omega_0,c,j)$ can be simplified to $m_{00}(\omega_0,m_0,0)$.
The pre-factor $\sigma_j$ is $\pm1$ denoting the parity of $\mu_j \in V^{ab}_c$.}
\label{bounding_SU26}
\end{center}
\end{table}

\begin{table}
\begin{center}
\begin{align}
\nonumber
\xymatrix @M=0mm @R=5mm @C=5mm  {
&\text{
\begin{tabu}{ c | c }
type&$\quad  $ twist $\quad \;$ \\ \hline
$m_{01}(\omega_0,c,j)$ &$-e^{-3i\pi/16}$ \\
$m_{11}(\omega_0,c,j)$ &$1$ \\
$m_{21}(\omega_0,c,j)$ &$-i e^{-3i \pi /16}$ \\
\end{tabu}
}
&
\text{
\begin{tabu}{ c | c }
type& $\quad  $ twist $\quad \;$ \\ \hline
$m_{01}(\omega_\psi,c,j)$ &$-e^{-3 i \pi /16}$ \\
$m_{11}(\omega_\psi,c,j)$ &$1$ \\
$m_{21}(\omega_\psi,c,j)$ &$-i e^{-3 i \pi /16}$ \\
\end{tabu}
}
&
\text{
\begin{tabu}{ c | c }
type& twist \\ \hline
$m_{31}(\omega_0,c,j)$ &$e^{3 i \pi /4}$ \\
$q_{33}(\omega_0,c,j)$ &$1$ \\
\end{tabu}
}
\\
&\text{
\begin{tabu}{ c | c }
type& twist \\ \hline
$q_{03}(\omega_0,c,j)$ &$-e^{-15 i \pi /16}$ \\
$q_{13}(\omega_0,c,j)$ &$e^{- 3 i \pi /4}$ \\
$q_{23}(\omega_0,c,j)$ &$-i e^{-15 i \pi /16}$ \\
\end{tabu}
}
&
\text{
\begin{tabu}{ c | c }
type& twist \\ \hline
$q_{03}(\omega_\psi,c,j)$ &$-e^{-15 i \pi /16}$ \\
$q_{13}(\omega_\psi,c,j)$ &$e^{- 3 i \pi /4}$ \\
$q_{23}(\omega_\psi,c,j)$ &$- i e^{- 15i \pi /16}$ \\
\end{tabu}
}&
}
\end{align}
\caption{Non-Bounding idempotents for $SU(2)_6/\psi$
}
\label{nonbounding_SU26}
\end{center}
\end{table}

\subsection{Minimal idempotents of $SO(3)_6/\psi$}

Finding the idempotents in the $SO(3)_6/\psi$ theory is more difficult, since the parent theory $SO(3)_6$ isn't modular.
However since $SO(3)_6$ is obtained from $SU(2)_6$ by discarding the elements in $SU(2)_6$ 
that braid nontrivially with $\psi$, $SU(2)_6$ is a modular extension of $SO(3)_6$ (in fact, it is the minimal modular extension). 
This fact will allow us to compute the idempotents in the $SO(3)_6/\psi$ theory using our knowledge of the $SU(2)_6$ theory. 

Applying \eqref{OneStrandIdempotent} directly to $SO(3)_6$ will not yield a complete set of minimal idempotents due to the lack of modularity. 
Instead we use  \eqref{OneStrandIdempotent} by taking pairs of simple objects $(a,b)$ from $SU(2)_6/\psi \times SU(2)_6/\psi$ whose tensor product is in $SO(3)_6/\psi$.
Additionally, within this subset we need to take an appropriate linear combination of the $SU(2)_6/\psi$ idempotents so that the resulting annulus only has strands labeled by objects in $SO(3)_6/\psi$.
This linear combination is given by taking $\tilde{e}_{ab}(\omega_\unit, c,j)+\tilde{e}_{ab}(\omega_\psi, c,j)$ which results in the minimal idempotent
\begin{align}
\widetilde{e}_{ab}( \omega_\unit + \omega_\psi,c,j) \in \tube(SO(3))_6).
\end{align}
where the labels $(a,b) \in I_0/\psi \times I_0/\psi \cup I_1/\psi \times I_1/\psi$.\footnote{
Note that a minimal idempotent of 
$\tube(SO(3)_6/\psi)$ has a trivial lift to an idempotent of $\tube(SU(2)_6/\psi)$ but does not remain minimal.}
One can show that the procedure creates a complete set of minimal idempotents by direct calculation. 
The bounding idempotents are found when $(a,b) \in I_0/\psi \times I_0/\psi$, 
these are the minimal idempotents found from a naive application of \eqref{OneStrandIdempotent}. 
The non-bounding idempotents are given by $(a,b) \in I_1/\psi \times I_1/\psi$

We now write down the minimal idempotents of $SO(3)_6/\psi$ and their twists using the same notation as in Section \ref{SU2psiIdempotents}. 
In the bounding sector $J = B$, we have $4$ m-type idempotents, given by
\begin{align}
\xymatrix{
\text{
{{\tabulinesep=1.2mm
\begin{tabu}{ c c  }
type& twist \\ \hline
$m_{00}(\omega_\unit + \omega_\psi,0,0)$ &$1$ \\
$m_{02}(\omega_\unit + \omega_\psi,2,0)$ &$ -i $\\
$m_{20}(\omega_\unit + \omega_\psi,2,0)$ & $i$\\
$m_{22}(\omega_\unit + \omega_\psi,c,j)$ &$ \sigma_j $\\
\end{tabu}
}}}
} 
\end{align}
with $c \in m_2 \tp m_2$ and $j$ labeling a vector in the fusion space $V^{m_2 m_2}_c(\mcc/\psi)$.
Meanwhile the non-bounding idempotents are given by
\begin{align}
\xymatrix{
\text{
{{\tabulinesep=1.2mm
\begin{tabu}{ c c }
type& twist \\ \hline
$m_{11}(\omega_\unit + \omega_\psi,c,j)$ &$1$ \\
$q_{13}(\omega_\unit + \omega_\psi,2,j)$ &$ e^{-3 i \pi/4} $\\
$q_{31}(\omega_\unit + \omega_\psi,2,j)$ & $e^{3 i \pi /4} $\\
$m_{33}(\omega_\unit + \omega_\psi,c,j)$ &$ 1 $\\
\end{tabu}
}}}
}
\end{align}
Similarly, $c \in m_1 \tp m_1$ and $j$ labels a vector in $V^{m_1 m_1}_c$ for $m_{11}$, 
and $c \in q_3 \tp q_3$ and $j$ labels a vector in $V^{q_3 q_3}_c(\mcc/\psi)$ for $m_{33}$.
Here we've made use of the above observation that only $a \tp b$ need be in $SO(3)_6/\psi$, 
and hence the pairs $(a,b)$ in these idempotents are labeled by $m_1$ and $q_3$.
As the notation suggests $m_{33}$ is not q-type:
 to see this we note that if it were, then there would 
be an odd endomorphism of $\tube(SO(3)_6/\psi)$ with trivial boundary conditions (since $m_0 \in q_3 \tp q_3$).
That would require $SO(3)_6/\psi$ to contain a q-type simple object, 
but as explained in \eqref{noq-type} this is not possible since $\psi$ is transparent in $SO(3)_6$.
One can explicitly construct an odd endomorphism for $q_{13}$ by putting a fermion on $q_3$ and fusing all strands into the annulus;
similarly for $q_{31}$.

\subsection{Modular transformations}
\label{SO36ModularTransformations}

The modular transformations can be computed directly by manipulating string diagrams.
We already have the twists (which are read off from the string-net labels of the idempotents), 
and the modular $T$-transformation can be obtained directly from the twists.
The modular $S$-transformation requires a little more work, 
and so we discuss this in a little more detail.

We first close the minimal idempotents onto the torus. 
Using the basis above, one finds the image under $\text{cl}_Y$ is given by
\begin{align}
\tilde{e}_{ab}(\omega_\unit + \omega_\psi, c, j) \; \xrightarrow{\;\;  \cl_Y \;\;  }\; \tilde{e}_{ab} \;=\;  \TorusBraidBasis{a}{b\;}{}{\omega}{X}{Y}
\end{align}
where $a$ and $b$ label the idempotent as before, $X$ denotes the spin structure of the idempotent which is fixed by $b$, $Y$ denotes the spin structure around the newly closed cycle, 
and $\omega = \omega_\unit + \omega_\psi$. 
Note that when $X$ is non-bounding, 
$a, b\in \{ m_1, q_3 \}$, which lies outside of $SO(3)_6/\psi$.
However, $\tilde{e}_{ab} \in \cl_Y \tube(SO(3)_6/\psi)$ since when the diagram is fused into the torus we find a torus string net labeled with objects only in $SO(3)_6/\psi$, as required.

In this graphical convention, the modular $S$-transformation acts as
\begin{align}
{ \TorusBraidBasis{a}{b\;}{}{\omega}{X}{Y}} \stackrel{S}{\longrightarrow} \sum_{r,w} S^{{XY \rightarrow YX}}_{(a,b),(r,w)} \;
{\TorusBraidBasis{r}{w}{}{\omega}{Y}{X} }\; .
\end{align}
The matrix elements can be computed explicitly using the $S$-matrix of $SU(2)_6$. 
Explicitly, the $S$-matrix acts on each of the three bounding spin tori as 
\begin{align}
\left( \begin{matrix}
m_{00}\\
m_{02}\\
m_{20}\\
m_{22}\\
\end{matrix} \right)_{BB}
\xrightarrow{S^{BB \rightarrow BB}}
\frac{1}{2 \sqrt{2}} \left( \begin{matrix}
\frac{1}{d}&1&1&d \\
1& -\frac{1}{d} & d &-1 \\
1& d& -\frac{1}{d} & -1 \\
d & -1& -1& \frac{1}{d}\\
\end{matrix} \right)
\left( \begin{matrix}
m_{00}\\
m_{02}\\
m_{20}\\
m_{22}\\
\end{matrix} \right)_{BB}
\end{align}

\begin{align}
\left( \begin{matrix}
m_{11}\\
q_{13}\\
q_{31}\\
m_{33}\\
\end{matrix} \right)_{NB}
\xrightarrow{S^{NB \rightarrow BN}}
\frac{1}{2} \left( \begin{matrix}
1&1&1&1\\
1&-1&1&-1\\
1&1&-1&-1\\
1&-1&-1&1\\
\end{matrix} \right)
\left( \begin{matrix}
m_{00}\\
m_{02}\\
m_{20}\\
m_{22}\\
\end{matrix} \right)_{BN}
\end{align}

\begin{align}
\left( \begin{matrix}
m_{00}\\
m_{02}\\
m_{20}\\
m_{22}\\
\end{matrix} \right)_{BN}
\xrightarrow{S^{BN \rightarrow NB}}
\frac{1}{2} \left( \begin{matrix}
1&1&1&1\\
1&-1&1&-1\\
1&1&-1&-1\\
1&-1&-1&1\\
\end{matrix} \right)
\left( \begin{matrix}
m_{11}\\
q_{13}\\
q_{31}\\
m_{33}\\
\end{matrix} \right)_{NB}
\end{align}
As mentioned earlier, 
the Dehn twists follow directly from the twists computed above; 
see \eqref{CmodPsiTwistsModular}.
Again for the bounding spin tori we find
\begin{align}
\left( \begin{matrix}
m_{00}\\
m_{02}\\
m_{20}\\
m_{22}\\
\end{matrix} \right)_{BB}
\xrightarrow{T^{BB \rightarrow BN}}
\left( \begin{matrix}
1&&&\\
&-i&&\\
&&i&\\
&&&1\\
\end{matrix} \right)
\left( \begin{matrix}
m_{00}\\
m_{02}\\
m_{20}\\
m_{22}\\
\end{matrix} \right)_{BN}
\end{align}

\begin{align}
\left( \begin{matrix}
m_{11}\\
q_{13}\\
q_{31}\\
m_{33}\\
\end{matrix} \right)_{NB}
\xrightarrow{T^{NB \rightarrow NB}}
 \left( \begin{matrix}
 1&&&\\
&e^{-3 i \pi /4}&&\\
&&e^{3 i \pi /4}&\\
&&&1\\
\end{matrix} \right)
\left( \begin{matrix}
m_{11}\\
q_{13}\\
q_{31}\\
m_{33}\\
\end{matrix} \right)_{NB}
\end{align}

\begin{align}
\left( \begin{matrix}
m_{00}\\
m_{02}\\
m_{20}\\
m_{22}\\
\end{matrix} \right)_{BN}
\xrightarrow{T^{BN \rightarrow BB}}
\left( \begin{matrix}
1&&&\\
&-i&&\\
&&i&\\
&&&1\\
\end{matrix} \right)
\left( \begin{matrix}
m_{11}\\
q_{13}\\
q_{31}\\
m_{33}\\
\end{matrix} \right)_{BB}
\end{align}
One can verify that $(TS)^3 = \text{id}$, which holds since all three of the spin tori discussed above have 
 even fermion parity (recall that the more general identity is $(ST)^3 = (-1)^F$). 

The modular $S$ transformations on the torus with $NN$ spin structure are a little more tedious to calculate. 
A complete basis 
of net configurations on the $NN$ spin torus is given by
\begin{align}
\underbrace{
\xymatrix{
\TorusBraidBasisd{\scale{.6}{1}}{\scale{.6}{1}\;}{}{\omega}{N}{N}{\TorusBasisMTCNoLabel} \quad 
 \TorusBraidBasisd{\scale{.6}{3} }{\scale{.6}{3} \;}{}{\omega}{N}{N}{\TorusBasisMTCdd}
 \\}
 }_{\text{even}}
 \quad \quad 
\underbrace{
\xymatrix{ \TorusBraidBasisd{\scale{.6}{1}}{\scale{.6}{3}\;}{}{\omega}{N}{N}{\TorusBasisMTCdr} \quad
 \TorusBraidBasisd{\scale{.6}{3} }{\scale{.6}{1} \;}{}{\omega}{N}{N}{\TorusBasisMTCdl}
 \\}
 }_{\text{odd}}
\end{align}
On the non-bounding torus, m-type idempotents always close up into even parity states, 
while q-type idempotents close up into odd parity states.
The first two vectors have even parity and provide a basis for $\cl_N(m_{11})$ and $\cl_N(m_{33})$, 
while the second two have odd parity and provide a basis for $\cl_N(q_{13})$ and $\cl_N(q_{31})$.
After some calculation one finds that the $S$ and $T$ modular matrices act as 
\begin{align}
\left( \begin{matrix}
m_{11}\\
m_{33}\\
\stackrel{\bullet}{q}_{13}\\
\stackrel{\bullet}{q}_{31}\\
\end{matrix} \right)_{NN}
\xrightarrow{S^{NN \rightarrow NN}}
 \left( \begin{matrix}
1&0&&\\
0&1&&\\
&&e^{-i\pi/4}&0\\
&&0&e^{i \pi /4}\\ 
\end{matrix} \right)
\left( \begin{matrix}
m_{11}\\
m_{33}\\
\stackrel{\bullet}{q}_{13}\\
\stackrel{\bullet}{q}_{31}\\
\end{matrix} \right)_{NN}
\end{align}
and 
\begin{align}
\left( \begin{matrix}
m_{11}\\
m_{33}\\
\stackrel{\bullet}{q}_{13}\\
\stackrel{\bullet}{q}_{31}\\
\end{matrix} \right)_{NN}
\xrightarrow{T^{NN \rightarrow NN}}
 \left( \begin{matrix}
1&0&&\\
0&1&&\\
&&e^{-3i\pi/4}&0\\
&&0&e^{3i \pi /4}\\ 
\end{matrix} \right)
\left( \begin{matrix}
m_{11}\\
m_{33}\\
\stackrel{\bullet}{q}_{13}\\
\stackrel{\bullet}{q}_{31}\\
\end{matrix} \right)_{NN}
\end{align}
As required, the odd part of the $S$-matrix satisfies $S^4 = -\text{id}$ and $(TS)^3 = -\text{id}$, 
while the even part satisfies the usual $S^4 = \text{id}$ and $(TS)^3 = \text{id}$. 


\section{Fermion condensation in $\halfesix$} \label{halfesix}

In this section we perform fermion condensation in the category $\halfesix$.
This provides an example of a super pivotal category with fusion multiplicity.
After condensation we will obtain a theory with one non-trivial q-type particle, 
which we will denote by $\rho$.
$\rho$ obeys the fusion rule
\begin{align}
\rho \tp \rho = \mathbb{C}^{1|1} \mathds{1} \oplus \mathbb{C}^{1|1} \rho,
\end{align}
which is similar to the Fibonacci fusion rule but with q-type objects.\footnote{A simpler generalization of the Fibonacci theory 
to the super pivotal case, with one non-trivial q-type particle and fusion space 
$V^{\tilde{\tau} \tilde{\tau}}_{\tilde {\tau}}\cong \mathbb{C}^{1|1}$, does not exist. 
Indeed, suppose $\text{End}(\tau) = \mathbb{C} \ell_1$ and $V^{\tau \tau \tau} \cong \mathbb{C}^{1|1}$.
$\gamma \in \text{End}(\tau)$ denote the odd endomorphism in $V^{\tau\tau\tau}$.
Then we can write down three anti-commuting operators $\gamma \tp \gamma \tp 1$, $\gamma \tp 1 \tp \gamma$ and $1\tp \gamma \tp \gamma$ 
which are each even and as such preserve the grading on $V^{\tau\tau\tau}$. 
However, the even and odd subspaces of $V^{\tau\tau\tau}$ are one dimensional, and so these operators cannot be represented.
Hence a theory with a single q-type particle with Fibonacci-like fusion rules must have nontrivial fusion multiplicity.
In the $\halfesix$ theory we have $V^{\tau\tau\tau} \cong \cc^{2|2}$, which is big enough to represent all three of the above operators.}
The nontrivial fusion spaces are $V^{\rho\rho}_\rho \cong \cc^{2|2}$ and $V^{\rho\unit}_\rho \cong \cc^{1|1}$, 
with the first telling us that the theory has nontrivial fusion multiplicity.

This theory is richer than the examples we have considered previously, and serves as a good 
case study for the features that appear in phases described by more general super pivotal 
categories. 
These more general features include:
\begin{itemize}
\item There is a quasiparticle excitation which has a non-bounding spin structure but which is m-type 
(this also occurs in the $SO(3)_6/\psi$ theory discussed previously) and is oddly self dual.
\item The ground state degeneracy on the three spin tori with a bounding cycle (with spin 
structures $BB$, $BN$, and $NB$) is $\mathbb{C}^{3|0}$, and the ground state degeneracy on the 
non-bounding torus (with $NN$ spin structure) is $\mathbb{C}^{1|2}$.
In particular, the fermion parity of a ground state on the torus is not uniquely determined by 
the torus' spin structure (this also occurs in the $SO(3)_6/\psi$ theory discussed previously). 
\item As a fusion category, there is a fusion rule that takes two q-type particles to another q-type particle.\footnote{
Note that fusion products of this type cannot appear in the tube category of a fermionic theory since two idempotents 
in the tube category with non-bounding spin structure must fuse to an idempotent with bounding spin structure, which can never be q-type.
}
\end{itemize}

Performing the condensation requires one additional step that did not appear in the previous examples. 
This is because the category $\halfesix$ is not braided, and so doesn't have a fermion to condense.
However, as described in Section \ref{lift_and_condense}, 
it suffices to lift a particle in $\halfesix$ to a fermion in the Drinfeld center of $\halfesix$.

In what follows, we will first introduce the fusion theory of $\halfesix$ 
and its properties that are pertinent to the rest of the section.
We will then compute the half braid for the emergent fermion, and condense it in the same way we did in the previous examples.
Following this, we will compute the idempotents in the condensed theory, 
as well as the modular $S$ and $T$ matrices.

\subsection{Fusion theory of $\halfesix$}
\label{halfesixFusionTheory}

\begin{figure}
\begin{align}
\vcenter{
\xymatrix @!0 @M=4mm @R=28mm @C=40mm{
&\EsixDynkin  &   &\EsixCondensePsi&  \\
&\HalfEsixDynkin &   &\HalfEsixDynkinCondensed&  \\
 	}} \nonumber
\end{align}
\caption{
On the upper left we have the $E_6$ Dynkin diagram. 
The $E_6$ fusion theory has two closed fusion subcategories whose simple objects are $\left \{ \mathds{1},\; \sigma,\; y \right\}$ and $\left \{ \mathds{1}, x,y \right \}$. 
The first satisfies the Ising fusion rules while the second satisfies those of $\halfesix$ given in (\ref{halfEsixFusionRules}).
The figure on the upper right denotes the principal graph of the theory after condensing $y$ (after first lifting $y$ to the Drinfeld center).
On the bottom left we have drawn the $\halfesix$ principal graph.
The figure on the bottom right is the principal graph of $\halfesix/y$ studied in this section.
The fermionic quotient of the $\halfesix$ fusion subcategory reduces the particle content to $\{\unit,\rho\}$.  
}
\label{EsixDynkin}
\end{figure}
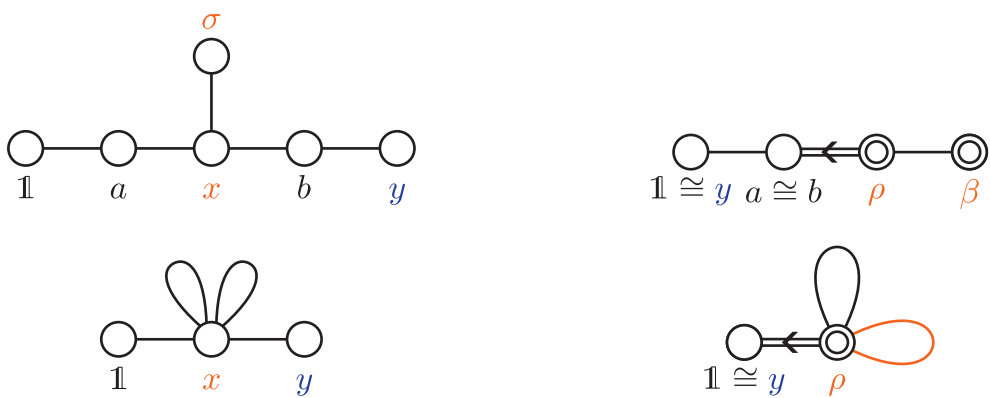

The $E_6$ fusion category is the fusion category whose principle graph is given by the $\text{E}_6$ Dynkin diagram, 
shown to the left in Figure \ref{EsixDynkin}.
The $E_6$ fusion category has two sub-categories: one subcategory has the fusion rules of the Ising theory, 
while the other is known as $\halfesix$ \cite{Hong2008} and has more complicated fusion rules. 

The fusion category $\halfesix$ has three particles, $\mathds{1}$, $x$, and $y$. 
The non-trivial fusion rules are
\begin{align}
y \tp y \cong \mathds{1} \quad \quad y \tp x \cong x \tp y \cong x \quad \quad x\tp x \cong \mathds{1}\oplus 2x \oplus y,
\label{halfEsixFusionRules}
\end{align}
and the quantum dimensions are given by
\begin{align}
d_{\mathds{1}} = 1 \quad \quad d_x = 1 + \sqrt{3} \quad \quad d_y = 1.
\end{align}
Note that one of the fusion spaces ($V_x^{xx}$) has dimension greater than 1.
Note also that $x$ 
is invariant under fusion with $y$, and that $y$ has quantum dimension $1$. 
If the theory were braided, and $y$ were fermionic, then condensing $y$ would lead to a super pivotal 
fusion theory with only two objects, $\mathds{1}$ and $\rho$, the image of $x$ under condensation of $y$. 
This theory however is not braided, and so we will have to do more work to condense $y$, 
as discussed in the next subsection 
(see also the discussion in Section \ref{lift_and_condense}).

\medskip

We now to lay out some of the basic data of $\halfesix$ which will be useful to us in the following sections. 
Specifically we will give all information required to manipulate the $y$ line, 
which will be useful knowledge to have on hand when we condense $y$. 

From looking at (\ref{halfEsixFusionRules}) we notice that all particles are self dual, 
and therefore we must specify their Frobenius-Schur indicators. 
In this case both Frobenius-Schur indicators are equal to 1.
These can be found from the associators $\kappa_x = d_x \left[ F^{xxx}_x \right]_{\mathds{1}\mathds{1}}$, and similarly for $y$. 
We list all the $F$-symbols (as found in \cite{okazaki2013,Wakui2002}) in App.~\ref{E6Fsymbols}.
Using the $F$-symbols in App.~\ref{E6Fsymbols} we can check that, in this gauge, $y$ has nice pivotal properties:
\begin{align}
\begin{matrix}
&{\xxypivotadual  = \xxypivotbdual  =  \Vxxydual} \\
&\\
&{\xxypivota =  \xxypivotb = \Vxxy } \\
\end{matrix}
&\;\;
\begin{matrix}
&{\yxxpivotadual = \yxxpivotbdual = \Vyxxdual }\\
&\\
&{\yxxpivota = \yxxpivotb=  \Vyxx }\\
\end{matrix}
&\;\;
\begin{matrix}
&{\xyxpivotadual = \xyxpivotbdual = \Vxyxdual } \\
&\\
&{\xyxpivota =  \xyxpivotb =  \Vxyx}\\
\end{matrix}
\end{align}
The fact that these diagrams are trivially pivotal is a reflection of the gauge choice used for the splitting spaces.

Next, we look at what happens when we slide a $y$ line past a $V^{xx}_x$ fusion space.
Since $\text{dim} V^{xx}_x = 2$ the fusion space requires a multiplicity index labeling the independent vectors spanning this vector space. 
We denote them $v_1$ and $v_2$, and diagrammatically label them with an index at the fusion vertex:
\begin{align}
V^{xx}_x \cong \mathbb{C} \left[\;\; \Vxxxa \;\;  \right  ] \qquad \qquad a = 1, 2.
\end{align}
The next three relations show what happens when $y$ shifts past the fusion space $V^{xx}_x$:
\begin{align}
\Vxyxxa = \sigma^x_{ab} \; \Vxyxxb
\quad \quad 
\Vyxxxa = \sigma^z_{ab} \;\; \Vyxxxb
\quad \quad 
\Vxxyxa =  \sigma^y_{ab} \;\Vxxyxb,
\label{yslide}
\end{align}
where the $\sigma^w$, $w = x,y,z$ are the standard Pauli matrices.\footnote{
Explicitly $\sigma^x = \left( \begin{matrix} 0 &1\\ 1&0 \end{matrix} \right) \quad  \sigma^y = \left( \begin{matrix} 0 &-i\\ i&0 \end{matrix} \right)  \quad \sigma^z = \left( \begin{matrix} 1 &0\\ 0&-1 \end{matrix} \right)$}
When we condense $y$, these sliding moves will determine the action of $\End(\rho) \tp \End(\rho) \tp \End(\rho)$ on $V^{\rho \rho}_\rho$.

Lastly, we have
\begin{align}
&\Vxxxva \; = \; \left( W_{\mathds{1}} \right)_{ab} \Vxxxvb \quad \quad 
W_{\mathds{1}} = \frac{e^{-7 i \pi /12}}{\sqrt{2}}\left( \begin{matrix} 1 &-i \\ 1 & i \end{matrix} \right) \\
\nonumber \\
&  \Vxxxya \; = \;  \left( W_{y} \right )_{ab} \; \Vxxxyb  \quad \quad   W_y = \frac{e^{-7 i \pi /12}}{\sqrt{2}}\left( \begin{matrix} 1 &-i \\ -1 & -i \end{matrix} \right)
\end{align}
These will be of use to us when we specify the pivotal properties of $\halfesix$ 
after condensing $y$; see Section \ref{ESixPivotal} for more details.
The only data left to specify is the associators for the $V^{xxx}_x$ fusion space, 
which we list in Appendix~\ref{E6Fsymbols}.

\subsection{Fermion condensation in $\halfesix$}
\label{condensey}

In this subsection we will describe the procedure for condensing the $y$ particle in $\halfesix$. 
As mentioned earlier, $\halfesix$ is not braided, and so when we say condense $y$, 
we actually mean that we lift $y$ to the Drinfeld center (where it is an emergent fermion), and condense the lift of $y$. 
Since the center of $\halfesix$ has been computed in several places \cite{Hong2008,Izumi2001, kawahigashi2001} we will not provide all details. 

The lift of $y$ to the Drinfeld center can be found by solving \eqref{halfbraid_resolution} 
subject to the constraints \eqref{identityBraid} and \eqref{junctionBraid}.
Using the fusion theory of $\halfesix$ defined above in Section~\ref{halfesixFusionTheory} one readily finds the unique solution,
\begin{align}
\halfbraid{\ebox}{y}{\unit +x+y}{} = \halfbraid{\eone}{y}{\mathds{1}}{} - \; i   \halfbraid{\ex}{y}{x}{x} -   \halfbraid{\ey}{y}{y}{\mathds{1}}.
\label{yhalfbraid}
\end{align}
The negative sign on the last term makes the statistics and twist
of (the lift of) $y$ fermionic. 

We are now in a position to condense $y$.
Since $x$ is invariant under fusion with $y$ after condensation it becomes a q-type simple object, which we will denote by $\rho$:
\begin{align}
x \xrightarrow{\;\; \text{condense $y$\;\;}} \rho \quad \quad \quad \text{End}(\rho) \cong \mathbb{C} \ell_1
\end{align}
$\rho$ is the only non-trivial simple object in the condensed theory. 
Furthermore, since $x$ has fusion multiplicity in the parent theory, $\rho$ has fusion multiplicity in the condensed theory.
This is captured by the fusion space
\begin{align}
V^{\rho \rho}_\rho \cong \mathbb{C}^{2|2}.
\end{align}
The nontrivial fusion rule of the condensed theory is
\begin{align}
\rho \tp \rho = \mathbb{C}^{1|1}\cdot \mathds{1} \oplus \mathbb{C}^{1|1} \cdot \rho.
\end{align}
The fusion rule coefficients $\Delta^{\rho\rho}_\rho = \Delta^{\rho\rho}_\unit \cong \cc^{1|1}$ 
appearing in the above formula are determined by the relation $V^{ab}_c \cong \Delta^{ab}_c \tp \End(c)$ and the 
knowledge of the fusion spaces $V^{\rho\rho}_\rho \cong \cc^{2|2}, V^{\rho \unit}_\rho \cong \cc^{1|1}$.\footnote{We could have 
also taken $\Delta^{\rho \rho}_\rho \cong \cc^{2|0}$ or $ \cc^{0|2}$.
}

\subsubsection{Pivotal structure}
\label{ESixPivotal}

Since $\text{End}(\rho) \cong \cliff_1$, $\End(\rho)$ possesses an odd endomorphism which we will denote as $f$. 
We will denote the even basis vectors of $V^{\rho \rho}_\rho$ as $v_1$ and $v_2$, so that the odd basis 
vectors are $f {v_1}$ and $f {v_{2}}$, where the $f v_i$ are obtained by acting with $f$ on the bottom leg of the fusion space.
Diagrammatically we can denote this vector space by
\begin{align}
V^{\rho \rho }_\rho \; \cong \; \mathbb{C} \left[ \VSeven , \quad \VSa \right] \quad \quad \quad a = 1, 2,
\end{align}
where $f$ is represented graphically by the blue dot. 
We can then use our knowledge of local relations in the parent $\halfesix$ theory and the lift of $y$ to 
derive the following relations in the condensed theory:
\begin{align} 
\VSa = \sigma^z_{ab} \VSb \qquad \VSa = \sigma^y_{ab} \VSc \qquad \VSd = -i \sigma^x_{ab} \VSc,
\label{esixdotslide}
\end{align}
where the $\sigma^{w}$ are the standard Pauli matrices (compare \eqref{yslide}). 
We can also obtain the following pivoting moves:
 \begin{align}
 \PivotEsixEven  = 
P_{ab}\Vrhorhorho \quad \quad P = \frac{e^{7 i \pi/12}}{\sqrt{2}}\left( \begin{matrix}
 1& 1\\ 
 i &- i
 \end{matrix} \right)  \\
 \PivotEsixOdd =
\overset{\bullet}{P}_{ab} \Vrhorhorhoodd \quad \quad \overset{\bullet}{P} =  \frac{e^{7 i \pi/12}}{\sqrt{2}}\left( \begin{matrix} 
 i& -i\\ 
 -1 & -1
 \end{matrix} \right) 
 \end{align}
 Note that we have $P^3=\unit,( \overset{\bullet}{P} )^3=-\unit$, which is consistent with the fact that $P$ acts on even vectors 
 while $\overset{\bullet}{P}$ acts on odd ones (so that $(\overset{\bullet}{P})^3$ rotates an odd vector by $2\pi$, which produces a minus sign).

\subsection{The tube category and the torus}

\subsubsection{Tube category morphism spaces}

\begin{table}
\centering
\scalebox{0.86}{
\begin{tabu}{ r  c c c }
&$e$&$h$&$\stackrel{\bullet}{h}$\\
\cline{2-4} 
&      &     &  \\[-2ex]
$\mor^B(e \ra e) $ 
&$\AddDat{\eTube}{B}{} $
&$\AddDat{\hTube}{B}{} $&\\
\\
$\mor^N(e \ra e) $ 
&$\AddDat{\eTube}{N}{} $
&&$\AddDat{\hTube}{N}{\dotb} $\\
\end{tabu}
}
\caption{A complete basis of morphisms for $\text{mor}^B(e \ra e) \in \tube(\halfesix/y)$.
The labels above each tube are short-hand for that tube: $e$ -- empty tube; $h$ -- tube with horizontal $\rho$ line. 
The dot above $h$ denotes that the morphism is an odd morphism.}
\label{etoe}

\bigskip
\bigskip

\scalebox{0.86}{
\begin{tabu}{ r  c c c }
&$k$&$\stackrel{\bullet}{k}$\\
\cline{2-3} 
&      &     \\[-2ex]
$\mor^B(\rho \ra e) $ 
&$\AddDat{\AddDat{\TaddownTubeprime{1}}{B}{}}{}{} $
&$\AddDat{\AddDat{\TaddownTubeprime{1}}{B}{\dota}}{}{} $\\
\\
$\mor^N(\rho \ra e) $ 
&$\AddDat{\AddDat{\TaddownTubeprime{1}}{N}{}}{}{} $
&$\AddDat{\AddDat{\TaddownTubeprime{1}}{N}{\dota}}{}{} $
\end{tabu}

\quad \quad \quad \quad

\begin{tabu}{ r  c c c }
&$\tilde{k}$&$\stackrel{\bullet}{\tilde{k}}$\\
\cline{2-3} 
&      &     \\[-2ex]
$\mor^B( e \ra \rho ) $ 
&$\AddDat{\AddDat{\TadupTube{1}}{B}{}}{}{} $
&$\AddDat{\AddDat{\TadupTube{1}}{B}{\dotc}}{}{} $\\
\\
$\mor^N(e \ra \rho ) $ 
&$\AddDat{\AddDat{\TadupTube{1}}{N}{}}{}{} $
&$\AddDat{\AddDat{\TadupTube{1}}{N}{\dotc}}{}{} $
\end{tabu}

}
\caption{\label{rhotoe} A complete basis of morphisms $\mor(\rho \ra e) \in  \tube(\halfesix/y) $ and $\mor(e \ra \rho) \in  \tube(\halfesix/y) $.
We have denoted these by $k$, $\stackrel{\bullet}{k}$ $\tilde{k}$, and $\stackrel{\bullet}{\tilde{k}}$.
}

\bigskip
\bigskip

\scalebox{0.86}{
\begin{tabu}{ r  c c c c c c c c }
&$v$&$t$&$X_{11}$&$X_{12}$&$\stackrel{\bullet}{v}$&$\stackrel{\bullet}{t}$&$\stackrel{\bullet}{X}_{11}$&$\stackrel{\bullet}{X}_{12}$ \\
\cline{2-9} 
&      &      &      &     &      &     &    &\\[-2ex]
$\mor^B(\rho \ra \rho) $ 
&$\AddDat{\vTube}{B}{} $
&$\AddDat{\tTube}{B}{} $
&$ \AddDat{\XTube{1}{1}}{B}{} $
&$ \AddDat{\XTube{1}{2}}{B}{}  $
&$\AddDat{\vTube}{B}{\dotb} $
&$\AddDat{\tTube}{B}{\dotc} $
&$ \AddDat{\XTube{1}{1}}{B}{\dotc} $
&$ \AddDat{\XTube{1}{2}}{B}{\dotc}$\\
\\
$\mor^N(\rho \ra \rho) $ 
&$\AddDat{\vTube}{N}{} $
&$\AddDat{\tTube}{N}{} $
&$ \AddDat{\XTube{1}{1}}{N}{} $
&$ \AddDat{\XTube{1}{2}}{N}{}  $
&$\AddDat{\vTube}{N}{\dotb} $
&$\AddDat{\tTube}{N}{\dotc} $
&$ \AddDat{\XTube{1}{1}}{N}{\dotc} $
&$ \AddDat{\XTube{1}{2}}{N}{\dotc}$\\
\end{tabu}
}
\caption{A complete basis of morphisms for $\mor(\rho \ra \rho) \in  \tube(\halfesix/y)$.
The labels above each tube are shorthand for that tube:
$v$ -- tube with vertical $\rho$ strand; 
$t$ -- tube with $\rho$ strand wrapping both cycles;
$X$ -- tube with all labels given by $\rho$. 
As before, a dot denotes an odd vector.}
\label{rhotorho}
\end{table}

In this subsection we compute bases for tube category morphism spaces. 
We will make use of the notation $s(X)$ defined in \eqref{sX_defn}.
Using the relations
\begin{align}
& \AddDat{\hTube}{X}{} \; =  s(X)  \AddDat{\hTube}{X}{} \quad \quad \quad \AddDat{\hTube}{X}{\dotb} \; =  -s(X)  \AddDat{\hTube}{X}{\dotb} 
\label{hlocalrelation}
\end{align}
we see that
a complete basis of morphisms from trivial (empty) boundary to the empty boundary is listed in Table \ref{etoe}. 
(The relations are found by taking two fermions out of the vacuum and sliding them around the annulus.)

Similarly, we have
\begin{align}
\AddDat{\TaddownTube{\mu }}{X}{}\;= s(X) \sigma^{x}_{\mu \nu} \AddDat{\TaddownTube{\nu}}{X}{} \quad \quad \quad
\AddDat{\TadupTube{\mu}}{X}{}\;= s(X) \sigma^y_{\mu \nu} \AddDat{\TadupTube{\nu}}{X}{}.
\end{align}
Hence some of these states are linearly dependent, and so a complete basis of morphisms is found if we fix the fusion space, 
and also consider the action of the odd endomorphism. 
The basis we choose is listed in Table \ref{rhotoe}.

The morphisms of $\text{mor}(\rho \ra \rho)$ satisfy
\begin{align}
& \AddDat{\XTube{\mu}{\nu}}{X}{} \;\;  =s(X) (\sigma^x \tp \sigma^y )_{\mu \nu; \kappa \tau}\AddDat{\XTube{\kappa}{\tau}}{X}{} 
\label{XlocalRelationAnnulus}
\end{align}
and also
\begin{align}
\AddDat{\XTube{\mu}{\nu}}{X}{\dotc} \;= (\sigma^y \tp \sigma^z)_{\mu \nu; \kappa \tau} \; \AddDat{\XTube{\kappa}{\tau}}{X}{\dota}
\end{align}
We can use these results to obtain a basis for $\text{mor}(\rho \ra \rho)$, which is given in Table \ref{rhotorho}.

\subsubsection{Bases for tori} \label{TorusLocalRelations}

In this subsection we compute bases for the Hilbert spaces of spin tori.
We could of course do this using knowledge of tube category idempotents and \ref{torus_basis_theorem}, but
it's an instructive exercise to also compute bases using more elementary means.
In addition, when computing $S$ and $T$ matrices, it will be useful to have a topologically simple basis at our disposal.

\medskip

On the torus there are four spin structures.
We will first investigate the local relations on the three bounding tori which have spin structure $(X,Y) = (B,B),(N,B),(B,N)$.
Then we will consider the $(N,N)$ torus separately.
We use the same notation as in Section~\ref{modulartforms}. 

Depending on the spin structure, some of the annular tubes become zero after identifying the boundaries to form tori.
Since there is always one cycle with bounding spin structure, an odd tube is 
identified with zero for the same reason as discussed in \ref{c2_stringnets_torus}.
For another example, due to (\ref{hlocalrelation}) we have
\begin{align}
 0 = \;\AddDatTorus{\hTube}{N}{B} \;=\; \AddDatTorus{\vTube}{B}{N}\; = \; \AddDatTorus{\tTube}{B}{B}
\end{align}

We also get another local relation from \eqref{XlocalRelationAnnulus} after closing up the annulus to a torus, namely
\begin{align}
\AddDatTorus{\XTube{a}{b}}{X}{Y}{} \; = M_{ab;\alpha \beta} \AddDatTorus{\XTube{\alpha}{\beta}}{X}{Y}{} \qquad M = s(X) \sigma^x \tp \sigma^y, \; \; s(Y) \sigma^y \tp \sigma^z 
\end{align}
The two relations above can be multiplied to find a third:
\begin{align}
\AddDatTorus{\XTube{1}{1}}{X}{Y}{} \; = -s(X)s(Y) \AddDatTorus{\XTube{1}{2}}{X}{Y}{}  \; = -is(Y) \AddDatTorus{\XTube{2}{1}}{X}{Y}{} \; = -is(X) \AddDatTorus{\XTube{2}{2}}{X}{Y}{} 
\label{XLocalRelation}
\end{align}
We take the state with $(a,b) = (1,1)$ as the representative of this set of linearly dependent vectors. 

There is one additional useful linear relation to be found. 
This relation 
comes from nucleating a $\rho$ loop and extending it around the torus before fusing it back into the canonical basis:
\begin{align}
d_\rho \AddDatTorus{\eTube}{X}{Y} \; = \;
\AddDatTorus{\TorusLocalRelationa}{X}{Y} \;=\;
\AddDatTorus{\TorusLocalRelationb}{X}{Y} \;=\; 
\AddDatTorus{\TorusLocalRelationc}{X}{Y} \;=\; 
\sum_\lambda \; c_\lambda \AddDatTorus{\TorusNoLabelsx{\lambda}}{X}{Y},
\label{rhoLoopRelation}
\end{align}
where $c_\lambda$ are coefficients that depend on the $F$-symbols.
The string of equalities gives an additional local relation, in particular, it allows us 
solve for the tubes in (\ref{XLocalRelation}) above in terms of the other non-zero tubes.
Explicitly in the three sectors $(B,B), (B,N)$, and $(N,B)$ we have,
\begin{align}
& \AddDatTorus{\XTube{1}{1}}{B}{B} \;= e^{-i \pi /12} \frac{d+1}{\sqrt{d}} \left( \;\AddDatTorus{\eTube}{B}{B}\; -  \frac{1}{d} \left(\AddDatTorus{\hTube}{B}{B}   +  \AddDatTorus{\vTube}{B}{B}  \right) \right)\\
&\\
 & \AddDatTorus{\XTube{1}{1}}{B}{N} \; = e^{-5i \pi /12} \frac{d+1}{\sqrt{d}} \left( \AddDatTorus{\eTube}{B}{N} -  \frac{1}{d} \left( \AddDatTorus{\hTube}{B}{N}   + \AddDatTorus{\tTube}{B}{N} \right) \right)\\
 &\\
& \AddDatTorus{\XTube{1}{1}}{N}{B}  \;= e^{3 i \pi /4} \frac{d+1}{\sqrt{d}} \left( \AddDatTorus{\eTube}{N}{B}-  \frac{1}{d} \left( \AddDatTorus{\vTube}{N}{B} + \AddDatTorus{\tTube}{N}{B} \right) \right)
\end{align}

We now move on to the torus where the spin structure is non-bounding along both cycles.
We first notice that
\begin{align}
 0 = \;\AddDatTorus{\hTube}{N}{N} \;=\; \AddDatTorus{\vTube}{N}{N}\; = \; \AddDatTorus{\tTube}{N}{N}\;,
\end{align}
which can be seen by nucleating two fermions out of the vacuum along each $\rho$ line and 
dragging one of them along the entire $\rho$ line before fusing them back to the vacuum.
Furthermore, the same calculation as \eqref{rhoLoopRelation} implies that the empty tube is identified with zero:
\begin{align}
\AddDatTorus{\eTube}{N}{N} \; = 0
\end{align}
The only non-zero tube with even parity is given by
\begin{align}
\AddDatTorus{\XTube{1}{1}}{N}{N}
\end{align}
and those which are proportional to it are given by \eqref{XLocalRelation}. 

As one may expect from the Ising example, there are odd tubes which are non-zero.
Indeed we can find four of them,
\begin{align}
\AddDatTorusDot{\hTube}{N}{N}{\dotb}\quad \AddDatTorusDot{\vTube}{N}{N}{\dotb} \quad \AddDatTorusDot{\tTube}{N}{N}{\dotc} \quad \AddDatTorusDot{\XTube{1}{1}}{N}{N}{\dotc}.
\end{align}
However, these four tubes are not linearly independent. 
There are two independent linear relations that can be found between them by multiplying the tadpole-like diagrams in two different ways,
\begin{align}
\AddDatTorusDot{\TaddownTube{\mu}}{N}{N}{} \cdot \AddDatTorusDot{\TadupTube{\nu}}{N}{N}{\dotc} \;=\;
\AddDatTorusDot{\TadupTube{\nu}}{N}{N}{\dotc} \cdot  \AddDatTorusDot{\TaddownTube{\mu}}{N}{N}{} .
\end{align}
This is an instance of the familiar relation $cl(a\cdot b) = cl(b\cdot a)$.
Despite the indices $\mu$ and $\nu$ varying over four distinct values, this yields only two linearly independent relations on the torus. 
They are given by
\begin{align}
\frac{de^{ -i \pi / 4 }}{\sqrt{2}}  \AddDatTorusDot{\hTube}{N}{N}{\dotb}\; =\;
 \AddDatTorusDot{\vTube}{N}{N}{\dotb}  \; + \; 
 \AddDatTorusDot{\tTube}{N}{N}{\dotc} \; - \frac{2 e^{i \pi /4}}{\sqrt{d}}  
 \AddDatTorusDot{\XTube{1}{1}}{N}{N}{\dotc}
\end{align}
and\footnote{One can check that the second relation follows from the first by performing an $S$ transformation.}
\begin{align}
\frac{d e^{i \pi/4}}{\sqrt{2}}  \AddDatTorusDot{\vTube}{N}{N}{\dotb}\; = \;
\AddDatTorusDot{\hTube}{N}{N}{\dotb}\; + i 
 \AddDatTorusDot{\tTube}{N}{N}{\dotc}\; + \frac{2 e^{11i \pi /12}}{\sqrt{d}} 
 \AddDatTorusDot{\XTube{1}{1}}{N}{N}{\dotc}
\end{align}
We can now solve for any two of the above four states.
We choose
\begin{align}
 \AddDatTorusDot{\tTube}{N}{N}{\dotc}  \;&=\; \AddDatTorusDot{\hTube}{N}{N}{\dotb} \; -i\;  \AddDatTorusDot{\vTube}{N}{N}{\dotb}  \\
 &\\
 \AddDatTorusDot{\XTube{1}{1}}{N}{N}{\dotc} \;&=\;\sqrt{\frac{d}{2}}\left( e^{i \pi /3}\;  \AddDatTorusDot{\hTube}{N}{N}{\dotb} \; -i \; \AddDatTorusDot{\vTube}{N}{N}{\dotb} \;\right) 
\end{align}

In summary, the Hilbert spaces on each of the different spin tori are
\begin{align}
\begin{split}
A(T^2_{BB})  &\cong \mathbb{C}^{3|0} = \mathbb{C}\left[ \AddDatTorus{\eTube}{B}{B}{},\AddDatTorus{\hTube}{B}{B}{},\AddDatTorus{\vTube}{B}{B}{}\right]\\
&\\
A(T^2_{BN})  &\cong \mathbb{C}^{3|0}  = \mathbb{C}\left[\AddDatTorus{\eTube}{B}{N}{}, \AddDatTorus{\hTube}{B}{N}{}, \AddDatTorus{\tTube}{B}{N}{} \right]\\
&\\
A(T^2_{(NB})  &\cong \mathbb{C}^{3|0}=   \mathbb{C}\left[\AddDatTorus{\eTube}{N}{B}{}, \AddDatTorus{\vTube}{N}{B}{}, \AddDatTorus{\tTube}{N}{B}{} \right] \\
&\\
A(T^2_{NN})  &\cong \mathbb{C}^{1|2} = \mathbb{C} \left[ \AddDatTorusDot{\hTube}{N}{N}{\dotb}\;,\AddDatTorusDot{\tTube}{N}{N}{\dotc}\;,  \AddDatTorusDot{\XTube{1}{1}}{N}{N}{} \right]
\end{split}
\label{TorusBasis}
\end{align}

\subsection{The tube category of $\halfesix / y$}   \label{he6mp-tube-cat}

In this subsection we compute the minimal idempotents of the tube category of $\halfesix / y$. 
The tube category is somewhat exotic, 
and highlights many of the non-trivial features which arise when studying fermionic theories. 
We start by analyzing the idempotents from a purely algebraic point of view using only the dimensions of the morphism spaces, which can be inferred from Tables \ref{etoe}--\ref{rhotorho}.
We also provide explicit representations of the idempotents, which are given in Tables \ref{MIdempotentsprime} -- \ref{QIdempotents}.

\medskip

We begin by examining the bounding tube category. 
A basis for the morphism space $\mor^B(e \to e)$ is given in Table \ref{etoe}. 
As a vector space $\mor^B(e \to e)$ is isomorphic to $\cc^{2|0}$, 
and (as we have seen before in \eqref{C2etoe}) there is only one possible super algebra structure on $\cc^{2|0}$, given by $\cc \oplus \cc$.
(As before $\cc$ is shorthand for the trivial 1-dimensional algebra.)
Therefore this subcategory contains two inequivalent minimal idempotents. 
One is the trivial idempotent which we denote $m_1$, 
the other we denote $m_2$.
Explicit representations of these idempotents are given in Table \ref{MIdempotentsprime}. 
The trivial idempotent is the usual one, just given by the quantum dimensions. 
The non-trivial idempotent $m_2$ is easily computed from the constraint $m_1 + m_2 = \text{id}_{e}$ and is also listed in Table \ref{MIdempotentsprime}.

Next we look at $\mor^B(\rho \to e)$, 
a basis for which is listed in Table \ref{rhotoe}. 
As a vector space $\mor^B(\rho \to e ) \cong \cc^{1|1}$. 
This implies that $\mor^B(\rho \to \rho)$ contains a two by two matrix algebra $M(1|1)$.
One may think that it could just have easily been $Q(1)\oplus Q(1)$, 
but we know that the bounding tube category does not admit q-type simple objects; see the discussion at the end of Section \ref{ground_states_on_torus}.
Thus we find two minimal idempotents in $\mor^B(\rho \to \rho)$, both isomorphic to $m_2$ (they cannot be isomorphic to $m_1$ due to the no tadpole axiom.).
One of these minimal idempotents is evenly isomorphic to $m_2$ which we denote $m_2^+$, the other is oddly isomorphic and denoted $m_2^-$.
The two isomorphic idempotents are explicitly written in Table \ref{MIdempotents}.
(They are proportional to $k \cdot m_2\cdot  \tilde{k}$ and $\overset{\bullet}{k} \cdot m_2 \cdot \overset{\bullet}{\tilde{k}}$.)

We now look at $\mor^B(\rho \to \rho)$ and ask which part of it has not been accounted for. 
We see from Table \ref{rhotorho} that $\mor^B(\rho \to \rho) \cong \cc^{4|4}$ as a vector space. 
An $M(1|1)$ matrix algebra has been accounted for by $m_2^{\pm}$. 
Thus the only consistent super algebra structure is $\mor^B(\rho \to \rho) \cong M(1|1) \oplus M(1|1)$.
The remaining $M(1|1)$ contains two equivalent oddly isomorphic minimal idempotents, 
which we label as $m_3^+$ and $m_3^-$.

\medskip

The super algebra structure of the non-bounding tube category is more exotic but equally well tamed by the classification of semisimple super algebras.
We begin by looking at $\mor^N(e\to e) \cong \cc^{1|1}$. 
There is a unique simple algebra structure on $\cc^{1|1}$ given by $Q(1)$. 
We denote the corresponding q-type minimal idempotent by $q_1$.

Next we look at $\mor^N(e \to \rho)$ which is isomorphic to $\cc^{1|1}$ as a vector space. 
It follows that $\mor^N(\rho \to \rho)$ contains a $Q(1)$ summand, 
and a minimal idempotent isomorphic to $q_1$ which we denote $q_1'$, and list in Table \ref{QIdempotents}. 

We now turn to $\mor^N(\rho \to \rho) \cong \cc^{4|4}$. 
A $\cc^{1|1}$ dimensional subalgebra has been accounted for by $q_1$. 
Hence we only need to look at complement of this subalgebra, which as a vector space is isomorphic to $\cc^{3|3}$.
There are two possible super algebra structures available: either three copies of $Q(1)$ or one copy of $Q(1)$ and one copy of $M(1|1)$.
However, $\tube^B$ and $\tube^N$ must contain the same number of non-isomorphic minimal idempotents, 
as shown at the end of Section \ref{ground_states_on_torus}. 
Since $\tube^B(\halfesix/y)$ has three non-isomorphic minimal idempotents, 
so must $\tube^N(\halfesix/y)$. 
Hence the algebra structure on the remaining $\cc^{3|3}$ must be $Q(1)\oplus M(1|1)$.
Correspondingly we find one q-type minimal idempotent which we label $q_2$, 
and two oddly isomorphic m-type minimal idempotents which we label $m_4^+$ and $m_4^-$.
These idempotents are given explicitly in Table \ref{QIdempotents}.

It is useful to write the isomorphisms between the idempotents in a standard form. 
Namely, if $e$ and $e'$ are isomorphic we can write $ e = u \cdot v$ and $e' = v \cdot u$ for some $u$ and $v$. 
In an obvious notation we have,
\begin{align}
m_2 = \frac{e^{i \pi /4}}{2} \sqrt{\frac{d}{3}} \AddDatTorusDot{\TaddownTube{1}}{B}{}{}  \cdot  \AddDatTorusDot{\TadupTube{1}}{B}{}{}  \; = \frac{e^{- i \pi /4}}{2} \sqrt{\frac{d}{3}} \AddDatTorusDot{\TaddownTube{1}}{B}{}{\dota} \cdot \AddDatTorusDot{\TadupTube{1}}{B}{}{\dotc} 
\end{align}
which can be used to track the isomorphisms across the three representatives $m_2$, $m_2^+$ and $m_2^-$.
Similarly we have,
\begin{align}
q_1 = \frac{e^{i \pi/4}}{\sqrt{d}} \AddDatTorusDot{\TaddownTube{1}}{N}{}{}  \cdot  \AddDatTorusDot{\TadupTube{1}}{N}{}{} .
\end{align}
Lastly we note that $m_3^+ \cdot \overset{\bullet}{v}  = \overset{\bullet}{v}  \cdot m_3^-$, and $m_4^+ \cdot \overset{\bullet}{v}  = \overset{\bullet}{v}  \cdot m_4^-$ where $\overset{\bullet}{v} $ is listed in Table \ref{rhotorho}.
Consequently we have,
\begin{align}
m_3^+ = \lambda^{-1}(m_3^+ \cdot \overset{\bullet}{v} ) \cdot (\overset{\bullet}{v}  \cdot m_3^+) \quad \quad \quad m_3^- = \lambda^{-1}(\overset{\bullet}{v}  \cdot m_3^+) \cdot  (m_3^+ \cdot \overset{\bullet}{v} )
\end{align}
and similarly, 
\begin{align}
m_4^+ = \lambda^{-1}(m_4^+ \cdot \overset{\bullet}{v} ) \cdot (\overset{\bullet}{v}  \cdot m_4^+) \quad \quad \quad m_4^- = \lambda^{-1}(\overset{\bullet}{v}  \cdot m_4^+) \cdot  (m_4^+ \cdot \overset{\bullet}{v} ).
\end{align}

The twists and quantum dimensions of the idempotents can be computed with the techniques developed 
in previous sections; we list the results in Table \ref{TubehalfesixQD}.
 
We also note that an idempotent of the parent tube category can always be included into the tube category of the condensed theory.
As in Section \ref{mtc_idem_subsect} the inclusion can be non non-trivial: some of the idempotents may become isomorphic, or even equal to zero.
In Appendix \ref{IdempotentsHalfESix}, we compute the minimal idempotents of $\tube(\halfesix)$ and track their images under the inclusion into $\tube(\halfesix/y)$. 
This also provides an additional crosscheck on the minimal idempotents given in tables \ref{MIdempotentsprime} -- \ref{QIdempotents}.

\begin{table}
\centering
{{\tabulinesep=1.2mm
\begin{tabu}{ c c | c c }
type &twist & $\AddDat{\eTube}{B}{}$ & $\AddDat{\hTube}{B}{}$ \\ \hline
${m}_{1}$ &$1$ & $\frac{1}{d \sqrt{3}}$ & $\frac{1}{2 \sqrt{3}}$\\ \hline
${m}_{2}$&$1$ & $ \frac{d}{2 \sqrt{3}}$ & $\frac{-1}{2 \sqrt{3}}$ 
\end{tabu}
}}
\caption{
\label{MIdempotentsprime}
Quasiparticles of $\halfesix$ with bounding spin structure and trivial boundary condition.
The particle is a linear combination of the tubes shown at the top of the table multipled by the coefficients in the row.}
\end{table}

\begin{table}
\centering
{\tabulinesep=1.2mm
\begin{tabu}{ c c | c c c c }
type &twist &  $\AddDat{\vTube}{B}{} $ & $\AddDat{\tTube}{B}{} $ & $\AddDat{\XTube{1}{1}}{B}{} $ & $\AddDat{\XTube{1}{2}}{B}{} $ \\ \hline
${m}_{2}^+$&$1$ & $\frac{1}{2 \sqrt{3}}$  & $\frac{1}{2 \sqrt{3}}$  & $\frac{e^{i \pi /4}}{2 \sqrt{d}}$ & $\frac{- e^{-i \pi/4} }{2 \sqrt{3d}}$ \\
${m}_2^{-}$&$-1$  & $\frac{1}{2 \sqrt{3}}$  & $\frac{-1}{2 \sqrt{3}}$  &  $\frac{ e^{-i \pi/4} }{2 \sqrt{3d}}$& $\frac{-e^{i \pi /4}}{2 \sqrt{d}}$ \\ \hline
${m}_3^+$&$e^{i \pi /3}$  & $\frac{1}{2 + d}$ & $\frac{e^{- i \pi /3}}{2+d}$ & $\frac{- e^{i \pi /12}}{\sqrt{3d}}$ &  \\
${m}_3^-$&$-e^{i \pi /3}$  & $\frac{1}{2 + d}$ & $\frac{-e^{- i \pi /3}}{2+d} $&  & $\frac{e^{i \pi /12}}{\sqrt{3d}}$  \\ 
\end{tabu}
}
\caption{\label{MIdempotents} Quasiparticles of $\halfesix$ with bounding spin structures, and boundary condition $\rho$.
The tube with a single $\rho$ line is a direct sum of four simple objects, 
two of which we name $m_2^+$ and $m_3^+$.
The other two are oddly isomorphic to $m_2^+$ and $m_3^+$, 
and we denote them $m_2^-$ and $m_3^-$ 
}
\end{table}

\begin{table}
{{\tabulinesep=1.2mm
\begin{tabu}{ c c | c }
type& twist & $\AddDat{\eTube}{N}{}$ \\ \hline
${q}_{1}$ &$1$&1
\end{tabu}
}
$\qquad$
{\tabulinesep=1.2mm
\begin{tabu}{c  c | c c c c }
type& twist & $\AddDat{\vTube}{N}{}$ & $\AddDat{\tTube}{N}{}$  & $\AddDat{\XTube{1}{1}}{N}{}$ & $\AddDat{\XTube{1}{2}}{N}{}$ \\ \hline
${q}_1'$&$1$ & $ \frac{1}{d}$&$ \frac{1}{d} $& $\frac{- e^{i \pi/4}}{d\sqrt{d}}$ & $\frac{- e^{i \pi/4}}{d\sqrt{d}}$ \\ \hline
${q}_2$&$-i$ &$ \frac{1}{2+d} $& $\frac{i}{2+d}$ & $\frac{i \gamma}{\sqrt{2+d}}$ & $\frac{i \gamma}{\sqrt{2+d}}$ \\ \hline 
${m}_4^+$&$e^{5i \pi/6}$&$ \frac{1}{2+d} $&$ \frac{e^{-5i\pi/6}}{2+d} $ & $\frac{- \alpha e^{5 i \pi/6}}{\sqrt{2+d}}$ & $\frac{\beta e^{5 i \pi /6}}{\sqrt{2+d}}$ \\ 
${m}_4^-$&$e^{5i \pi/6}$&$ \frac{1}{2+d} $&$ \frac{e^{-5i\pi/6}}{2+d} $ &  $\frac{\beta e^{5 i \pi /6}}{\sqrt{2+d}}$  &$\frac{- \alpha e^{5 i \pi/6}}{\sqrt{2+d}}$ 
\end{tabu}
 }
\caption{  \label{QIdempotents} Quasiparticles of $\frac{1}{2} E6$ with vortex (periodic) spin structures. Two are q-type, and one is m-type. The m-type particle is two-dimensional, consisting of of two smaller simple modules.  $\Pi$ is an odd isomorphism, and 
$\alpha = \frac{1}{2} \left( 1+ 1/\sqrt{2d+1} \right)$, and $\beta = \frac{1}{2} \left( 1- 1/\sqrt{2d+1} \right)$. 
}}
\end{table}

\begin{table}
\begin{flalign*} & \begin{array}{r@{ \quad \quad \quad}  c @{\quad \quad} c @{\quad \quad} c @{\quad \quad \quad \quad } c  @{\quad \quad} c @{\quad \quad} c  }
			\text{particle}				&m_1		&m_2		&m_3^+	&q_1	&q_2	&m_4 \\[.5ex] \hline \\ [-2ex]
			\text{quantum dimension}		&1			&1+d		&d		&2+d	&d		&d\\ [.5ex]
			\text{twist}				&1			&1		&e^{- i \pi/3}	&1	&i	&e^{-5 i \pi /6} \\
						\end{array} & \end{flalign*}
	\caption{The $\halfesix$ quantum dimensions and twists. 
	We have normalized the quantum dimensions by the quantum dimension of the trivial idempotent. 
	 The total quantum dimension is given by $\mcd = d\sqrt{6}$.}
	\label{TubehalfesixQD}
\end{table}

\medskip

The fusion rules in the condensed theory can be calculated using \eqref{PplusQ} and \eqref{PminusQ} as well as the S-matrix which we compute in the next section. 
We list the fusion rules in Table \ref{TubehalfesixFusionRules}. 
A particularly noteworthy fusion rule is:
\begin{align}
m_4^+ \tp m_4^+ \cong \cc^{0|1} m_1 \oplus m_2 \oplus m_3^+.
\end{align}
The odd vector space coefficient of the trivial object $m_1$ implies that $m_4^+$ is oddly self dual. 
Let's show explicitly that $\mor(\unit \ra m_4^+ \tp m_4^+) \cong \cc^{0|1}$.
We first note the following linear relations,
\begin{align}
\label{sphericity_for_twist}
\underset{m_\unit}{\Pantsmsixa} &= \underset{m_\unit}{\Pantsmsixb}\\
\label{sphericity_for_twistprime}
\underset{m_\unit}{\Pantsmsixc} &= \sum_{\alpha \beta} P_{a\alpha} (P^2)_{b \beta} \underset{m_\unit}{\Pantsmsixd}.
\end{align}
The boundary is labeled by $m_\unit$ and allows us to treat the pair of pants 
as if it were a sphere with two disks removed each with one marked point labeled $\rho$.
Indeed \eqref{sphericity_for_twist} follows directly from this observation, while \eqref{sphericity_for_twistprime} also requires some pivots. 
After plugging in the coefficients and using \eqref{XlocalRelationAnnulus}, one finds that
\begin{align}
\underset{m_\unit}{\Pantsmsixe} &= \underset{m_\unit}{\Pantsmsixh}  \\
\underset{m_\unit}{\Pantsmsixf} &= \underset{m_\unit}{ \Pantsmsixg}.
\end{align}
Under interchanging $X_{11}$ with $X_{12}$, we see that $m_4^+$ is interchanged with $m_4^-$. 
It follows that inserting the $m_4^+$ idempotent on the right hand side of the pair of pants is equivalent to inserting $m_4^-$ on the left hand side of the pair of pants.
Using that $\overset{\bullet}{v}\cdot  m_4^- = m_4^+ \cdot \overset{\bullet}{v}$, a basis for $V^{m_4^+ m_4^+}_\unit$ is generated by a single odd vector:
\begin{align}
\nonumber
V^{m_4^+ m_4^+}_\unit =   
&\left \langle \underset{m_\unit}{\Vmsixa}
+e^{-5 i \pi /6} \underset{m_\unit}{\Vmsixb} \right. \\ 
-& \left. \alpha \sqrt{2+d}  e^{5 i \pi /6}\underset{m_\unit}{\Vmsixc}
+\beta \sqrt{2 + d} e^{5 i \pi /6} \underset{m_\unit}{\Vmsixd}
\right \rangle,
\end{align}
and so $V^{m_4^+ m_4^+}_\unit \cong \cc^{0|1}$.

			\begin{table}
\begin{flalign*} & \begin{array}{c@{ \quad}  | @{\quad \quad} l @{\quad \quad} l @{\quad \quad} l}
			\mca \tp \mca 		&m_1		&m_2		&m_3^+		\\[.5ex] \hline \\ [-2ex]
			m_1		 		&m_1		&m_2		&m_3^+		\\
			m_2		 		&m_2		&m_1\oplus \cc^{1|1}m_2	\oplus \cc^{1|1}m_3^+	&\cc^{1|1}  m_2 \oplus \cc^{0|1}m_3^+		\\
			m_3^+		 		&m_3^+		&\cc^{1|1} m_2 \oplus \cc^{0|1}m_3^+		&m_1 \oplus \cc^{0|1}m_2 \oplus m_3^+		\\
			\end{array} & \end{flalign*}
	
\begin{flalign*} & \begin{array}{c@{ \quad}  | @{\quad \quad} l @{\quad \quad} l @{\quad \quad} l}
			\mcv \tp \mca 		&m_1		&m_2									&m_3^+		\\[.5ex] \hline \\ [-2ex]
			q_1				&q_1		&\cc^{1|1}q_1 \oplus q_2 \oplus \cc^{1|1}m_4^+		&q_1 \oplus q_2 \oplus \cc^{1|1}m_4^+		\\
			q_2				&q_2		&q_1 \oplus \cc^{1|1} m_4^+					&q_1 \oplus  q_2		\\
			m_4^+		 		&m_4^+		&q_1 \oplus q_2 \oplus \cc^{0|1}m_4^+			&q_1 \oplus \cc^{0|1} m_4^+		\\
			\end{array} & \end{flalign*}		

\begin{flalign*} & \begin{array}{c@{ \quad}  | @{\quad \quad} l @{\quad \quad} l @{\quad \quad} l}
			\mcv \tp \mcv 		&q_1		&q_2		&m_4^+		\\[.5ex] \hline \\ [-2ex]
			q_1		 		&\cc^{1|1}m_1\oplus \cc^{2|2} m_2 \oplus \cc^{1|1}m_3^+		&\cc^{1|1}m_2 \oplus \cc^{1|1}m_3^+		&\cc^{1|1}m_2 \oplus \cc^{1|1}m_3^+		\\
			q_2		 		&\cc^{1|1}m_2 \oplus \cc^{1|1}m_3^+		&\cc^{1|1}m_1 \oplus m_3^+		&\cc^{1|1}m_2		\\
			m_4^+		 		&\cc^{1|1}m_2 \oplus \cc^{1|1}m_3^+ 		&\cc^{1|1}m_2 		&\text{\textcolor{red}{$\cc^{0|1}$}}m_1 \oplus m_2 \oplus m_3^+		\\
			\end{array} & \end{flalign*}
	\caption{ \label{TubehalfesixFusionRules} 
$\halfesix$ fusion rules. We define $\mca = \{ m_1, m_2, m_3^+\}$ and $\mcv = \{ q_1, q_2, m_4^+ \}$ as the set of non-vortex and vortex quasiparticles, respectively. 
	The $\cc^{p|q}$ denote the vector space associated with $\Delta^{ab}_c$ which is related to the fusion space through $V^{ab}_c =\Delta^{ab}_c \tp \text{End}(c)$.
	Notice that $m_4^+$ is oddly self-dual (the relevant fusion channel is marked in red), and hence it has a Frobenius-Schur indicator of $\pm i$.
			} 
\end{table}

\subsection{Modular transformations}

\subsubsection{Topological and idempotent bases}
There are two natural bases on the torus. 
One is the topological basis, corresponding to (\ref{TorusBasis}), the other is the 
idempotent basis (or quasiparticle basis) given in Tables \ref{MIdempotents} and \ref{QIdempotents}.
We will compute the modular transformations in the topological basis first, and then change over to the idempotent basis.

We define the shorthand notation for the tubes as in Tables \ref{etoe}--\ref{rhotorho}
We will then denote the spin structure by a subscript, for example,
\begin{align}
h_{BN} =\; \AddDatTorus{\hTube}{B}{N}{}
\end{align}

Every state on the torus can be expanded in terms of the above states, as long 
as we are also careful to use the relations provided in Section \ref{TorusLocalRelations}.
Hence one can directly compute the change of basis by taking the particles in 
Tables \ref{MIdempotentsprime}, \ref{MIdempotents},  and \ref{QIdempotents} and 
projecting them onto the torus, and then modding out by the relations described in Section \ref{TorusLocalRelations}.
The change of basis matrices in the spin sectors with at least one bounding cycle are 
\begin{align}
\label{VAA}
\left( \begin{matrix}
{m}_1\\
{m}_2\\
m_3^+\\
\end{matrix} \right)_{BB} 
&= \left( \begin{matrix}
\frac{1}{d\sqrt{3}} & 0 & \frac{1}{2 \sqrt{3}} \\
\frac{d}{2 \sqrt{3}} & 0 & - \frac{1}{2 \sqrt{3}} \\
- \frac{d}{2 \sqrt{3}} & \frac{1}{2} & \frac{1}{2 \sqrt{3}}
\end{matrix} \right)
\left( \begin{matrix}
e\\
v\\
h\\
\end{matrix} \right)_{BB}\\
\label{VAP}  
\left( \begin{matrix}
{m}_1\\
{m}_2\\
m_3^+\\
\end{matrix} \right)_{BN}
&= \left( \begin{matrix}
\frac{1}{d\sqrt{3}} & 0 & \frac{1}{2 \sqrt{3}} \\
\frac{d}{2 \sqrt{3}} & 0 & - \frac{1}{2 \sqrt{3}} \\
\frac{d e^{2 \pi i/3}}{2 \sqrt{3}} & \frac{e^{- i \pi /3}}{2} & \frac{e^{- i \pi /3}}{2 \sqrt{3}}\\
\end{matrix} \right)
\left( \begin{matrix}
e\\
t\\
h\\
\end{matrix} \right)_{BN} \\
\label{VPA} 
\left( \begin{matrix}
{q}_1\\
{q}_2\\
m_4^+\\
\end{matrix} \right)_{NB}
&= \left( \begin{matrix}
1 & 0 & 0 \\
\sqrt{\frac{1+d}{3}} e^{-3i \pi /4} & \frac{e^{i \pi /6}}{\sqrt{3}} & \frac{e^{i \pi /3}}{\sqrt{3}} \\
\frac{e^{7 i \pi /12}}{\sqrt{6}} & \frac{e^{- i \pi / 6}}{2 \sqrt{3}} & \frac{e^{-2 i \pi /3}}{2 \sqrt{3}}\\
\end{matrix} \right)
\left( \begin{matrix}
e\\
v\\
t\\
\end{matrix} \right)_{NB} \\
\end{align}

For the non-bounding $NN$ spin structure the q-type idempotents need to be closed up into a torus with an odd isomorphism
We define $\left[ \stackrel{\bullet}{q}_{i}\right]^2  = q_i$. 
This has a $\pm$ ambiguity, we denote the $\pm$ signs by $\sigma_i$. 
We can also change to the idempotent basis with,
\begin{align}
\label{VPP}
\left( \begin{matrix}
\stackrel{\bullet}{q}_{1}\\
\stackrel{\bullet}{q}_{2}\\
m_4^+ \\ 
\end{matrix} \right) \; =\;
\left( \begin{matrix}
\sigma_1 e^{- i \pi /4} &0&0\\
0&\sigma_2 e^{- i \pi /4} &0\\
0&0&1 \\
\end{matrix} \right)
\left( \begin{matrix}
\frac{e^{- i \pi /4}}{\sqrt{2}} & 0&0 \\
- \frac{e^{- i \pi /4}}{\sqrt{2}} & 1 & 0 \\
0 & 0& - \frac{e^{5 i \pi /6}}{\sqrt{2 + d}}\\
\end{matrix} \right)
\left( \begin{matrix}
\stackrel{\bullet}{h} \\
\stackrel{\bullet}{v} \\
X \\
\end{matrix} \right)
\end{align}

The q-type idempotents have norm square of $2$ (due to their two-dimensional endomorphism algebras), 
as opposed to the m-type idempotents that have norm square of $1$. 
To ensure that the modular matrices are unitary, we adjust for this by defining $\widehat{q} = q/\sqrt{2}$. 
When written in terms of the $\widehat{q}$, the modular matrices are unitary.

\subsubsection{$S$ transformation}
\label{E6S_matrix_section}

The $S$ transformation exchanges the longitudinal and meridional cycles of the torus.
Since we are drawing the tori as annuli with their boundaries identified, the $S$ transformation looks like
\begin{align}
\xymatrix @!0 @M=1mm  @C=20mm{
S^{XY \rightarrow \widetilde{X}\widetilde{Y}}:& A(T^2_{XY}) \ar[rr] &&A(T^2_{\widetilde{X}\widetilde{Y}}) & \\
&&&&\\
& \AnnularTubex{\AnnularTubeNoIndex}{}{}{\psi}{X}{Y}\ar@{|->}[rr] &&\SAnnulusx{\psi}{\widetilde{X}}{\widetilde{Y}}\; & 
 }
 \label{STopologicalBasis}
\end{align}
with the transformed spin structure $\widetilde{X}\widetilde{Y}$ being found with the aid of Figure  \ref{spin_str_mapping_class_group}. 
In terms of the matrix elements of $S$, we have
\begin{align}
\SAnnulusx{\psi}{\widetilde{X}}{\widetilde{Y}} \;= \sum_{\lambda \in A(T^2,\widetilde{X}\widetilde{Y})}  S^{XY \rightarrow \widetilde{X}\widetilde{Y}}_{\psi \lambda} \AnnularTubex{\AnnularTubeNoIndex}{}{}{\lambda}{\widetilde{X}}{\widetilde{Y}} \\
\end{align}
Where we have taken $\psi,\lambda  \in \bigoplus_{ab} V^{aba^*}_b$ modulo local relations.

We can now work out the $S$-matrix for each spin structure in the topological basis, and then change over to the idempotent basis. 
The calculation is the same in each case, we find the linear map on the in to the 
topological basis based on (\ref{STopologicalBasis}), and then change back to the particle basis using (\ref{VPA}--\ref{VPP}).

For the $BB$ spin structure one simply finds that $v$ and $h$ are interchanged so that,
\begin{align}
\left( \begin{matrix}
e\\
v\\
h\\
\end{matrix} \right)_{BB}
\xrightarrow{S^{BB \rightarrow BB}}
\left(\begin{matrix}
1& 0& 0 \\
0& 0&1  \\
0&1 &0 \\ 
\end{matrix} \right)
\left( \begin{matrix}
e\\
v\\
h\\
\end{matrix} \right)_{BB}
\end{align}
We can now write down the $S$-matrix in the idempotent basis using the change of basis in (\ref{VAA})
\begin{align}
\left( \begin{matrix}
m_1\\
m_2\\
m_3^+\\
\end{matrix} \right)_{BB} \xrightarrow{S^{BB \rightarrow BB}} 
\frac{1}{\sqrt{3}}\left( \begin{matrix}
\frac{1}{d} & \frac{d}{2} & 1\\ 
\frac{d}{2} & \frac{1}{d} & -1\\
1 & -1 & 1\\
\end{matrix} \right)
\left( \begin{matrix}
m_1\\
m_2\\
m_3^+\\
\end{matrix} \right)_{BB}
\end{align}

Next we compute the $S$-matrix elements that transition between the $BN$ and $NB$ tori.
We first find the action of S on the $BN$ torus:
\begin{align}
\left( \begin{matrix}
e\\
h\\
v\\
\end{matrix} \right)_{BN} 
 \xrightarrow{S^{BN\rightarrow NB}}
\left( \begin{matrix}
1&0&0\\
0&1&0\\
d e^{-i \pi/6} & e^{5 i \pi/6}  & e^{2 \pi i /3}\\
\end{matrix} \right)
\left( \begin{matrix}
e\\
v\\
t\\
\end{matrix} \right)_{NB}
\end{align}
which can be written in the idempotent basis as
\begin{align}\left( \begin{matrix}
{m}_1\\
{m}_2\\
m_3^+\\
\end{matrix} \right)_{BN}
\xrightarrow{S^{BN \rightarrow NB}}
\left( \begin{matrix}
\frac{1}{2} & \frac{1}{2 \sqrt{3}} &  \frac{1}{\sqrt{3}} \\
\frac{1}{2} & - \frac{1}{2\sqrt{3}} & -\frac{1}{\sqrt{3}} \\
0& \frac{1}{\sqrt{3}} & -\frac{1}{\sqrt{3}} \\
\end{matrix} \right)
\left( \begin{matrix}
{q}_1\\
{q}_2\\
m_4^+\\
\end{matrix} \right)_{NB}.
\end{align}
Similarly we can work out the $S$-matrix in the topological basis for the $(N,B)$ torus,
\begin{align}
\left( \begin{matrix}
e \\ 
v\\ 
t\\ 
\end{matrix} \right)_{NB}
  \xrightarrow{S^{NB \rightarrow BN}}
\left( \begin{matrix}
1&0&0\\
0&1&0\\
de^{i \pi/6} & e^{-5 i \pi /6} & e^{-2 i \pi /3}
\end{matrix} \right)
\left( \begin{matrix}
e \\
h\\ 
t\\ 
\end{matrix} \right)_{BN} 
\end{align}
And again we can write this in the idempotent basis:
\begin{align}
\left( \begin{matrix}
{q}_1\\
{q}_2\\
m_4^+\\
\end{matrix} \right)_{NB}
\xrightarrow{S^{NB \rightarrow BN}}
\left( \begin{matrix}
1& 1& 0 \\
\frac{1}{\sqrt{3}} & - \frac{1}{\sqrt{3}} & \frac{2}{\sqrt{3}} \\
\frac{1}{\sqrt{3}} & - \frac{1}{\sqrt{3}} & - \frac{1}{\sqrt{3}} \\
\end{matrix} \right)
\left( \begin{matrix}
{m}_1\\
{m}_2\\
m_3^+\\
\end{matrix} \right)_{BN}
\end{align}
Notice that the $S$-matrix is invertible, but not unitary. 
This is because we didn't normalize our idempotents appropriately. 
As mentioned earlier, we write the normalized $Q$ idempotents with a hat, $\widehat{Q}_i = Q_i /\sqrt{2}$.
Once doing so we find the appropriately normalized $S$-matrix is given by
\begin{align}\left( \begin{matrix}
{m}_1\\
{m}_2\\
m_3^+\\
\end{matrix} \right)_{BN}
\xrightarrow{S^{BN \rightarrow NB}}
\left( \begin{matrix}
\frac{1}{\sqrt{2}} & \frac{1}{\sqrt{6}} &  \frac{1}{\sqrt{3}} \\
\frac{1}{\sqrt{2}} & - \frac{1}{\sqrt{6}} & -\frac{1}{\sqrt{3}} \\
0& \sqrt{\frac{2}{3}} & -\frac{1}{\sqrt{3}} \\
\end{matrix} \right)
\left( \begin{matrix}
\widehat{{q}}_1\\
\widehat{{q}}_2\\
m_4^+\\
\end{matrix} \right)_{NB}
\end{align}
\begin{align}
\left( \begin{matrix}
\widehat{{q}}_1\\
\widehat{{q}}_2\\
m_4^+\\
\end{matrix} \right)_{NB}
\xrightarrow{S^{NB \rightarrow BN}}
\left( \begin{matrix}
\frac{1}{\sqrt{2}}& \frac{1}{\sqrt{2}}& 0 \\
\frac{1}{\sqrt{6}} & - \frac{1}{\sqrt{6}} & \sqrt{\frac{2}{3}} \\
\frac{1}{\sqrt{3}} & - \frac{1}{\sqrt{3}} & - \frac{1}{\sqrt{3}} \\
\end{matrix} \right)
\left( \begin{matrix}
{m}_1\\
{m}_2\\
m_3^+\\
\end{matrix} \right)_{BN}
\end{align}
Notice that the matrix is symmetric, and unitary.

Lastly, we can work out the $S$-matrix on the non-bounding torus.
\begin{align}
\left( \begin{matrix}
\stackrel{\bullet}{h} \\
\stackrel{\bullet}{v} \\
X \\
\end{matrix} \right)_{NN}
\xrightarrow{S^{NN\rightarrow NN}} 
\left( \begin{matrix}
0&i  &0 \\ 
1&0 &0 \\
0&0& -i \\
\end{matrix} \right)
\left( \begin{matrix}
\stackrel{\bullet}{h} \\
\stackrel{\bullet}{v} \\
X \\\end{matrix} \right)_{NN}
\end{align}
In the idempotent basis, this is
\begin{align}   \label{hE6_S_NN}
\left( \begin{matrix}
\stackrel{\bullet}{q}_{1}\\
\stackrel{\bullet}{q}_{2}\\
m_4^+ \\ 
\end{matrix} \right)_{NN}
 \xrightarrow{S^{NN \rightarrow NN}}
\frac{e^{i \pi /4}}{\sqrt{2}}\left( \begin{matrix} 
1&\Sigma &0 \\
\Sigma &-1&0\\
0&0& -\sqrt{2} e^{i \pi /4}\\
\end{matrix} \right)
\left( \begin{matrix}
\stackrel{\bullet}{q}_{1}\\
\stackrel{\bullet}{q}_{2}\\
m_4^+ \\ 
\end{matrix} \right)_{NN}
\quad \quad \text{where $\Sigma = \sigma_1 \sigma_2$.}
\end{align}
Note that $S^{NN\ra NN}$ splits as $S^{NN\ra NN}_q \oplus S^{NN\ra NN}_m$ into blocks which operate 
on q-type and m-type particles, as it must: 
the basis vectors coming from m-type (q-type) idempotents are even (odd), and $S$ preserves fermion parity.

In summary, we have \eqref{hE6_S_NN} together with 
\begin{align}
\left(\begin{matrix}
\left( \begin{matrix}
{m}_1\\
{m}_2\\
m_3^+\\
\end{matrix} \right)_{BN} \\
\\
\left( \begin{matrix}
\widehat{{q}}_1\\
\widehat{{q}}_2\\
m_4^+\\
\end{matrix} \right)_{NB}\\
\\
\left( \begin{matrix}
{m}_1\\
{m}_2\\
m_3^+\\
\end{matrix} \right)_{BB} \\
\end{matrix} \right)
\xrightarrow{S} \left( \begin{matrix}
&&&			\frac{1}{\sqrt{2}} & \frac{1}{\sqrt{6}} &  \frac{1}{\sqrt{3}} &			&&\\
&&&			\frac{1}{\sqrt{2}} & - \frac{1}{\sqrt{6}} & -\frac{1}{\sqrt{3}}& 			&&\\
&&&			0& \sqrt{\frac{2}{3}} & -\frac{1}{\sqrt{3}}& 			&&\\
\frac{1}{\sqrt{2}}& \frac{1}{\sqrt{2}}& 0&			&&& 			&&\\
\frac{1}{\sqrt{6}} & - \frac{1}{\sqrt{6}} & \sqrt{\frac{2}{3}}&			&&& 			&&\\
\frac{1}{\sqrt{3}} & - \frac{1}{\sqrt{3}} & - \frac{1}{\sqrt{3}}&			&&& 			&&\\
&&&			&&&			\frac{1}{d\sqrt{3}} & \frac{d}{2\sqrt{3}} & \frac{1}{\sqrt{3}}\\
&&&			&&& 			\frac{d}{2\sqrt{3}} & \frac{1}{d\sqrt{3}} & -\frac{1}{\sqrt{3}}\\
&&&			&&& 			\frac{1}{\sqrt{3}} & -\frac{1}{\sqrt{3}} & \frac{1}{\sqrt{3}}\\
\end{matrix} \right)
\left(\begin{matrix}
\left( \begin{matrix}
{m}_1\\
{m}_2\\
m_3^+\\
\end{matrix} \right)_{BN} \\
\\
\left( \begin{matrix}
\widehat{{q}}_1\\
\widehat{{q}}_2\\
m_4^+\\
\end{matrix} \right)_{NB}\\
\\
\left( \begin{matrix}
{m}_1\\
{m}_2\\
m_3^+\\
\end{matrix} \right)_{BB} \\
\end{matrix} \right)
\end{align}

\subsubsection{$T$ transformation}

The Dehn twist ($T$-transformation) corresponds to cutting the torus open along one cycle, 
applying a full $2\pi$ rotation and then gluing the torus back together along that cycle.
\begin{align}
\xymatrix @!0 @M=1mm  @C=20mm{
T^{XY \rightarrow \widetilde{X}\widetilde{Y}}:& A(T^2_{XY}) \ar[rr] &&A(T^2_{\widetilde{X}\widetilde{Y}}) & \\
&&&&\\
& \AnnularTubex{\AnnularTubeNoIndex}{}{}{\psi}{X}{Y}\ar@{|->}[rr] &&\TAnnulusx{\psi}{\widetilde{X}}{\widetilde{Y}}\; & 
 }
 \label{TTopologicalBasis}
\end{align}
with the spin structure transforming according to Figure \ref{spin_str_mapping_class_group}. 
In terms of the matrix elements of $T$, 
\begin{align}
\TAnnulusx{\psi}{\widetilde{X}}{\widetilde{Y}} \;= \sum_{\lambda \in A(T^2_{\widetilde{X}\widetilde{Y}})}  T^{XY \rightarrow \widetilde{X}\widetilde{Y}}_{\psi \lambda} \AnnularTubex{\AnnularTubeNoIndex}{}{}{\lambda}{\widetilde{X}}{\widetilde{Y}} \\
\end{align}

Aside from spin structure considerations, all idempotents are eigenstates of the Dehn twist,
with $T$ acting diagonally within each spin-structure block.

To find the eigenvalues, we compute the Dehn twist in the topological basis and then change back to the idempotent basis as usual. 
We find
\begin{align}
\left( \begin{matrix}
\stackrel{\bullet}{q}_{1}\\
\stackrel{\bullet}{q}_{2}\\
m_4^+ \\ 
\end{matrix} \right)_{NN} \xrightarrow{T^{NN\rightarrow NN}}
\left( \begin{matrix} 
1 & 0&0 \\
0 & -i & 0 \\
0 & 0& e^{5 i \pi /6}\\
\end{matrix} \right) 
\left( \begin{matrix}
\stackrel{\bullet}{q}_{1}\\
\stackrel{\bullet}{q}_{2}\\
m_4^+ \\ 
\end{matrix} \right)_{NN}
\end{align}
for the $NN$ torus, and 
\begin{align}
\left(\begin{matrix}
\left( \begin{matrix}
{m}_1\\
{m}_2\\
m_3^+\\
\end{matrix} \right)_{BN} \\
\\
\left( \begin{matrix}
\widehat{{q}}_1\\
\widehat{{q}}_2\\
m_4^+\\
\end{matrix} \right)_{NB}\\
\\
\left( \begin{matrix}
{m}_1\\
{m}_2\\
m_3^+\\
\end{matrix} \right)_{BB} \\
\end{matrix} \right)
\xrightarrow{T} \left( \begin{matrix}
&&&			&&&			1&0&0	\\
&&&			&&&			0&1&0	\\
&&&			&&&			0&0&e^{i \pi/3}	\\
&&&			1&0&0&			&&	\\
&&&			0&-i&0&			&&	\\
&&&			0&0&e^{5 i \pi /6}&			&&	\\
1&0&0&			&&&			&&	\\
0&1&0&			&&&			&&	\\
0&0&e^{i \pi /3}&			&&&			&&	\\		
\end{matrix} \right)
\left(\begin{matrix}
\left( \begin{matrix}
{m}_1\\
{m}_2\\
m_3^+\\
\end{matrix} \right)_{BN} \\
\\
\left( \begin{matrix}
\widehat{{q}}_1\\
\widehat{{q}}_2\\
m_4^+\\
\end{matrix} \right)_{NB}\\
\\
\left( \begin{matrix}
{m}_1\\
{m}_2\\
m_3^+\\
\end{matrix} \right)_{BB} \\
\end{matrix} \right)
\end{align}
for the spin tori with at least one bounding cycle.
One can check that the modular matrices defined above satisfy $S^4 = (-1)^F$ and $(ST)^3 = (-1)^F$.


\section{Super pivotal categories}  \label{def_sect}

In this section we give a more formal (though not completely formal) definition of super pivotal categories.
Since the usual bosonic case is well covered in the literature, we will concentrate on
the differences between the bosonic case and the fermionic/super case.
We will also fix some notation and conventions used elsewhere in the paper.

There are various ways of axiomatizing string nets, including Kuperberg spiders \cite{kup_spider}, 
planar algebras \cite{jones_pa},
disk-like 2-categories \cite{blob_paper}, and pivotal tensor categories \cite{Joyal1991}.

The first three are better suited to our applications, but the last one is likely the most familiar to a majority or our readers,
so we will describe a fermionic/super version of pivotal tensor categories.

So far as we know, the earliest definition of a super pivotal tensor category was given in \cite{blob_paper}.
In the higher category definition given in that paper, one of the parameters was the type of balls used to 
specify morphism spaces.
If we take those balls to be 2-dimensional and equipped with spin structures, then one has a definition of a super pivotal 2-category.
We can then take a super pivotal tensor category to be a super pivotal 2-category with only one 0-morphism.
The more traditional-style definition given below is reverse-engineered to be
equivalent with the definition already contained in \cite{blob_paper}.

\medskip

Recall from 
\cite{Joyal1991}
that the data of a pivotal category includes:
\begin{itemize}
\item A set of objects $\spc^1$.
\item A set of morphisms $\spc^2$.
\item A binary operation $\otimes$ (horizontal composition) on objects and morphisms.
\item A binary operation $\circ$ on morphisms.
\item A pivotal structure $*$ defined on both objects and morphisms.
\end{itemize}

The definition of a super pivotal category differs from the usual bosonic case in the following ways:

\begin{enumerate}
	\item The space of morphisms between two objects has the structure of 
	a super vector space.
	Morphisms also satisfy the {\em super interchange law} \cite{brundan2016}:
	\begin{align}
	(f_1\tp f_2) \circ (g_1\tp g_2) = (-1)^{|f_2| |g_1|} (f_1\circ g_1)\tp (f_2\circ g_2)
	\end{align}
	where $|f|$ is the parity of the morphism $f$. 
	\item There are two distinct types of simple object, ``m-type" and ``q-type".
	m-type simple objects have trivial endomorphism algebras $\cc$, as is the case in bosonic theories.  
	q-type simple objects have endomorphism algebras isomorphic to $\cliff_1$, 
	and so their endomorphism algebras contain odd elements in addition to scalars. (See Section \ref{def_sob_ss}.)
	\item In order to keep track of Koszul signs arising from exchanging fermions, 
	we must keep track of a sign-ordering of individual fusion spaces (See \ref{koszul_signs}).
	\item Fusion spaces $V^{abc}$, $V^{ab}_c$, $V_{abcd}$, etc.\ are not merely supervector spaces; they come equipped with an action of the
	endomorphism algebras of the objects being fused. 
	For example, $V^{abc}$ possess an action of $\End(a)\tp \End(b)\tp \End(c)$.
	(See \ref{fusion_spaces}.)
	\item When combining basic 3-valent fusion spaces $V^{ab}_c$ to form fusion spaces of higher valence, we must take
	tensor products over the endomorphism algebras of the simple objects which connect two fusion spaces. 
	For example, we form the fusion space $V^{ab}_{cd}$ as $V^{ab}_{cd} \cong \bigoplus_e V^{ab}_e \tp_{\End(e)} V^e_{cd}$.
	If $e$ is m-type, this is just the usual tensor product over $\cc$, as in the bosonic case.
	But if $e$ is q-type, then we must take a non-trivial tensor product over $\cliff_1$.
	(See \ref{modified_tensor_product}.)
	\item The square of the pivotal anti-automorphism is 
	the fermion parity functor $(-1)^F$, rather than the identity functor (see \ref{pivotal_structure}). 
	If $*$ is the pivotal anti-automorphism, then 
	\be f^{**} = (-1)^{|f|}f.\ee
	In order to keep track of minus signs that result from rotating fermions by $2\pi$, we must keep track of a spin-framing at each fusion space. 
	\item The coherence equations for the basic data of the theory (e.g.\ the pentagon relations) are modified to incorporate  
	 Koszul signs resulting from reordering various fusion spaces. 
	 They are also modified to incorporate the tensor products over endomorphism algebras mentioned above. (see \ref{Fsymbols})
	\item In order to define an inner product, we need to equip the manifold on which our string-nets are 
	defined with a $pin_\pm$ structure, which is discussed in Appendix \ref{fermion_line_bundle}.
\end{enumerate}

\subsection{Simple objects}  \label{def_sob_ss}

We will assume that our category $\spc$ is {\it idempotent complete} -- 
every idempotent is the identity morphism of an associated object.
We also assume that $\spc$ is {\it additive} -- we can take direct sums of objects.

An object $a$ of $\spc$ is called {\it simple} if any homogeneous non-zero endomorphism of $a$ is an isomorphism.
Equivalently, $a$ is simple if it has no quotient objects.
(We also stipulate that the zero object is not a simple object.)

In the usual bosonic, non-super case, the only possible endomorphism algebra for a simple object
is the trivial algebra $\cc$ (scalars).
In the fermionic/super case, there is a second possibility: the complex Clifford algebra $\cliff_1$, 
which is the only nontrivial $\zt$-graded division algebra other than $\cc$.
$\cliff_1$ is generated over $\cc$ by the identity (which is even) and an odd element $f$ such that $f^2 = \lambda \cdot \id$
(or $f^2 = \lambda$ for short) for some non-zero complex number $\lambda$. 
Note that by rescaling the odd generator $f$ we can make $\lambda$ in the definition of $\cliff_1$ any nonzero complex number. 

It follows that simple objects in a super pivotal category fall into two classes, according to whether their endomorphism algebras are $\cc$ or $\cliff_1$. 
These are the m-type and q-type objects discussed earlier. 
A simple object is {m-type} if its endomorphism algebra
is $\cc$, and {q-type} if its endomorphism algebra is $\cliff_1$:
\begin{align}
\vcenter{\xymatrix @!0 @M=1mm @C=25mm{
& \text{End}(x) = \mathbb{C} \ar@{<->}[rr] &   &\text{$x$ is a simple m-type object}&  \\
&\text{End}(x) = \mathbb{C} \ell_1 \ar@{<->}[rr]  &  &\text{$x$ is a simple q-type object}&
	}}
\end{align}
This terminology comes from the notation of \cite{jozefiak1988}, which classifies simple super algebras over $\cc$ as either
$M(p|q) = \End(\cc^{p|q})$ or $Q(n)$ (see appendix \ref{superstuff}).
Note that we are using ``simple" here in two different (and well-established) senses: 
any $M(p|q)$ or $Q(n)$ is a simple super algebra
(because it has no non-trivial ideals), but the endomorphism algebra of a simple object must be either
$M(1|0) \cong M(0|1) \cong \cc$ or $Q(1) \cong \cliff_1$,
because all of the larger $M(p|q)$ or $Q(n)$ contain non-invertible elements.

The existence of q-type particles is responsible for much of the novel physics present after performing fermion condensation. 
q-type objects were also discussed in \cite{usher2016,gaiotto2016}, where they were referred to as ``Majorana objects''. 
(We prefer the m-type/q-type terminology, since it makes clearer the relationship to the Morita classification of 
simple super algebras.)

\subsection{Fusion spaces} \label{fusion_spaces}

Arbitrary morphism spaces in a super pivotal category can be built out of basic fusion spaces
$V^{ab}_c = \mor(c\to a\tp b)$,
where $a$, $b$ and $c$ are simple objects (equivalently, minimal idempotents).
This is a super vector space of dimension
$N^{ab}_c = \dim V^{ab}_c = p|q$, where $p$ is the dimension of the even part
of $V^{ab}_c$ and $q$ is the dimension of the odd part of $V^{ab}_c$.

Alternatively, we can treat $a$, $b$ and $c$ more symmetrically and define
$V^{abc} = \mor(\unit\to a\tp b\tp c)$, a super vector space of dimension $N^{abc}$.
In most of this paper we use $V^{ab}_c$, but in Sections \ref{Super_pivotal_Hamiltonian} and \ref{state_sums} we find it more convenient
to use $V^{abc}$.
Elements of the morphism spaces $V^{abc}$ and $V^{ab}_c$ are depicted by
\begin{align}
\PitchFork{a}{b}{c}{\mu} \quad \quad \quad \text{and} \quad \quad \quad \Nonpitchfork{a}{b}{c}{\mu}.
\end{align}
We call the fusion spaces $V^{abc}$ ``pitchforks" because of their graphical depiction.
Of course we have $V^{ab}_c \cong V^{abc^*}$ (see \ref{pivotal_ss} below).

More generally, we define $V^{ab}_{cd} = \mor(c\tp d\to a\tp b)$, $V^{ab}_{cde} = \mor(c\tp d\tp e\to a\tp b)$,
$V_{abcd} = \mor(a\tp b\tp c\tp d\to \unit)$, and so on.
In general, we do not require that the objects $a$, $b$ etc. be simple.

\medskip

It is very important to note that $V^{ab}_c$ is not merely a super vector space -- it also comes equipped with an action
of (i.e.\ module structure for) the endomorphism algebras of $a$, $b$ and $c$, and hence admits an action of $\End(a)\tp \End(b)\tp \End(c)$. 
It is impossible to construct the full super pivotal category without knowing this module structure (see \ref{modified_tensor_product} below), 
so the module
structure is part of the input data.
Note that the module structure implies that $N^{ab}_c = n|n$ if any of $a$, $b$ or $c$ is q-type.
This is because any representation of $\cliff_1$ has equal even and odd dimensions.
Acting with the odd (and invertible) element of $\cliff_1$ gives an isomorphism between the even and odd parts of $V^{ab}_c$, 
and hence they must have the same dimension.

The explicit matrix elements of these isomorphisms can be defined in the following way. 
Let $\psi \in V^{ab}_c$ and suppose that $b$ is a q-type simple object, and let $\Gamma_b \in \End(b)$ be an odd endomorphism.
Then 
\begin{align}
\Gamma_b \left( \Vertexa \right) = \sum_{\eta \in V^{ab}_c} \left[ \Gamma_b \right]_{\psi \eta }\Vertexe, 
\end{align}
where the matrix elements are obtained from 
\begin{align}
\label{Gamma_y_def}
\Gamma_b \left( \Vertexa \right) \; = \; \Vertexb \; =\; \sum_{\eta \in V^{ab}_c} \; \Vertexe \; \left( 
{\Vertexc}\middle/ {\Vertexd} \right), 
\end{align}
where the $\eta^*$ provide a complete orthogonal basis (with respect to the pairing \eqref{reflection_pairing_defn}) for $V^{b^*a^*}_{c^*}$.
If either $a$ or $c$ are q-type, then $\Gamma_a$ and $\Gamma_c$ matrices can be defined in a similar way.

\medskip

If at least one of $a,b,c$ is q-type, we can simplify the description of $V^{ab}_c$ slightly, 
which we have done when working out the examples considered earlier.
Suppose $c$ is q-type, and that $\{|\psi_i\rangle\} \in [V^{ab}_c]^0, i =1,\dots, r$ are the even basis vectors in $V^{ab}_c$. 
Then we can define a complete set of odd basis vectors $\{ |\eta_i\rangle \}$ for $[V^{ab}_c]^1$ by $|\eta_i\rangle = f|\psi_i\rangle$, where $f$ is the odd element of $\End(c) \cong \cc^{1|1}$.
When we write $|\eta_i\rangle$ graphically, we will write it as $f |\psi \rangle$, which allows us to  ``shift the oddness out of the vertex onto the edge'' by transferring the fermion residing on the fusion space to the q-type particle $c$. 
Graphically, this means that we are allowed to ``displace'' dots from trivalent vertices onto q-type worldlines:
\begin{align} \label{dot_displacement} 
\NonPitchforkDot{a}{b}{c}{\eta_i} \;\; = \;\; \NonPitchforkDotDisplaced{a}{b}{c}{\psi_i}
\end{align}
where the picture on the left is $|\eta_i\rangle$ and the one on the right is $f|\psi_i\rangle$, and $c$ is assumed to be q-type. 
Although this is not a deep fact, it proves to be helpful when doing graphical manipulations, 
and operators implementing transformations like \eqref{dot_displacement} will be crucial for writing down the lattice 
Hamiltonian which realizes the
super pivotal version 
of the Levin-Wen Hamiltonian. 

\medskip

Lastly, we will define a non-degenerate bilinear pairing between vectors in the vector space assigned to a disk with $n$ marked points. 
We will focus on fusion spaces of the form $V^{x_1\dots x_n}$, but the construction
for different types of fusion spaces is analogous. 
The pairing is defined by 
\begin{align} \label{reflection_pairing_defn}	
\mcb:\;  &V^{x_n^* \cdots x_2^* x_1^*} \tp V^{x_1 x_2 \cdots x_n}  \ra \cc \\
\nonumber &\Bilineara \tp \Bilinearb \mapsto \Bilinearc
\end{align} 
where $\mu\in V^{x_n^* \cdots x_2^* x_1^*}$ and $\nu  \in V^{x_1 x_2 \cdots x_n}$. 
We are working with the convention that the Koszul ordering of the tensor product increases 
in a left-to-right fashion (indicated by the numbers $1,2$ in the bottom right of \eqref{reflection_pairing_defn}); we will elaborate on this convention in Section \ref{koszul_signs}. 
This bilinear pairing is just the evaluation map, and is $\cc$-linear in both its arguments. 
It is non-degenerate, meaning that if $\nu_j$ is a complete basis for the fusion space $V^{x_1 x_2 \cdots x_n}$ and $\mu_i$ is a complete basis for the dual fusion space $V^{x_n^* \cdots x_2^* x_1^*} $, then the matrix $\mcb_{ij} = \mcb(\mu_i \tp \nu_j)$ is invertible. 
Hence we can define a set of vectors $\mu_j^* = \sum_i  (\mcb^{-1} )_{ji}\nu_i $ so that
$\mcb( \mu_j^* \tp \mu_i)=\delta_{ij}$.
Alternatively, we will can choose the normalization convention
\begin{align}  \label{b_pairing_defn}
\mcb( \mu_j^* \tp \mu_i )  = \sqrt{d_a d_b d_c} \, \delta_{ij},
\end{align} 
with $\mu_i \in V^{abc}$ and $\mu_j^* \in V^{c^* b^* a^*}$.

\subsection{Pivotal structure}  \label{pivotal_ss}   \label{pivotal_structure} 

The pivotal structure assigns to each object $a$ a dual object $a^*$.
It also provides linear isomorphisms
\be
	P_L : \mor(a \to b\tp c) \to \mor(b^*\tp a \to c)
\ee
and
\be
	P_R : \mor(a\tp b \to c) \to \mor(a \to c\tp b^*)
\ee
for any objects $a$, $b$ and $c$.
These isomorphisms are required to be functorial with respect to $a$ and $c$.
In addition, they are required to be twisted-functorial with respect to $b/b^*$, where we use the 
$*$ functor (defined below) to relate morphisms with domain $b$ to morphisms with range $b^*$.

For objects $a$, we require that $a^{**} = a$ on the nose (strict pivotal).
For morphisms $f : a\to b$, we define $f^* : b^* \to a^*$ by $f^* = P_L(P_R(f))$, diagrammatically by,
\begin{align}
\fstar \; = \; \fPLPR
\end{align}
(We have implicitly added and then removed some tensor units (trivial objects) here appearing 
in fusion spaces like $V^{aa^*}_\unit$.)
We think of $f^*$ as a $+\pi$ rotation of the morphism $f$.
We have
\be
	(f\cdot g)^* = (-1)^{|f||g|} g^* \cdot f^* .
\ee
In other words, $*$ is a contravariant functor if one takes Koszul signs into account.

For (strict) pivotal bosonic categories, one requires that $**$ is the identity functor, but
in the fermionic case one requires that $**$ is the spin-flip functor $(-1)^F$.
More specifically, we require
\be
\label{spin_flip_functor}
	f^{**} = (-1)^{|f|} f ,
\ee
since $f^{**}$ is a $2\pi$ rotation of $f$.

The part of the pivotal structure most used in calculations is the $+2\pi/3$ rotation on the 
basic trivalent fusion spaces.
We define the ``pivot" $P^{ab}_c : V^{ab}_c \to V^{bc^*}_{a^*}$
as $P_R\circ P_L$.
In terms of diagrams and matrices, this looks like
\begin{align}
P^{ab}_c \left( \;  \PivotYa \; \right) = \PivotYb
\end{align} 

The Frobenius-Schur indicator $\kappa_a$ can be computed in terms of pivot maps.
If $a\cong a^*$, then $\kappa_a$ is the eigenvalue of the composite map
\be
	V^{a}_{a^*} \stackrel{U_r}\longrightarrow V^{a1}_{a^*} \stackrel{P}\longrightarrow V^{1a}_{a^*} \stackrel{U_l^{-1}}\longrightarrow V^{a}_{a^*} ,
\ee
where $U_r$ and $U_l^{-1}$ are given by post-composition with the canonical isomorphisms (a.k.a.\ unitors)
$a \stackrel{\sim}{\to} a\tp \unit$ and $a \stackrel{\sim}{\to} \unit\tp a$.
(If $a$ is q-type then we take the eigenvalue for the even part of $V^{a}_{a^*}$.)

We also note that the modular $S$ matrix gives the square of the Frobenius-Schur indicator.
If $a \cong a^*$ we have $\kappa_a^2 = (S^2)_{aa}$.
For a bosonic theory the Frobenius-Schur indicators are $\pm1$, and this provides no new information. 
But in fermionic theories the oddly self-dual simple objects have Frobenius-Schur indicators of $\pm i$,
and this is detected by the diagonal entries of $S^2$.
The $m_4$ particle in the $\halfesix/y$ theory is an example of this; see \eqref{hE6_S_NN}.

We can similarly define $P^{abc} : V^{abc}\to V^{bca}$. 
In terms of diagrams, 
\begin{align}
\label{Pitchfork_pivot}
P^{abc} \left(  \Pitchforkabc \right ) \;=\;  \Pitchforkabcrot.
\end{align}
We will usually write simply $P$, since the $a$, $b$ and $c$ are typically clear from context.

Since $P^3$ acts as a $2\pi$ rotation, we have $P^3 = (-1)^F$. Diagrammatically, when acting on $V^{abc}$ this is written 
\begin{align}
P^3 \left(  \Pitchforkabc \right )\; = \; \PivotThreeTimes \; = \;(-1)^F \Pitchforkabc.
\end{align}

\subsection{Fusion rules and fusion spaces} \label{fusion_rules_and_fusion_spaces}

In this section we elaborate on the differences arising in fermionic theories between fusion spaces and the super vector spaces appearing in the fusion rules. 

We assume that our categories are additively complete, which means that it makes sense to
multiply objects by super vector spaces. 
(This is a categorified version of multiplying vectors by scalars.
Vectors are promoted to objects and scalars are promoted to vector spaces.)
For any collection of super vector spaces $\{W_a\}$ indexed by 
a finite set of objects $\{a\}$ in our category, 
we therefore have an object of the form
\be 
	\bigoplus_a W_a \cdot a .
\ee

Morphisms between these more general objects are calculated as  
\be  \label{amordef}
	\mor(\bigoplus_a W_a\cdot a \to \bigoplus_b W'_b\cdot b) = \bigoplus_{a,b} \Hom(W_a \to W'_b)\otimes_\cc \mor(a\to b) .
\ee

Because our category is semisimple, there exists a finite collection $\sob(\spc)$ of mutually non-isomorphic simple objects $x$ such
that any object $a$ is isomorphic to one of the form
\be \label{asumx}
	a \cong \bigoplus_{x\in \sob(\spc)} W_x\cdot x .
\ee
If we want the isomorphism to be canonical, we can take $W_x = \mor(x\to a)$ if $x$ is m-type, or $W_x$ to be the even morphisms in $\mor(x\to a)$
if $x$ is q-type.

Combining \eqref{asumx} and \eqref{amordef}, we can compute endomorphisms of objects by
\be
	\End(a) \cong \bigoplus_{x\in \sob(\spc)} \End(W_x) \otimes_\cc \End(x).
\ee

We are now ready to discuss fusion rules. 
For any $a$ and $b$, define the vector spaces $\Delta^{ab}_c$ by
\be
	a \otimes b \cong \bigoplus_{c\in \sob(\spc)} \Delta^{ab}_c \cdot c .
\ee
The $\Delta^{ab}_c$ are the fusion rule coefficients. 

The fusion spaces $V^{ab}_c$ are defined as the vector space of morphisms from $c$ to $a\tp b$:
\be \label{defn_of_V_by_Delta}
	V^{ab}_c = \mor(c \to a\tp b),
\ee
where $a,b,c$ are simple objects. 
Decomposing the tensor product and using the simplicity of $c$, we see that 
\be V^{ab}_c \cong \Delta^{ab}_c \tp \rm{mor}(c \ra c) \cong \Delta^{ab}_c \otimes \End(c).\ee
Thus, the fusion spaces can be larger than the vector spaces appearing in the fusion rules 
(in contrast to bosonic theories, where the fusion spaces and fusion rule coefficients are always equal).
As examples, in the $C_2$ theory studied earlier, we have 
\be \Delta^{q_\sigma q_\sigma}_{m_\psi} \cong \cc^{1|1},\quad\Delta^{q_\sigma m_\unit}_{q_\sigma} \cong \cc,\ee
while 
\be V^{q_\sigma q_\sigma}_{m_\psi} \cong V^{q_\sigma m_\unit}_{q_\sigma} \cong \cc^{1|1}.\ee

$V^{ab}_x$ is cyclically symmetric (up to isomorphism) in $a,b,x$ (if $a$ and $b$ are simple).
Explicitly, this is because 
\be \mor(c \ra a\tp b) \cong \mor(\unit \ra a\tp b \tp c^*) \cong \mor(a^* \ra b\tp c^*),\ee 
which allows us to cyclicly permute the indices of $V$, so long as we take the duals of any objects that move from subscripts to superscripts, and vice versa. 
For example, we have $V^{ab}_c \cong V^{bc^*}_{a^*} \cong V^{c^*a}_{b^*}$. 

On the other hand, $\Delta^{ab}_c$ is {\it not} cyclically symmetric in $a,b,c$, as the $C_2$ theory example shows. 
Additionally, while $V^{ab}_c$ has an action of $\End(a)\otimes\End(b)\otimes\End(c)$ (as mentioned earlier), $\Delta^{ab}_c$
only has an action of $\End(a)\otimes\End(b)$.

\subsection{Koszul sign rule and unordered tensor products} \label{koszul_signs}

We will treat Koszul signs as in \cite[Section 1.2]{deligne1999}.
This approach doesn't really do away with Koszul signs.
Rather, it pushes them to the background, where they don't need to be mentioned as frequently.
For explicit calculations, they must again be brought to the foreground.

Let $I$ be a finite (and unordered) index set.
For each $i\in I$, let $W_i$ be a super vector space.
We define the unordered tensor product,
\be
	\bigotimes_{i\in I} W_i ,
\ee
as follows.
For each bijection $f: \{1, \ldots,m\} \to I$ (i.e.\ for each ordering of $I$),
we have the ordered tensor product
\be
	T_f = W_{f(1)}\otimes\cdots\otimes W_{f(m)} ,
\ee
generated by elements
\be
	w_{f(1)}\otimes\cdots\otimes w_{f(m)}.
\ee
For any two orderings $f$ and $g$, there is a Koszul isomorphism
\be
	K_{fg} : T_f \to T_g ,
\ee
characterized by\footnote{
We should stress that in this section, 
the left-to-right ordering of tensor factors appearing in equations is tied to their Koszul ordering only, and
is independent of the order in which they appear when written down in diagrams. 
This is in contrast to several other points in the 
paper, where $a\tp b$ translates graphically into placing $a$ horizontally next to $b$ on the page.
} 
\begin{multline}
	K_{fg} : w_{f(1)}\otimes\cdots\otimes w_{f(k)} \otimes w_{f(k+1)} \otimes \cdots \otimes w_{f(m)} \mapsto \\
				(-1)^{|w_{f(k)}||w_{f(k+1)}|} w_{f(1)}\otimes\cdots\otimes w_{f(k+1)} \otimes w_{f(k)} \otimes \cdots \otimes w_{f(m)}
\end{multline}
when $g$ differs from $f$ by a simple transposition at $k$, and
\be
	K_{gh} \circ K_{fg} = K_{fh} .
\ee
An element of the unordered tensor product is then defined as an assignment to each ordering $f$ of an element $t_f\in T_f$
such that
\be
	K_{fg}(t_f) = t_g
\ee
for all orderings $f$ and $g$.
In other words, an element of the unordered tensor product is a collection of elements in all 
possible ordered tensor products which are related by the usual Koszul sign rules.

Note that to specify an element of the unordered tensor product, it suffices to give an element $t_f$ of one
particular ordered tensor product $T_f$.
All of the other $t_g$ are uniquely determined by $t_f$.

When writing equations involving particular ordered tensor products of fusion spaces, we will adopt the 
convention that the Koszul ordering is left-to-right on the page, unless explicitly 
indicated otherwise. If we want to indicate an ordering which departs from this left-to-right convention, 
we will indicate the ordering explicitly with numerical subscripts, 
e.g. $V^{ab}_{c,1} \tp V^{cd}_{e,2}$.
This explicit notation is often better suited to our diagrammatic calculus, where 
we frequently label the Koszul ordering of fermion dots in a way that is not tied to the 
left-to-right order in which we write down tensor products (this was done throughout Sections
\ref{C2_condense_sect} and \ref{C2_quasiparticles}, for example). 

When drawing diagrams with a particular ordering in mind, we always indicate the ordering explicitly by numbers near each fusion space (i.e.\ near
each vertex in the string net).
Another possible convention would be to use the ordering corresponding to (say) bottom-to-top on the page,
but this creates opportunities for error when changing diagrams by isotopies,
and is not really workable for diagrams drawn on higher genus surfaces.

A map between unordered tensor products
\be
	\bigotimes_i W_i \; \to \; \bigotimes_j V_j
\ee
is defined to be a collection of maps between
all possible pairs of ordered tensor products.
We will call such a collection an ``unordered map".
These maps are required to commute with the Koszul isomorphisms on either side.
To specify such a map, it suffices to give a single map between one particular pair of ordered tensor products.
All other maps in the collection are uniquely determined by this choice and the commutativity requirement.
This map will be called an ``ordered representative" of the unordered map.
See \ref{coherence_ss} for a further discussion of the distinction between unordered maps and
their ordered representatives.

\subsection{Modified tensor product} \label{modified_tensor_product}

Let $\sob(\spc)$ be a complete collection of simple objects (minimal idempotents) in some 
input super fusion category $\spc$, one from each equivalence class.
(In this subsection, as in most of the paper, we are assuming that our category $\spc$ is semisimple with finitely many equivalence
classes of simple objects.)
For arbitrary objects $x$ and $y$, we have
\be \label{sobdecomp}
	\mor(x\to y) \: \cong \: \bigoplus_{a\in\sob(\spc)} \mor(x\to a) \tp_{\End(a)} \mor(a\to y) .
\ee
Recall that the relative tensor product $\tp_{\End(a)}$ on the RHS above is defined as the usual tensor product over scalars, modulo elements
of the form $\alpha \cdot f \tp \beta - \alpha\tp f \cdot \beta$ with $\alpha \in \mor(x \to a)$, $\beta \in \mor(a \to y)$ and $f\in \End(a)$.
Clearly such elements are in the kernel of the composition map $\mor(x\to a) \tp \mor(a\to y) \to \mor(x\to y)$.
Our semisimplicity assumption implies that the composition map (summing over all $a\in \sob(\spc)$) is surjective and that such
elements generate all of the kernel.

In terms of diagrams, the relative tensor product is responsible for allowing fermionic dots to move across edges labeled by q-type simple objects.
Loosely, taking a tensor product over $\End(a)$ when $a$ is q-type allows us to identify diagrams that 
differ only by the position of a fermionic dot on an $a$ strand.

It follows (though not quite directly) from \eqref{sobdecomp} that we have isomorphisms
\be \label{pfi1}
	V^{abc}_d \cong \bigoplus_{x\in \sob(\spc)} V^{ab}_x \tp_{\End(x)} V^{xc}_d
\ee
and also
\be \label{pfi2}
	V^{abc}_d \cong \bigoplus_{y\in \sob(\spc)} V^{ay}_d \tp_{\End(y)} V^{bc}_y .
\ee
Diagrammatically these read,
\begin{align} 
\bigoplus_{x}  \Fusionspaceb \cong \Fusionspacea \cong  \bigoplus_{y}  \Fusionspacec
\end{align} 
where the unlabeled trivalent vertices denote the fusion spaces $V^{ab}_x$, $V^{xc}_d$, $V^{ay}_d$, $V^{bc}_y$, 
and the unlabeled tetravalent vertex in the middle diagram denotes the fusion space $V^{abc}_d$.
The tensor product over endomorphisms is implicit in the diagram.
Similarly, 
\be
	V^{abcd} \cong \bigoplus_{x\in \sob(\spc)} V^{ab}_x \tp_{\End(x)} V^{xcd} ,
\ee
and using the isomorphism $P_R: V^{ab}_x \to V^{abx^*}$ this becomes
\be  \label{pfpfi1}
	V^{abcd} \cong \bigoplus_{x\in \sob(\spc)} V^{abx} \tp_{\End(x)} V^{x^*cd} .
\ee
(We are implicitly using the $*$ functor to convert an $\End(x)$ action into an $\End(x^*)$ action.)
Alternatively,
\be  \label{pfpfi2}
	V^{abcd} \cong \bigoplus_{x\in \sob(\spc)} V^{bc}_x \tp_{\End(x)} V^{axd} \cong \bigoplus_{x\in \sob(\spc)} V^{axd}  \tp_{\End(x)} V^{x^*bc}.
\ee
Diagrammatically we have,
\begin{align}
\vcenter{\xymatrix @!0 @M=4mm @R=14mm @C=20mm{
 \bigoplus_x \TensorProductaprime &\quad \cong& \TensorProducteprime  &\cong\quad&\bigoplus_x \TensorProductcprime   \\
\quad \quad \rotatebox[origin=c]{90}{$\cong$}   &          &                                      &         &\quad\quad\rotatebox[origin=c]{90}{$\cong$}\\
 \bigoplus_x \TensorProductbprime  &          &                                      &         &\bigoplus_x \TensorProductdprime  
	}}
\end{align}

\subsection{F-symbols} \label{Fsymbols}

It follows from \eqref{pfi1} and \eqref{pfi2} that there is an isomorphism
\be  \label{fdef21}
	F^{abc}_d : \bigoplus_{x\in \sob(\spc)} V^{ab}_x \tp_{\End(x)} V^{xc}_d \;\to\; \bigoplus_{y\in \sob(\spc)} V^{ay}_d \tp_{\End(y)} V^{bc}_y .
\ee
For the pitchfork version, we instead use \eqref{pfpfi1} and \eqref{pfpfi2} to obtain
\be  \label{fdef30}
	F^{abcd} : \bigoplus_{x\in \sob(\spc)} V^{abx^*} \tp_{\End(x)} V^{xcd}   \;\to\;   \bigoplus_{x\in \sob(\spc)} V^{bcx^*} \tp_{\End(x)} V^{axd} .
\ee
The tensor products appearing in the above isomorphisms are unordered tensor products. 
For numerical applications, particular ordered representatives of the tensor products need to be chosen. 
For the ordered $F^{abc}_d$ isomorphism, 
we will adopt the convention
\be \label{Fmove_ordering_conventions}
	F^{abc}_d : \bigoplus_{x\in \sob(\spc)} V^{xc}_{d} \tp_{\End(x)} V^{ab}_{x} \;\to\; \bigoplus_{y\in \sob(\spc)} V^{ay}_{d} \tp_{\End(y)} V^{bc}_{y} ,
\ee
with implicit sign ordering which increases from left to right on the page.
Graphically, and written as a matrix equation, we thus have
\begin{align}
 \label{graphical_Fmove} 
\FusionSpaceLeftOrdered =\sum_y \sum_{\sigma \rho}\left( F^{abc}_d \right)_{(x; \mu \nu)(y;\sigma \rho)}   \FusionSpaceRightOrdered
\end{align}
where the greek indices label particular fusion space basis vectors, with $\mu\in V^{ab}_x,\nu\in V^{xc}_d,\alpha\in V^{bc}_y,$ and $\beta\in V^{ay}_d$, 
and where the first sum is over $y\in \sob(\spc)$. 
We also stick with the left-to-right ordering convention for the $F^{abcd}$ move:
\be  
	F^{abcd} : \bigoplus_{x\in \sob(\spc)} V^{abx} \tp_{\End(x)} V^{x^*cd}   \;\to\;   \bigoplus_{y\in \sob(\spc)} V^{ayd} \tp_{\End(x)} V^{y^*bc} .
\ee
Which written as a matrix equation is
\begin{align}
\label{PitchforkFMove}
\PitchforkFLeftprime = \sum_{y} \sum_{\rho \sigma} \left( F^{abcd} \right)_{(x; \mu \nu)(y; \rho \sigma)} \PitchforkFRightprime .
\end{align}

We will also find the following identity helpful:
\begin{align}
\label{idpitchfork}
\Idabba = \sum_{x} \frac{d_x}{\mcb(\mu^* \tp \mu )} \IdPitchfork, 
\end{align}
where the pairing $\mcb$ is defined in \eqref{reflection_pairing_defn}.

\subsection{Coherence relations} \label{coherence_ss}

We will not list all coherence relations here.
Instead we will give a few examples, in order to highlight how the bosonic case
must be changed to take account of Koszul signs and relative tensor products.

\medskip

We will start with the well-known pentagon equation, in the version that uses the basic fusion spaces $V^{ab}_c$.
If we work in terms of {\it unordered maps}, with \eqref{fdef21} interpreted as an unordered map between direct sums of
unordered tensor products, then the fermionic pentagon equation looks just like the bosonic case, namely
that the following diagram commutes:
\begin{align}
\newcommand{\A}{\bigoplus_{p,q}V^{xy}_{p}\tp_p V^{pz}_{q}  \tp_q V^{qw}_{u}}
\newcommand{\AB}{\ar[rrrru]^{F^{pzw}_u} \ar[rrd]^{F^{xyz}_q}}
\newcommand{\B}{\bigoplus_{p,t} V^{xy}_{p} \tp_p V^{pt}_{u} \tp_t V^{zw}_{t}}
\newcommand{\BC}{\ar[rrrrd]^{F^{xyt}_u}}
\newcommand{\D}{\bigoplus_{r,q}V^{xr}_{q}\tp_r V^{yz}_{r}  \tp_q V^{qw}_{u}}
\newcommand{\DE}{\ar[rrrr]^{F^{xrw}_u} }
\newcommand{\E}{\bigoplus_{s,r}V^{xs}_{u}\tp_s V^{yz}_{r}  \tp_r V^{rw}_{s}}
\newcommand{\EF}{\ar[rru]^{F^{yzw}_s}} 
\newcommand{\F}{\bigoplus_{t,s} V^{xs}_{u} \tp_s V^{yt}_{s} \tp_t V^{zw}_{t}}
\vcenter{\xymatrix @!0 @M=5mm @R=28mm @C=16mm{
&&&&\B\BC&&&&\\
\A\AB&&&&&&&&\F\\
&&\D\DE&&&&\E\EF&&
	}} 
		\label{bosonPentagon}
\end{align}
where all the sums are over a representative set of simple objects and we have used the notation $\tp_x \equiv \tp_{\text{End}(x)}$.

However, if we peer under the hood and look at {\it ordered representatives} (as we would need to do if, for example,
we were checking the pentagon equation on a computer), then we
see that a Koszul sign appears:

\begin{align}
\newcommand{\A}{\bigoplus_{p,q}\Penta}
\newcommand{\AB}{\ar[rru]^{F^{pzw}_u} \ar[rrd]^{F^{xyz}_q}}
\newcommand{\B}{\bigoplus_{p,t} \Pentb}
\newcommand{\BC}{\ar[rrr]^{K_{23}}}
\newcommand{\C}{\bigoplus_{p,t} \Pentc}
\newcommand{\CF}{\ar[rrd]^{F^{xyt}_u}}
\newcommand{\D}{\bigoplus_{r,q}\Pentf}
\newcommand{\DE}{\ar[rrr]^{F^{xrw}_u} }
\newcommand{\E}{\bigoplus_{s,r}\Pente}
\newcommand{\EF}{\ar[rru]^{F^{yzw}_s}} 
\newcommand{\F}{\bigoplus_{t,s} \Pentd}
\vcenter{\xymatrix @!0 @M=2mm @R=28mm @C=19mm{
&&\B \BC &&&\C \CF &&\\
\A \AB &&&&&&&\F \\
&&\D \DE&&&\E \EF&&
	}} 
	\label{endoPentagonDiagram}
\end{align}
or equivalently, 
\begin{align}
\newcommand{\A}{\bigoplus_{p,q}V^{qw}_{u}\tp_q V^{pz}_{q}  \tp_p V^{xy}_{p}}
\newcommand{\AB}{\ar[rru]^{F^{pzw}_u} \ar[rrd]^{F^{xyz}_q}}
\newcommand{\B}{\bigoplus_{p,t} (V^{pt}_{u} \tp_t V^{zw}_{t}) \tp_p V^{xy}_{p}}
\newcommand{\BC}{\ar[rrr]^{K_{23}}}
\newcommand{\C}{\bigoplus_{p,t} (V^{pt}_{u} \tp_p V^{xy}_{p}) \tp_t V^{zw}_{t}}
\newcommand{\CF}{\ar[rrd]^{F^{xyt}_u}}
\newcommand{\D}{\bigoplus_{r,q}V^{qw}_{u}\tp_q V^{xr}_{q}  \tp_r V^{yz}_{r}}
\newcommand{\DE}{\ar[rrr]^{F^{xrw}_u} }
\newcommand{\E}{\bigoplus_{s,r}V^{xs}_{u}\tp_s V^{rw}_{s}  \tp_r V^{yz}_{r}}
\newcommand{\EF}{\ar[rru]^{F^{yzw}_s}} 
\newcommand{\F}{\bigoplus_{t,s} V^{xs}_{u} \tp_s V^{yt}_{s} \tp_t V^{zw}_{t}}
\vcenter{\xymatrix @!0 @M=2mm @R=22mm @C=19mm{
&&\B \BC &&&\C \CF &&\\
\A \AB &&&&&&&\F \\
&&\D \DE&&&\E \EF&&
	}} 
	\label{endoPentagon}
\end{align}
Here, $K_{23}$ denotes the Koszul isomorphism associated to transposing the second and third tensor factors.
Again, we are using the implicit left-to-right Koszul ordering of each tensor product.
In terms of the matrix elements of the $F$-symbols, this reads 
\be  
\begin{aligned}
 \sum_{r\in \sob(\spc)}
 \sum_{\sigma\in V^{xr}_q}
 \sum_{\omega \in V^{yz}_r}
 \sum_{\eta\in V^{rw}_s}
 & 
 [F^{xyz}_q]_{(p;\mu\nu)(r;\omega\sigma)}
 [F^{xrw}_u]_{(q;\sigma\lambda)(s;\eta\gamma)}
 [F^{yzw}_s]_{(r;\omega \eta)(t;\alpha\delta)} \\ 
 & = \sum_{\beta \in V^{pt}_u}
 [F^{pzw}_u]_{(q;\nu\lambda)(t;\alpha\beta)}
 (-1)^{|\mu||\alpha|}
 [F^{xyt}_u]_{(p;\mu\beta)(s;\delta \gamma)}, 
\end{aligned} 
\ee
where $\mu\in V^{xy}_p,\nu\in V^{pz}_q,\lambda\in V^{qw}_u,\gamma\in V^{xs}_u,\delta\in V^{yt}_s,$ and $\alpha\in V^{zw}_t$. 
The Koszul sign $K_{23}$ appearing in \eqref{endoPentagon} appears in the above formula as $(-1)^{|\mu||\alpha|}$.

Other coherence relations are modified to take into account Koszul 
signs.
For example, requiring consistency between $F$-moves and the pivot means that the following 
diagram must commute:
\begin{align}
\newcommand{\A}{\bigoplus_x \PitchforkFLeftprimesmall}
\newcommand{\B}{\bigoplus_x  \PivotCoherenceb}
\newcommand{\C}{\bigoplus_x\;  \PivotCoherencec  }
\newcommand{\D}{\bigoplus_y   \PivotCoherencea }
\newcommand{\E}{\bigoplus_y \PitchforkFRightprimesmall }
\vcenter{\xymatrix @!0 @M=1.5mm @R=25mm @C=30mm{
&\A  \ar[rr]^{F^{abcd}} &&\E \ar[rd]^{(P^{ayd})^{-1}}&\\
\B\ar[ru]^{K_{12}} &&&&\D \ar[lld]^{F^{dabc}}   \\
&&\C \ar[llu]^{P^{dxc} \tp P^{x^*ab} } &&
	}} 
	\label{pivotconsistent}
\end{align}

\subsection{Reflection structure} \label{reflection_ss}

A {\it reflection structure} on $\spc$ is an antilinear anti-automorphism $r$ from $\spc$ to itself which preserves (rather than reverses) 
the tensor product:
\begin{align}
	a & \; \mapsto \; r(a) \\
	\alpha: a\to b & \; \mapsto \; r(\alpha): r(b) \to r(a) \\
	r(\lambda \alpha) & \;=\;  \bar\lambda r(\alpha) \\
	r(\alpha\beta) & \;=\; r(\beta)r(\alpha) \\
	r(a\tp b) & \;=\;  r(a) \tp r(b) \\
	r(\alpha\tp \beta) & \;=\;  r(\alpha) \tp r(\beta)
\end{align}
for objects $a,b$ and morphisms $\alpha,\beta$. Diagrammatically, the action of  $r$ 
reflects diagrams about the horizontal axis (while acting as complex conjugation on $\cc$). 
Outside of this section, we usually denote $r$ by a bar: $\bar a = r(a)$ and $\bar\alpha = r(\alpha)$.

For objects, we require that $r(r(a)) = a$.\footnote{
We also frequently restrict our attention to reflection-invariant objects which satisfy
$r(a) = a$, but this is not a requirement for all objects.}
For morphisms, we have two choices.
In a pin+ reflection structure, we require $r^2$ to be the identity functor:
\be
	r^2 = \id .
\ee
In a pin$-$ reflection structure, we require $r^2$ to be the spin flip functor,
\be
	r^2 = (-1)^F .
\ee
The main examples of this paper all have pin+ reflection structures.

We require $r$ to be compatible with the other structure maps of $\spc$ (pivots, $F$, etc.).
For example, 
we require the following diagrams to be commutative:
\begin{align}
\vcenter{
\xymatrix @!0 @M=2mm @R=34mm @C=19mm{
\refa \ar[rr]^r \ar[d]_P&& \refb\\
\refd \ar[rr]^r&&\refc \ar[u]^P\\
	}}
\end{align}
and
\begin{align}
\vcenter{
\xymatrix @!0 @M=2mm @R=34mm @C=19mm{
\refcoha\ar[rr]^{F^{abc}_d} \ar[d]_r &&\refcohb \ar[rr]^r && \refcohc \ar[d]^{P_L^{-1} \tp P_R} \\
\refcohh\ar[rr]^{P_R \tp P_L^{-1}} &&\refcohg \ar[d]_{K_{12}} && \refcohd \ar[d]^{K_{12}} \\
    &&\refcohf && \refcohe \ar[ll]^{F^{\bar{a}^* \bar{d} \bar{c}^*}_{\bar{b}}} \\
	}}
\end{align}

A pin+ reflection structure on $\spc$ allows us to define the action of pin+ diffeomorphisms on $\spc$ string nets.

It follows from the ``back wall" line bundle construction of Appendix \ref{flb_appendix} that super pivotal categories $\mcc/\psi$
obtained via fermion condensation will have pin+ reflection structures whenever the parent category has an
ordinary bosonic reflection structure.

Our main use for pin+ reflections is to define a sesquilinear inner product on the string net space $A(Y; c)$.
Let $Y$ be a spin surface and let $-Y$ denote the same underlying surface but with the reversed spin structure.
The ``identity" map from $Y$ to $-Y$ is not a spin diffeomorphism (as it reverses orientation), but it is a pin+ diffeomorphism.
Using the reflection structure on $\spc$, we can use this pin+ diffeomorphism to map string nets in $A(Y; c)$ to 
string nets in $A(-Y; r(c))$.
If $r(c) = c^*$, then string nets in $A(Y;c)$ and $A(-Y;r(c))$ can be glued together to get a string net on the closed
spin surface $Y\cup_{\bd Y} -Y = \bd(Y\times I)$.
Using the path integral $Z(Y\times I) : A(Y\cup_{\bd Y} -Y) \to \cc$ now yields a sesquilinear inner product on $A(Y;c)$.
Since $A(Y;c)$ is finite-dimensional, we also get an inner product on the dual space $Z(Y;c)$.

Let $M$ be a spin 3-manifold.
Then the path integrals $Z(M): A(\bd M)\to \cc$ and $Z(-M): A(-\bd M)\to \cc$ are related by
\be \label{3man-pi-orev}
	Z(-M) = Z(M)\circ R ,
\ee
where $R:A(-\bd M) \to A(\bd M)$ is the antilinear map induced by the orientation-reversing identity map from $-\bd M$ to $\bd M$.
Equivalently, $Z(M)\in Z(\bd M)$ and $Z(-M)\in Z(- \bd M)$ and
\be \label{3man-pi-orevz}
	Z(M) = R(Z(-M)) .
\ee

Let $Y_1$ and $Y_2$ be spin surfaces and let $M$ be a cobordism from $Y_1$ to $Y_2$
(i.e.\ $\bd M = Y_2\cup -Y_1$).
Then $-M$ is a cobordism from $Y_2$ to $Y_1$.
The path integrals can be viewed as maps $Z(M): Z(Y_1)\to Z(Y_2)$ and $Z(-M):Z(Y_2)\to Z(Y_1)$.
It follows from \eqref{3man-pi-orevz} that $Z(-M)$ is the adjoint of $Z(M)$ with respect to the inner products on $Z(Y_1)$
and $Z(Y_2)$ defined above.


 \section{Super pivotal Hamiltonian}
 \label{Super_pivotal_Hamiltonian}

In this section we will write down a commuting projector Hamiltonian 
for a generic fermionic topological phase. 
Since our goal in this section is to be rather general, we will put a fair amount of effort into making our construction 
mathematically precise -- readers who are only interested in the final result may skip to \ref{terms_in_Hamiltonian}. 

 \begin{figure}
\begin{center}
      \includegraphics{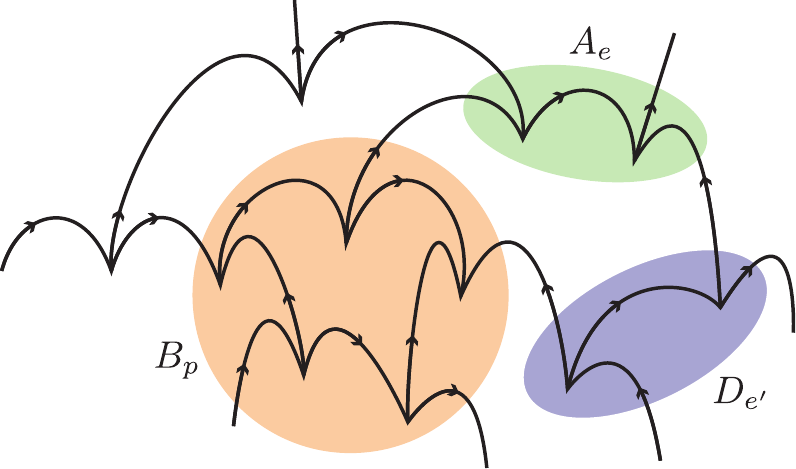}
 \caption{A cartoon of the support of the operators appearing in a super pivotal Hamiltonian on a section of a generic graph. 
The motivation for the strange-looking vertices is explained in Section \ref{standardized_handles}.
Dashed ellipses indicate the support of various terms in the Hamiltonian, on edges $e,e'$ and a plaquette $p$.
The vertex terms $A_e$ act on edges and project onto states with edge colorings that are consistent between adjacent vertices. 
The edge terms $D_e$ are responsible for sliding fermions along q-type edges. 
The plaquette terms $B_p$ involve all of the vertices and edges neighboring $p$.
They project onto graph configurations that contain no quasiparticles within the plaquette $p$.
}
 \label{example_lattice}
 \end{center}
 \end{figure}

We will follow the same basic construction as in \cite{levin2005}, with modifications to take into account 
the fermionic nature of the phases under consideration. 
The most important modifications are as follows:
First, we will need to fix a spin structure on the manifold on which we are working.
This spin structure affects the details of the local projections which constitute the Hamiltonian, and is a necessary feature of any fermionic lattice model.
Additionally, we will need to allow the local degrees of freedom that constitute the 
Hilbert space for our lattice model to be super vector spaces, rather than the regular vector spaces in bosonic models. 
Finally, we will need to add a new term in the Hamiltonian with support on the edges and pairs of neighboring vertices in the lattice that allows fermions to fluctuate across edges 
that host q-type strings.

To begin the construction of the lattice model, we will thus need the following data:
\begin{enumerate}
\item a super pivotal fusion category ${\cal C}$,
\item a spin surface $\Sigma$,
\item and a graph $\mcg$ embedded in $\Sigma$ (more precisely, a cell decomposition of $\Sigma$) which inherits information about the spin structure.
\end{enumerate}
In the following subsections we define the Hamiltonian explicitly in terms of the above data.

We will write a frustration-free Hamiltonian as a sum of local projectors, which fall into
three classes.
Two of these classes of projector are the fermionic analogues of the plaquette and vertex terms from the usual Levin-Wen Hamiltonian, 
while the third is an edge term which allows fermions to fluctuate across edges hosting q-type strings.
The Hamiltonian takes the form
\begin{align} \label{ham}
H = \lambda_p \sum_{p \in \mcf} (1-B_p) + \lambda_e \sum_{e \in \mce} (1-D_e) + \lambda_v \sum_{e \in \mce} (1-A_e),
\end{align}
where the sums are over the plaquettes (faces) $\mcf$ and edges $\mce$ of the graph, 
and the $\lambda_p,\lambda_e,\lambda_v$ are positive constants.
As in the bosonic case, we require a hierarchy in the couplings of $\lambda_v \gg \lambda_e \gg \lambda_p$.\footnote{
This makes the low energy spectrum consist purely of ``flux" excitations: 
the ``charged" idempotents always violate both the plaquette and $A_e$ term, which with this hierarchy of coefficients costs a large amount of energy. 
One way to put charge and flux excitations on an energetically more equal footing is to add ``tails'' to each 
plaquette that host the charge degrees of freedom, as in \cite{Hu2015}.
Alternatively, we could modify the Hamiltonian so that $B_p \rightarrow B_p \prod_{e \in p} (1-D_e) \prod_{v \in p} (1-A_v)$, and $D_e \rightarrow D_e \prod_{v \in e} (1-A_v)$.
}
This is because the operators appearing in the plaquette terms are not well-defined unless the edge term energies are minimized, and in turn the terms appearing in the edge operators are not well-defined unless the terms involving $A_e$ are minimized.
Thus, the projectors associated with $k$-cells are only defined on the ground states of the projectors associated with $l$-cells, for all $l<k$.  
In Figure~\ref{example_lattice} we illustrate the support of each operator appearing in the Hamiltonian with dashed circles, where we have drawn a section of the graph $\mcg$ embedded in the plane for the sake of visualization. 
For simplicity, we will assume that all vertices in $\mcg$ are trivalent.\footnote{
Any surface possesses a cell decomposition with trivalent vertices.
Such cell decompositions are Poincar\'e dual to triangulations.}

\subsection{Hilbert space} \label{hilbertspacesect}

We will locate all all degrees of freedom (spins) at the vertices of the graph $\mcg$, so
the big Hilbert space on which the Hamiltonian is defined is
\begin{align}	\label{GraphHilbertSpace}
 \mch_\mcg =\bigotimes_{v \in \mcv} \mch_v. 
\end{align}
We are making use of the unordered tensor product defined in Section \ref{koszul_signs}.

The vertex Hilbert spaces $\mch_{v}$ will depend on three sets of choices.
First, we must choose an orientation of each edge of $\mcg$.
This is because of the possibility of non-trivial Frobenius-Schur indicators; see below.
Second, we must choose an ordering of the edges incident to each vertex, consistent with with the intrinsic cyclic ordering of those edges
coming from the embedding of $\mcg$ in the oriented surface
(so if the vertex is $r$-valent, there are $r$ possible orderings).
Finally, we must choose a spin framing at each vertex.

In order to make the choice of cyclic ordering manifest in diagrams, and in order to simplify spin-structure-related
aspects of our construction, we will choose to draw our graphs in a way such that all of the 
edges at each trivalent vertex are ``on the same footing''. This is in contrast to the conventions in the physics literature
and in the earlier sections of this paper, where vertices are drawn with one edge extending below the vertex and two 
edges extending above, since in this convention the edge extending below is distinguished from the two other edges.
We will choose a convention in which all three edges at each trivalent vertex extend above the vertex: with this
choice, each vertex resembles a pitchfork. 

We must also choose a spin framing at each vertex, consistent with the ``pitchforkization'' convention.
The pitchforkization and spin framing are needed in order to define an unambiguous
isomorphism between local degrees of freedom at $v$ is standard vector spaces $V^{abc}$.
Without this standardization, the local Hilbert spaces would be ambiguous up to automorphisms. 
The standardization procedure is nothing more than fixing a convenient choice of gauge in the way we present our fusion diagrams. 

If all the edges incident to a vertex $v$ point out of the vertex, we define the vertex Hilbert space at $v$ by 
\be
	\mch_v = \bigoplus_{a,b,c} V^{abc}  \quad\quad\quad\quad\quad\quad \PitchFork{a}{b}{c}{},
	\label{pitchfork_basis}
\ee
where the sum is over the simple objects of $\spc$.
If the first edge points in and the other two point out, then we define
\be
	\mch_v = \bigoplus_{a,b,c} V^{a^*bc}  \quad\quad\quad\quad\quad\quad \PitchForkas{a}{b}{c}{} =  \PitchFork{a^*}{b}{c}{},
\ee
and so on for all eight possible patterns of in/out of the three incident edges.

\medskip

This completes the definition on the big Hilbert space for the Hamiltonian.
Impatient readers should now skip to the next subsection, but readers who are puzzled by some of the choices we made above 
are encouraged to read on.

\medskip

Why are there no spins on edges, as in the original Levin-Wen Hamiltonian?
Levin and Wen explicitly assume that $V^{abc}$ is at most 1-dimensional.
This is true for theories based on Temperley-Lieb or $Rep_q(sl_2)$, but it is not true in general,
so we need to add spins on vertices.
But each basis vector in $\bigoplus_{a,b,c} V^{abc}$ ``knows" the labels on the adjacent edges, so once we have these vertex degrees of
freedom the edge degrees of freedom become redundant and can be eliminated.\footnote{
An argument in favor of edge degrees of freedom is that they can lead to smaller local Hilbert spaces, at least for simple theories.
For example, one can write a Hamiltonian for the $C_2$ theory which has a $(2|0)$-dimensional Hilbert space at each edge
and a $(1|1)$-dimensional Hilbert space at each vertex.
The general Hamiltonian we are now discussing, specialized to the $C_2$ theory, 
would assign a $(4|3)$-dimensional Hilbert space to each vertex (but no Hilbert spaces for edges).}

Why must we choose an orientation of each edge?
The short answer: because of the possibility of non-trivial Frobenius-Schur indicators.
Now for the longer answer.
If the edges are not oriented, then we would assign vertex Hilbert spaces as above, but with all edges point out at each vertex.
This means that each edge sees two inward pointing edges.
\begin{align}
\PitchForkWithEdge
\end{align}
If the two labels coming from the two adjacent vertices are $a$ and $b$
(which we will assume are both m-type, for simplicity), then the
associated vector space for the edge is $V_{ab}$, which is 0-dimensional unless $a \cong b^*$.
If $a$ is not self-dual, or if $a$ is evenly self-dual with FS indicator 1, then there is a canonical identification of $V_{ab}$ with 
$\cc = \cc^{1|0}$ and we can ignore it.

But if $a$ is evenly self-dual with FS indicator $-1$, then there is a sign ambiguity in identifying $V_{aa}$ with $\cc$, and we will have to
keep careful track of this sign when defining the Hamiltonian.
Even worse, if $a$ is {\it oddly} self-dual (and $a$ is m-type), with FS indicator $\pm i$ (as occurs in the $\halfesix$ theory studied earlier), then $V_{aa}$ is an odd vector space, 
non-canonically isomorphic to $\cc^{0|1}$.
Keeping track of these odd vector spaces would entail even more bookkeeping.

Overall, we think the least annoying solution to the above problems is to orient each edge of the graph.
This allows us to treat $a$ and $a^*$ as distinct objects, even when they happen to be isomorphic.
Now, instead of $V_{aa}$, we have $V_{aa^*} \cong \End(a)$, which has a canonical element $id: a\to a$,
even when Frobenius-Schur indicators are nontrivial.

Why the pitchforks? As alluded to earlier, rotations by $2\pi/3$ can act non-trivially on $V^{aaa}$, and so it is helpful to choose a vertex configuration where every outgoing edge is placed on the same footing. 
This is true even in the bosonic case.

Why the spin framings?
Because $V^{abc}$ has a spin-flip automorphism, and also because we need to 
enhance the graph $\mcg$ with information related to the spin structure of the ambient manifold 
in order to write the edge terms, as explained below.

\subsection{Spin structure considerations and the standardization of the graph} \label{standardized_handles}

To derive the Hamiltonian \eqref{ham} and explain the nature of the $B_p$, $D_e$, and $A_v$ operators, we will first need to describe how the graph inherits spin structure data from the ambient spin manifold on which it is defined. 

\begin{figure}
  \includegraphics{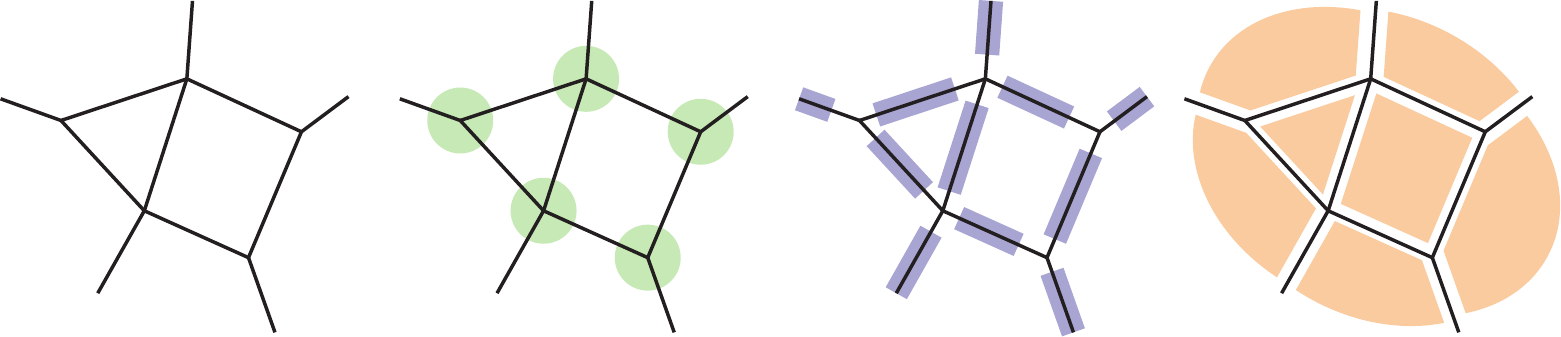}
  \caption{A handle decomposition obtained from the graph on the far left.
  The $0$ handles (green disks) are neighborhoods of the vertices, the $1$-handles (purple strips) are neighborhoods of the edges, 
  and the $2$ handles (orange polygons) which are the compliment of the union of the $0$- and $1$-handles.}
  \label{HandleDecomposition}
\end{figure}

Recall that we have a graph $\mcg$ embedded in an orientable spin surface $\Sigma$.
In order to define the super pivotal Hamiltonian we will need to equip $\mcg$ with information about the spin structure $\sigma$. 
In order to talk about the spin structure data at the vertices and edges of $\mcg$, we will thicken the cell decomposition to a handle decomposition of $\Sigma$.
A handle decomposition is essentially a fattened version of a graph; see Fig.~\ref{HandleDecomposition} for an illustration.
The handle decomposition can be obtained by expanding each vertex of the graph into a disk ($0$-handle) and each edge into
a thickened strip (1-handle).
The remaining faces constitute the $2$-handles, which are homeomorphic to disks.

Recall from Section \ref{hilbertspacesect} that 
in order to define the local vertex degrees of freedom we choose an orientation for each edge, a ``pitchforkization" for each vertex, and a spin framing at each vertex.
These choices are analogous to choosing a gauge -- different choices lead to isomorphic Hamiltonians and ground states.

The choices of pitchforkization and spin framing are equivalent to choosing a spin diffeomorphism from a 0-handle to a standard model for a 0-handle.
We define a spin diffeomorphism $\varphi_v$ from a generic 0-handle $v$ to a disk in $\rr^2$ with its standard spin structure, so that the attaching regions for the 1-handles terminating on $v$ are all located on the top part of the disk. 
That is, we use the spin diffeomorphisms to turn each 0-handle into a ``standardized 0-handle'', where the configuration of the 1-handles terminating on each 0-handle means that that each 0-handle looks like a pitchfork.
The spin diffeomorphism $\varphi_v$ that maps a generic 0-cell $v$ to a standardized 0-cell (thereby implementing the pitchforkization procedure) is defined pictorially by
\begin{align} \label{std0handle}
\xymatrix @!0 @M=1mm @R=14mm @C=35mm{
 \varphi_v: \; \; D_v\ar[r]            & D(n) \\
\;\;\;\;\;\;\; \Dv \ar@{|->}[r] & \Dfv
	} 
\end{align} 
where $n$ is the number of edges which terminate at $v$ and the black arrow in the picture on the right hand side denotes the fermion framing of the 0-handle, which is constant throughout the 0-handle.  
When writing down the Hamiltonian we always take $n=3$ without loss of generality, but when discussing tensor network constructions of these phases it is helpful to let $n$ be unspecified. 
Using the pitchforkization map $\varphi_v$ 
we can pull back the standard spin framing of $\rr^2$ to the 0-handle.
This results in a spin framing which is parallel to the outgoing edges at the top of the 0-handle. 

\begin{figure}\centering{
  \includegraphics{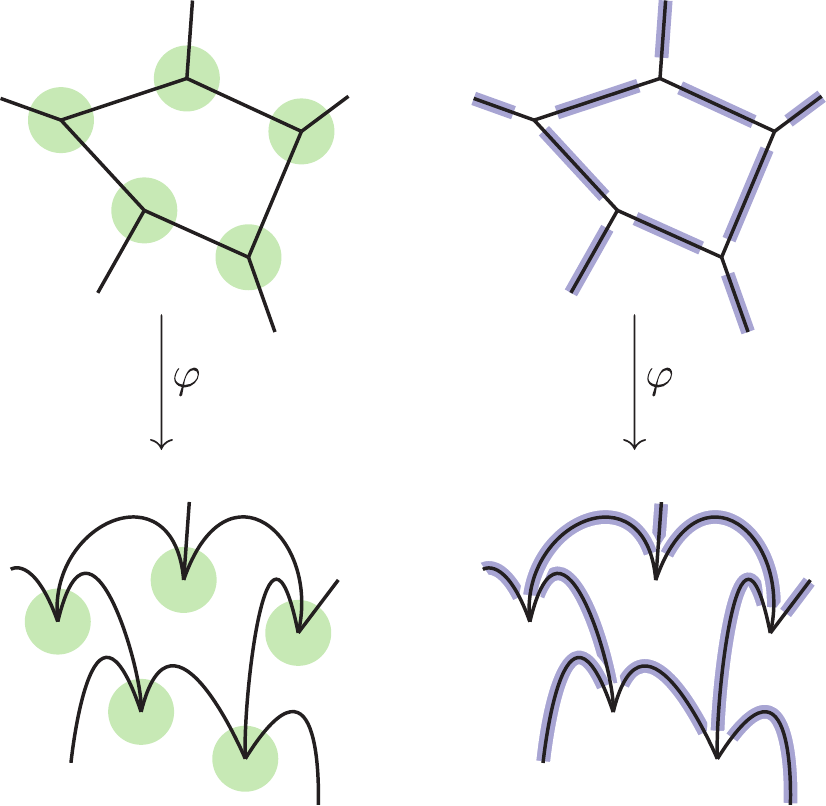}}
  \caption{The mapping $\varphi$ that maps a generic handle decomposition (top row) onto a ``standardized'' handle decomposition in which each 0-handle (green disk) has an identical pitchfork configuration (bottom row).
    }
  \label{SpinIsomorphisms}
\end{figure}

Just as we do for the 0-handles, we will ``standardize'' the 1-handles so that they all assume the same form. 
After standardizing our 0-handles, 1-handles will always enter/exit from a 0-handle ``vertically'' (see Figure \ref{SpinIsomorphisms}), 
and so our standardized 1-handles will look like 
\begin{align}
\Horseshoe
\end{align}

For theories with q-type particles, the Hamiltonian will contain terms that allow fermions to fluctuate across a 1-handle from vertex to vertex. 
The spin structure on a 1-handle (relative to the two attaching intervals) will determine what phase factor a fermion picks up when it moves across a 1-handle. 

\begin{figure}
\begin{center}
\includegraphics{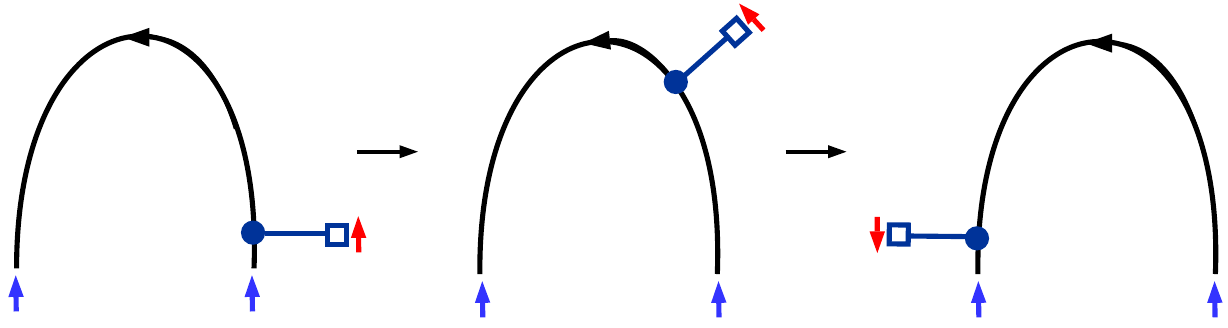}
\end{center}
\caption{\label{framing_rot} Parallel-transporting a fermion along a q-type edge, which is oriented as shown. 
The red arrow keeps track of the fermion framing, which rotates by $\pi$ when proceeding along the direction of the edge's orientation.
The blue arrows at the ends of the edge indicate the fixed framing at each 0-handle at the endpoints of the edge. 
}
  \label{framing_rotation}
\end{figure}

Once we have chosen coordinates at each vertex, we can associate a spin rotation of $+\pi$ or
$-\pi$ to each 1-handle.
First, we choose a standard spin framing at the incoming side (recall that each edge is directed) of the 1-handle.
A standard choice exists because we have chosen a standard spin framing for the 0-handle at the incoming end of the 1-handle. 
We then parallel transport the spin framing along the 1-handle, keeping the first basis vector of the spin framing tangent to 
the core of
the 1-handle during the transporting process.
This procedure is illustrated in Figure \ref{framing_rotation}, where the red arrow denotes the first basis vector of the spin framing. 
When we arrive at the end of the 1-handle, the spin framing we have transported will not agree with the standardized spin framing at the second 0-handle. 
We can see this from Figure \ref{framing_rotation}; we have chosen the spin framing to point upwards at each 0-handle, but the red arrow points downward when it reaches the end of the 1-handle, which disagrees with the framing at the 0-handle. 
These two framings are related by either a $+\pi$ or $-\pi$ spin rotation in $Spin(2)$.
\begin{figure}
\begin{center}
\includegraphics{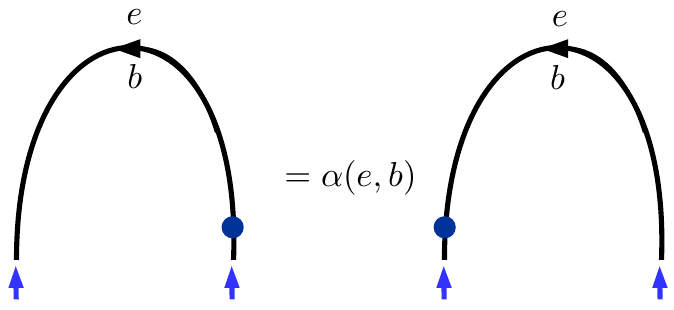}
\caption{\label{alphae_defn} The action of sliding a fermion across an edge $e$ labeled by the q-type object $b$ is given by $\alpha(e,b)$. 
The spin framings of the 0-handles on either end of $e$ are denoted by the blue arrows, 
and  the spin framing of the fermion on the right hand side is taken to match that of the left 0-handle. }
\end{center}
\end{figure}
We will denote this element of $Spin(2)$ by $\alpha(e)$, where $e$ is the edge corresponding to the 1-handle.

Note that the collection $\{\alpha(e)\}$ is determined by the spin structure of $\Sigma$ and
the choice of spin framings at each 0-handle.
Conversely, any collection $\{\alpha(e)\}$ determines a spin structure on $\Sigma\setminus\{\mbox{2-handles}\}$.
In order for this spin structure to extend to all of $\Sigma$, $\{\alpha(e)\}$ must satisfy the constraint that the 
boundary of each 2-handle has a bounding spin structure; see below.

Let $b$ be a q-type simple object.
We want to analyze the effect of sliding a fermionic dot over the edge $e$ when $e$ is labeled by $b$.
At the outset, we will choose an odd element $\gamma_b\in\End(b)$ such that $\gamma_b^2 = \id_b$, and also
$\gamma_{b^*}\in\End(b^*)$ such that $\gamma_{b^*}^2 = \id_{b^*}$.
The requirement that $\gamma_b^2 = \id_b$ determines $\gamma_b$ up to sign.
Let $r(b)\in\cc$ be such that $R_\pi\cdot\gamma_b = r(b)\gamma_{b^*}$, where
\be
	R_\pi : \End(b) \to \End(b^*)
\ee
denotes the spin rotation by $+\pi$.
It is easy to see that $r(b) = \pm i$.
The exact value will depend on the choices of standard generators $\gamma_b$ and $\gamma_{b^*}$.
If $b$ is not isomorphic to $b^*$, then we can always choose $\gamma_b$ and $\gamma_{b^*}$ so that $r(b) = i$.
But if $b = b^*$ (as happens in the $C_2$ theory, for example), then $\gamma_b = \gamma_{b^*}$ and the value of $r(b)$
is forced upon us, independent of the choice of $\gamma_b$.\footnote{
The condition that $b$ is equal to, rather than merely isomorphic to, $b^*$ is in some sense pathological.
But for theories build out of unoriented strands, like $C_2$, it is convenient to allow this pathology.}

We can now, finally, describe the effect of sliding a standard fermionic endomorphism (dot) over an edge $e$ labeled by 
a q-type particle $b$.
Let $\gamma_b * e$ denote the edge with the standard generator $\gamma_b$ placed at the incoming end (left side of Figure \ref{alphae_defn}).
Let $e* \gamma_{b^*}$ denote the edge with the standard generator $\gamma_{b^*}$ placed at the outgoing end (right side of Figure \ref{alphae_defn}).
Then
\be
	\gamma_b * e = \alpha(e, b) \cdot e* \gamma_{b^*} ,
	\label{fermion_slide}
\ee
where
\be
	\alpha(e, b) =  \left\{   \begin{array}{ll}  
		r(b) & \mbox{if $\alpha(e) = R_\pi$} \\
		-r(b) & \mbox{if $\alpha(e) = R_{-\pi}$} \\
	\end{array}  \right. .
\ee
This is illustrated in Figure \ref{alphae_defn}.

Fermionic dots can also be ``absorbed'' into vertices using the action of $\cliff_1$ on 
fusion spaces involving q-type objects, as discussed in Section \ref{fusion_spaces}. 
We can do this in analogy with \eqref{Gamma_y_def} by constructing odd operators $\Gamma_i$, $i\in \{1,2,3\}$, which map a pitchfork with a standard dot 
on the $i$-th outgoing pitchfork edge to a pitchfork without a fermion dot on the $i$-th leg. 
Graphically, $\Gamma_1$ is defined by 
\be \label{gamma1_defn}
\underset{\psi}{\Pitchforkdotone} = \sum_{\eta}[ \Gamma_1 ]_{\psi \eta} \underset{\eta}{\PitchforkLarge}
\ee
where $\psi,\eta$ are basis vectors for $V^{abc}$, and the fermionic dot has a higher sign ordering than the basis vector it sits on.
$\Gamma_2$ and $\Gamma_3$ are defined similarly:
\be \label{gamma2gamma3_defn}
\underset{\psi}{\Pitchforkdottwo} = \sum_{\eta}[ \Gamma_2 ]_{\psi \eta} \underset{\eta}{\PitchforkLarge}, \quad \quad
\underset{\psi}{\Pitchforkdotthree} = \sum_{\eta}[ \Gamma_3 ]_{\psi \eta} \underset{\eta}{\PitchforkLarge}
\ee
Since only q-type objects can host fermionic dots, $\Gamma_i$ is only defined when the $i$-th 
leg of the vertex is labeled by a q-type object. 
Since only q-type objects have odd endomorphisms, 
$\Gamma_i$ is only defined when the i-{th} label of the pitchfork is q-type.
The $\Gamma_i$ are all odd matrices, reversing fermion parity.

\medskip

\begin{figure}
\begin{center}
\includegraphics{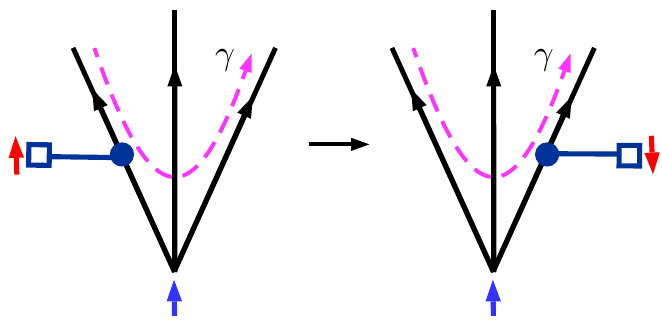}
\caption{ \label{spin_rot_through_0handle} An illustration of a $+\pi$ spin framing rotation picked up as the path $\gamma$ (drawn in dashed purple) passes through a 0-handle. 
If we were to proceed along $\gamma$ in the opposite direction, we would pick up a $-\pi$ spin framing rotation instead.   }
\end{center}
\end{figure}

We mentioned above that the collection of $\{\alpha(e)\}$ must satisfy some constraints if it is 
to extend over the 2-cells and give a specified spin structure.
Here are the details:
let $\gamma$ be an oriented framed loop in $\Sigma$.
For simplicity, we will assume that (a) $\gamma$ is embedded, (b) the framing is the natural one coming from the tangent 
space of $\gamma$, and (c) $\gamma$ is contained in the union of the 0- and 1-handles.
We want to compute the spin rotation, either 0 or $2\pi$, which $\gamma$ picks up from the spin structure on the 0- and 1-handles.
When $\gamma$ goes over a 1-handle $e$, it picks up a rotation of $\alpha(e)$ or $-\alpha(e)$, depending on whether it goes over $e$
with or against the orientation of $e$.
Each time $\gamma$ passes though a 0-handle, it also picks up a rotation of $\pm\pi$.
Suppose $\gamma$ enters the 0-handle at the $i$-th 1-handle and exits the 0-handle at the $j$-th 1-handle.\footnote{
Note that $i$ and $j$ refer to the ordering of 1-handle attachments local to the 0-handle, not to some global ordering. The $i$ and $j$ are assigned in the same way as the $\Gamma_i$ and $\Gamma_j$ in \eqref{gamma1_defn} and \eqref{gamma2gamma3_defn}.}
With our ordering conventions, this gives a rotation of $+\pi$ if $i<j$ and a rotation of $-\pi$ if $i > j$ (see Figure \ref{spin_rot_through_0handle}).
Combining all of these $\pm\pi$ rotations results in a rotation of 0 or $2\pi$.
A bounding spin structure on $\gamma$ yields a $2\pi$ rotation (since a $2\pi$ rotation acts as 
multiplication by $-1$ and produces the requisite anti-periodic boundary conditions), and a non-bounding 
spin structure corresponds to no rotation.
In particular, if $\gamma$ is the boundary of a single 2-cell/plaquette, then it must correspond to a spin 
framing rotation of $2\pi$, since we assume that the spin structure around each 2-cell is bounding (which 
allows the spin structure on the 0- and 1-handles extends to a spin structure on all of $\Sigma$). 
More generally, $\gamma$ will always get a spin framing rotation by $2\pi$ if it is in the trivial homology 
class of $H_1(\Sigma)$.

\subsection{Terms in the Hamiltonian} \label{terms_in_Hamiltonian}

In this section we will finally write down the Hamiltonian, and then explain the terms appearing in it in detail. 
As mentioned earlier, the Hamiltonian consists of three kinds of mutually commuting projections:
\begin{align} \label{terms_in_ham}
H = \lambda_p \sum_{p \in \mcf} (1-B_p)  + \lambda_e \sum_{e \in \mce} (1-D_e) + \lambda_v \sum_{e\in \mce} (1-A_e)
\end{align}
We'll start with a discussion of the ``vertex'' term $A_e$ and then address the new edge term $D_e$, 
finally ending by describing the plaquette term $B_p$.

\subsubsection{Vertex term}   \label{VertexHamiltonian}

Let $v_1$ and $v_2$ be two vertices joined by an edge $e$.
We define
\begin{align}
A_e(\ket{\psi_1} \otimes \ket{\psi_2}) = 
\left\{
                \begin{array}{ll}
                   \ket{\psi_1} \otimes \ket{\psi_2} & \text{if $\psi_1$ and $\psi_2$ assign the same label to $e$} \\
                  0 & \text{otherwise}
                \end{array}
              \right.
\end{align}
In other words, $A_e$ forces the labels on each end of an edge to agree.

Why do we call these ``vertex terms" when they are indexed by edges, and the support is a pair of adjacent vertices, joined by an edge?
Because it does the same work as the vertex term of the usual bosonic LW Hamiltonian.
If our fermionic category happens to be bosonic (i.e.\ it lacks fermions), then the ground state of our ``vertex" term is isomorphic to the ground state
of the vertex term in the usual LW Hamiltonian.\footnote{Note that the ``admissibility'' condition that requires $V^{abc}$ to be in the ground state vector space only if $\unit\in a\tp b \tp c$ is already satisfied for us, since our local Hilbert spaces at the vertices already have this condition built in.}
If we had chosen to put spins on edges as well as vertices, then we could write a vertex term that was actually indexed by vertices.
But its ground state would be isomorphic to the above edge-like vertex term.

Note that vectors in the ground state of the vertex term can be interpreted as string nets.
If $K$ is the ground state of the vertex term, we have maps
\be
	K \to \mch(\Sigma \setminus \mbox{2-cells}) \to \mch(\Sigma) .
	\label{ground_state_projection}
\ee
where $\mch(\Sigma \setminus \mbox{2-cells})$ and $\mch(\Sigma)$ are the ground-state Hilbert spaces of $\Sigma \setminus \mbox{2-cells}$ and $\Sigma$, respectively.
The job of the edge term (below) will be to pick out a subspace of $K$ on which the first map is an isomorphism. 
The job of the plaquette term will be to further reduce to a subspace such that the composite map to $H(\Sigma)$ is an isomorphism.

\subsubsection{Edge term} 
\label{edge_term}

The edge terms are the qualitatively new feature of this Hamiltonian, and are a necessary ingredient for the Hamiltonian of any theory possessing q-type particles. 
They are only well-defined on ground states of the vertex term. 
They allow fermions to fluctuate across edges of the graph which are labelled by q-type particles,
and provide a way of energetically implementing the isotopy relations associated with sliding fermions along the worldlines of q-type particles. 
In Section \ref{modified_tensor_product} we solved the same problem in
a different context by replacing a tensor product over scalars with a tensor product
over the endomorphism algebra of a q-type simple object.
The edge term of the Hamiltonian is a stand-in for the tensor product over a non-trivial endomorphism algebra.

The edge term will coherently add and remove fermions at the end points of the q-type bonds, as well as tunnel them across.\footnote{Heuristically we can think of two adjacent vertices as islands which can hold a number of fermions and whose fermion parity is well defined. 
If these vertices are connected by a q-type edge, we can think of that edge as a 1D superconductor which coherently couples the two islands together.}
This term favors an equal-weight (meaning equal up to a phase factor) 
superposition of fermions across all vertices connected by 
q-type simple objects with a fixed fermion parity.
Since the edge term is responsible for allowing fermions to fluctuate (``hop'') across edges labeled by q-type objects, 
it will be absent in any theory with no q-type objects.

Fermion hopping across q-type edges is implemented by the $\Gamma$ operators. 
To do this for an edge $e$, we can create a pair of fermions near the vertex at the beginning of $e$, slide one of the fermions along $e$ to the vertex at the end of $e$, and then use the $\Gamma$ operators to ``absorb'' each fermion into their respective vertices. 
For example, if $e$ hosts a q-type edge label $x$, then we have 
\begin{align}
\xymatrix @!0 @M=1mm @R=18mm @C=30mm{
\Edgea\ar[rr]^{\lambda^{-1}}&&\Edgeb\ar[dd]^{\alpha(e,x)} \\
&&\\
\Edgea&&\ar[ll]_{\Gamma_3 \tp \Gamma_1} \Edgec
	} 
	 \label{dot_slide_gamma} 
\end{align}
where we have defined tensor products of $\Gamma$ operators to act so that operators located further to the left in tensor products absorb fermions with higher order than the operators to their right.
Note that although the diagrams in the first and last steps in the above sequence look the same, they are not: the fermion parity of the vectors in the two vertex Hilbert spaces $V^{abx^*}$ and 
$V^{xcd}$ has been switched. 

For a generic edge $e$ oriented from $v_1$ to $v_2$ and colored by a fixed object $x$ the edge term
can be written as the projector
\begin{align} \label{De_defn}
D_e = \frac{1}{2}(1+J_e),
\end{align}
where $J_e$ acts on vectors $|\psi_1\rangle \tp |\psi_2\rangle \in \mch_{v_1}\tp \mch_{v_2}$ 
by implementing the local relations coming from $\End(x)$:
\be J_e(|\psi_1\rangle\tp |\psi_2\rangle) = 
\begin{cases}
  \lambda^{-1}\alpha(e,x)\sum_{\eta_1,\eta_2}[\Gamma_i]_{\psi_1\eta_1} |\eta_1\rangle \tp [\Gamma_j]_{\psi_2\eta_2}|\eta_2\rangle \quad &\text{if $x$ is q-type}\\ 
 |\psi_1\rangle\tp |\psi_2\rangle \quad &\text{if $x$ is m-type} \end{cases}
\ee
Here, we have taken $e$ to be the $i$th leg of the pitchfork at $v_1$ and the $j$th leg at $v_2$, and $|\eta_1\rangle \tp |\eta_2\rangle \in \mch_{v_1} \tp \mch_{v_2}$.
Since $D_e$ acts as the identity on edges colored by m-type objects, edges with m-type edges will automatically lie in the ground state of the edge term.

In summary, including $D_e$ in the Hamiltonian provides a way of energetically enforcing the conditions that the ground states of theories containing q-type particles are realized by superpositions of string-net configurations possessing all possible ways of arranging fermions on the q-type strings. 
Ensuring that ground states are superpositions of different fermion configurations is tantamount to projecting from the Hilbert space $\mch_\mcg$ to the physical Hilbert space $\mch_{\rm phys}$, in which redundant degrees of freedom created by different fermion configurations are modded out.

As mentioned earlier, it can be conceptually helpful to note that a 1D Kitaev wire in the topological phase also exhibits the same behavior as the q-type strings in our theories.
However, since generic theories (like the $\halfesix$ example considered earlier) can have fusion rules in which two q-type simple objects to fuse to a third q-type simple object, this analogy is not perfect, since at the junction of three Kitaev wires a zero mode is left behind, which does not occur in the $\halfesix$ theory.

\subsubsection{Plaquette term}  \label{ss_plaquette_term}

As with the vertex term, the plaquette term is essentially the same as the plaquette term in the usual string-net Hamiltonian: 
it inserts an $\omega$ loop 
into each plaquette, and uses local relations to fuse these $\omega$ loops into the boundaries of the plaquettes. 
Physically it is responsible for the dynamics of net configurations which reside in the ground space of the vertex and edge terms, 
and it is designed so that two string nets that correspond to the same state vector receive the same amplitude. 

Using the definition of the $\omega$ loop, we can write $B_p$ as 
\begin{align} \label{plaquette_term_defn}
B_p =\frac{1}{\mcd^2} \sum_{a \in \sob(\spc)} \frac{d_a}{\text{dim} \; \text{End}(a)} B^a_p, 
\end{align}
where the operator $B_p^a$ fuses a loop labeled $a$ into the edges and vertices neighboring the plaquette $p$ (note that 
as discussed earlier, the operator $B_p^a$ is only defined when all vertices and edges neighboring the plaquette $p$ satisfy the corresponding vertex and edge terms).

\begin{figure}
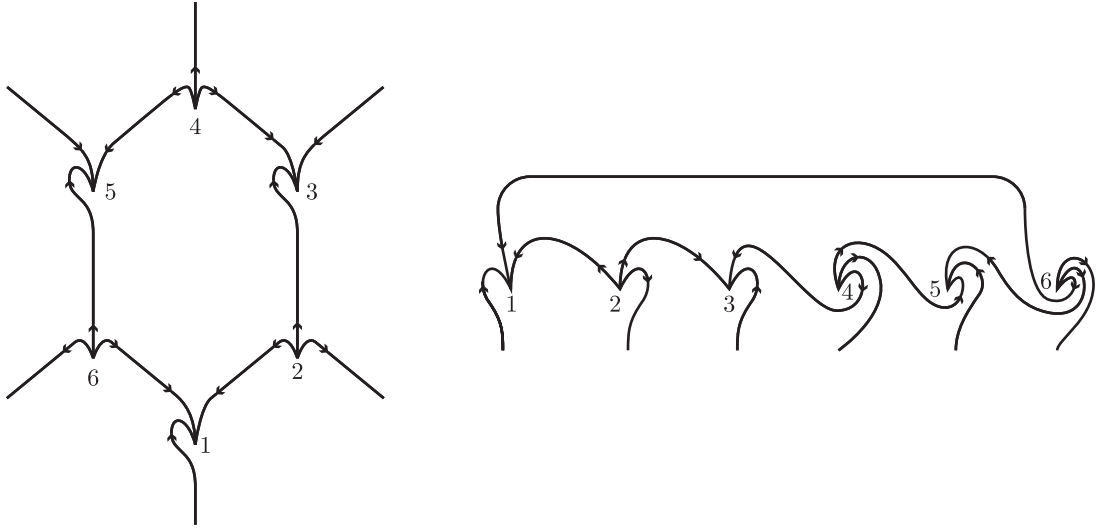

\begin{center}
\begin{align}
\nonumber 
{\HexPlaquette} \quad \quad \quad
{\HexPlaquetteStretchedPrime}
\end{align}
\caption{\label{HexagonCell}
On the left we have single 2-cell along with its neighboring edges and vertices.
Tiling the this 2-cell results in a honeycomb latter.
Each vertex has been standardized in to a pitchfork.
On the right we have stretched out the 2-cell in preparation for applying the plaquette operator.
We can transform it into our standard basis \eqref{plaquett_std_basis} via $R = P \tp \id \tp \id \tp P^{-1} \tp P^{-1} \tp P^{-2}$. 
}
\end{center}
\end{figure}

The matrix elements of $B_p^a$ depend on the choice of cell decomposition and pitchforkization procedure. 
For a generic cell decomposition it is somewhat tedious to write down these matrix elements, 
although the procedure is straightforward. 
To expedite this process we apply yet another standardization procedure.
Suppose we are given a plaquette $p$ with $n$ neighboring vertices labeled from $1, \cdots, n$ in a counterclockwise fashion with respect to the orientation of $\Sigma$, see Figure \ref{HexagonCell}.
The Hilbert space associated to the plaquette $p$ is the subspace of
\begin{align}
\mch_p = \mch_{v_1} \tp \mch_{v_2} \tp \cdots \tp \mch_{v_n}
\end{align}
which satisfies the vertex and edge terms. 
Diagrammatically, 
states in this space take the form of the figure on the right hand side of \ref{HexagonCell}.
It is useful to apply a number of pivots to vectors in $\mch_p$ so that the vector space takes on a standard form which will facilitate the application of $B_p^a$.
Define the operator $R_p$ by
\begin{align}
R_p: \; \mch_p \ra \bigoplus_{\{ x_i, z_i \} } V^{x_n^* x_1^* z_1^*} \tp V^{x_1 x_2 z_2} \tp \cdots \tp V^{x_{n-1} x_n z_n}
\end{align} 
which can be explicitly written as 
\begin{align}
R_p = P_1 \tp P_2 \tp \cdots \tp P_n
\end{align}
with each $P_j = \bigoplus_{abc} (P^{abc} )^{l_j}$ defined in \eqref{Pitchfork_pivot} and with $l_j = -2,-1,0,1,2$ controlling the angle by which the $j$th vertex is pivoted (if the edges aren't all oriented out of the vertex then the appropriate object in $P^{abc}$ needs to be replaced by its dual). 
We choose the pivots so that a vector in the image of $R_p$ takes the form
\begin{align} 
\label{plaquett_std_basis}
\Plaquettea
\end{align}
We now write down the matrix elements of $B_p^a$ in this basis, 
which we call the {\em standardized} basis for the plaquette $p$.
We choose an implicit sign ordering which increases from left to right (in particular, the numerical 
subscripts on $x,\mu,z,$ etc. denote string-net labels and not Koszul orders). 
As with \eqref{pitchfork_basis} if some of the edges have a different orientation then the object is replaced by its dual.

The following is identical to the standard Levin-Wen plaquette term written in the notation of Section \ref{def_sect}.
To find the action of $B_p^a$ in the standardized basis of \eqref{plaquett_std_basis}
we first insert a closed strand labeled $a$ into the interior of $p$:
\begin{align}
\Plaquetteb
\end{align}
Next, we begin fusing the strand into the plaquette. We start with the strand labeled by $x_6$ and use the resolution of the identity \eqref{idpitchfork} on the $a$ and $x_6$ strands, which gives the picture
\begin{align}
\label{plaquette_first_F_move}
\sum_{y_6, \sigma_0} \frac{d_{y_6}}{\mcb(\sigma_0^* \tp \sigma_0)}\Plaquettec
\end{align}
Next we use the associator \eqref{PitchforkFMove} to pull the $a$ string over $x_6$ in the diagram on the right to obtain
\begin{align}
\sum_{y_1, \sigma_1} &(F^{y_6^*a x_1^*z_1^*} )_{(x_6; \sigma_0^* \mu_1) (y_1^*; \nu_1 \sigma_1 )} \times \\ 
&\times  \Plaquetted.
\end{align} 
It will be helpful to use the short hand $(F)_{(\sigma_0^* \mu_1) (  \nu_1\sigma_1 )}$ for $(F^{y_6^*a x_1^*z_1^*} )_{(x_6; \sigma_0^* \mu_1) (y_1^*; \nu_1 \sigma_1 )}$, 
which is well-defined so long as $\sigma_0,\mu_1, \sigma_1, \nu_1$ are defined, 
which will be clear from context.
We keep applying F-moves until we are left with
\begin{align}
 &\sum_{\substack{\nu_1 \cdots \nu_6 \\ \sigma_1 \cdots \sigma_6}}F_{(\sigma_0^* \mu_1)( \nu_1\sigma_1)} 
F_{(\sigma_1 \mu_2)( \nu_2 \sigma_2)} \cdots 
F_{(\sigma_5 \mu_6)( \nu_6 \sigma_6)} 
  \times \\
 &\times \Plaquettee
\end{align} 
We note that the sign ordering is increasing from left to right.
We now perform one last isotopy and remove the ``bubble": 
\begin{align}
&\Plaquettee =\\ 
&\nonumber \\
&=\Plaquettef \\ 
&\nonumber \\
& = (-1)^{|\sigma_6|} (-1)^{|\sigma_6| ( |\sigma_0|+|\nu_1| + \cdots+ |\nu_6|)}\frac{\mcb(\sigma_6 \tp \sigma_0)}{d_{y_6}} \times \\
&\times \Plaquetteg.
\end{align}
In the last step, the factor of $(-1)^{|\sigma_6|}$ comes from applying a $2\pi$ pivot 
to the $\sigma_6$ vertex and the factor of $(-1)^{|\sigma_6| ( |\sigma_0|+|\nu_1| + \cdots+ |\nu_6|)}$
is the Koszul sign coming from ensuring that $\sigma_6$ is located immediately before $\sigma_0$ in the ordering (recall that the pairing $\mcb$ is only defined on diagrams with a specific Koszul ordering; see \eqref{reflection_pairing_defn}). 
$\mcb(\sigma_6 \tp \sigma_0)$ is zero unless $\sigma_6$ is dual to $\sigma_0$,
and thus the $\mcb(\sigma_6\tp \sigma_0)/d_{y_6}$ factor is cancelled by the factor of $d_{y_6}/\mcb(\sigma_0^*\tp\sigma_0)$ introduced in \eqref{plaquette_first_F_move}.
Noting that $|\sigma_6| = |\sigma_0|$, we can write (recall \eqref{plaquett_std_basis}),
\begin{align}
\nonumber
[B_p^a]_{(\mu_1 \cdots \mu_6)(\nu_1 \cdots \nu_6)} =  \frac{1}{\mcd^2} \frac{d_a}{\dim \End(a)} \sum_{\sigma_1 \cdots \sigma_6} &F_{(\sigma_6 \mu_1)( \nu_1\sigma_1)} 
F_{(\sigma_1 \mu_2)( \nu_2 \sigma_2)} \cdots 
F_{(\sigma_5 \mu_6)( \nu_6 \sigma_6)} 
\\
& \times (-1)^{|\sigma_6| (|\nu_1| + \cdots+ |\nu_6|)}
\label{plaquettematrix}
\end{align} 
The plaquette term in the basis $\mch_p$ is given by conjugating $B_p^a$ with $R_p$.
Note that due to \eqref{pivotconsistent} the pivots commute with the $F$-moves, 
and so the computation can be carried out in either basis.

\medskip

We now briefly remark on surfaces with boundary.
Let $Y$ be such a surface and let $c$ be a boundary condition for $Y$, i.e.\ a collection of labeled string net endpoints on $\bd Y$.
Choose a graph $G$ in $Y$ such that the ``boundary" of $G$ is $c$ and each component of $Y\setminus G$ which does not meet $\bd Y$ is a disk.
Given this input, we construct a Hamiltonian similarly to above.
There are plaquette terms only for the interior 2-cells.
There are edge terms only for interior edges of $G$.
Vertices of $G$ which are adjacent to the labeled points of $c$ will have one or more of their edge labels fixed.
The ground state of this Hamiltonian can be identified with $Z(Y;c)$.

We will see in Section \ref{kitaev_wire} that one instance of this construction (with $Y$ a disk and $c$ consisting of two points labeled by a q-type particle
and its dual) yields the Kitaev chain Hamiltonian.

\medskip

Before moving on, we briefly mention a 
high-level way of understanding the plaquette operator in terms of the tube category.
Let $\widehat\Sigma$ denote the surface $\Sigma$ with a disk removed from each 2-cell.
Each boundary component of $\widehat\Sigma$ corresponds to a plaquette term in the Hamiltonian. 
There is an obvious bijection between the boundary components of $\widehat\Sigma$ and the plaquette terms of the Hamiltonian.
For any collections $c$ of labeled string endpoints on the boundary of $\widehat\Sigma$, we have
the string net Hilbert space $Z(\widehat\Sigma; c)$.
If $c$ is the empty boundary condition (no labeled endpoints), then this Hilbert space can be identified with
the ground state of the vertex and edge terms of the Hamiltonian.
The tube categories of each boundary component act on the collection of vector spaces $\{Z(\widehat\Sigma; c)\}$,
and consequently we can decompose these spaces according to the simple objects of the collective tube category.
The summands of this decomposition correspond to the labelings of each boundary component of $\widehat\Sigma$ with an
anyon of the tube category.
Now consider gluing disks to each boundary component of $\widehat\Sigma$ to obtain $\Sigma$.
This has the effect of projecting to the summand corresponding to placing the trivial anyon at each boundary component.
Another way of achieving the same effect is to place a copy of the of the trivial tube category idempotent $e_0$
on an annulus at each boundary component.
This is exactly what the plaquette term of the Hamiltonian does.
This point of view also explains why the elementary excitations of the Hamiltonian
correspond to placing tube category idempotents at 2-cells~\cite{kirillov2011}.

\subsection{Excitation spectrum} \label{excitations_of_H}

Finally, we briefly comment on the spectrum of the Hamiltonian \eqref{ham} and the types of excitations it model supports. 
Each of the terms in the Hamiltonian enforces one of the local relations of the vector space assigned to $\Sigma$ by $\spc$. 
As such, the zero-energy ground space is just the vector space assigned to $\Sigma$ by $\spc$.
The deconfined anyonic excitations in the model correspond to violations of the plaquette and vertex terms.
By construction, these are in one-to-one correspondence with the bounding idempotents of the tube category.
This is simply because as discussed earlier, the $B_p$ operator in the plaquette term is the projector $\Pi_\unit$, which projects onto states containing no quasiparticles in the interior of $p$. 
The fusion rules of the excitations can be computed with the tube category methods developed in earlier sections. 

\begin{figure}
\begin{center}
\includegraphics{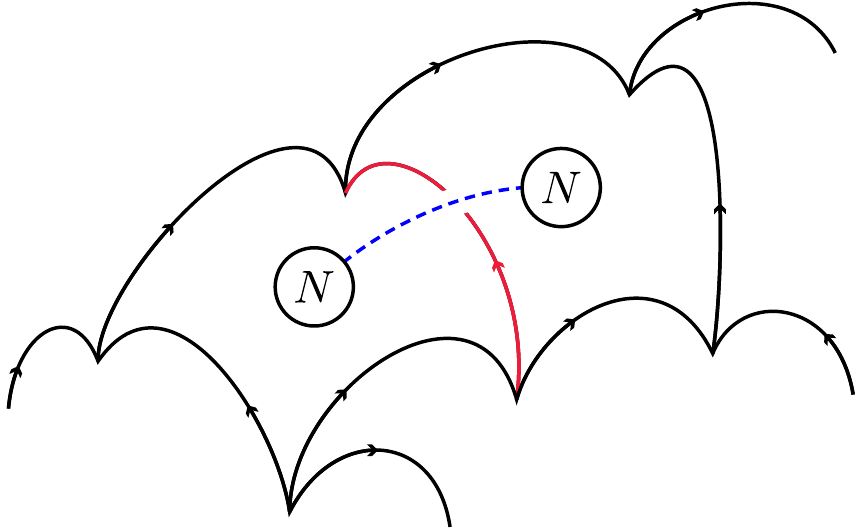}
\caption{\label{lattice_w_vortices} A region of the graph $\mcg$ containing two vortex excitations. 
The two punctures hosting the vortex excitations are marked as circles, with the label $N$ denoting their non-bounding spin structure. 
The dashed blue line is the spin structure defect connecting the two punctures.
The edge colored red intersects the spin structure defect and is excited, violating the edge term $D_e$ in the Hamiltonian. 
}
\end{center}
\end{figure}

One can also consider the excitations corresponding to violations of the edge terms 
(note that these will only be present in theories with q-type objects).
If a given state has a $+1$ eigenvalue under an edge term $(1-D_e)$, 
then the fermions traveling across $e$ pick up an additional minus sign relative to the background spin structure.
This is illustrated in Figure \ref{lattice_w_vortices}, where the violated edge $e$ is shown in red, and the two circles marked $N$ denote the vortices created on either side of $e$. 
Diagrammatically we denote the additional minus sign with a branch cut (the dashed line in Figure \ref{lattice_w_vortices}).
This implies that violating a single edge term nucleates a pair of vortices (the set of which are in one-to-one correspondence with the non-bounding idempotents of the tube category) on the plaquettes adjacent to the edge $e$ (recall that the plaquette terms are only non-zero in the ground space of the edge and vertex terms).
This means that the vortex excitations can only be separated
at the expense of a linear increase in energy, and so are linearly confined.

Alternatively, by modifying the Hamiltonian we can introduce vortices by hand. 
We remove the plaquette terms where we wish the vortices to reside, 
and require the corresponding plaquette boundaries
to have a non-bounding spin structure.
Relative to the un-modified Hamiltonian, 
the vortices will be connected by spin structure branch cuts.
(The edge terms of the modified and unmodified Hamiltonians will differ for edges which intersect these branch cuts.)
A ground state of the modified Hamiltonian will be an excited state of the unmodified Hamiltonian, 
whose energy depends on the choice of branch cut.
To deconfine the vortices one needs to give the spin structure dynamics; 
we leave the study of this possibility to future work.


\section{Super pivotal state sums and tensor networks} \label{state_sums}

In this section we describe a version of the Turaev-Viro-Barrett-Westbury (TVBW) state sum \cite{Turaev1992,Barrett1996}
for super pivotal fusion categories 
and a tensor network for the ground state wave function of the Hamiltonian constructed in Section \ref{Super_pivotal_Hamiltonian}.
Related 
work was presented in~\cite{bhardwaj2016}, 
see also \cite{Bultinck2017}.
We first review the TVBW construction for bosonic spherical fusion categories. 
We then show how to write the state-sum as a tensor contraction on a tensor 
network.
Next we detail the modifications needed for the fermionic versions of the state sum and tensor network.
Lastly we use the state sum to write down an explicit wave function for the ground ground state of \eqref{ham}.

\medskip

\begin{figure}
\begin{center}
\includegraphics[scale = 1]{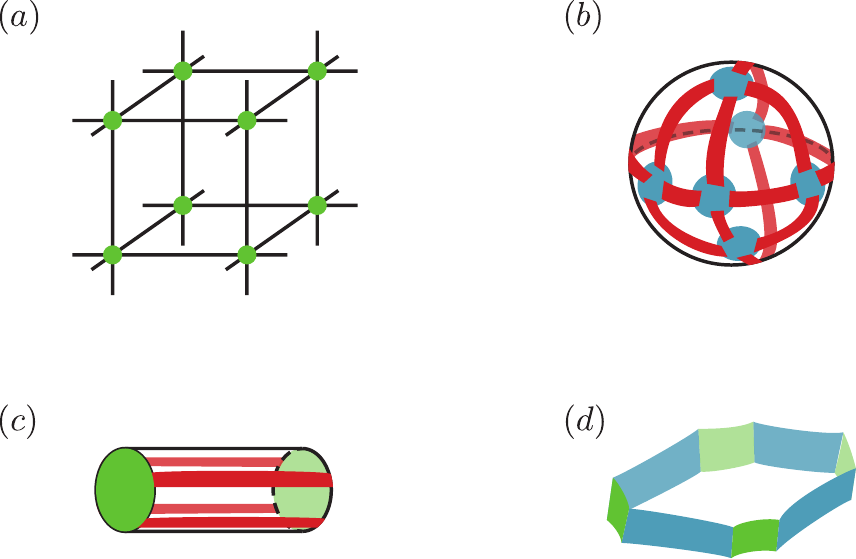}
\end{center}
\caption{\label{HandleDecompFig}%
The handles corresponding to a standard cubic cell decomposition (a).
The 0-, 1- and 2-handles are shown in (b), (c) and (d), with colors green, blue and red.
In (b) the blue disks on the 0-handle indicate where the 1-handles attach to the 0-handle, and the red rectangles
indicate where the 2-handles attach to the 0-handle.
In (c) the green disks indicate where the 0-handles attach to the 1-handle and the red rectangles indicate where the 2-handles attach to the 1-handle.
Similarly, in (d) the blue and green rectangles indicate where the 2-handles attach to the 0- and 1-handles.
We have omitted the 3-handles from the figure.}
\end{figure}

Before we begin, we need to establish some terminology regarding cell and handle decompositions. 
Recall that a handle decomposition for a 
3-manifold $M$ is built from a series of $k$-handles, with $k=0,1,2,3$, each of which is identified with $D^k\times D^{3-k}$. 
Handle decompositions can be obtained from cell decompositions by thickening each $k$-cell into a $k$-handle.
Conversely, each handle decomposition determines a cell decomposition by taking the cores of the handles.
(See Section \ref{standardized_handles} for more details.)
We will often refer to a $k$-cell and its associated $k$-handle with the same letter, since
it will be convenient for us to be able to describe things in terms of both handle decompositions 
and their corresponding cell decompositions. 
We call $S^{k-1} \times D^{3-k}$ the attaching region (or attaching boundary) of the $k$-handle,
and $D^k\times S^{3-k-1}$ the non-attaching boundary.
The attaching map of a $k$-handle is a homeomorphism from the attaching region to 
a submanifold of the boundary of the union of the lower-dimensional handles.
The topology of $M$ is encoded by the various attaching maps.
The different types of $k$-handles are illustrated in Figure 
\ref{HandleDecompFig}.

\subsection{Bosonic TVBW state sum}

\subsubsection{Definition of the state sum}

Our first task is to describe the TVBW (bosonic) state sum.
The original references are \cite{Turaev1992,Barrett1996}.
We will use the form for a general cell/handle decomposition, as described in \cite{Walker2006}.

Let $M$ be a closed oriented 3-manifold equipped with a handle decomposition $\mch$.
Choose auxiliary orientations of the 1- and 2-cells of the cell decomposition corresponding to $\mch$.
Let $\mch_i$ denote the set of $i$-handles ($i = 0,1,2,3$).
The state sum has the form
\be \label{bos_tv_sum}
	Z(M) = \sum_{\beta\in\mcl(\mch)}
		\prod_{c\in\mch_3} \mcd^{-2}
		\prod_{f\in\mch_2} d(f, \beta)
		\prod_{e\in\mch_1} \widetilde\Theta(e, \beta)^{-1}
		\prod_{v\in\mch_0} \text{Link}(v, \beta) .
\ee
The next few paragraphs define the notation used in \eqref{bos_tv_sum}.

We use the 2-cell orientations to define an oriented graph (unlabeled string net) on the boundary of each 0-, 1- and 2-handle,
as shown in Figure \ref{TwoHandleToGraph}.
String-net graphs are assigned to the $k$-handles as follows:
\begin{figure}
\begin{center}
\includegraphics[scale = 1]{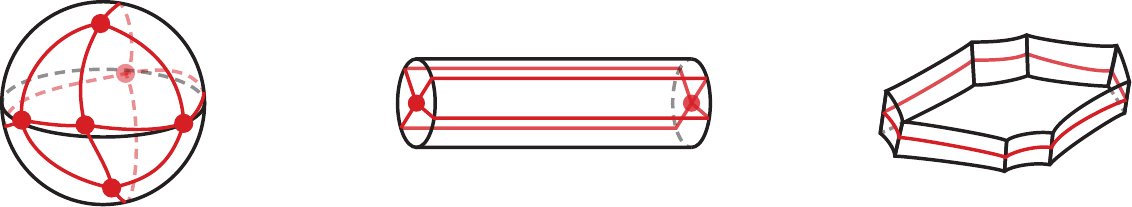}
\end{center}
\caption{\label{TwoHandleToGraph}%
Graphs on the boundaries of 0-, 1- and 2-handles, in the case of a cubic cell decomposition.
Compare Figure \ref{HandleDecompFig}.}
\end{figure} 
\begin{itemize}
\item On 2-handles, the graph is a single loop along the core of the attaching annulus of a 2-handle.
The orientation of the loop is determined by the orientation of the 2-cell. 
\item On 1-handles, the graph is a generalized $\Theta$ graph, which we will call a $\widetilde\Theta$ graph.
The graph has one edge for each 2-handle adjacent to the 1-handle.
The middle part of each edge of the graph corresponds to where the cores of the 2-handles meet the boundary of the 1-handle.
The two vertices of the graph are on the two attaching disks of the 1-handle.
The edges are oriented opposite to the orientations used in the 2-handle loops above.
\item On each 0-handle, the graph is determined by the pattern of 2- and 1-handles adjacent to the 0-handle.
The graph has one edge for each adjacent 2-handle and one vertex for each adjacent 1-handle.
The orientations of the edges are opposite to the orientations of the 2-handle loops.
We denote this graph $\text{Link}(v)$.
\end{itemize}

\begin{figure}
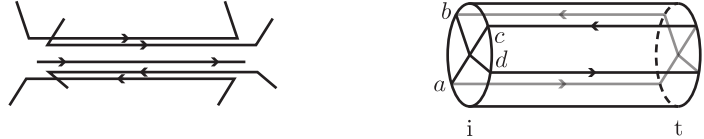

\centering
\begin{align}
\nonumber
\CellDecompNearOneHandle \quad \quad \quad \quad \quad \OneHandlePrime 
\end{align}
\caption{\label{OneHandlePrime}
On the left, we have an illustration of four 2-cells meeting a 1-cell.
For clarity we have put a small gap between the 1-cell and the four 2-cells.
On the right we have the corresponding 1-handle and a particular labeling. 
We have denoted the corresponding attaching disks by `i' for initial, and `t' for terminal. 
}
\end{figure}

Recall that we have an orientation of each 1-cell.
This allows us to distinguish an ``initial" and ``terminal" attaching disk for each 1-handle; see Figure \ref{OneHandlePrime}.
On the initial disk we see a graph with a single vertex in the interior of the disk and $k$ edges connecting the central vertex
to the boundary of the disk (where $k$ is the number of 2-handles which cross the 1-handle).
For each labeling $\ell$ of these edges by simple objects in $\sob(\mcc)$, we have an associated vector space $V(\ell)$.
For example, in the case of Figure \ref{OneHandlePrime} the vector space is isomorphic to $V^{ab^*c^*d}$.
Let $B(\ell)$ be some chosen basis of this vector space.
For each 1-handle $e$ define $B(e)$ to be the union over all labelings $\ell$ of $B(\ell)$,
and also define $V(e)$ to be the direct sum of all the vector spaces $V(\ell)$.

We define the set of all labelings $\mcl(\mch)$ to be the product over all 1-handles $e$ of the basis sets $B(e)$.
In other words, we choose (independently, without any compatibility constraints) a labeling by simple objects of the edges of each initial
disk graph, then choose a basis vector for each associated vector space.

We also associate a vector space $V^*(\ell)$ to the terminal disk of each labeled 1-handle.
In the example, this is isomorphic to $V^{d^*cba^*}$.
There is a nondegenerate bilinear pairing between $V(\ell)$ and $V^*(\ell)$, 
given by evaluating the labeled string net on the boundary of the 1-handle (which is a 2-sphere).
We will choose a basis of $V^*(\ell)$ such that the pairing matrix is diagonal.
(It is sometimes convenient to not insist that the diagonal entries be $\delta_{ij}$.)
We also define $V^*(e)$ to be the direct sum (over all labelings $\ell$) of $V^*(\ell)$.

We are now ready to define the weights appearing in the state-sum $Z(M)$. 
Let $\beta\in\mcl(\mch)$ and let $f$ be a 2-handle.
The labeling $\beta$ associates a simple object to each intersection of $f$ with a 1-handle.
If these simple objects are not all the same, we define $d(f, \beta) = 0$.
If they are all equal to the same simple object $a\in\sob(\mcc)$, we define the weight $d(f, \beta)$
appearing in \eqref{bos_tv_sum} by $d(f,\beta) = d_a$.

Let $e$ be a 1-handle.
The labeling $\beta$ associates a basis vector $\mu$ to the initial disk of $e$.
Define $\widetilde\Theta(e, \beta)$ to be the value of the bilinear pairing evaluated on $\mu^*$ and $\mu$.
Diagrammatically, $\widetilde\Theta(e, \beta)$ is found by connecting the open strings in $V(\ell)$ to their dual counterparts in $V^*(\ell)$ and evaluating 
the resulting diagram.
Continuing with our example in Figure \ref{OneHandlePrime}, 
we have 
\be \label{four_banana}
\widetilde \Theta (e, \beta) =  \Bananafourmu.
\ee

\begin{figure}
\centering
\includegraphics{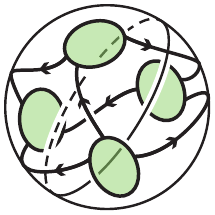}
\caption{ \label{TetSphere_Fig} 
An example of the surface of a 0-handle on which four 1-handles are attached to. 
The attaching regions for the 1-handles are marked in green. The coloring $\beta$ assigns 
objects to the oriented strands and fusion space basis vectors to the green regions.
}
\end{figure}

Let $v$ be a 0-handle.
The labeling $\beta$ determines a labeling of the graph $\text{Link}(v)$ as follows.
Near each vertex of $\text{Link}(v)$ we place the basis element $\mu^* \in V^*(e)$ (or $\mu \in V(e)$) assigned by $\beta$ to the 
corresponding 1-handle if $v$ is attached to the initial (terminal) end of the 1-handle.
If these vertex labels are incompatible along edges of $\text{Link}(v)$, we define $\text{Link}(v, \beta) = 0$.
If they are all compatible then we define $\text{Link}(v, \beta)$ to be the evaluation of the resulting labeled graph (string net).
For cell decompositions dual to a triangulation, 
the labeled graph is a tetrahedral string net on a sphere.
This is illustrated in Figure \ref{TetSphere_Fig}, which shows an example 0-handle on which 
four 1-handles terminate. 

This completes the definition of the state sum.

\medskip

It follows from Section 8.2 of \cite{Walker2006} that the state sum computes $Z(M)$, independently of the choice of handle decomposition
and choice of orientations of 1- and 2-cells.

\medskip

If $M$ has non-empty boundary,
then we choose the cell decomposition so that $\bd M$ lies in the union of the 2-skeleton (union of 0-1, 1- and 2-cells).
(An alternative choice would be to require that $\bd M$ is transverse to the 2-skeleton.
The two conventions each have strengths and weaknesses.)

The 0- and 1-cells on $\bd M$ will do double duty as the underlying graph of a string net on $\bd M$.
Choose an orientation of each 1-cell on $\bd M$.
(This is analogous to choosing an orientation of the boundary of a 2-cell in the interior of $M$.)
Choose a labeling of these oriented edges by simple objects in $\sob(\mcc)$.
For each vertex (0-cell) on $\bd M$, choose an element of the appropriate disk vector space.

We now have a labeled string net $g$ on $\bd M$.
The state sum will evaluate the path integral $Z(M)(g)$ 
(i.e.\ the path integral of $M$ with boundary condition given by $g$).
The labelings and weights are defined as before, except that some of the labels are already determined by the string net $g$ on the boundary.

\subsubsection{The state sum as a tensor network}

Our goal in this subsection is to reinterpret \eqref{bos_tv_sum} as a tensor network.
We will first discuss that case when $M$ is closed, then consider the case when $\bd M$ is non-empty.

\medskip

If we (temporarily) ignore the factors of $d(f, \beta)$ and $\mcd^{-2}$ in \eqref{bos_tv_sum}, 
it is easily seen to compute the contraction of a tensor network.
The underlying graph of the tensor network is the union of 0- and 1-cells of the cell decomposition.
The vector space associated to each edge $e$ is $V(e)$ as defined above.
The matrix elements of the tensor associated to a 0-cell $v$ are the numbers $\text{Link}(v, \beta)$ defined above.
The factors of $\widetilde\Theta(e, \beta)^{-1}$ arise from the pairing of dual tensor indices.

To incorporate the factors of $d(f, \beta)$ and $\mcd^{-2}$, we make some ad hoc choices.
We consider ``dressed" 
0-handle weights that incorporate the factors of $d(f,\beta)$ and $\mcd^{-2}$ for certain adjacent 2- and 3-cells.
For each 2-handle $f$ we choose an adjacent 0-handle $v_f$.
For a 0-handle $v$ we modify the associated weight ${\text{Link}}(v,\beta)$ by multiplying factors of $d(f,\beta)$ for each 2-handle $f$ such that $v = v_f$.
Similarly, for each 3-handle we choose an adjacent 0-handle and multiply the associated weight by $\mcd^{-2}$.
Denoting the modified 0-handle weights by $\widetilde{\rm Link}(v,\beta)$, 
we define the 0-handle tensors $T_v$ as follows.
Let $e_1, \ldots, e_k$ be the 1-handles adjacent to the 0-handle $v$.
Let 
\begin{align}
\label{0_handleVectorspaces}
V_i = \begin{cases}
V(e_i) & \text{if $v$ is adjacent to the terminal end of $e_i$}\\
V^*(e_i) & \text{ if $v$ is adjacent to the initial end of $e_i$}.\\
\end{cases}
\end{align}
We define
\be
	T_v \in V^*_1\tp\cdots\tp V^*_k
\ee
by
\be
\label{vertex_tensor}
	T_v(w_1\tp\cdots\tp w_k) = \widetilde{\text{Link}}(v, w_1\tp\cdots\tp w_k),
\ee
where $w_i\in V_i$.
In other words, $T_v$ evaluates the link graph with labels determined by $w_1,\ldots,w_k$ and multiplied by the factors of $d(f,\beta)$ and $\mcd^{-2}$ as described above.
To obtain the partition function, we trace out the tensor product of all the 0-handle tensors constructed in this way.
Because the vector space associated to the region on a $0$-handle attached to the 
terminal end of a 1-handle $e$ is dual to the vector space associated to the corresponding region on the 
0-handle attached to the initial end of $e$, there are precisely as many dual vectors as vectors in the tensor product, and 
contracting each vector with its associated dual vector computes the complex number $Z(M)$. 
That is, we have 
\be Z(M) = \tr \left(\bigotimes_{v\in \mch_0} T_v\right),\ee
where the trace $\tr$ denotes the tensor contraction.
It is easy to see that this tensor network gives the same state sum as \eqref{bos_tv_sum} and 
is independent of the way we assign factors of $d(f,\beta)$ and $\mcd^{-2}$ to the 
vertex tensors. 

\medskip

In the case where $\partial M \neq \emptyset$, we define the 0-handle tensors $T_v$ as before,
but in this case some of the legs of the tensors are unpaired (not contracted).
Specifically, there is one unpaired leg for each 0-cell on $\bd M$.
If $W_1, \ldots, W_n$ are the vector spaces associated to the boundary 0-cells, we have
\be \label{trace_pM_nonempty}
	Z(M) = \tr \left( \bigotimes_{v\in \mch_0} T_v \right) \in W_1^*\tp\cdots\tp W_n^* .
\ee
Each string net on $\bd M$ (i.e. each labeling of 0- and 1-cells as described above)
determines an element of $W_1\tp\cdots\tp W_n$ (the vertex labels).
By evaluating \eqref{trace_pM_nonempty} on this element we obtain the amplitude of the wave function for the string net.

\subsubsection{Standardization procedures} \label{bosonic_standardization} 

The above tensor network construction is irregular, in the sense that (potentially) every edge vector space is different 
and every 0-handle tensor is different.
There are several standardization procedures that reduce this irregularity. 

One standardization procedure is to start with a cell decomposition that is dual to a triangulation.
This ensures that there are exactly three 2-handles meeting each edge, and that the vertex graphs $\text{Link}(v,\beta)$ are all tetrahedral.
However, this still leaves us with several different types of edge vector spaces and vertex tensors.
For example, depending on the choice of 1- and 2-cell orientations, there will be eight possible vector spaces associated to 1-handles,
corresponding to $V^{abc}$, $V^{ab^*c}$, $V^{a^*b^*c^*}$, etc.
(Note that if all Frobenius-Schur indicators are equal to 1, then we can ignore these distinctions.)
Similarly, there will be many different tetrahedral vertex tensors, depending on the orientations of the edges of the tetrahedral graph.

We can further standardize the tensor network by choosing a global ordering of the 3-cells of the handle decomposition.
(This is equivalent to a global ordering of the vertices of the dual triangulation.)
We then choose 2-cell orientations so that the orientation of a 2-cell together with a normal vector pointing toward the higher-ordered of the two
adjacent 3-cells agrees with the orientation of $M$.
We can now choose 1-cell orientations so that the graph on the initial disk of each 1-handle has a trivalent vertex with two
outgoing and one incoming edge.
With these orientation choices, we have, for every 1-handle $e$,
\be
	V(e) = \bigoplus_{a,b,c\in\sob(\mcc)} V^{abc^*} .
\ee
Furthermore, all of the tetrahedral graphs have the same pattern of edge orientations, so all of the 0-handle tensors
in the tensor network are of the same form. 

However, it is not always convenient to choose a global ordering.
For example, our main application is a tensor network associated to $Y\times I$.
If $Y$ is a torus, we might hope that the network has translational symmetry, but this is not compatible with the global ordering trick.
For this reason, in what follows we will work with a cell decomposition dual to a triangulation, but we will not employ the global ordering trick.

We will find it useful to employ a standardization 
procedure in which all fusion spaces in the string nets assigned to the 0-handles 
assume the ``pitchfork'' form introduced 
in our treatment of the Hamiltonian. 
(These vertices are all trivalent since we are now 
assuming a cell decomposition dual to a triangulation.)
In this convention, the $T_v$ tensor weights are all computed by the evaluation of tetrahedral diagrams:
\begin{align} 
\label{Bosonic_tet}
	T_v(\alpha \tp \beta \tp \gamma \tp \delta) = {\rm Tet}(v,\alpha \tp \beta \tp \gamma \tp \delta) = \TetrahedronPrime , 
\end{align}
where we have defined the tensor weight by its evaluation of the picture on the right with 
$\alpha \in \bigoplus_{abc} V^{abc}$, $\beta \in \bigoplus_{abc} V^{a^* bc}$, $\gamma \in \bigoplus_{abc} V^{a^* b^* c^*}$, and $ \delta \in \bigoplus_{abc} V^{a^* b c^*}$.
The pitchforkization procedure does not entirely fix the form of the $T_v$ tensors above, 
since the tetrahedral nets associated with each 0-handle will in general have different 
edge orientations.  
Since there are $2^6 = 64$ possible choices of edge orientations near a 0-handle, there will be 64 
different types of $T_v$ tensors, given by the appropriate diagram evaluation with $\alpha,\beta,\gamma,\delta$ chosen from the appropriate fusion spaces.%
\footnote{Not all these $64$ tetrahedra are independent however, as 
some of them can be transformed into one another by using the pivotal and spherical structure of the input category.}

In this procedure, we choose standardizations (pitchforkizations) at each 0-handle independently.
This means that the two ends of each 1-handle are standardized independently 
of each other, and the pairing induced from the $\widetilde\Theta$ graph on the 1-handle 
\eqref{four_banana}
will not necessarily agree with the standard pairing \eqref{reflection_pairing_defn}.
Instead, the 1-handle pairing and the standard pairing will be 
related by a pivot operation $P_e=P^{l_e}$ for $l_e = 0,1,2$, where $P$ is the pivot defined in \eqref{Pitchfork_pivot}.
We define $P_e$ to rotate diagrams counterclockwise relative to the orientation of $e$. 
If $l_e \neq 0$, then the 
pitchforks at the
initial and terminal vertices of $e$ are twisted by $2\pi l_e / 3$ relative to 
one another, and this twisting data needs to be incorporated into the 1-handles. 
The 1-handles now look like
\begin{align} \label{horshoe_resln} 
\HorshoeIdentity, 
\end{align}
where the double arrows designate the orientation of $e$. 

\begin{figure} 
\begin{centering} 
\includegraphics[scale=0.6]{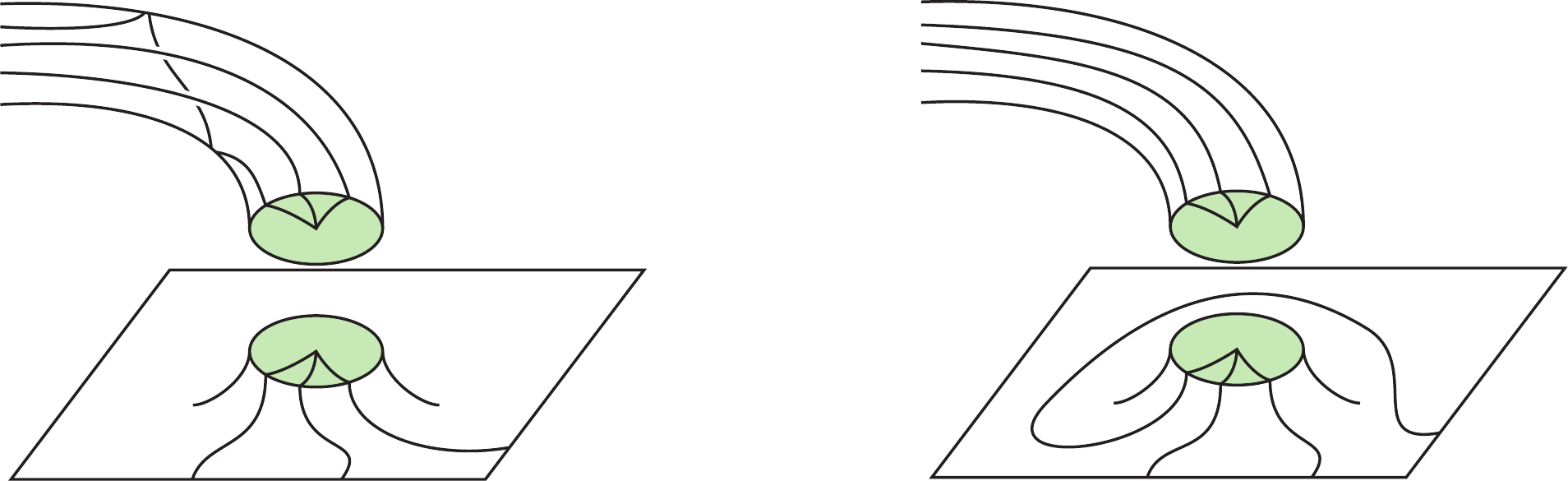} 
\caption{\label{OneHandlevsTetPivot}
 A 1-handle and its attaching region on an adjacent 0-handle (the plane denotes a section of the surface of the 0-handle).
Standardizing the 0-handle attaching regions to be ``pitchforks" requires that we either have to add pivots to the 1-handle (left, where one of the strands on the 1-handle is wrapped around the 1-handle) or to the corresponding 0-handle attaching region (right, where one strand on the 0-handle is pivoted around the attaching region). 
In \eqref{bos_tv_sum_std} we choose to include the pivots within 1-handles.}
\end{centering} 
\end{figure}

In summary, by standardizing each 0-handle independently, we have managed to make each
0-handle tensor isomorphic to a standard (up to edge orientations) tetrahedral tensor.
The price we pay for this is that we have to keep track of the pivots which relate the two ends of each 1-handle.
In terms of the tensor network, this means that we either need to insert two-legged pivot/1-handle tensors between each pair
or adjacent 0-handle tensors, or we need to further ``dress" the 0-handle tensors by merging each such pivot tensor
into one of the two adjacent 0-handle tensors.
See Figure \ref{OneHandlevsTetPivot}.

\medskip

These standardizations also serve to make the original state sum \eqref{bos_tv_sum} more uniform.
We can now write
\be \label{bos_tv_sum_std}
	Z(M) = \sum_{\beta\in\mcl(\mch)}
		\prod_{c\in\mch_3} \mcd^{-2}
		\prod_{f\in\mch_2} d(f, \beta)
		\prod_{e\in\mch_1} \Theta(P_e, \beta)^{-1}
		\prod_{v\in\mch_0} \text{Tet}(v, \beta) .
\ee
Here $\Theta(P_e, \beta)$ is a standard pairing as in \eqref{reflection_pairing_defn}, but modified by the pivot isomorphism $P_e$.
The weight $\text{Tet}(v, \beta)$ is a standard tetrahedral symbol (though there are still variants which depend on
the orientations of the edges of the tetrahedron).
The point is that we have now written the state sum for an arbitrary 3-manifold in terms of a finite number
of standard weights.

\subsection{The fermionic state sum}

\subsubsection{Definition of the fermionic state sum}

We now extend the bosonic state sum to the fermionic case. 
We start with two pieces of data: a super pivotal fusion category $\spc$, and a spin
3-manifold $M$ possessing a cell decomposition with orientations of 1- and 2-cells.
The fermionic version of the state sum is similar to the bosonic version:
\begin{align}
\label{fermionic_tv_sum}
	Z(M) = \sum_{\beta\in\mcl(\mch)}(-1)^{\kappa_\beta}
		\prod_{c\in\mch_3} \mcd^{-2}
		\prod_{f\in\mch_2} \frac{d(f, \beta)}{n(f,\beta)}
		\prod_{e\in\mch_1}  \widetilde \Theta(e, \beta)^{-1}
		\prod_{v\in\mch_0} \text{Link}(v, \beta) .
\end{align}
The fermionic version of the state sum differs from the bosonic version in the following ways:
\begin{itemize} 
\item 
The string nets corresponding to the 0- and 1-handle weights require a sign-ordering of their string net vertices. 
This in turn requires that the partition function is weighted by a Koszul sign $(-1)^{\kappa_\beta}$
which measures the difference between the global sign ordering coming from the 1-handles and the global sign ordering coming from the 0-handles.
\item The weights assigned to the 2-handles need to be properly normalized, 
resulting in a factor of $n(f,\beta) = \dim \End(a)$ if $a$ is the simple object labeling the core of the attaching annulus for the 2-handle $f$ by $\beta$.
\item The spin structure on $M$ determines how the basis elements which make up the labeling $\beta$ are inserted into
the graphs $\text{Link}(v)$.
\end{itemize} 
As in the previous section, we will now explain 
the factors appearing in \eqref{fermionic_tv_sum}. 

As before, we use the 2-cell orientations to define an oriented graph (unlabeled string net) on the boundary of each 0-, 1- and 2-handle.
String net graphs are assigned to the $k$-handles in the same way as in the bosonic case. 
The set of all labelings $\mcl(\mch)$ is defined as the product over all 1-handles $e$ of the basis sets $B(e)$.
For a fixed labeling $\beta \in \mcl(\mch)$, the weights are determined as follows.

The 2-handle weight $d(f, \beta)$ is defined in the same way as before. 
However we now divide by the factor $n(f, \beta) = n_a = \dim\End(a)$,
where $a$ is the simple object labeling the boundary of the core of the 2-handle.
This factor is necessary because the norm-square of the $a$-labeled loop is $n_a$.

The 1-handle weights are determined by the bilinear pairings given by each 1-handle $e$. 
When evaluating the graph on the boundary of $e$, we choose the Koszul ordering which puts the terminal vertex
immediately before the initial vertex in the ordering (similar to the convention in \eqref{reflection_pairing_defn}).

For each 0-handle $v$, $\text{Link}(v,\beta)$ is defined in the same way as in the bosonic case:
we evaluate a string net determined by the 1- and 2-handles incident on $v$ and the labeling $\beta$.
There are two subtleties here.
First, when mapping a vertex label $\mu$ from the initial (terminal) end of a 1-handle to the 0-handle adjacent to the terminal (initial) end of the 1-handle, 
we must employ the attaching map which connects the terminal (initial) end of the 1-handle to the 0-handle.
This attaching map is a spin diffeomorphism, 
and changing the attaching map by a spin flip changes the sign
of the label on the 0-handle by $(-1)^{|\mu|}$.
It is here (and only here) that the state sum is sensitive to the spin structure on $M$.
The second subtlety concerns Koszul orderings.
In order (pun noticed but not intended)
to evaluate the string net on the boundary of the 0-handle, we must choose an (arbitrary) ordering
of the string net vertices of the graph on the boundary of the 0-handle.
Thus the evaluation $\text{Link}(v, \beta)$ is arbitrary up to a sign.
However, we will see that a change of Koszul ordering which changes the sign of $\text{Link}(v, \beta)$ also produces a compensating
change in the factor $(-1)^{\kappa_\beta}$, and so the overall state sum is well defined.

\begin{figure} 
\centering
\includegraphics{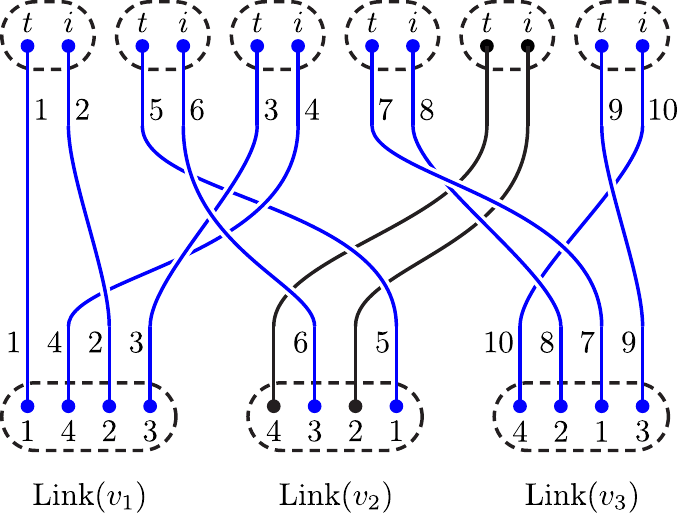} 
\caption{\label{KoszulFig} 
An example of how to compute the Koszul sign $(-1)^{\kappa_\beta}$ diagrammatically  
for a graph consisting of three 0-handles and six 1-handles. 
The lower dashed regions represent the 0-handles, while the upper dashed regions 
represent the 1-handles (with initial and terminal ends marked as $i$ and $t$, respectively). 
The Koszul orderings indicated by the numbers are explained in the text, and 
$\kappa_\beta$ is given by the number of crossings of the blue strands (in this example, 
$(-1)^{\kappa_\beta}=-1$).
}
\end{figure}

The Koszul sign $(-1)^{\kappa_\beta}$ is defined as follows.
Consider, for fixed $\beta$, the tensor product of the all the super vector spaces
associated to the attaching disks on all the 0-handles.
If there are $k$ 1-handles, then there are $2k$ tensor factors in this tensor product, one 
for each 1-handle end. 
We will compare two different orderings of the tensor factors.
In the first ordering, we place each terminal disk immediately before each initial disk in the ordering.
Such an ordering is well-defined up to even permutations.
In the second ordering, 
we choose a global ordering of the 0-handles and then
use the above choices of local ordering for the factors associated to
each 0-handle.
This is again well-defined up to even permutations, since each 0-handle graph evaluates to zero when the total
parity at that 0-handle is odd.
(It depends on the choice of local orderings but not on the choice of global ordering of the 0-handles.)
We then define $(-1)^{\kappa_\beta}$ to be Koszul sign relating these two orderings. 

\medskip

We now describe a convenient way to compute $(-1)^{\kappa_\beta}$ graphically, with an 
example shown in Figure \ref{KoszulFig} for a graph consisting of six 1-handles 
(upper dashed regions) and 3 0-handles $v_1,v_2,v_3$ (lower dashed regions). 
Each 0-handle has four 1-handle attaching regions, which are indicated by the lower small dots and 
which possess a local ordering relative to one another (indicated by the numbers within the lower dashed regions). 
Each attaching region can either be even (black) or odd (blue). 
If it is odd, we assign a Koszul ordering to the attaching
region, consistent with the local ordering at the 0-handle.
This ordering is indicated by the numbers appearing just outside the lower dashed regions.

Each 1-handle either has two even ends or two odd ends: if it has two odd ends, we assign an ordering 
to the ends by placing the terminal end immediately before the initial end in the ordering. This ordering
is denoted by the top row of numbers below the 1-handles in Figure \ref{KoszulFig}. 

To evaluate $(-1)^{\kappa_\beta}$, we draw a fermion line connecting each odd 0-handle attaching region
with Koszul order $k$ to the respective 1-handle end with Koszul order $k$. $(-1)^{\kappa_\beta}$ is then 
simply $(-1)^{n_c}$, where $n_c$ is the number of crossing between fermion lines in the resulting diagram. 
In the example of Figure \ref{KoszulFig} we have $n_c=11$, and so $(-1)^{\kappa_\beta}=-1$. 

\medskip

The case of non-empty $\bd M$ presents one new issue not present in the bosonic version:
we must pick a Koszul ordering of the labels corresponding to string net vertices on $\bd M$.
Once this has been done, we can combine that ordering with the ordering coming from the 1-handles.
The Koszul sign $(-1)^{\kappa_\beta}$ is now defined to be the sign arising from comparing 
the 0-handle ordering with the combined $\bd M$ and 1-handle ordering.

\subsubsection{The fermionic state sum as a tensor network}

We now turn to the task of reinterpreting \eqref{fermionic_tv_sum} as a tensor network.

\medskip

We incorporate the factors of $d(f, \beta)/n(f,\beta)$ and $\mcd^{-2}$ into the 0-handle 
weights 
in the same way as in the bosonic case. 
As in the bosonic case we denote the dressed 0-handle weights by $\widetilde{\rm Link}(v,\beta)$.
We define the vertex tensor in a similar fashion to \eqref{vertex_tensor}. 
Let $e_1, \cdots, e_k$ be the 1-handles adjacent to the 0-handle $v$ with the same ordering as the vertices of the graph $\text{Link}(v)$.
Let $V_i$ be defined in the same way as \eqref{0_handleVectorspaces} (with the modification that $V_i$ is a super vector space).
We define 
\begin{align} \label{Tv_defn}
T_v  \in V_1^* \tp \cdots \tp V_k^*
\end{align}
by 
\begin{align}
T_v(w_1 \tp \cdots \tp w_k) = \widetilde{\text{Link}}(v, w_1 \tp \cdots \tp w_k)
\end{align} 
where $w_i \in V_i$.
In words, $T_v$ evaluates the string-net graph determined by $\text{Link}(v)$ with the ordered vertices labeled by $w_1, \cdots, w_k$ (in the same order), and multiplied by the appropriate factors of $d(f,\beta)/n(f,\beta)$ and $\mcd^{-2}$.

As in the bosonic case, the partition function 
is computed as a trace of the $T_v$ tensors:
\be \label{fermion_Z_as_tr} Z(M) = \tr \left(\bigotimes_{v\in \mch_0} T_v\right).\ee
where the $\tr$ denotes contracting dual indices.
In practice, to perform the trace, we require that the pair of vectors to be contracted be adjacent in the Koszul ordering (with terminal preceding initial). 
To make the pair of vectors adjacent in the Koszul ordering we need to apply a number of Koszul isomorphisms. 
After contracting all vectors we pick up the appropriate factor of $(-1)^{\kappa_\beta}$. 
Again, $Z(M)$
is independent of the way we assign factors of $d(f,\beta)/n(f,\beta)$ and $\mcd^{-2}$ to the 
vertex tensors. 

If $\bd M$ is non-empty, we again follow the bosonic prescription in \eqref{trace_pM_nonempty} 
to obtain $Z(M) = \tr \left( \bigotimes_{v\in \mch_0} T_v \right) \in W_1^*\tp\cdots\tp W_n^*$, where 
$W_1, \ldots, W_n$ are the vector spaces associated to the boundary 0-cells and we 
are implicitly making use of the undordered tensor product. When using $Z(M)$ to compute 
amplitudes of different string-net boundary conditions, care must be taken when performing the 
tensor contraction on ordered representatives because of Koszul sign issues.

\subsubsection{Fermionic standardization procedures} \label{fermionic_standardization}

As in the bosonic case, we can standardize the tensor network by
choosing a generic cell decomposition (dual to a triangulation)
and ``pitchforkizing" all trivalent vertices which appear on the boundaries of 0-handles.
Note that in this fermionic setting, pitchforkizing includes choosing a spin framing at each trivalent vertex.
This standardization procedure results in string net vertices on 0-handles which are standardized independently of their partners at the opposite
ends of the 1-handles: 
the form of a given trivalent vertex at the initial edge of a 1-handle $e$
and the form of the associated vertex on the terminal edge of the 1-handle may be related by a pivot operation.
Properly accounting for this requires inserting pivot operators $P_e=P^{l_e}$ into the 
1-handles, as in \eqref{horshoe_resln}. 
Rather than tacking the spin-structure signs onto the 0-handle weights 
we incorporate them into the pivots.
Indeed, we now have $P_e^3=(-1)^F$, so that $l_e$ is 
valued in $\zz_6$ as opposed to $\zz_3$.
(Alternatively, we could keep $l_e \in \zz_3$  but insert $(-1)^{F} \cdot P^{l_e}$ where appropriate.)
Note that the spin structure of the underlying 3-manifold is encoded in the edge pivots $P_e$ (and the standardized 0-handles).

The standard tetrahedral string net on each 0-handle must of course incorporate a Koszul ordering in the fermionic case: 
\begin{align} 
	T_v(\alpha \tp \beta \tp \gamma \tp \delta) = {\rm Tet}(v,\alpha \tp \beta \tp \gamma \tp \delta) = \Tetrahedron.
\end{align}
Note that there are still multiple versions of the standardized tetrahedral weights ${\rm Tet}$ differing by choices of the edge orientations.
The fermionic analogue of \eqref{bos_tv_sum_std} can now be written:
\begin{align}
\label{ferm_tv_sum_std}
	Z(M) = \sum_{\beta\in\mcl(\mch)}(-1)^{\kappa_\beta}
		\prod_{c\in\mch_3} \mcd^{-2}
		\prod_{f\in\mch_2} \frac{d(f, \beta)}{n(f,\beta)}
		\prod_{e\in\mch_1}  \Theta(P_e, \beta)^{-1}
		\prod_{v\in\mch_0} \text{Tet}(v, \beta) .
\end{align}
Again, $\Theta(P_e, \beta)$ is a standard pairing as in \eqref{reflection_pairing_defn} but modified by the pivot isomorphism $P_e = P^{l_e}$.

\subsection{The shadow world and ground state wave functions}\label{shadowworld}

In this subsection we construct a state sum and corresponding tensor network that produces the ground state wave function of the 
Hamiltonian defined in Section \ref{Super_pivotal_Hamiltonian}.
In a nutshell, the idea is to apply the general tensor network construction of the previous subsection to the spin 3-manifold
$\Sigma\times I$, where $\Sigma$ is the spin surface which hosts the Hamiltonian.

Recall that the big Hilbert space for the Hamiltonian defined on a cell decomposition $\mcg$ is
\begin{align} \label{bhs_redef}
	\mch_\mcg =\bigotimes_{v \in \mcv} \mch_v,
\end{align}
where $\mch_v $ is defined to be $\bigoplus_{a,b,c} V^{abc}$ if all edges point away from the vertex, 
with similar definitions of $\mch_v$ in the case of other edge orientation arrangements.
If a basis vector of $\mch_\mcg$ satisfies edge label compatibility for all adjacent pairs of vertices
(equivalently, if the basis vector lies in the ground state of the vertex term of the Hamiltonian),
then it can be interpreted as defining a string net on $\Sigma$.
A wave function (not ``the" wave function unless the ground state is 1-dimensional) $\Psi$ assigns a weight $\Psi(w)$ to each such basis vector $w$, 
in such a way that if $\sum_i c_i w_i$ is equal to zero in $A(\Sigma)$, then $\sum_i c_i \Psi(w_i) = 0$.
For basis vectors $w$ which violate edge label compatibility, we have $\Psi(w) = 0$.

Given a string net $g$ on $\Sigma$, we can define a wave function $\Psi_g$ via
\be  \label{wfg-def}
	\Psi_g(w) = Z(\Sigma\times I)(\bar g \cup w) .
\ee
In other words, we evaluate the path integral $Z(\Sigma\times I)$ with boundary condition $\bar g$ on $\Sigma\times\{0\} \cong -\Sigma$
and boundary condition $w$ on $\Sigma\times\{1\} \cong \Sigma$.
(Recall that $\bar g$ is the reflected version of the string net $g$ on the orientation-reversed surface $-\Sigma$.)
Note that as $g$ runs through a basis of $A(\Sigma)$, $\Psi_g$ runs through a basis of the wave functions for the ground state of the Hamiltonian.
Our task now reduces to using the techniques of the previous subsection to construct a tensor network which evaluates the RHS of \eqref{wfg-def}.

\medskip

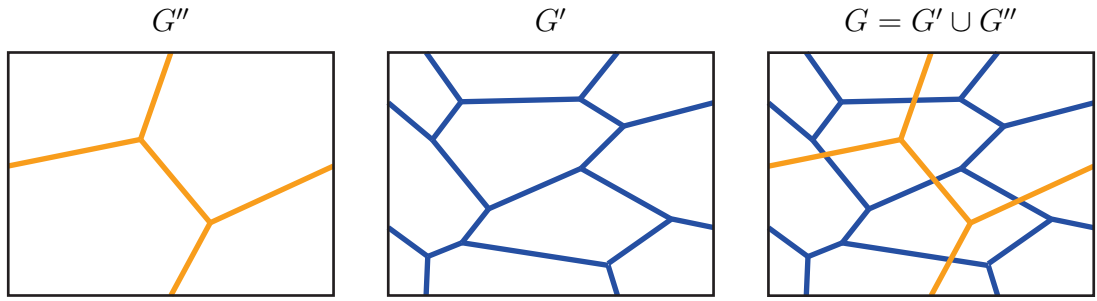
\begin{figure}
\begin{align}
\nonumber
\vcenter{
\xymatrix @!0 @M=2mm @R=20mm @C=50mm{
G'' & G' & G = G' \cup G'' \\
\GraphGprimeprime&\GraphGprime& \GraphG
	}}
\end{align}
\caption{\label{slab_decomp}
The cell decompositions described in the text. $G''$ (left) is the cell 
decomposition on which the input string-net state is defined, and $G'$ (center) is the decomposition on 
which the Hamiltonian acts. The union $G=G'\cup G''$ (right) is the cell decomposition we use to 
construct the ground state wavefunctions. }
\end{figure}

First we must specify a handle decomposition of $\Sigma\times I$.
Let $G'$ be the 1-skeleton of the cell decomposition of $\Sigma$ associated with $\mch_\mcg$, and let $G''$ be the 1-skeleton of the
cell decomposition of $\Sigma$ underlying the input string net $g$.
(In practice, $G'$ will be as fine a lattice as our computer can handle, while $G''$ will be as simple as possible
subject to the constraint that $g$ can represent a basis of $A(\Sigma)$.)
We stipulate that $G'$ and $G''$ are transverse (i.e., only their 1-cells intersect, and all the intersections are transverse), and are each dual to a triangulation, so that their vertices are all 3-valent.
We define the cell decomposition $G$ to be the union of $G'$ and $G''$.
The graph $G$ has three types of vertices: vertices of $G'$,
vertices of $G''$, and vertices corresponding to the points of $G'\cap G''$ which
are 4-valent; see Figure~\ref{slab_decomp}.

\begin{figure}
\begin{centering}
\includegraphics[scale=.7]{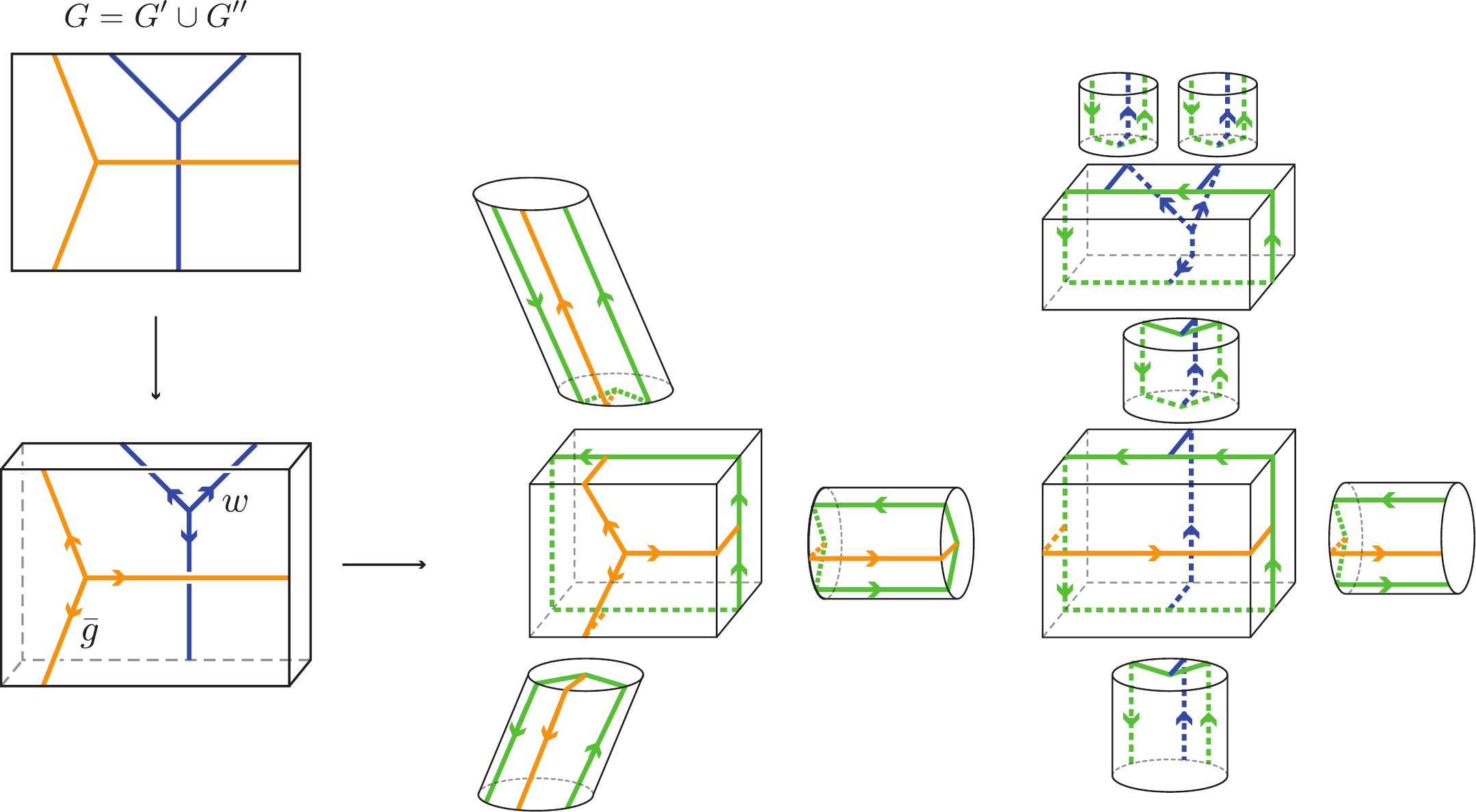}
\caption{\label{VertexTypes}
An example cell decomposition on $\Sigma \times I$. 
In the upper left figure, we show the cell decomposition $G$ for a given simple choice of $G'$ (blue) and $G''$ (orange). 
In the lower left figure, we include the ``time direction'' in the picture, which thickens it into a box with boundary conditions 
set by $\bar g$ on the initial boundary $\Sigma \times \{0\}$ and boundary conditions set by $w$ on the terminal boundary $\Sigma \times \{1\}$. 
The right figure shows the full handle decomposition for this setup. Each box shows a 0-handle in the composite cell decomposition $G$, 
while each cylinder shows a 1-handle. The green lines denote string nets in the interior of $\Sigma \times I$, which are not fixed 
by either of the boundary conditions $\bar g, v$.
}
\end{centering}
\end{figure}

Our handle decomposition for $\Sigma\times I$ will be a thickened version of $G$.
We have a 0-handle for each vertex of $G$, a 1-handle for each edge of $G$, and a 2-handle for each 2-cell of the complement of $G$.
There are no 3-handles.
Figure \ref{VertexTypes} illustrates this handle decomposition, and also shows how the string nets
$v$ and $\bar g$ are situated on its boundary.

\begin{figure}
\begin{center}
\includegraphics[scale=0.6]{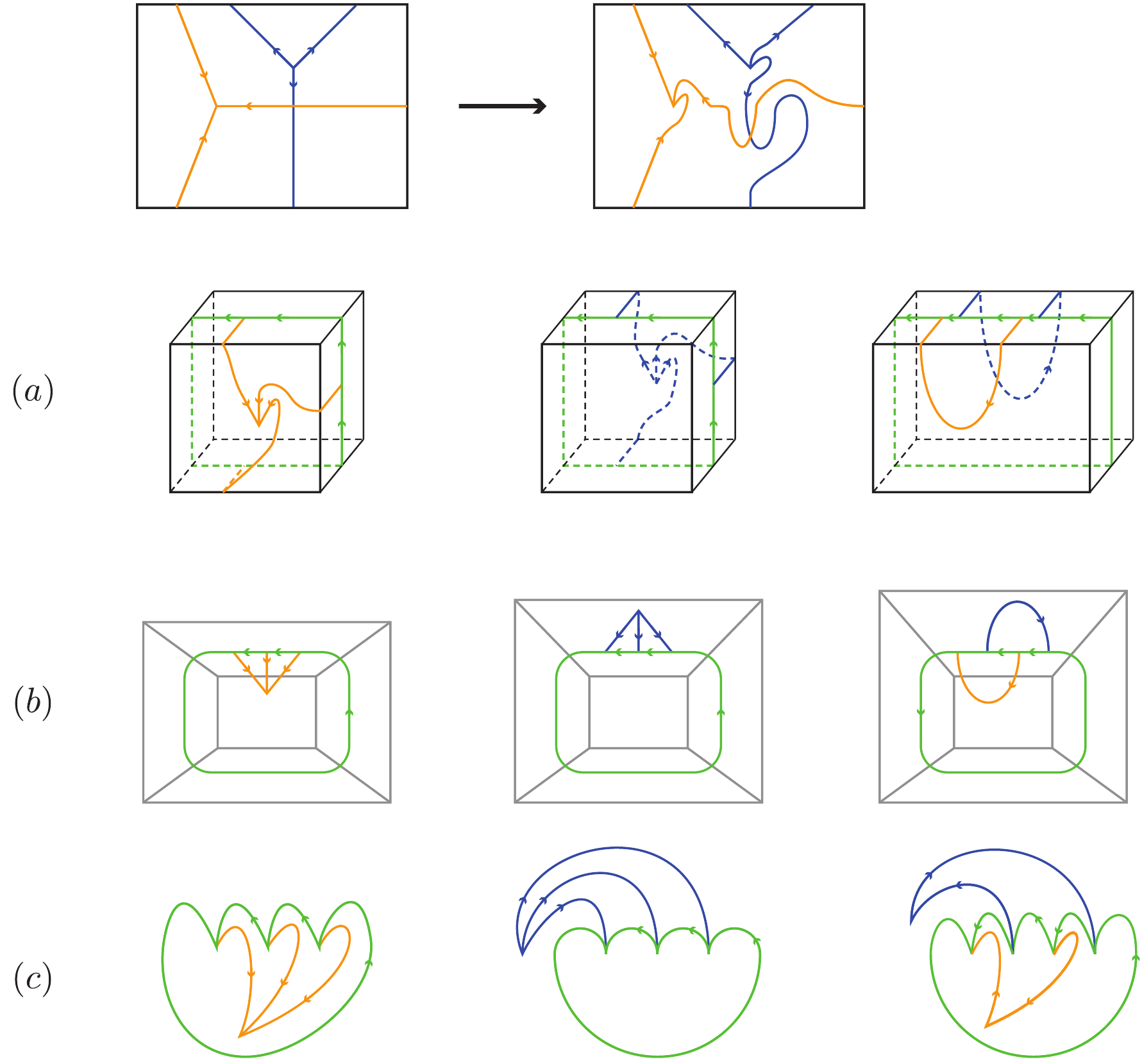}
\caption{\label{StandardizedSlabTensorsprime} \label{StandardizedSlabTensors}
The different types of 0-handles that appear in the cell decomposition of $G$. 
The top picture shows how a simple cell decomposition is standardized by putting 
each of the trivalent vertices in pitchfork form. 
In row (a) we show the three different types of standardized 0-cells that can appear in $G$. The cubes are drawn so that the 0-cells are located in their centers. 
Orange (blue) lines represent 1-handles in $G''$ ($G'$), and 
green lines represent intersections of 2-cells in $G$ with the cube. 
In row (b) we show the standardized string nets projected into the plane, and in  
row (c) we finish the standardization of the diagrams by making each trivalent vertex a pitchfork. The appropriate 0-cell 
tensors are found by evaluation of these diagrams. 
}
\end{center}
\end{figure}

The next step is to standardize the string nets on the boundary of each 0-handle, by following the procedures outlined in Sections \ref{bosonic_standardization} and \ref{fermionic_standardization}.
This is illustrated in Figure \ref{StandardizedSlabTensorsprime} for the three different types of 0-handle in our handle decomposition $G$ (one of each type of 0-handle is shown in the rightmost picture of Figure \ref{VertexTypes}).
Note that our conditions on the handle decompositions $G',G''$ ensure that all three types of 0-handle string nets are tetrahedral.

\medskip

We are now in a position to apply the state sum and tensor network constructions of the previous subsection.
The state sum turns out to be a version of the ``shadow world" state sum of \cite{kirillow1989,turaev2016quantum}. 
In other words, the shadow world state sum is a special case of the Turaev-Viro state sum.
If we fix the (labeled) string net $\bar g$ at the outset, the tensor network has an output corresponding to \eqref{bhs_redef}.

To compute the tensor network, we just need to know how to assign tensors to the 0-handles in $G$. 
This is done by computing evaluations of tetrahedral string-net diagrams, as in previous sections. 
Explicitly, for a 0-cell $v$ of $G$ consisting of three 1-cells of $G'$ (middle column of row (a) of Figure \ref{StandardizedSlabTensorsprime}), 
we assign the tensor $T_v$ as follows: 
\begin{align}
\Tensora \; \ra T_v(\mu_0 \tp s_1 \tp s_2 \tp s_3) &= \;  \Tensorcc\,,
\end{align}
In the diagrams, the green letters $A,B,C$ denote the labels of the 2-cells in the ``interior'' of the cell decomposition on $\Sigma \times I$ (those drawn in green in Figure \ref{VertexTypes}), which will 
be summed over when computing the amplitude $Z(\Sigma\times I)(\bar g \cup v)$. The labels of 
the blue lines are fixed, and are determined by the labels of the 1-cells 
in the string-net graph $v$ on $\Sigma \times \{1\}$. 

Tensors for the other types of 0-cells in $G$ are determined similarly. For the type of 
0-cell in the left column of row (a) in Figure \ref{StandardizedSlabTensorsprime} involving three 1-handles from $G''$, we assign the tensor
\begin{align}
\Tensorc \;\ra  T_v(\nu_0 \tp s_1 \tp s_2 \tp s_3) &= \; \Tensoraa\,, \\
\end{align}
where the labels of the orange lines are fixed by the input string-net state $\bar g$.
Similarly, for the type of 0-cell in the right column (involving the intersection of 1-handles in $G'$ and $G''$) we assign the tensor
\begin{align}
\Tensorb \; \ra T_v(s_0 \tp s_1 \tp s_2 \tp s_3) &= \;  \Tensorbb\,, \\
\end{align}
which corresponds to a string operator. 

Now that we know how to assign tensors to each 0-handle, we can construct the partition function $Z(\Sigma \times I)(\bar g \cup v)$ in the same way as in previous sections, namely by performing a tensor contraction over $\bigotimes_{v\in \mch_0}T_v$, 
where the tensor product runs over all 0-cells of $G$.
(The evaluation of this tensor contraction involves the same 
treatment of Koszul signs and 1-handle pivots as the fermionic state sum discussed in the previous section.)

\medskip

Instead of fixing a particular input string net $g$, we can put $g$ and $v$ on more equal footing by constructing
a tensor network which computes an operator from $\mch_{G''}$ to $\mch_{G'}$, where $\mch_{G''}$ and $\mch_{G'}$ are versions of \eqref{bhs_redef} corresponding to the vertices of $g$ and $v$, respectively. 
In particular, we can take $G''$ to be isotopic\footnote{
We can't take $G'' = G'$ because we require that $G'$ and $G''$ be transverse.}
to $G'$ and compute a projection from the big Hilbert space to itself.
This projection is, of course, the projection onto the ground state of the Hamiltonian of Section \ref{Super_pivotal_Hamiltonian}.

There is one small technical hurdle to overcome before constructing this operator.
Previously we adopted the convention that boundaries of 3-manifolds are contained in the 2-skeletons of cell decompositions 
corresponding to handle decompositions.
This is convenient for many purposes, but if we want to glue 3-manifolds along their boundaries (and perform analogous operations with
tensor networks), then it would have been more convenient to take the boundaries to be transverse to the cell decompositions.
In practice, this means that we must assign some additional factors of $\mcd^{-2}$ and $d_a/n_a$ to our 0-handle tensors (as described above \eqref{0_handleVectorspaces}), corresponding to
3- and 2-cells which straddle the surface along which we are gluing 3-manifolds.
Specifically, for each 2-cell of $G''$ we choose an adjacent 0-cell and assign a factor of $\mcd^{-2}$ to the corresponding 0-handle tensor,
and to each 1-cell of $G''$ we choose an adjacent 0-cell and assign a factor of $d_a/n_a$ to the corresponding 0-handle tensor.
These 2- and 1-cells in $\Sigma$ correspond to the 3- and 2-cells which straddle the gluing surface when we glue two copies
of $\Sigma\times I$ together.

Let $H$ denote the resulting tensor network operator. 
The fact that $(\Sigma\times I) \cup (\Sigma\times I) \cong \Sigma\times I$ implies that
$H\circ H = H$.
The fact that $\Sigma\times I \cong -(\Sigma\times I)$ 
(via a homeomorphism which is the identity on the $\Sigma$ factor and reverses the $I$ factor)
implies that $H$ is self-adjoint
(see the end of Section \ref{reflection_ss}).


\section{Kitaev chain} \label{kitaev_wire}

In this section, we show how the graphical formalism developed in previous sections
can be used to capture the salient features of the ``Kitaev wire", 
Kitaev's toy model of a one-dimensional spinless $p$-wave superconductor \cite{kitaev2001}. 
This highlights the connection between Majorana zero modes and Ising anyons and serves 
as a nice application of the graphical calculus of the $C_2$ theory.
Most of what we say applies beyond the $C_2$ theory and can be carried out for any theory containing at least one $q$-type object.
The Hamiltonian we write down is a special case of
the one constructed in Section \ref{Super_pivotal_Hamiltonian}, for particular choices of cell decompositions of a disk (annulus) with fixed boundary conditions.
The associated wavefunctions we construct
are the same as those found in, e.g., \cite{fidkowski2011}, but presented in a more graphical formalism 
that serves as a simple example of the techniques discussed in Section \ref{state_sums}.

Recall that the $C_2$ theory has two simple objects, $\unit$ and $\beta$,
with $\beta\tp \beta\cong \cc^{1|1}\unit$,
and $\End(\beta) \cong \cliff_1$.
We will focus on the $\beta$ object in the $C_2$ 
theory for concreteness, but the analysis can be applied q-type objects $q$ in any theory.

In what follows, 
we will show that a single strand of $\beta$ string is a diagrammatic description for the zero correlation length limit of the Kitaev chain. 
This means that the string-net Hamiltonian in Section \ref{Super_pivotal_Hamiltonian} 
based on the $C_2$ theory describes a phase of fluctuating Kitaev wires, 
an idea previously investigated in \cite{tarantino2016,ware2016,kapustin2017}.

The basic strategy is to cut a single $\beta$ strand into pieces and analyze how to glue those pieces back together to recover the uncut strand. 
Physically, we will implement the gluing by requiring the vectors to be in the ground space of a particular Hamiltonian, which is similar to what we did in Section \ref{Super_pivotal_Hamiltonian}. 
We first note that the vector space associated to a single interval $I$ with boundary conditions labeled by $\beta$ can be written graphically as 
\be\label{VIbetabeta}
 A(I;\beta,\beta) = \cc \left[ \halfchain\;, \; \halfchaindot \right] \cong \cc^{1|1}.\ee
 
 Now we can consider splitting the interval $I$ into two smaller intervals $I_1,I_2$, such that $I_1\cup I_2 = I$.
 We then can reconstruct the vector space $A(I;\beta,\beta)$ from the vector 
 spaces $A(I_1;\beta,\beta),A(I_2;\beta,\beta)$ by gluing the two intervals $I_1,I_2$ together. 
Algebraically, this gluing is implemented by the tensor product. However, 
we must be sure to make the proper choice 
of tensor product to ensure that we don't produce any extra degrees of freedom during the gluing. 
The standard tensor product $\tp_\cc$ doesn't work, since then $A(I_1;\beta,\beta) \tp_\cc A(I_2;\beta,\beta) \cong \cliff_2 \not\cong A(I;\beta,\beta)$. 

The correct tensor product to use is the relative tensor product $\tp_{\End(\beta)}$ (a.k.a $\tp_{\cliff_1}$) discussed in Section \ref{modified_tensor_product}. 
With this tensor product, we (rather trivially) have 
\be A(I;\beta,\beta) \cong A(I_1;\beta,\beta) \tp_{\End(\beta)} A(I_2;\beta,\beta),\ee
which tells us how to split apart the $I$ interval correctly. 

Graphically, the relative tensor product $\tp_{\End(\beta)}$  is needed to mod out by local relations involving the sliding of fermions along $\beta$ lines, 
as was discussed Section \ref{modified_tensor_product}. 
Utilizing $\tp_{\End(\beta)}$ is equivalent to performing the regular tensor product $\tp_\cc$ and modding out by the equivalence relations
\begin{align}
\label{graphical_equiv_reln} 
\halfchain \tp_\cc \halfchaindot \; &=\; \halfchaindot \tp_\cc \halfchain\; ,\\
\nonumber
\halfchaindot \tp_{\cc} \halfchaindot \;  &= -A^4\; \halfchain \tp_{\cc}  \halfchain.
\end{align}
where we have assumed a Koszul ordering for the fermions which increases from left to right (see Table \ref{C2_data_table} for the origin of the phase $A^4$).

As we did with the string-net Hamiltonian in Section \ref{Super_pivotal_Hamiltonian}, 
we can implement these equivalence relations energetically, via an appropriately defined Hamiltonian, which 
will be the same as the edge term in the lattice Hamiltonian defined in \eqref{De_defn}. 

Consider an interval $I$ of $\beta$ string cut into $n$ segments: $I = I_1\cup I_2\cup\dots\cup I_n$.
Each segment $I_i$ will end up mapping to a single physical site in the Kitaev chain. 
The local Hilbert space at each $I_i$ segment is generated by two basis vectors $v_e,v_o$, 
which for convenience we draw as
\begin{align} \label{vevodefn}
v_e \; = \;\LocalHilba\,,\qquad v_o \; = \; \LocalHilbb\;.
\end{align}
The upward-curved ends on each $\beta$ segment are drawn purely for aesthetic purposes, and exist solely to make drawing the Kitaev chain slightly easier. 
The local Hilbert space is then
\begin{align}
A(I_i; \beta, \beta) = \cc \left[ \LocalHilba, \LocalHilbb \right] \cong \cc^{1|1}.
\end{align}

The total Hilbert space of the chain is given by tensoring each local Hilbert space together:
\begin{align} 
\label{HilbInterval}
\mch_I  = A(I_1; \beta, \beta)  \tp_\cc A(I_2; \beta, \beta)  \tp_\cc \cdots \tp_\cc A(I_n; \beta, \beta).  
\end{align} 
States in this Hilbert space are expressed graphically as
\begin{align} \label{example_hilb_vectors}
\quarterpiperprime \tp  \LocalHilba \tp  \LocalHilba \tp  &\cdots \tp  \LocalHilba \tp  \quarterpipelprime, \\
\quarterpiperdotprime \tp  \LocalHilbb \tp  \LocalHilba \tp  &\cdots \tp  \LocalHilba \tp  \quarterpipelprime, \\
\quarterpiperdotprime \tp  \LocalHilba \tp  \LocalHilbb \tp  &\cdots \tp  \LocalHilba \tp  \quarterpipelprime, \\
&\;\; \vdots
\end{align}
and so on. 
Instead of using the relative tensor product $\tp_{\End(\beta)}$ to mod out by the equivalence relations \eqref{graphical_equiv_reln} we use $\tp_\cc$ (abbreviated as $\tp$ above) and define a Hamiltonian so that the ground space is isomorphic to $A(I; \beta, \beta)$. 
The Hamiltonian can be written as:
\begin{align} \label{kitaev_chain_ham_sitei}
H_i =  \frac{1}{2} \left( \Id \;\; + \;\; \TwoLinedotdot \right),
\end{align}
which have a non-trivial action only on the adjacent Hilbert spaces associated to the intervals $I_i$ and $I_{i+1}$.%
\footnote{Note that there are no terms that act on both the right and left ends of a single strand/interval; such a term would perturb us away from the zero correlation length limit.
In the language of the Kitaev chain, this corresponds to tuning the chemical potential to zero and the 
magnitude of the superconducting gap to the hopping amplitude.}

The image of this projector on a pair of adjacent string endpoints is  
\begin{align} 
\quarterpiper \tp_\cc \quarterpipel + \quarterpiperdot \tp_\cc \quarterpipeldot\,,
\end{align}
so that using $\tp_\cc$ and projecting with $H$ is equivalent to using $\tp_{\End(\beta)}$. 
Thus, $H_i$ is responsible for gluing together ends of $\beta$ strands. 
In terms of electronic operators, $H_i$ implements hopping and pairing between electrons in nearest neighbor sites $i$ and $i+1$.

We form our Hamiltonian from a sum of projectors, $H_i$, acting between each pair of strands:
\begin{align}
\label{KWHam}
H = t \sum_{i = 1}^{n-1} (1- H_i).
\end{align}
This Hamiltonian describes the zero correlation length limit of the Kitaev chain, albeit in a slightly unconventional language.
To understand this, we proceed to investigate the ground state wave functions.

We note that the nontrivial term in the Hamiltonian is proportional to the fermion parity measured between adjacent physical sites. 
Indeed, acting with that term on the vectors $v_e,v_o$ defined in \eqref{vevodefn}, we find
\begin{align}
\TwoLinedotdot\; \circ \; v_e =\; v_e \quad \text{and} \quad \TwoLinedotdot\; \circ\; v_o \; =\;   - \; v_o.
\end{align}
which is precisely the action of $(-1)^F$ (which can be identified with $i\gamma_1\gamma_2$ in the conventional Clifford algebra language).

It is straightforward to find the ground states of \eqref{KWHam}.
Noting that the non-trivial term in the Hamiltonian is just measuring the parity shared between adjacent ``physical" sites, 
we can do a change of basis using an F-move so that the Hamiltonian is diagonal and annihilates the ground state.
In this basis the (un-normalized) ground state wavefunctions take the form
\begin{align} \label{kitaev_wire_ground_states}
\Psi_e = \;\StaggaredGSEvenprime \; \cdots \; \StaggaredGSEvenRprime\;, 
\qquad \text{and} \qquad 
\Psi_o =\; \StaggaredGSOddprime \; \cdots  \; \StaggaredGSEvenRprime\;.
\end{align}
In this basis the Hamiltonian acts as $(1-(-1)^F)$ on each pair of vertical strands, which is clearly zero.

To better understand the wavefunctions $\Psi_e,\Psi_o$, we can apply a series of F-moves to change to the physical ``on-site" basis.\footnote{Recall the physical Hilbert space is associated to $A(I_i;\beta, \beta)$, 
whereas the wavefunctions $\Psi_e,\Psi_o$ are expressed in a basis that's ``shared" between adjacent sites.}
In a Kitaev chain with $n$ physical sites (i.e., $n$ intervals), 
we recover the well known result
\begin{align}
\Psi_e = \frac{1}{d^{n-1}} \sum_{\substack{ \{ v_i\} \\  n_f = \text{even} }} (A^4)^{n_f/2} \; v_1 \tp v_2 \tp \cdots \tp v_n
\end{align}
where the sum is over all configurations of $v_i =v_o,v_e$ such that only an even number $n_f$ of odd vectors $v_o$ appear 
in the tensor product, and where $v_e,v_o$ are defined as in \eqref{vevodefn}.
For the odd wavefunction $\Psi_o$, we find 
\begin{align}
\Psi_o = \frac{1}{d^{n-1}}\sum_{\substack{ \{ v_i\} \\  n_f = \text{odd} }}  (A^4)^{(n_f-1)/2}\; v_1 \tp v_2 \tp \cdots \tp v_n
\end{align}
where the sum is now restricted so that only an odd number $n_f$ of odd vectors $v_o$ vectors in the tensor product.
From the expressions for $\Psi_e,\Psi_o$ in this basis, we see that they are given by configurations 
that are coherent sums over all fermion parity even and fermion parity odd states, respectively.  

Note that the fermion dot appearing in $\Psi_o$ of \eqref{kitaev_wire_ground_states} has zero energy (since the Hamiltonian does not act on either the beginning of the first strand in the chain or the end of the last strand), while this is not true for fermion dots 
appearing on the interior cups. 
Physically, this is due to the presence of a pair of Majorana zero modes localized at the ends of the chain. 
One can explicitly construct the zero mode operators by considering the odd operators acting on either end of the chain; 
they commute with the Hamiltonian, anti-commute with one another, anti-commute with $(-1)^{F_\text{tot}}$ (the total fermion parity), and up to a pre-factor each square to the identity.
These are exactly the properties of a Majorana zero mode. 
A nice feature of the diagrammatic notation we use is that one can easily see that acting on the left end of the chain with one zero mode operator is equivalent to acting on the right end with the other zero mode operator (up to a phase).
To see this, one simply slides the fermionic dot appearing from the zero mode operator along the bottom $\beta$ strand appearing in the presentation of the wavefunction (see \eqref{kitaev_wire_ground_states}).
Physically, this means the ends of the wire share a fermionic mode, 
and no information about the occupancy of this mode can be detected by local measurements.

By considering the same spin chain but using the regular Ising fusion category $A_3$ (rather than the condensed $A_3/\psi$ theory), one finds exactly the transverse field Ising model. 
Fermion condensation provides a map between these two models in the same way the Jordan-Wigner transformation does.
Of course, we have only discussed the zero correlation length limit, but on-site terms can be added as well, and the analysis carries through, except that the zero modes are exponentially localized to the boundary (for a small perturbation).
In the zero correlation length limit, excited states are easily constructed by putting dots on the intermediate cups. 

We now turn our attention to the Kitaev chain defined on a circle. 
The bulk of the Hamiltonian is constructed from a sum of projectors defined in \eqref{kitaev_chain_ham_sitei}. 
To ``glue" the end points of the interval into a circle, we need to add an additional term across the boundary.
There are two ways to do the gluing, differing from one another by a $2\pi$ rotation of the spin framing
(i.e.\ a spin flip). 
These choices correspond to the two spin structures on the circle, $S^1_B$ and $S^1_N$, corresponding to anti-periodic and periodic fermionic boundary conditions, respectively.
To define periodic boundary conditions we define $H_{n+1}$, the Hamiltonian term which glues the 
two endpoints of the interval together, by
\begin{align}
H_{n+1}^N = \frac{1}{2}
\left(   \; \HambdR  \cdots \HambdL \quad  +  \quad 
   \HambdLdot \cdots \HambdRdot  \;\right)    
\end{align}
where the leftmost string acts on the left side of $I_1$ and the rightmost string on the right side of $I_n$.
When closing the interval with anti-periodic boundary conditions (to form the bounding spin structure) we need to apply a spin flip twist to $H_n$, resulting in an additional minus sign multiplying the nontrivial term in the projector:
\begin{align}
H_{n+1}^B = \frac{1}{2}
\left(   \; \HambdR  \cdots \HambdL \quad  -  \quad 
   \HambdLdot \cdots \HambdRdot  \;\right)    
\end{align}
We note that these give us explicit matrices for the even linear maps $\cl_B: A(I; \beta, \beta) \to A(S^1_B, \beta)$ and $\cl_N: A(I;\beta, \beta) \to A(S^1_N,\beta)$.
In this case we are led 
to identify $\cl_B$ with $H_{n+1}^B$ and $\cl_N$ with $H_{n+1}^N$.
To substantiate these identifications we will explicitly compute the ground state wavefunctions.

In Section \ref{C2excitations} we noted that closing a q-type object along a bounding (non-bounding) spin structure results in an even (odd) parity vector; if the identifications made above are correct, then the ground state of the bounding Hamiltonian should have even parity, and the ground state of the non-bounding Hamiltonian should have odd parity. 
To see that this is indeed the case, we note that that when acting with the non-trivial term in $H^{X}_{n+1}$ for $X\in \{B,N\}$
on the subspace spanned by \eqref{kitaev_wire_ground_states} (with the outer legs now turned up), 
we need to slide one of the fermions around the full $S^1_{X}$.
Using this, the fact that the structure of the Hamiltonian is the same as in \eqref{kitaev_chain_ham_sitei}, and using our earlier results \eqref{kitaev_wire_ground_states}, we can write the (unnormalized) wavefunctions on the $B$ and $N$ sectors as 
\begin{align} \label{kitaev_wire_circle_ground_states}
\Psi_B =  \;\StaggaredGSEven \; \cdots \; \StaggaredGSEvenR_B, 
\qquad \qquad 
\Psi_N =\; \; \StaggaredGSOdd \; \cdots  \; \StaggaredGSEvenR_N \; ,
\end{align}
where the subscripts denote the spin structure. 
Although our graphical presentation may give the impression that these pictures are drawn on an interval, they are not: the presence of the $H_{n+1}$ terms, which act on the left-most and right-most strands
in the graphical presentation of $\Psi_e$ and $\Psi_o$ are responsible for gluing the interval into a circle. 

Note that the other possible candidates for ground-state wavefunctions (an odd-parity version of $\Psi_B$ or an even-parity version of $\Psi_N$) are identically zero, which can be seen by using the graphical 
calculus of the $C_2$ theory (see the discussion around \eqref{BoundingNullVector}).

\medskip

\begin{figure}
\centering
\includegraphics[scale=2.5]{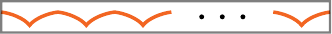}
\caption{\label{celldecompKitWire}
A cell decomposition of $I \times I$ with two marked points on the boundary, each labeled $\beta$. 
The interior graph contains $n$ ``pitchforkized" 2-valent vertices.
}
\end{figure}

The above Hamiltonian can be viewed as a special case of the Hamiltonian in Section \ref{Super_pivotal_Hamiltonian} as follows.
We take the ambient 2-manifold to be a long, thin rectangle (i.e.\ a disk).
We fix a $\beta$ strand boundary condition at each of the short sides of the rectangle.
On the long sides we impose empty boundary conditions.
In the interior, the ``lattice" contains only 2-valent vertices, as shown in Figure \ref{celldecompKitWire}.
Applying the general prescription in Section \ref{Super_pivotal_Hamiltonian} 
to this case yields essentially the same Kitaev wire Hamiltonian as defined above.
The spins at each 2-valent vertex are $V^{\beta\beta} \cong \cc^{1|1}$.
The vertex terms of the general Hamiltonian do nothing interesting, and there are no plaquette terms.
The edge terms of the general Hamiltonian \eqref{ham}, which we recall serve the purpose of allowing fermion dots to fluctuate along q-type strings, are the same as \eqref{kitaev_chain_ham_sitei}.
Thus, when acting on single strands of q-type string, the general Hamiltonian \eqref{ham} reduces to the Kitaev chain Hamiltonian. 

We can similarly glue the two ends of the rectangle together (either periodically or antiperiodically) to obtain
the Kitaev chain Hamiltonians for spin circles.

\medskip

We now use the Hamiltonian to construct matrix product operators (MPOs) and their related matrix product states (MPSs) for the ground state wavefunctions \eqref{kitaev_wire_ground_states} of \eqref{KWHam}.
This is a well known result, see e.g. \cite{fidkowski2011,turzillo2016,bultinck2017b}.
We write it here as it in some sense gives a ``gentler'' version of the tensor network discussed in Section \ref{state_sums}, and provides a nice application of the graphical calculus developed in the body of the paper.

We seek an MPO that projects a given state into the image of the projectors $H_i$ defined in \eqref{kitaev_chain_ham_sitei}. 
For convenience, we write $H_i =\frac{1}{2}( e_{i} + f_{i})$, with $e_{i}$ proportional to the identity operator on the junction between the intervals $I_i$ and $I_{i+1}$, and $f_{i}$ proportional to the fermion parity operator $(-1)^F$ across the junction.
Temporarily putting aside the issue of boundary conditions, to find the MPO we simply act with $\prod_i H_i$ on 
a given initial vector $V = v_1\tp \cdots \tp v_n$, where $v_i=v_e,v_o$ (graphically, these input vectors look like those appearing in \eqref{example_hilb_vectors}).
After expanding the product $\prod_i H_i$, we find an operator which is a sum over all possible configurations of the operators $e_i$ and $f_i$ straddling the junctions between intervals $I_i$ and $I_{i+1}$. 
We will denote the resulting state by $\Psi$.

We now just need to simplify the resulting state $\Psi$ using the local relations of the $C_2$ theory. 
Each physical site $v_i$ can be acted on by two terms in the Hamiltonian: $H_i$ (acting on the right strand of $v_i$) and $H_{i-1}$ (acting on the left strand). 
Hence, when we simplify $\Psi$, the phase factor associated with a given physical site $v_i$ will depend on a pair of indices $(e,e)$,$(e,f)$,$(f,e)$, and $(f,f)$, where 
the left (right) index denotes the term in $H_{i-1}$ $(H_i)$ that contributes to the phase.

Focusing on a single site with input vector $v_i$, 
we can succinctly write the action of the Hamiltonian as a matrix $(W^{v_i \to v_i'})_{xy}$, where $v_i'$ denotes the output vector obtained after acting with the Hamiltonian and where $x,y \in \{ e, f \}$.
These matrices are straightforward to compute using the rules of the $C_2$ graphical calculus.
If the input vector is $v_e$, we find
\begin{align}
W^{v_e \to v_e} = 
\frac{1}{2}\left( \begin{matrix} 
1 & 0\\
0 & 1 \\
\end{matrix} \right) \quad \quad \quad \quad 
W^{v_e \to v_o} = 
\frac{1}{2}\left( \begin{matrix} 
0& A^4\\
1 & 0 \\
\end{matrix} \right),
\end{align}
while we get
\begin{align}
W^{v_o \to v_e} = 
\frac{1}{2}\left( \begin{matrix} 
0 & 1\\
A^4 & 0\\
\end{matrix} \right) \quad \quad \quad \quad 
W^{v_o \to v_o} = 
\frac{1}{2}\left( \begin{matrix} 
1& 0\\
0 & -1 \\
\end{matrix} \right)
\end{align}
if the initial input vector is $v_o$.

We thus obtain an MPO on the interval:
\begin{align}
W_{b_l b_r}: \; \mch_I &\to \mch_I\\
v_1 \tp \cdots \tp v_n& \mapsto  \sum_{\{v_i' \}} \left( W^{v_1 \ra v_1'} W^{v_2 \ra v_2'} \cdots W^{v_n \ra v_n'} \right)_{b_l b_r} \; v_1' \tp \cdots \tp v_n' ,
\end{align}
where $b_l,b_r \in \{e,f\}$ and $b_r$ are boundary conditions for the interval.
If $b_l=f$ ($b_r=f$), an additional fermionic dot is present on the left side of the first interval (right side of the last interval). 
For example, the diagrammatics allows one to check that $W_{ef}$ is an odd operator which satisfies $W_{ef} = A^4 W_{fe} \circ (-1)^F$.This is a consequence of $\beta$ having an odd endomorphism.
The additional $(-1)^F$ accounts for sliding a fermion past an odd operator.

To obtain an MPS for the ground states, we simply fix an initial vector $v_1\tp \dots \tp v_n$ and boundary conditions for the MPO. 
On the interval we can construct the even parity ground state by
\begin{align}
\Psi_{e} = \sum_{\{v_i' \}} \left( W^{v_e \ra v_1'} W^{v_e \ra v_2'} \cdots W^{v_e \ra v_n'} \right)_{ee} \; v_1' \tp \cdots \tp v_n' 
\end{align} 
and the odd parity ground state with
\begin{align}
\Psi_{o} = \sum_{\{v_i' \}} \left( W^{v_o \ra v_1'} W^{v_e \ra v_2'} \cdots W^{v_e \ra v_n'} \right)_{ee} \; v_1' \tp \cdots \tp v_n'.
\end{align} 
Setting the boundary conditions to be $b_l=b_r=f$ in both cases would provide the same wavefunction up to an overall phase.

Similarly using \eqref{bounding_trace}, one can find the (unnormalized) MPS on a bounding spin circle by
\begin{align} 
\Psi_B = \sum_{\{ v_i ' \} } \tr \left( W^{v_e \ra v_1'} W^{v_e \ra v_2'} \cdots W^{v_e \ra v_n'} \right) v_1' \tp \cdots \tp v_n' 
\end{align}
and by using \eqref{nonbounding_trace} we find the (unnormalized) MPS on the non-bounding circle:
\begin{align} 
\Psi_N = \sum_{\{ v_i ' \} } \str \left( W^{v_o \ra v_1'} W^{v_e \ra v_2'} \cdots W^{v_e \ra v_n'} \right) v_1' \tp \cdots \tp v_n'.
\end{align}
In order to produce a non-zero state we need to choose a even parity initial vector for the bounding spin structure, 
and an odd parity initial vector for the non-bounding spin structure.

To summarize, we showed that strands of q-type objects are intimately related to the Kitaev chain. 
One can think of the string net Hamiltonian (defined in Section~\ref{def_sect}) for the $C_2$ theory as describing a phase of fluctuating Kitaev wires, a point of view adopted in \cite{tarantino2016,ware2016,kapustin2017}.
We also noted that fermion condensation is closely tied with the Jordan-Wigner transformation, as one 
With the one-dimensional Hamiltonian at hand, we showed how to explicitly construct the MPS for the ground state wavefunctions in this graphical language, recovering the same MPS as \cite{fidkowski2011,turzillo2016,bultinck2017b}.

\medskip

We also note that the same MPS can be found by employing the shadow world construction outlined in Section~\ref{shadowworld}: we work on the 3-manifold $D\times I$ where $D$ is a disk,  
and fix the string-net graph $w$ on the $D \times \{1\}$ boundary to have two marked $\beta$ points on the disk boundary and $n$ interior 2-valent vertices (this string-net state is illustrated in Figure \ref{celldecompKitWire}, where $w$ is illustrated as being made up of a union of intervals).



\section{Outlook} \label{discussion}

One potentially interesting aspect of the fermionic topological orders we have studied in this paper is their possible quantum information 
applications, which we now briefly speculate on. 
We consider a hybrid system with spin structure defects 
and deconfined anyonic excitations.
Each spin structure defect harboring a q-type vortex admits an action by $\text{End}(q)$, 
and so $n$ such defects admit an action of $\End(q_1) \tp \End(q_2) \tp \cdots \tp \End(q_n) \cong \cliff_1^{\tp n} \cong \cliff_n$.
One could then imagine utilizing this action to perform quantum computations. 
Physically, the action of $\cliff_n$ is implemented by choosing pairs of vortices $q_i$ and $q_j$ and pumping a charge into $q_i$ and out of $q_j$.
A natural platform for pumping charge through the q-type vortices is a Kitaev chain.
Of course, in addition to the action of $\cliff_n$, computations can also be performed with the conventional braiding of the deconfined quasiparticles appearing in $q_1 \tp q_2 \tp \cdots \tp q_n$.

It would be useful to make more precise connections between physical entities and some of the mathematical 
devices we have used to construct the fermionic theories 
we have studied. 
In particular, it would be useful to clarify the precise physical meaning of the complex line bundle 
and the ``back wall'' that we use to perform fermion condensation. 
In simple examples like the $C_2$ theory, the most natural interpretation for these constructs 
seems to be that they constitute a topological p-wave superconductor.
Indeed, the way we treat the physical fermions we use to perform condensation 
in our models is identical to the behavior of 
superconductors: they are 
free fermion states, 
where wavefunctions that differ by pairs of 
fermions are related by a phase. 
The specialization to the p-wave pairing channel is made because of the spinlessness
of the fermions we use to induce the condensation, which we assumed from the very 
beginning.  
The superconducting nature of the devices we use to perform condensation is forced on us
by our assumption that the emergent fermion $\psi$ we condense possesses $\zt$
fusion rules, and that in the complex line bundle we construct, pairs of
$\psi$ worldline endpoints can be created or destroyed in pairs. 
Evidence for the presence of a p-wave superconductor is clearly seen in the $C_2$ theory: 
restricting our attention to the non-bounding torus with $NN$ spin structure, both the modular $S$ and $T$ matrices 
factorize as $S=S_{Ising}\tp S_{p\pm ip}, T = T_{Ising} \tp T_{p\pm ip}$ (where the 
choice of $\pm$ is determined by the ``angular momentum'' of the fermionic dots 
in our graphical calculus, i.e. the choice of $\pm A^4$ when removing a semicircular fermion
worldline), suggesting a possible interpretation of this sector 
as a stack of the original Ising theory with a topological superconductor. 
Indeed, this was noticed recently in \cite{ware2016}.
Furthermore, the fact that the parity of the ground states on the torus with $(N,N)$ spin structure is always 
odd agrees with this interpretation, since the fermion parity of a topological 
p-wave superconductor on such a torus is always odd \cite{you2015}.

In our discussion of the modular $S$ and $T$ matrices in each of the examples we've worked out, 
we have focused on the modular transformation perspective, rather than on the braiding statistics perspective. 
For example, we have computed the $S$-matrix by considering the way it acts to exchange the two 
cycles of the torus and 
we have not focused on the statistical picture behind the $S$-matrix, in which matrix elements $S_{ab}$ correspond to double braids between $a$ and $b$ particles. 
While we have checked that the computation of double braids reproduces the correct $S$-matrix for $\tube(C_2)$,
some subtleties involving relative spin structures rear their heads when trying to compute particular braiding data in more general settings. 
We plan to address these subtleties in future work. 

 \paragraph{Acknowledgements}

Ethan Lake and Dave Aasen are grateful to Nick Bultinck, Nicolas Tarantino, Ryan Thorngren, Brayden Ware, and Dominic Williamson for helpful discussions.
Dave Aasen thanks Parsa Bonderson for explaining his unpublished work at early stages of this project. 
Dave Aasen gratefully acknowledges support from the KITP Graduate Fellows Program, the National Science Foundation through grant DMR- 1723367
and the Caltech Institute for Quantum Information and Matter, 
an NSF Physics Frontiers Center with support of the Gordon and Betty Moore Foundation through Grant GBMF1250.
Ethan Lake is supported by the Fannie and John Hertz Foundation.
Dave Aasen and Ethan Lake acknowledge support by the 2016 Boulder Summer School for Condensed
Matter and Materials Physics through NSF grant DMR-13001648.
Kevin Walker thanks Zhenghan Wang and Scott Morrison for helpful conversations, and
thanks the Aspen Center for Physics and  the Mathematisches Forschungsinstitut Oberwolfach
for providing stimulating research environments.


\clearpage
\appendix


\section{Spin and pin structures} \label{spin_and_pin}

%
In this appendix we review basic definitions and properties of spin and pin structures.

\medskip

Roughly, a spin structure on an oriented $n$-manifold $M$ is a specification for how fermions pick up phases of $-1$ as they move around $M$.
Locally, we of course require that fermions pick up a minus sign when rotated through $2\pi$,
or when two fermions are exchanged within a small neighborhood of $M$.
But if a fermion moves along a non-contractible loop in $M$, it is not clear what sign we should assign.
A spin structure on $M$ is a consistent set of answers to all possible questions of this form.

More formally, we can define a spin structure on $M$ to be a double covering of the frame bundle $F(M)$, such that on each fiber of $F(M)$
the covering is isomorphic to the standard double covering $Spin(n) \to SO(n)$.
Such double coverings correspond to cohomology classes in $H^1(F(M), \zz/2)$ which restrict to the generator of $H^1(SO(n), \zz/2)$ on each fiber.
It follows that the difference between any two spin structures is canonically identified
with an element of $H^1(M, \zz/2)$; spin structures on $M$ form a $H^1(M, \zz/2)$-torsor.
In particular, the number of distinct spin structures on $M$ is given by the number of elements in $H^1(M, \zz/2)$.
(Things need to said differently when $n<2$.
One way around this problem is to work with the stabilized frame bundle for 0- and 1-manifolds.)

It is important to note that there is no canonical correspondence between spin structures on $M$ and $H^1(M, \zz/2)$; simply naming
a cohomology class does not pick out a spin structure.

One way to specify a spin structure is to specify a framing on the 1-skeleton of a cell decomposition of $M$.
We can think of this framing as an embedded graph in the frame bundle $F(M)$, and the spin structure is uniquely determined by requiring
that the double covering of $F(M)$ be trivial when restricted to this graph.
We can also think of the 1-skeleton framing as specifying a collection of possible fermion paths which do not pick up a factor of $-1$.

In this paper, most of the diagrams we draw are embedded in the page/blackboard/$\rr^2$.
$\rr^2$ has a standard framing, so the 1-skeleton of any such diagram inherits a framing, and unless stated otherwise we work in the spin
structure associated to that framing.
In practice, then means that fermions pick up a minus sign only when their framing rotates with respect to the page.

If we have designated a reference spin structure on $M$ (for example, the blackboard spin struture), 
then any other spin structure can be specified by giving an
element of $H^1(M, \zz/2)$, or equivalently by giving the Poincar\'e dual element in $H_1(M, \zz/2)$.
In this context, we refer to the Poincar\'e dual homology class as a ``branch cut".
Fermions obey the rules of the reference spin structure, except that they pick up a $-1$ whenever they cross the branch cut.

\medskip

For unoriented manifolds, we must replace $SO(n)$ with $O(n)$ and replace $Spin(n)$ with a $\zz/2$ extension of $O(n)$.
There are two such extensions, called $Pin_+(n)$ and $Pin_-(n)$.
In $Pin_+(n)$, lifts of reflections in $O(n)$ square to the identity, while in $Pin_-(n)$ such lifted reflections square to the ``spin flip"
in $Spin(n)$.
(When $n=1$, we have $O(1)\cong\zz/2$, $Pin_+(1)\cong\zz/2\times\zz/2$, and $Pin_-(1)\cong\zz/4$.)
Roughly speaking, in pin+ manifolds reflecting a fermion twice returns us to the same state, while in pin$-$ manifolds reflecting a fermion twice
picks op a factor of $-1$.

None of the examples in this paper have a pin$-$ reflection structure; we only work with pin+ structures.

Specifying pin+ structures in terms of framings (as we did above for spin structures) is a little awkward.
It is usually more convenient to use lifts of classifying maps or, when $n=2$, quadratic refinements of intersection pairings.


\section{Constructing the fermion line bundle}
\label{fermion_line_bundle} \label{flb_appendix}

Recall our set-up from the end of Section \ref{condensing_psi}: 
We have a back wall $B$, which is a spin (and therefore oriented) 2-manifold.
Associated to $B$ is the configuration space of $\psi$-ribbon endpoints, $\mcr(B)$.
This configuration space is a disjoint union of pieces $\mcr(B)_k$, where $k$ is number of ribbon endpoints
in a configuration.

Our goal in this subsection is to construct a complex line-bundle-with-flat-connection $F(B)$ over $\mcr(B)$,
satisfying the following six conditions alluded to in Section \ref{condensing_psi}:
\begin{enumerate}
\item $F$ is functorial with respect to spin diffeomorphisms. That is,
if $f: B \to B'$ is spin diffeomorphism, then there is a corresponding bundle isomorphism $F(B) \to F(B')$ which
preserves the flat connections and complex structure.
\item $F$ is functorial with respect to orientation-reversing $\mbox{pin}_+$ diffeomorphisms. That is, 
if $f: B \to B'$ is an orientation-reversing $\mbox{pin}_+$ diffeomorphism, then there is a corresponding map
$F(B) \to F(B')$ which preserves the flat connections and is complex antilinear on the fibers.
(Recall that any spin manifold has an associated $\mbox{pin}_+$ structure.
By ``orientation-reversing $\mbox{pin}_+$ diffeomorphism", we mean a $\mbox{pin}_+$ diffeomorphism
of the associated $\mbox{pin}_+$ manifolds which reverses the orientations of the underlying oriented manifolds.)
This condition is needed in order to define Hermitian/unitary structures.
\item The holonomy around a loop in $\mcr(B)$ corresponding to a $2\pi$ rotation of a ribbon endpoint is $-1$.
This condition is needed to compensate for fermionic twist of $\psi$.
\item The holonomy around a loop in $\mcr(B)$ corresponding to an exchange of two ribbon endpoints 
(inside a fixed disk) is $-1$.
This condition is needed to compensate for the fermionic statistics of $\psi$.
\item $F$ is local in the following sense.
Given a decomposition $B = B'\cup B''$, there is an obvious map $u: \mcr(B')\times\mcr(B'')\to\mcr(B)$, 
and a corresponding pull-back bundle $u^*(F(B))$ over $\mcr(B')\times\mcr(B'')$.
We require an isomorphism $u^*(F(B)) \cong F(B')\otimes F(B'')$ which is natural with respect to spin diffeomorphisms.
\item $F$ satisfies a cancellation property.
Given a configuration $r \in \mcr(B)_k$ and a point $x\in B$ distinct from the ribbon endpoints of $c$,
we can create a new configuration $c_+ \in \mcr(B)_{k+2}$ by inserting a pair of endpoints in a standard configuration
near $x$.
We require an isomorphism of fibers $F(B)_c \cong F(B)_{c_+}$ which is compatible with the flat connection as explained below.
This condition is needed to allow for well-defined creation and annihilation of pairs of $\psi$ 
endpoints in line with the fusion rule $\psi \tp \psi \cong \unit$.
\end{enumerate}

\medskip

To represent an element of $F(B)$, we will choose spin framings at each ribbon endpoint and also assign
an ordering to the ribbon endpoints.
The main idea is fairly simple, but making this construction compatible with orientation reversal
and ribbon endpoint cancellations requires a bit of fussiness with the details.

\medskip

Recall the group $\pin_+(1) \cong \zz/2 \times \zz/2$.
We will call the non-identity elements of $\pin_+(1)$ the ``spin flip" 
(the non-trivial element of the kernel of the covering map $\pin_+(1) \to O(1)$), 
the ``reflection", and the ``other reflection"
(the latter two reflections map to the single reflection in $O(1)$).

To construct $F(B)$, we will first construct a principal $\pin_+(1)$ bundle $P(B)$ over $\mcr(B)$.
The construction of $P(B)$ will be independent of reversing the orientation of $B$.
We define the action of $\pin_+(1)$ on $\cc$ as follows:
the spin flip sends $z\in \cc$ to $-z$;
the reflection sends $z$ to the complex conjugate $\bar z$;
and the other reflection sends $z$ to $-\bar z$.
Using this action, we can now define $F(B)$ to be
\be \label{fermbundef}
	F(B) = P(B) \times_{\pin_+(1)} \cc ,
\ee
the associated $\cc$ bundle over $\mcr(B)$.
(Recall that this means that elements of $F(B)$ are represented by pairs $(f, z) \in P(B) \times \cc$,
and that for each element $a \in \pin_+(1)$, we identify $(f\cdot a, z)$ with $(f, a \cdot z)$.)
Since $P(B)$ is a bundle with discrete fibers, it has a canonical flat connection.
This induces a flat connection on $F(B)$.

Note that $F(B)$, as defined above, has two different complex-linear structures, one the conjugate of the other.
We will see below that the orientation of $B$ picks out one of the two possible complex structures.

\medskip

We are now ready, finally, to construct $P(B)$.
Let $r \in \mcr(B)_k$ be a configuration of $k$ ribbon endpoints.
At each endpoint of $r$, there is a distinguished unit tangent vector $v \in TB$
pointing in the direction of the front of the ribbon.
There are two unit tangent vectors $w_1$ and $w_2$ in $TB$ orthogonal to the distinguished vector.
The orientation of $B$ allows us to designate one of these two orthogonal vectors as ``positive"
and the other as ``negative".
(We call $w_i$ positive if $(v, w_i)$ is a positively oriented frame with respect to the orientation of $B$.)
We will call a collection of framings $(v, w_i)$ at each endpoint ``consistent" if they are all positive
or all negative.
Note that there are exactly two possible consistent collections of framings for each fixed configuration $r$.
We will denote this set of two elements by $cf(r)$.
Note that reversing the orientation of $B$ does not change $cf(r)$.\footnote{It is tempting 
to say that $cf(r)$ depends only on the underlying unoriented manifold of $B$, but this
is not true if $B$ has more than one path component.
Reversing the orientation of some but not all of the path components of $B$ would change $cf(r)$.}

\newcommand{\tcf}{\widetilde{cf}}

Now define $\tcf(r)$ to be the set of all collections of $\pin_+(2)$-framings (one at each endpoint)
which cover an element of $cf(r)$.
At each endpoint, there are two possible lifts of an $O(2)$ framing, 
so $\tcf(r)$ is a set with $2^{k+1}$ elements.
Again, $\tcf(r)$ does not change if we reverse the spin structure on $B$.\footnote{
Reversing a spin structure is analogous to reversing the orientation of an oriented manifold.
One way to define it is to extend the Spin bundle $E$ to a Pin+ bundle $E_+$;
the reversed Spin bundle is $E_+ \setminus E$.
}

Let $S(r)$ denote the set of all orderings of the endpoints of $r$.
It is a set with $k!$ elements.
Now consider $\tcf(r) \times S(r)$, yet another set associated to a ribbon endpoint configuration $r$.
This set has a group of symmetries $G$ which is generated by (a) spin flips acting on any single endpoint, 
and (b) the symmetric group $S_k$, acting on the $S(r)$ component.
Let $G_e \subset G$ is the subgroup with even total parity, where 
parity in this context is defined by the homomorphism from $G$ to $\zz/2$
characterized by the condition that is odd for single spin flips and transpositions.
Now, finally, define
\be
	P(B)_r = (\tcf(r) \times S(r))/G_e .
\ee

We claim that $P(B)_r$ is naturally a torsor for $\pin_+(1)$.
First, let's check the cardinality:
$\tcf(r) \times S(r)$ has $2^{k+1} k!$ elements,
$G_e$ has $2^{k-1} k!$ elements, and $G$ acts freely on $\tcf(r) \times S(r)$.
Therefore $P(B)_r$ has four elements.

Now we define the action of $\pin_+(1)$ on $P(B)_r$.
The spin flip acts by changing (any) one of the spin framings of $\tcf(r)$ by a spin flip.
The reflection acts diagonally on all of the spin framings of $\tcf(r)$
(i.e.\ each spin framing is reflected).
These two actions are well-defined and commute, so we have
an action of $\pin_+(1) \cong \zz/2\times\zz/2$.
(If $k = 0$ and $r \in \mcr(B)_0$ is the unique configuration of zero ribbon endpoints,
we define $P(B)_r = \pin_+(1)$ and let $\pin_+(1)$ act in the obvious way.)

\medskip

In summary, an element of $F(B)_r$ is represented by a triple $(f, o, z) \in \tcf(r) \times S(r) \times \cc$.
Spin flips, permutations of ribbon endpoints, and reflections act as
\begin{itemize}
\item If $f$ and $f'$ differ by a spin flip at a single ribbon endpoint, then $(f, o, z) = (f', o, -z)$.
\item If $o$ and $o'$ differ by an odd permutation, then $(f, o, z) = (f, o', -z)$.
\item If $f$ and $f'$ differ by reflecting all of the spin framings, then $(f, o, z) = (f', o, \bar z)$.
\end{itemize}

To define complex multiplication by $a\in\cc$ on $F(B)_r$, we choose a collection of framings $f$ which is positive
with respect to the orientation of $B$ and then define $a\cdot (f, o, z) = (f, o, az)$.
If we were to reverse the orientation of $B$, then we would get the conjugate complex structure on $F(B)$.
In other words, $F(B)_r = F(-B)_r$ as sets (and even as vector spaces over $\rr$), but the identity map from
$F(B)_r$ to $F(-B)_r$ is complex antilinear.

\medskip

We began this subsection with a list of several desiderata for $F(B)$.
It is more or less obvious that $F(B)$ has the right sort of functoriality for both orientation-preserving
and orientation-reversing spin/pin maps.
It should also be clear that $F(B)$ has the desired holonomies for rotations and exchanges.
So all that's left to discuss is locality (gluing) and ribbon endpoint cancellation.

\medskip

We consider locality first.
Let $B = B_1 \cup B_2$.
Let $r_i \in \mcr(B_i)$ and let $(f_i, o_i, z_i)$ represent an element of $F(B_i)_{r_i}$.
If the spin framing collections $f_1$ and $f_2$ are either both positive or both negative, then
$f_1 \cup f_2$ is a consistent spin framing in $\tcf(r_1 \cup r_2)$, and the triple
$(f_1 \cup f_2, o_1\cup o_2, z_1 z_2)$ represents an element of $F(B)_{r_1\cup r_2}$.
It is easy to check that this map gives a well-defined  isomorphism between $u^*(F(B_1)$ 
and $F(B')\otimes F(B_2)$, both thought of as line bundles over $\mcr(B_1)\times\mcr(B_2)$.

\medskip

Now for cancelations.
We want a relation of the type 
\be \label{eval_psi_semicirc}
\PsiEnd  = \lambda \times \text{(vaccuum)},
 \ee
for some $\lambda \in \cc$. 
On the left hand side we have two ribbon endpoints in a disk $D \subset B$ connected by a ribbon in $D\times I$.
We have chosen coordinates so that the front of the ribbon always points in the same direction.
The spin framings at the two endpoints are chosen to be related by a translation in these coordinates
and to both be positive.
We have chosen the ordering so that the second vector at endpoint 1 points toward endpoint 2;
we will call this a ``standard configuration".
As indicated, we want this standard picture to be equal to $\lambda$ times the empty picture.

We will show that in order for this relation to be compatible with reflections, 
we must have that $\lambda = -\bar\lambda$, i.e.\ $\lambda$ must be pure imaginary.
Note that in order to define the action of a reflection on string-nets, it is essential that we have defined an 
action of $\pin_+(1)$ on $P(B)_r$ (rather than merely an action of $\spin(1)$). 
(A $\pin_-(1)$ structure would also work, but our examples happen to have $\pin_+$ rather than $\pin_-$ structures.)
The existence of a reflection structure also allows us to define inner products of diagrams.

Consider first the RHS of \eqref{eval_psi_semicirc}. 
Reflections take the empty picture to the empty picture, and so, since reflections act antilinearly on $F(B)$,
the RHS of \eqref{eval_psi_semicirc} maps under reflection to $\bar\lambda$ times the empty picture.

Now for the LHS of \eqref{eval_psi_semicirc}.
After a reflection (by which we mean an orientation-reversing map), 
the framings are no longer positive, and so we must reflect them in order to compare to a standard configuration
in the target manifold.
After the framings are reflected, the second vector at endpoint 1 points away from endpoint 2,
so we must swap the ordering in order for it to be a standard configuration.
This change of orderings means that under a reflection, a standard configuration map to $-1$ times a standard
configuration.
It follows that we must have $\lambda = -\bar\lambda$, and so $\lambda$ must be purely imaginary.

One can show that this cancellation relation satisfies the necessary coherence relations.


\section{Basic facts about super algebras} \label{superstuff}

In this appendix we briefly recall some of the key mathematical facts about semisimple super algebras; 
details can be found in \cite{jozefiak1988}.

There are two distinct classes of simple super algebras over $\cc$. One class is the set of 
super algebras $M(r|s)$ for $r,s\in\zz_{\ge 0}$, which are $(r+s)\times(r+s)$ matrices whose even and odd subspaces take the form
\be 
	M(r|s)_0 = \text{matrices of the form\ } \begin{pmatrix} A & 0 \\ 
	0 & B \end{pmatrix},\qquad M(r|s)_1 = \text{matrices of the form\ } \begin{pmatrix} 0 & C \\ D & 0 \end{pmatrix}.
\ee
In these expressions, $A$ is an $r\times r$ matrix, $B$ an $s\times s$ matrix, $C$ an $r\times s$ matrix, and $D$ an $s\times r$ matrix. 

The other class of simple super algebras are denoted by $Q(n)$ for $n\in \zz_{>0}$, 
which are $(2n)\times(2n)$ matrices with even and odd subspaces of the form
\be 
	Q(n)_0 = \text{matrices of the form\ } \begin{pmatrix} A & 0 \\ 0 & A \end{pmatrix},
	\qquad Q(n)_1 = \text{matrices of the form\ } \begin{pmatrix} 0 & B \\ B & 0 \end{pmatrix},
\ee
where both $A$ and $B$ are $n\times n$ matrices. 
In particular, $Q(1) = \langle \unit, \sigma^x\rangle$ is the first complex Clifford algebra $\cliff_1$. 

Note that all of the $M(r|s)$ are Morita equivalent to the trivial algebra $M(1|0) \cong \cc$.
All of the $Q(n)$ are Morita equivalent to $Q(1) \cong\cliff_1$.

If $x$ is an object in a super pivotal category, then $\End(x)$ will be isomorphic to a direct sum of instances of $M(r|s)$ and $Q(n)$.

The form of a general simple super algebra $A$ can be deduced by computing the center $Z(A)$.
(An element $a$ is in $Z(A)$ if it super commutes with everything in $A$, 
i.e.\ if $ax = (-1)^{|a||x|}xa$ for all $x\in A$, with $|x|$ denoting the parity of $x$.)
Since the super algebras $Q(n)$ treat even and odd vectors symmetrically, 
they expect that they will have odd elements in their center, while this will not be true for the super algebras $M(r|s)$. 
Indeed, we have that if $Z(A) \cong \cc^{1|0}$ then $A \cong M(r|s)$ for some $r,s$, while if $Z(A) \cong \cc^{1|1}$ then $A\cong Q(n)$ for some $n$. 

If $A,B$ are super algebras, their tensor product $C=A\tp B$ is defined as the super algebra such that $C_0 = A_0\tp B_0 + A_1\tp B_1$, $C_1 = A_0\tp B_1 + B_1\tp A_0$. 
The simple super algebras presented above can be tensored together by using the following rules:
\be \begin{aligned} 
	M(r|s) \tp M(p|q) & \cong M(rp+sq|rq+sp) \\ 
	M(r|s) \tp Q(n) & \cong Q(rn+sn) \\ 
	Q(n) \tp Q(m) & \cong M(nm | nm). 
\end{aligned} \ee
Note that all the $Q(n)$ can be generated from $M(n|0)$ and $Q(1)$ according to $Q(n) \cong M(n|0)\tp Q(1)$.


\section{$\halfesix$ data}

\subsection{Associators}   \label{E6Fsymbols}

There are four solutions to the pentagon equation for $\frac{1}{2}$E6 fusion rules \cite{Hagge2007}.
They split into two sets, one pair has all positive quantum dimensions, while the other has negative quantum dimension on the $x$ particle. 
The solutions in each set are related by complex conjugation. 
Here we present one of the solutions with positive quantum dimensions on all particles (they have been extracted from \cite{Wakui2002}) . 
Several of the $F$ symbols are trivial,
\begin{align}
F^{yyy}_y = F^{xyy}_x = F^{yyx}_x = F^{xxy}_y  = F^{yxx}_y = F^{xyx}_{\mathds{1}} = F^{xxy}_{\mathds{1}} = F^{yxx}_{\mathds{1}}= 1
\end{align}
Let $v_1, v_2$ be orthogonal unit vectors for the two-dimensional splitting space $V^{xx}_x$. 
We define the $F$ symbols acting on these vectors by $V^{xy}_x \tp V^{xx}_x  = F^{xyx}_x V^{xx}_x \tp V^{yx}_x$ in the basis $(v_1, v_2)^{T}$.
Explicitly we have,
\begin{align}
&F^{xyx}_x = \left(\begin{matrix}
0&1\\
1&0
\end{matrix} \right) \quad
F^{xxy}_x = \left(\begin{matrix}
0&-i\\
i&0
\end{matrix} \right) \quad
F^{yxx}_x = \left(\begin{matrix}
1&0\\
0&-1
\end{matrix} \right) 
\\
\\
&F^{xxx}_{\mathds{1}} = c_2^* \left(\begin{matrix}
1&-i\\
1&i
\end{matrix} \right) \quad
F^{xxx}_y = c_2^*\left(\begin{matrix}
1&-i\\
-1&-i
\end{matrix} \right), \quad \quad c_2  = \frac{e^{7 i \pi/12}}{\sqrt{2}}
\end{align}

Lastly we have the $F$ symbol whose four external legs are all labeled by $x$.
We write this $F$ symbol down in the basis $(0,y, v_1\tp v_1,v_1 \tp v_2, v_2 \tp v_1, v_2 \tp v_2)$, where we have 
\begin{align}
F^{xxx}_x  = 
 \left(\begin{matrix}
\frac{1}{d} & \frac{1}{d} & \frac{c_1^*}{\sqrt{2}} & \frac{c_1^*}{\sqrt{2}} & \frac{c_1^*}{\sqrt{2}} & - \frac{c_1^*}{\sqrt{2}} \\
\frac{1}{d} & - \frac{1}{d} & \frac{c_1^*}{\sqrt{2}} & -\frac{c_1^*}{\sqrt{2}} & \frac{c_1^*}{\sqrt{2}} & \frac{c_1^*}{\sqrt{2}} \\
\frac{e^{- i \pi/4}}{\sqrt{d}} & 0 & -\frac{1}{d} & 0 & -i c_4^* & 0 \\
0 & \frac{e^{i \pi /4}}{\sqrt{d}} & 0 &c_4^* & 0 & \frac{ i}{d} \\
0 & \frac{e^{- i \pi /4}}{\sqrt{d}} & 0 & -\frac{1}{d} & 0 & i c_4^* \\ 
\frac{e^{i \pi /4}}{\sqrt{d}} & 0 & c_4^* & 0 & - \frac{i}{d} & 0
\end{matrix} \right) \\
d = 1 + \sqrt{3} \quad \quad c_1 = \frac{e^{-5 i \pi/6} }{\sqrt{d}} \quad \quad c_4 = \frac{e^{-i \pi/4}}{\sqrt{2}}
\end{align}

\subsection{Idempotents}
\label{IdempotentsHalfESix}

In this appendix we provide the minimal idempotents of $\tube(\halfesix)$, and give there images under condensation of $y$.
The inclusion is performed by simply condensing fermions off the tubes in $\tube(\halfesix)$ to get tubes in $\tube(\halfesix / y)$. 
Special care must be taken with respect to spin structure issues, 
since removing $y$ lines may force a pair of fermions to traverse a cycle of the tube. 

The minimal idempotents of $\tube(\halfesix)$ are listed in tables \ref{mor0to0}, \ref{mor1to1}, and \ref{mor2to2}.
They were found by brute force on a computer.
We also identify the minimal idempotents of $\tube(\halfesix)$ with simple objects of the Drinfeld center $\mcz(\halfesix)$ listed in \cite{Hong2008}.%
\footnote{Disclaimer: 
At the level of fusion rules and spins,
idempotents $X_2$ and $X_3$ are identical and so there is an ambiguity in identifying these minimal idempotents with the simple objects in $Z(\mcc)$ of \cite{Hong2008}.
The spins here differ from those in \cite{Hong2008} by complex conjugation. } 
Under condensation of $y$ described in Section \ref{halfesix} we find the following maps from idempotents in $\tube(\halfesix)$ to those in in $\tube(\halfesix / y)$:
\begin{align}
\xymatrix @!0 @M=1mm @C=10mm{
& &m_1 && m_2 && &&& m_3^+ &&  \\
 \\
 \quad&\mathds{1} \ar[ruu] \quad & Y \ar[uu] & W\ar[dd] & U\ar[uu] & V \ar[uul] & X_1 \ar[dd] & X_{2}\ar[ddr] &X_{3} \ar[dd] & X_4\ar[uu] & X_5\ar[uul] &  \\
\\
&&&q_1& &&q_2 &&m_4^+& &&  \\
},
\end{align}
where the center line lists the idempotents in $\tube(\halfesix)$ and the upper and lower objects are the objects in $\tube(\halfesix / y)$.
The identifications are made by taking a minimal idempotent in $\tube(\halfesix)$ and using the inclusion to $\tube(\halfesix/y)$, 
as discussed in Section \ref{double_fermionic_quotient}.

\begin{table}
{\tabulinesep=1.2mm
\begin{tabu}{ c c | c c c c }
particle & spin &$e$&$l_x$&$l_y$& \\ \hline
$\unit$ & $
       1       

$ &
$
\frac{1}{2+d^2}
$
 &
$
\frac{1}{2 \sqrt3}
$
 &
$
\frac{1}{2+d^2}
$
 &
\\
$W$ & $
       1       

$ &
$
\frac{1}{2}
$
 &
$
$
 &
$ -
\frac{1}{2}
$
 &
\\
$U$ & $
       1       

$ &
$
\frac{d}{4 \sqrt{3}}
$
 &
$ -
\frac{1}{2 \sqrt3}
$
 &
$
\frac{d}{4 \sqrt{3}}
$
 &
\end{tabu}
\caption{
Minimal idempotents for $\tube_{\unit \to \unit}$.
 We have used the notation $e$ = empty diagram, $l_x = \cl(x) $, and $l_y = \cl(y)$}
\label{mor0to0}}
{\tabulinesep=1.2mm
\begin{tabu}{ c c | c c c c }
particle & spin &$t_y$&$t_y l_x$&$v_y$& \\ \hline
$W$ 
& $       1       $ 
&$\frac{1}{2}$
 &$$
 &$\frac{1}{2}$
 &
\\
$Y$ & $
       1/2     

$ &
$ -
\frac{1}{2+d^2}
$
 &
$ e^{\frac{6i\pi}{12}}
\frac{1}{2 \sqrt3}
$
 &
$
\frac{1}{2+d^2}
$
 &
\\
$V$ & $
       1/2     

$ &
$ -
\frac{d}{4 \sqrt{3}}
$
 &
$ e^{\frac{-6i\pi}{12}}
\frac{1}{2 \sqrt3}
$
 &
$
\frac{d}{4 \sqrt{3}}
$
 &
\end{tabu}
}
\caption{
Minimal idempotents for $\tube_{y \to y}$.
We have used the notation $v_y = \text{id}_y \in \tube_{y \to y}$, $t_y = t_{y\unit y y; 11}$, and $t_y l_x = t_{yxyx; 11}$.
\label{mor2to2}}
\end{table}

\begin{table}
\resizebox{\textwidth}{!}{
{\tabulinesep=1.2mm
\begin{tabu}{ c c | c c c c c c c c c }
particle & spin &$t_x$&$v_x$&$X_{11}$ & $X_{12}$ &$X_{21}$ & $X_{22}$ & $v_x l_y$ & $t_x h_y$ \\ \hline
$W$ & $
       1       
$ &
$
\frac{1}{2d}
$
 &
$
\frac{1}{2d}
$
 &
$ e^{\frac{-9i\pi}{12}}
\frac{\sqrt{3d-8}}{4}
$
 &
$ e^{\frac{-9i\pi}{12}}
\frac{\sqrt{3d-8}}{4}
$
 &
$ e^{\frac{9i\pi}{12}}
\frac{\sqrt{3d-8}}{4}
$
 &
$ e^{\frac{-3i\pi}{12}}
\frac{\sqrt{3d-8}}{4}
$
 &
$ e^{\frac{6i\pi}{12}}
\frac{1}{2d}
$
 &
$ e^{\frac{-6i\pi}{12}}
\frac{1}{2d}
$
 &
\\
$U$ & $
       1       

$ &
$
\frac{1}{4 \sqrt{3}}
$
 &
$
\frac{1}{4 \sqrt{3}}
$
 &
$ e^{\frac{3i\pi}{12}}
\frac{1}{4 \sqrt{d} }
$
 &
$ e^{\frac{9i\pi}{12}}
\beta
$
 &
$ e^{\frac{-9i\pi}{12}}
\beta
$
 &
$ e^{\frac{-3i\pi}{12}}
\frac{1}{4 \sqrt{d} }
$
 &
$ e^{\frac{-6i\pi}{12}}
\frac{1}{4 \sqrt{3}}
$
 &
$ e^{\frac{6i\pi}{12}}
\frac{1}{4 \sqrt{3}}
$
 &
\\
$V$ & $
       1/2     

$ &
$ -
\frac{1}{4 \sqrt{3}}
$
 &
$
\frac{1}{4 \sqrt{3}}
$
 &
$ e^{\frac{-3i\pi}{12}}
\beta
$
 &
$ e^{\frac{-9i\pi}{12}}
\frac{1}{4 \sqrt{d} }
$
 &
$ e^{\frac{-3i\pi}{12}}
\frac{1}{4 \sqrt{d} }
$
 &
$ e^{\frac{-9i\pi}{12}}
\beta
$
 &
$ e^{\frac{-6i\pi}{12}}
\frac{1}{4 \sqrt{3}}
$
 &
$ e^{\frac{-6i\pi}{12}}
\frac{1}{4 \sqrt{3}}
$
 &
\\
$X_2 $ & $
      -5/12    

$ &
$ e^{\frac{-10i\pi}{12}}
\frac{1}{2+d^2}
$
 &
$
\frac{1}{2+d^2}
$
 &
$ e^{\frac{-2i\pi}{12}}
\gamma
$
 &
$ e^{\frac{10i\pi}{12}}
\alpha
$
 &
$ e^{\frac{4i\pi}{12}}
\alpha
$
 &
$ e^{\frac{4i\pi}{12}}
\gamma
$
 &
$ e^{\frac{6i\pi}{12}}
\frac{1}{2+d^2}
$
 &
$ e^{\frac{8i\pi}{12}}
\frac{1}{2+d^2}
$
 &
\\
$X_1$ & $
       1/4     

$ &
$ e^{\frac{6i\pi}{12}}
\frac{1}{2+d^2}
$
 &
$
\frac{1}{2+d^2}
$
 &
$ e^{\frac{6i\pi}{12}}
\frac{1}{2 \sqrt{6d}}
$
 &
$ e^{\frac{6i\pi}{12}}
\frac{1}{2 \sqrt{6d}}
$
 &
$
\frac{1}{2 \sqrt{6d}}
$
 &
$ -
\frac{1}{2 \sqrt{6d}}
$
 &
$ e^{\frac{6i\pi}{12}}
\frac{1}{2+d^2}
$
 &
$
\frac{1}{2+d^2}
$
 &
\\
$X_{3}$ & $
      -5/12    

$ &
$ e^{\frac{-10i\pi}{12}}
\frac{1}{2+d^2}
$
 &
$
\frac{1}{2+d^2}
$
 &
$ e^{\frac{10i\pi}{12}}
\alpha
$
 &
$ e^{\frac{-2i\pi}{12}}
\gamma
$
 &
$ e^{\frac{-8i\pi}{12}}
\gamma
$
 &
$ e^{\frac{-8i\pi}{12}}
\alpha
$
 &
$ e^{\frac{6i\pi}{12}}
\frac{1}{2+d^2}
$
 &
$ e^{\frac{8i\pi}{12}}
\frac{1}{2+d^2}
$
 &
\\
$X_5$ & $
       1/3     

$ &
$ e^{\frac{8i\pi}{12}}
\frac{1}{2+d^2}
$
 &
$
\frac{1}{2+d^2}
$
 &
$
$
 &
$ e^{\frac{1i\pi}{12}}
\frac{1}{2 \sqrt{3d} }
$
 &
$ e^{\frac{7i\pi}{12}}
\frac{1}{2 \sqrt{3d} }
$
 &
$
$
 &
$ e^{\frac{-6i\pi}{12}}
\frac{1}{2+d^2}
$
 &
$ e^{\frac{-10i\pi}{12}}
\frac{1}{2+d^2}
$
 &
\\
$X_4$ & $
      -1/6     

$ &
$ e^{\frac{-4i\pi}{12}}
\frac{1}{2+d^2}
$
 &
$
\frac{1}{2+d^2}
$
 &
$ e^{\frac{-11i\pi}{12}}
\frac{1}{2 \sqrt{3d} }
$
 &
$
$
 &
$
$
 &
$ e^{\frac{7i\pi}{12}}
\frac{1}{2 \sqrt{3d} }
$
 &
$ e^{\frac{-6i\pi}{12}}
\frac{1}{2+d^2}
$
 &
$ e^{\frac{2i\pi}{12}}
\frac{1}{2+d^2}
$
 &
\end{tabu}
}
}
\caption{Minimal idempotents for $\tube_{x \to x}$. 
We have used the notation $t_x =t_{x\unit xx; 11}$, $v_x = \text{id}_x \in \tube_{x \to x}$, $X_{ij} = t_{xxx; ij}$, $v_x l_y = t_{xxxy; 11}$, and $t_x h_y = t_{xyxx; 11}$. 
Where $\alpha = \frac{1}{2} \left( 1+ 1/\sqrt{2d+1} \right)$, and $\beta = \frac{1}{2} \left( 1- 1/\sqrt{2d+1} \right)$, $\gamma/\alpha =1/(2\sqrt{d} 3^{1/4})$, and $d = 1+ \sqrt{3} $. 
\label{mor1to1}}
\end{table}

Some of the idempotents are isomorphic. 
For example, $W$ appears in all three tables \ref{mor0to0}, \ref{mor1to1} and \ref{mor2to2} (with its boundary condition $\unit$, $y$ or $x$ implicit in each table).
As usual, if $e$ and $e'$ are isomorphic idempotents, then we can find morphisms $u$, $v$ such that $e = u\cdot v$ and $e' = v \cdot u$.
In the following we denote the boundary condition of each idempotent by a subscript and similarly for the morphisms, so that, e.g., $W_x = w_{x \unit} \cdot w_{\unit x} = w_{xy} \cdot w_{yx}$, and $W_{\unit} = w_{\unit x} \cdot w_{x \unit}  = w_{\unit y}\cdot w_{y \unit} $ and so on. 
We have:
\begin{align}
w_{\unit x} &= \frac{i}{\sqrt{2 \sqrt{2d}}}(t_{\unit xxx; 11} - t_{\unit xxx;12}) \\
w_{x \unit}  &= \frac{-i}{\sqrt{2\sqrt{2d}}} (t_{xx \unit x;11}-t_{xx\unit x ;21}) \\
w_{xy} & = \frac{1}{(8d)^{\frac14}}(t_{xx y x ;11} + t_{xxyx;21}) \\
w_{yx} & = \frac{1}{(8d)^{\frac14}}(t_{yxxx;11} + t_{yxxx;12}) \\
w_{\unit y } & \frac{-e^{-i\pi/4}}{\sqrt{2}} t_{\unit x y x;11}\\
w_{y \unit} & \frac{-e^{i\pi/4}}{\sqrt{2}} t_{yx \unit x ;11}
\end{align}
\begin{align}
u_{\unit x}  & = \frac{1}{2}\sqrt{\frac{d}{6}}( t_{\unit xxx;11}+t_{\unit xxx;12})\\
u_{x \unit} & = \frac{1}{2}\sqrt{\frac{d}{6}}(t_{xx\unit x;11}+t_{xx \unit x;21})\\
v_{xy } & = \frac{i}{2}\left( \frac{d}{6} \right)^{\frac14} (t_{xxy x;11}-t_{xxyx;21})\\
v_{yx} & = \frac{i}{2}\left( \frac{d}{6} \right)^{\frac14} (t_{yxxx;11}-t_{yxxx;21})
\end{align}
with
\begin{align}
t_{abcd; \mu \nu } = \AnnularTubexp{\AnnularTubeNoIndex}{a}{b}{c}{d}{\mu}{\nu}{}{}.
\end{align}
In terms of diagrams, we have
\begin{align}
\xymatrix @!0 @M=2mm @R=22mm @C=19mm{
W_{\unit} \ar@/^2pc/[rr]^{w_{\unit y }} \ar@<.5ex>[r]^{w_{\unit x}}
& W_x \ar@<.5ex>[l]^{w_{x \unit }} \ar@<.5ex>[r]^{w_{xy}}
& W_y\ar@<.5ex>[l]^{w_{yx}} \ar@/^2pc/[ll]^{w_{y\unit}}
}\qquad
\xymatrix @!0 @M=2mm @R=22mm @C=19mm{
U_\unit\ar@<.5ex>[r]^{u_{\unit x}}
& U_x \ar@<.5ex>[l]^{u_{x \unit}} 
}\qquad
\xymatrix @!0 @M=2mm @R=22mm @C=19mm{
V_x\ar@<.5ex>[r]^{v_{xy}}
& V_y \ar@<.5ex>[l]^{v_{yx}} 
}.
\end{align}
Composing the morphisms in various ways constructs all isomorphic idempotents.

\phantomsection
\addcontentsline{toc}{section}{References}

\bibliographystyle{utphys}
\bibliography{references}

\end{document}